\documentclass[journal=jpcbfh,manuscript=article]{achemso}
\usepackage{amsmath, physics, graphicx, txfonts, bm}
\SectionNumbersOn

\title{Helical Organic and Inorganic Polymers}

\author{So Hirata}
\email{sohirata@illinois.edu}
\affiliation{Department of Chemistry, University of Illinois at Urbana-Champaign, Urbana, Illinois 61801, USA}

\author{Yasuteru Shigeta}
\affiliation{Center for Computational Sciences, University of Tsukuba, Tsukuba, Ibaraki 305-8577, Japan}

\author{Sotiris S. Xantheas}
\affiliation{Advanced Computing, Mathematics and Data Division, Pacific Northwest National Laboratory, Richland, Washington 99352, USA}
\affiliation{Department of Chemistry, University of Washington, Seattle, Washington 98195, USA}

\author{Rodney J. Bartlett}
\affiliation{Quantum Theory Project, University of Florida, Gainesville, Florida 32611, USA}

\begin{document}

\begin{abstract}
Despite being a staple of synthetic plastics and biomolecules, helical polymers are scarcely studied with 
Gaussian-basis-set {\it ab initio} electron-correlated methods on an equal footing with molecules. 
This article introduces an {\it ab initio} second-order many-body Green's-function [MBGF(2)] method with nondiagonal, frequency-dependent Dyson self-energy for infinite helical polymers using screw-axis-symmetry-adapted
Gaussian-spherical-harmonics basis functions. Together with the Gaussian-basis-set density-functional theory for energies, analytical atomic forces, translational-period force, and helical-angle force, it can compute correlated energy, quasiparticle energy bands, structures, and vibrational frequencies of an infinite helical polymer, which smoothly converge at the corresponding oligomer results.
These methods can handle incommensurable structures, which have an infinite translational period and are hard to characterize by any other method, just as efficiently as commensurable structures.
We apply them to polyethylene ($2/1$ helix), polyacetylene (Peierls' system), and polytetrafluoroethylene ($13/6$ helix) to establish the quantitative accuracy 
of MBGF(2)/cc-pVDZ in simulating their (angle-resolved) ultraviolet photoelectron spectra, and of B3LYP/cc-pVDZ or 6-31G** 
in reproducing their structures,
infrared and Raman band positions, phonon dispersions, and (coherent and incoherent) inelastic neutron scattering spectra.
We then predict the same properties for infinitely catenated chains of nitrogen or oxygen and discuss their possible metastable existence under ambient conditions. They include planar zigzag polyazene (N$_2$)$_x$  (Peierls' system), $11/3$-helical isotactic polyazane (NH)$_x$,  $9/4$-helical isotactic polyfluoroazane (NF)$_x$, and $7/2$-helical polyoxane (O)$_x$ as potential high-energy-density materials. 
\end{abstract}

\maketitle 

\section{Introduction} 

In both chemistry and physics, there is a strong, persistent interest in catenated forms of nitrogen and oxygen, which are 
only metastable thermodynamically and thus extremely hard to synthesize, detect, or isolate.\cite{HEDM1996,Howlong,Bartlett2000,Steele_bookchapter2019,ReviewN,AQC2014} There are multiple reasons for this interest. 
These elements, which are abundant as atmospheric diatoms, 
may exist in extended covalently bonded structures under high pressure of planetary interior, which are often predicted to be metallic or even superconducting.\cite{Shimizu1998,Neaton2002,Ma2007,MengPNAS2008}
The fact that they are thermodynamically unstable yet have an appreciable lifetime makes them an ideal candidate for high-energy-density materials (HEDM)\cite{Bartlett2000,Steele_bookchapter2019,ReviewN,AQC2014}
used as explosives and propellants. They also serve as essential synthetic reagents.\cite{Kimball2002} However, the interest may ultimately be spurred by 
the dramatic change\cite{Wright1999} in the nature of chemical bonds upon going from stable organic polymers to their isoelectronic inorganic counterparts.\cite{Manners1996} 

The rapidly decreasing stability of longer catenated nitrogen and oxygen bonds can be rationalized by their bond strengths.
The N--N and N=N bonds have binding energies much less than one third and two thirds, respectively, of the N$\equiv$N bond, and the O--O bond much less than one half of the O=O bond.\cite{ReviewN} These are, in turn, caused by the lone-pair-lone-pair repulsions, which are stronger than the lone-pair-bond-pair or bond-pair-bond-pair repulsions.\cite{Gaucheeffect} Consequently, catenated nitrogen and oxygen bonds tend to favor the {\it gauche} conformation,\cite{Mizushima_book,Mizushima1954,Gaucheeffect} which separates neighboring lone pairs 
often at the expense of the extra stabilization brought to by the  
$\pi$-electron conjugation of planar structures. 

The preferred {\it gauche} conformation may, in turn, suggest the
helical polymeric forms of nitrogen\cite{Criton_cispolyN2021} and oxygen\cite{MartinsCostaCPL2009,Zhu2012} or the {\it cubic gauche} (cg) form\cite{McMahan1985,Mailhiot1992} of nitrogen allotrope,  in the infinite concatenation limit. The latter was  achieved synthetically under high pressure by Eremets {\it et al.}\cite{Eremets2004,Eremets2007}
Furthermore, intrachain hydrogen bonds may\cite{Schlegel1993} or may not\cite{Zhao1994} stabilize the helical chains. 

However, infinite helical chains have scarcely been computationally explored by {\it ab initio} electron-correlated methods\cite{DelRe1967,Andre1969,MintmireSabin1980,MintmireSabin1982,Kertesz1982,Sun1999,Hirata2009_PCCP} in spite of the ubiquity of helical conformations
in both synthetic polymers (e.g., polypropylene, polytetrafluoroethylene, and carbon nanotubes) and biopolymers (e.g., DNA, collagens, and $\alpha$ helix).\cite{Helicalreview1,Helicalreview2}  This, in turn, is caused by the dearth of {\it ab initio} computational machinery capable of treating infinite chains 
under helical periodic boundary conditions (see, however, pioneering empirical or semiempirical methods\cite{Imamura1970,Fujita1970,Blumen1977} and {\it ab initio} method\cite{Teramae1983,Karpfen1984,Andre1984,Teramea1984,SpringborgSulfur1986,Teramae1989,Mintmire1991,Mintmire1992,Mintmire1993,Hirata1997,Hirata1998,ZhangMintmire1999,Mintmire2006,Mintmire2022,Hirata_MP2022} including Gaussian-basis-set DFT).

On the one hand, molecular methods and software can be applied to longer helical
oligomers to extract bulk properties.\cite{Jovanovic2019} However, not only are such calculations saddled with slowly decaying terminal effects and inefficient, but they have difficulties
determining the equilibrium helical angle or quasiparticle energy bands as a function of wave vector.
On the other hand, three-dimensional solid-state methods and software can handle a crystalline phase of helical polymers.
This too is not ideal, as it imposes severe constraints on the helical angle so that the translational repeat unit does not become too large. For example, polytetrafluoroethylene 
adopts a $13/6$ helical conformation\cite{Bunn_Nature} in the translational notation,\cite{Clark1999} meaning that the three-dimensional translational periodic unit  must contain at least 13 CF$_2$ groups per chain (see, e.g., Ref.\ \cite{DAmore2006} for a calculation using such a large translational unit) as opposed to
just one CF$_2$ rototranslational repeat unit under helical periodic boundary conditions. 
It would be exceedingly difficult to optimize the helical angle using a method exploiting linear translational symmetry.
Furthermore, there is no {\it a priori} reason for a helix to adopt a commensurable structure,\cite{Weeks1981} and it is possible that the translational period is infinity. 
Also, solid-state and molecular methods often use disparate basis sets
and incompatible approximations, creating a gap in computational characterization between the short- and long-chain limits.

The objective of this study is three-fold: 

First, we introduce an {\it ab initio} second-order many-body Green's-function [MBGF(2)] method\cite{Linderberg65,Hedin,Ohrn65,Linderberg67,GoscinskiLukman,Doll,pickup,Yaris,linderbergohrn,Freed74,paldus,Ceder,cederbaumacp,simonsrev,herman,BakerPickup,ohrnborn,jorgensensimons,schirmer1982, schirmer,vonniessen,Prasad,GWLouie,oddershede,Kutzelnigg,GW1,ortiz_aqc,GW2,Wire,szabo,Hirata2017} with nondiagonal, frequency-dependent
Dyson self-energy for infinite helical polymers. By using screw-axis-symmetry-adapted Gaussian-spherical-harmonics basis functions, 
the method expands the Bloch factor of a crystal orbital as a linear combination of these functions centered in the single smallest rototranslational (i.e., physical) repeat unit.\cite{Bower1989}
Together with the screw-axis-symmetry-adapted Gaussian-basis-set density-functional theory (DFT) 
for analytical gradients,\cite{Hirata_MP2022} it permits
the determination of smooth, continuous quasiparticle energy bands and density of states (DOS), following 
the optimization of all structural parameters including helical angle and translational period.
These results can be directly compared with those obtained from the Gaussian-basis-set {\it ab initio} electron-correlated and DFT calculations for molecules.

Second, we apply these methods to three experimentally most thoroughly  
characterized hydrocarbon or fluorocarbon polymers:\ polyethylene (CH$_2$)$_x$, polyacetylene (C$_2$H$_2$)$_x$, and polytetrafluoroethylene (CF$_2$)$_x$. These calculations establish the  
quantitative accuracy of the MBGF(2) method with the cc-pVDZ basis set for 
valence  bands and electronic DOS and  of DFT using the Becke3--Lee--Yang--Parr (B3LYP) hybrid functional with the cc-pVDZ or 6-31G** basis set for the structure parameters, phonon dispersion, and phonon DOS, as judged by the 
comparison with their crystallographies, (angle-resolved) ultraviolet photoelectron spectra, infrared (IR) and Raman band positions, and (coherent or incoherent) inelastic neutron scattering (INS) spectra. 

Third, we predict the structure parameters, phonon dispersion, phonon DOS, valence  bands, and electronic DOS of infinite nitrogen and oxygen polymers. 
The polymers considered in this study are linear, zigzag, planar all-{\it trans} polyazene (N$_2$)$_x$, helical isotactic polyazane (NH)$_x$, helical
isotactic polyfluoroazane (NF)$_x$, and helical polyoxane (O)$_x$. They are nitrogen or oxygen analogues of polyacetylene, polyethylene, polytetrafluoroethylene, and polyethylene,
respectively. We discuss their thermodynamic and kinetic stability and thus possible metastable existence under ambient conditions. We present the predicted
IR and Raman band positions, INS spectra, and photoelectron spectra of these hypothetical polymers with the aim of assisting in their identification.

\section{Methods\label{sec:Methods}}

\subsection{Hybrid density-functional method for helical polymers\label{sec:DFT}}

One of the present authors recently reported a Gaussian-basis-set hybrid DFT method for energies and analytical energy gradients of a closed-shell infinite helical polymer,
in which the quadrature derivatives were shown to persist in the 
translational-period and helical-angle gradients even in the limit of an infinitely dense grid.\cite{Hirata_MP2022} See
Refs.\ \cite{Teramae1983,Teramea1984,SpringborgSulfur1986,Mintmire1991,Hirata1997,Hirata1998,Mintmire2006,Mintmire2022} for earlier Gaussian-basis-set crystal-orbital methods for helices. 
The formalism encompasses the spin-restricted Hartree--Fock (HF) method for helical polymers,
furnishing a reference wave function for MBGF(2) described in Sec.\ \ref{sec:GF2}.
We succinctly summarize the salient parts of the formalism.

The $n$th rototranslational (physical) repeat unit\cite{Bower1989} is axially translated by period or ``rise''\cite{Hauser} $a$ as it is rotated counterclockwise by angle or ``twist''\cite{Hauser} $\varphi$ to form the ($n$+1)th unit cell. 
An $m/n$ helix in the translational notation\cite{Clark1999} (an $m_n$ helix in the rotational notation\cite{Clark1999}) means that $m$ repeat units make $n$ complete turns and, therefore, $\varphi = 2\pi n / m$.
The $I$th atomic coordinates of the $n$th unit ($X_{I(n)},Y_{I(n)},Z_{I(n)}$) are then related to those in the zeroth unit cell ($X_{I(0)},Y_{I(0)},Z_{I(0)}$) by 
\begin{eqnarray}
X_{I(n)} &=& X_{I(0)}+na, \label{X} \\
Y_{I(n)} &=& Y_{I(0)} \cos n\varphi - Z_{I(0)} \sin n\varphi, \label{Y} \\
Z_{I(n)} &=& Y_{I(0)} \sin n\varphi + Z_{I(0)} \cos n\varphi, \label{Z}
\end{eqnarray}
where the chain is along the $x$ axis. 

In our formalism, we use three sets of atom-centered Gaussian basis functions, $\{\chi_{\nu(n)}(\bm{r})\}$, $\{\tilde{\chi}_{\nu(n)}(\bm{r})\}$, and $\{\bar{\tilde{\chi}}_{\nu(n)}(\bm{r})\}$, where $\nu(n)$ denotes
the $\nu$th contracted Gaussian basis function in the $n$th unit cell. (In practice, however, only one Gaussian basis set, such as 6-31G** or cc-pVDZ, needs to be specified.)
The first set, $\{\chi_{\nu(n)}(\bm{r})\}$, consists of Gaussian basis functions with Cartesian six $d$ and ten $f$ functions (if applicable) defined in the laboratory Cartesian coordinates; its members are not 
reoriented according to the screw-axis symmetry, but only their centers are translated-rotated along the chain.  The second set, $\{\tilde{\chi}_{\nu(n)}(\bm{r})\}$, is composed of 
screw-axis-symmetry-adapted (i.e., reorienting) Gaussian basis functions with Cartesian six $d$ and ten $f$ functions; its members in the $n$th unit cell are oriented 
in the $n$th cell's Cartesian coordinates, whose $y$ and $z$ axes are rotated counterclockwise by angle $n\varphi$ (relative to the zeroth cell). The third set, $\{\bar{\tilde{\chi}}_{\nu(n)}(\bm{r})\}$, is the same as the second set 
except that it has spherical five $d$ and seven $f$ functions (if applicable). The corresponding molecular integrals and expansion coefficients are decorated by a tilde and/or overbar. 

Since the Bloch theorem\cite{Kittel} should hold only for a screw-axis-symmetry-adapted basis set, 
the generalized eigenvalue equation to be solved for the crystal-orbital coefficients, $\bar{\tilde C}_{\nu p}(k)$, and orbital energy, $\epsilon_{p}(k)$,
is given by
\begin{eqnarray}
\sum_{\nu} \bar{\tilde F}_{\mu\nu}(k) \bar{\tilde C}_{\nu p}(k) = \sum_{\nu} \bar{\tilde S}_{\mu\nu}(k) \bar{\tilde C}_{\nu p}(k) \epsilon_{p}(k), \label{Roothaan}
\end{eqnarray}
where $p$ is an energy band index, $k$ is a wave vector, $\bar{\tilde F}_{\mu\nu}(k)$ and $\bar{\tilde S}_{\mu\nu}(k)$ are the $\mu\nu$th element of the Kohn--Sham (KS) Hamiltonian (or Fock) dynamical matrix 
and overlap dynamical matrix, respectively, whereas $\mu$ and $\nu$ label screw-axis-symmetry-adapted Gaussian basis functions with spherical $d$ and $f$ functions (the third basis set).
These matrices are formed by transformation of the corresponding matrices, ${\tilde F}_{\mu\nu}(k)$ and ${\tilde S}_{\mu\nu}(k)$, in the screw-axis-symmetry-adapted Gaussian basis functions
with Cartesian $d$ and $f$ functions (the second basis set).
\begin{eqnarray}
\bar{\tilde F}_{\mu\nu}(k) &=& \sum_{\mu^\prime,\nu^\prime} Q_{\mu\mu^\prime}Q_{\nu\nu^\prime}\tilde F_{\mu^\prime\nu^\prime}(k), \\
\bar{\tilde S}_{\mu\nu}(k) &=& \sum_{\mu^\prime,\nu^\prime} Q_{\mu\mu^\prime}Q_{\nu\nu^\prime}\tilde S_{\mu^\prime\nu^\prime}(k).
\end{eqnarray}
See Ref.\ \cite{Hirata_MP2022} for the transformation matrix, $Q_{\mu\mu^\prime}$. 
The crystal-orbital coefficients in the the second basis set are obtained as
\begin{eqnarray}
\tilde C_{\nu m}(k) = \sum_{\nu^\prime} \bar{\tilde C}_{\nu^\prime m}(k) Q_{\nu^\prime\nu}.
\end{eqnarray}

The dynamical (i.e., reciprocal-space) matrices are constructed from the corresponding, real-space matrices by
\begin{eqnarray}
\tilde F_{\mu\nu}(k) &=& \sum_{n=-S}^S \tilde F_{\mu(0)\nu(n)} e^{2\pi ink/K}, \\
\tilde S_{\mu\nu}(k) &=& \sum_{n=-S}^S \tilde S_{\mu(0)\nu(n)} e^{2\pi ink/K},
\end{eqnarray}
for wave vector index (integer) in the range of $0 \leq k < K$, summing over unit cells in the range of $-S \leq n \leq S$, where $K$ is 
the number of wave vector sampling points in the reciprocal unit cell and $S$ is 
the so-called short-range Namur lattice-sum cutoff.\cite{Delhalle} 
$\tilde F_{\mu(0)\nu(n)}$ and $\tilde S_{\mu(0)\nu(n)}$ stand for, respectively, the KS Hamiltonian (or Fock) and overlap matrix elements
between the $\mu$th Gaussian basis function in the zeroth unit cell and the $\nu$th Gaussian basis function in the $n$th unit cell, the latter being reorientated by angle $n\varphi$. 

The latter matrices in the basis of the screw-axis-symmetry-adapted (i.e., reorienting) Gaussian basis functions are, in turn, obtained by transformation of those in the basis 
of non-reorienting Gaussian basis functions, $\{\chi_{\nu(n)}(\bm{r})\}$, in the first basis set:
\begin{eqnarray}
\tilde F_{\mu(0)\nu(n)} &=& \sum_{\nu^\prime} R_{\nu\nu^\prime}(n) F_{\mu(0)\nu^\prime(n)}  , \\
\tilde S_{\mu(0)\nu(n)} &=& \sum_{\nu^\prime} R_{\nu\nu^\prime}(n) S_{\mu(0)\nu^\prime(n)}  ,
\end{eqnarray}
where, for instance, the overlap matrix element in the right-hand side is defined by
\begin{eqnarray}
 S_{\mu(0)\nu(n)} &=&  \int d\bm{r}\,  \chi^*_{\mu(0)}(\bm{r})  \chi_{\nu(n)}(\bm{r}),
\end{eqnarray}
with $\chi_{\nu(n)}(\bm{r})$ being the $\nu$th Gaussian basis function in the $n$th unit cell in the laboratory Cartesian coordinates.
Such an integral can be computed by the standard algorithms.\cite{Obara1986} See Ref.\ \cite{Hirata_MP2022} for the evaluation
of $F_{\mu(0)\nu^\prime(n)}$, which can include the multipole-expansion correction.\cite{Delhalle}
The transformation matrix, $R_{\nu\nu^\prime}$, is also given in the same article.\cite{Hirata_MP2022}

With $R_{\nu\nu^\prime}(n)$, the crystal-orbital coefficients in the first basis set are then obtained by transformation,
\begin{eqnarray}
C_{\nu p}(k;n) = \sum_{\nu^\prime} {\tilde C}_{\nu^\prime p}(k) R_{\nu^\prime\nu}(n). \label{C}
\end{eqnarray}
Note that $C_{\nu p}(k;n)$ are no longer translationally invariant and vary with the unit-cell index $n$. 

See Ref.\ \cite{Hirata_MP2022} for a complete formulation of the hybrid DFT energies and analytical gradients for infinite helical polymers.
Quadrature derivatives are essential for analytical gradients with respect to the translational period and helical angle. 

\subsection{Second-order many-body Green's-function method for helical polymers\label{sec:GF2}}

In a closed-shell infinite helical polymer, whose Bloch orbitals are expanded by $n$ basis functions per rototranslational (physical) repeat unit,\cite{Bower1989} 
its one-particle many-body Green's function\cite{Linderberg65,Hedin,Ohrn65,Linderberg67,GoscinskiLukman,Doll,pickup,Yaris,linderbergohrn,Freed74,paldus,Ceder,cederbaumacp,simonsrev,herman,BakerPickup,ohrnborn,jorgensensimons,schirmer1982, schirmer,vonniessen,Prasad,GWLouie,oddershede,Kutzelnigg,GW1,ortiz_aqc,GW2,Wire,szabo,Hirata2017} is an $n$-by-$n$ matrix defined by its elements as
\begin{eqnarray}
{G}_{pq}(\omega; k) &=& \sum_{\mu} \frac{\langle \Psi_{N,0} |\hat{p}_k^\dagger|\Psi_{N-1,\mu}\rangle\langle\Psi_{N-1,\mu}|\hat{q}_k |\Psi_{N,0} \rangle}{\omega - E_{N,0} + E_{N-1,\mu}}
\nonumber\\&&
+ \sum_\mu \frac{\langle \Psi_{N,0} |\hat{q}_k|\Psi_{N+1,\mu}\rangle\langle\Psi_{N+1,\mu}| \hat{p}_k^\dagger |\Psi_{N,0} \rangle}{\omega - E_{N+1,\mu} + E_{N,0}}.  \label{def2}
\end{eqnarray}
Here, $\Psi_{N,0}$ is the exact, ground-state wave function with $N$ electrons having the exact energy $E_{N,0}$, $\Psi_{N\pm1,\mu}$ is the exact, $\mu$th-state wave function 
with $N\pm1$ electrons and exact energy $E_{N+1,\mu}$, and $\hat{p}_k^\dagger$ and $\hat{q}_k$ are electron creation and annihilation operators, respectively, in the $p$th and $q$th energy bands
with wave vector $k$. The Green's function diverges whenever $\omega$ coincides with an exact electron-binding energy.

Replacing the exact wave functions and energies in Eq.\ (\ref{def2}) by their corresponding quantities of a mean-field method, such as the HF method, we obtain
the zeroth-order Green's function. It simplifies to
\begin{eqnarray}
{G}^{(0)}_{pq}(\omega; k) &=& \frac{\delta_{pq}}{\omega-\epsilon_p(k)},  \label{def3}
\end{eqnarray}
where $\epsilon_p(k)$ is the mean-field one-electron energy of the $p$th energy band with wave vector $k$ obtained by solving Eq.\ (\ref{Roothaan}).
Starting with this reference Green's function as the zeroth order, we expand the exact Green's function of Eq.\ (\ref{def2}) in the so-called Feynman--Dyson perturbation series.\cite{Hirata2017}

Before doing so, let us define an $n$-by-$n$ matrix known as the Dyson self-energy, $\bm{\Sigma}(\omega;k)$, by the equation,
\begin{eqnarray}
\bm{G}(\omega;k) = \bm{G}^{(0)}(\omega;k) +  \bm{G}^{(0)}(\omega;k) \bm{\Sigma}(\omega;k) \bm{G}(\omega;k).
\end{eqnarray}
We seek $\omega$'s that make the exact Green's function divergent, which is where the determinant of its inverse becomes zero.
\begin{eqnarray}
0 &=& \left|\{ \bm{G}(\omega;k) \}^{-1}\right| \\ 
&=& \left| \{ \bm{G}^{(0)}(\omega;k) \}^{-1} -  \bm{\Sigma}(\omega;k) \right|  \\
&=& \left| \omega\bm{1} - \bm{\epsilon}(k) -  \bm{\Sigma}(\omega;k) \right|. \label{invDyson1}
\end{eqnarray}
The roots of this equation are, in turn, the solutions of the $n$-by-$n$ matrix eigenvalue equation,
\begin{eqnarray}
\{ \bm{\epsilon}(k) + \bm{\Sigma}(\omega_q;k)\}\bm{U}_q(k) = \omega_q(k) \bm{U}_q(k), \label{invDyson2}
\end{eqnarray}
where $\bm{\epsilon}(k)$ is the diagonal matrix of $\epsilon_p(k)$, $\omega_q(k)$ is the $q$th eigenvalue, and $\bm{U}_q(k)$ is the $q$th eigenvector.
One can view this equation, known as the inverse Dyson equation, as an effective one-electron equation, akin to the Hartree--Fock--Roothaan or Kohn--Sham equation, with $\bm{\Sigma}(\omega_q;k)$ serving as the one-electron correlation potential,
which is nonlocal and frequency-dependent. Owing to the frequency-dependence, there are much more than $n$ roots in this $n$-by-$n$ matrix eigenvalue equation, and
these roots, $\omega_q(k)$, form the exact quasiparticle energy bands
as a function of wave vector $k$ for Koopmans' as well as non-Koopmans' (satellite or shake-up) states. The corresponding eigenvector $\bm{U}_q(k)$ defines a Dyson orbital.\cite{OrtizDyson} 

There are widely used approximations to the self-energy. 
In the diagonal approximation,\cite{Hirata_GF2} we neglect all the off-diagonal elements of $\bm{\Sigma}(\omega_q;k)$, turning the matrix eigenvalue equation into a polynomial equation,
\begin{eqnarray}
\epsilon_q(k) + \Sigma_{qq}(\omega_q;k) = \omega_q(k), \label{diag}
\end{eqnarray}
whose Dyson orbital becomes just the orbital of the reference mean-field method; i.e., $\bm{U}_q(k)$ is a unit vector. 
In the diagonal, frequency-independent approximation,\cite{suhai_qp,SunBartlett1996,sun_qp,Hirata_GF2} the electron-binding energy is obtained by a one-time evaluation of the left-hand side of the following equation:
\begin{eqnarray}
\epsilon_q(k) + \Sigma_{qq}(\epsilon_q;k) = \omega_q(k), \label{indep}
\end{eqnarray}
at the expense of losing roots for non-Koopmans' states. This approximation is equivalent to the $\Delta$MP$n$ method for $1 \leq n \leq 3$.\cite{szabo,deltamp,Hirata2017}
The use of neither approximation is denoted the ``full'' self-energy in this article. 

The residue of the Green's function gives the weight of one-electron character in the electron-detachment or attachment transition, to which photoelectron cross section 
is proportional. Using the following relationship implied by Eq.\ (\ref{invDyson1}), 
\begin{eqnarray}
G_{qq}(\omega;k) = \frac{1}{\bm{U}^\dagger_q(k) \{ \omega\bm{1} - \bm{\epsilon}(k) -  \bm{\Sigma}(\omega;k) \} \bm{U}_q(k)},
\end{eqnarray}
we can evaluate the residue as\cite{OrtizDyson}
\begin{eqnarray}
&& \text{Res}_{\omega_q} G_{qq}(\omega;k) \nonumber\\
&&= \lim_{\omega \to \omega_q} (\omega - \omega_q) \,G_{qq}(\omega;k) \\
&&= \lim_{\omega \to \omega_q}  
\frac{\omega - \omega_q}{\bm{U}^\dagger_q(k) \{ \omega\bm{1} - \bm{\epsilon}(k) -  \bm{\Sigma}(\omega;k) \} \bm{U}_q(k)} \\
&&= \lim_{\omega \to \omega_q}  
\frac{(\partial/\partial\omega)(\omega - \omega_q)}{(\partial/\partial\omega) \bm{U}^\dagger_q(k) \{ \omega\bm{1} - \bm{\epsilon}(k) -  \bm{\Sigma}(\omega;k) \} \bm{U}_q(k)} \label{lhopital} \\
&& = \left\{ 1 - \bm{U}^\dagger_q(k) \frac{\partial \bm{\Sigma}(\omega;k)}{\partial \omega}\Big|_{\omega=\omega_q} \bm{U}_q(k) \right\}^{-1}. \label{HellmannFeynman} 
\end{eqnarray}
In Eq.\ (\ref{lhopital}), L'H\^{o}pital's rule was used. The derivatives of $\bm{U}_q$ need not be taken in Eq.\ (\ref{HellmannFeynman})
by virtue of the Hellmann--Feynman theorem.

In the diagonal approximation, Eq.\ (\ref{HellmannFeynman}) simplifies\cite{OrtizDyson} to
\begin{eqnarray}
\text{Res}_{\omega_q} G_{qq}(\omega;k) 
= \left\{ 1 - \frac{\partial \Sigma_{qq}(\omega;k)}{\partial \omega}\Big|_{\omega=\omega_q} \right\}^{-1},  \label{diagpole}
\end{eqnarray}
which can still discern Koopmans' roots from non-Koopmans' ones. 
In the diagonal, frequency-independent approximation, only Koopmans' roots exist, and therefore,
\begin{eqnarray}
\text{Res}_{\omega_q} G_{qq}(\omega;k) =1. 
\end{eqnarray}

We adopt the second-order Feynman--Dyson perturbation approximation to the self-energy,\cite{Hirata2017}
\begin{eqnarray}
\bm{\Sigma}(\omega;k) \approx \bm{\Sigma}^{(1)}(\omega;k) + \bm{\Sigma}^{(2)}(\omega;k).
\end{eqnarray}
With the canonical HF reference, which corresponds to the M\o ller--Plesset partitioning of the Hamiltonian, 
the first-order correction vanishes:
\begin{eqnarray}
\Sigma_{pq}^{(1)}(\omega;k) = 0.
\end{eqnarray}
The second-order correction is given by
\begin{scriptsize}
\begin{eqnarray}
\Sigma_{pq}^{(2)}(\omega;k) &=& \sum_j^{\text{occ.}} \sum_{a,b}^{\text{vir.}} \sum_{k_j,k_b} 
\frac{2 \langle q(k)a(k_a) | j(k_j)b(k_b) \rangle \langle p(k)a(k_a) | j(k_j)b(k_b) \rangle^*
-  \langle q(k)a(k_a) | j(k_j)b(k_b) \rangle \langle j(k_j)a(k_a) | p(k)b(k_b) \rangle^* }{\omega + \epsilon_j(k_j) - \epsilon_a(k_a) - \epsilon_b(k_b)} \nonumber\\
&& +  \sum_{i,j}^{\text{occ.}} \sum_{b}^{\text{vir.}} \sum_{k_j,k_b} 
\frac{2 \langle i(k_i) p(k) | j(k_j)b(k_b) \rangle \langle i(k_i) q(k) | j(k_j)b(k_b) \rangle^*
-  \langle i(k_i) p(k) | j(k_j)b(k_b) \rangle \langle j(k_j)q(k) | i(k_i) b(k_b) \rangle^* }{\omega + \epsilon_b(k_b) - \epsilon_i(k_i) - \epsilon_j(k_j)}, \label{mbgf2}
\end{eqnarray}
\end{scriptsize}
where $i$ and $j$ run over occupied (valence) energy bands, $a$ and $b$ over virtual (conduction) energy bands, $p$ and $q$ stand for any energy band, and
$k_j$ and $k_b$  run over $K$ evenly spaced wave vector sampling points in the reciprocal unit cell, whereas $k_a$ and $k_i$ are, in turn, determined by the momentum conservation law:
\begin{eqnarray}
k + k_j &\equiv& k_a + k_b \,\,\,(\text{mod}\,K), \\
k_i + k_j &\equiv& k + k_b \,\,\,(\text{mod}\,K). 
\end{eqnarray}
Each bracket denotes a two-electron integral (in the Mulliken notation) in the basis of crystal (Bloch) orbitals, which is obtained by transformation of the same in the Gaussian basis set. Using $\mu$, $\nu$, $\kappa$, and $\lambda$
as labels for the Gaussian basis functions in the first basis set (i.e., defined in the laboratory Cartesian coordinates), we can write the transformation as
\begin{eqnarray}
&& \langle p(k_p) q(k_q) | r(k_r)s(k_s) \rangle \nonumber\\
&&= \frac{1}{K} \sum_{\mu,\nu,\kappa,\lambda} 
\sum_{m_1=-S}^{S}\sum_{m_2=-L}^{L}\sum_{m_3=m_2-S}^{m_2+S} 
\nonumber\\&&\times\,
{C}_{\mu p}^*(k_p;0){C}_{\nu q}(k_q;m_1){C}_{\kappa r}^*(k_r;m_2){C}_{\lambda s}(k_s;m_3)
\nonumber\\&&\times\,
e^{2\pi i (m_1k_q-m_2k_r+m_3k_s)/K}
\langle \mu(0) \nu(m_1) | \kappa(m_2)\lambda(m_3) \rangle, \label{transform}
\end{eqnarray}
where $S$ and $L$ stand for the short- and long-range lattice-sum cutoffs according to the Namur protocol,\cite{Delhalle}
and $K$ is the number of wave vector sampling points, and
\begin{eqnarray}
&& \langle \mu(0) \nu(m_1) | \kappa(m_2)\lambda(m_3) \rangle = 
\nonumber\\
&& \iint  d\bm{r}_1d\bm{r}_2\,  \frac{\chi^*_{\mu(0)}(\bm{r}_1)\chi_{\nu(m_1)}(\bm{r}_1)   \chi^*_{\kappa(m_2)}(\bm{r}_2)\chi_{\lambda(m_4)}(\bm{r}_2)}{|\bm{r}_1 - \bm{r}_2|} .
\end{eqnarray}
The above integral can be evaluated by the standard algorithms\cite{Obara1986} since the Gaussian basis functions $\chi_{\nu(m)}(\bm{r})$ are defined in the laboratory Cartesian coordinates, whereas
${C}_{\nu q}(k_q;m)$ is given by Eq.\ (\ref{C}). 

The derivative of the self-energy needed for computing the residue [Eq.\ (\ref{HellmannFeynman})] is obtained analytically as 
\begin{scriptsize}
\begin{eqnarray}
\frac{\partial \Sigma_{pq}^{(2)}(\omega;k)}{\partial \omega} &=&
- \sum_j^{\text{occ.}} \sum_{a,b}^{\text{vir.}} \sum_{k_j,k_b} 
\frac{2 \langle q(k)a(k_a) | j(k_j)b(k_b) \rangle \langle p(k)a(k_a) | j(k_j)b(k_b) \rangle^*
-  \langle q(k)a(k_a) | j(k_j)b(k_b) \rangle \langle  j(k_j)a(k_a) | p(k)b(k_b) \rangle^* }{\{ \omega + \epsilon_j(k_j) - \epsilon_a(k_a) - \epsilon_b(k_b)\}^2} \nonumber\\
&& -  \sum_{i,j}^{\text{occ.}} \sum_{b}^{\text{vir.}} \sum_{k_j,k_b} 
\frac{2 \langle i(k_i) p(k) | j(k_j)b(k_b) \rangle \langle i(k_i) q(k) | j(k_j)b(k_b) \rangle^*
-  \langle i(k_i) p(k) | j(k_j)b(k_b) \rangle \langle j(k_j)q(k)  | i(k_i) b(k_b) \rangle^* }{\{\omega + \epsilon_b(k_b) - \epsilon_i(k_i) - \epsilon_j(k_j)\}^2}. \label{pole2}
\end{eqnarray}
\end{scriptsize}

Also, as we compute $\bm{\Sigma}^{(2)}(\omega;k)$, we simultaneously obtain the second-order many-body perturbation [MBPT(2)] correction to energy per unit cell as
\cite{Kunz1972,Liegener1988,SuhaiCPL,suhai_qp,Ye1993,SuhaiJCP1994,SuhaiPRB1994,Suhai1995,sun_qp,SunBartlett1996,HirataIwata1998,Ayala2001,HirataCC2004,Shimazaki2009,hirata_qp,Ohnishi2010}
\begin{scriptsize}
\begin{eqnarray}
E^{(2)} &=& \frac{1}{K} \sum_{i,j}^{\text{occ.}} \sum_{a,b}^{\text{vir.}} \sum_{k_i,k_j,k_a} 
\frac{2 \langle i(k_i)a(k_a) | j(k_j)b(k_b) \rangle \langle i(k_i)a(k_a) | j(k_j)b(k_b) \rangle^*
-  \langle i(k_i)a(k_a) | j(k_j)b(k_b) \rangle \langle j(k_j)a(k_a)|i(k_i)b(k_b)  \rangle^* }{\epsilon_i(k_i) + \epsilon_j(k_j) - \epsilon_a(k_a) - \epsilon_b(k_b)} , \label{mbpt2}
\end{eqnarray}
\end{scriptsize}
where $k_b$ is determined by the momentum conservation law:
\begin{eqnarray}
k_i + k_j &\equiv& k_a + k_b \,\,\,(\text{mod}\,K). 
\end{eqnarray}
Equations (\ref{mbgf2}), (\ref{pole2}), and (\ref{mbpt2}) are expressed in the forms that are conducive to parallelized summation loops over the crystal-orbital (band) index $b$ and wave-vector index $k_b$. 

The computational cost is dominated by the integral transformation of Eq.\ (\ref{transform}), which is carried out as four consecutive quarter transformations. 
Their nominal cost scales as $O(N^5S^2LK)$, where $N$ is the number of basis functions, $S$ and $L$ are the short- and long-range lattice-sum cutoffs of the Namur protocol, and $K$ is the number of wave vectors.
However, this cost function may only serve as an upper limit because integrals in the Gaussian basis functions are aggressively distance-based screened with only a few percent of the total number evaluated and stored.
Furthermore, the value of $K$ used in the MBGF(2) step is smaller than that in the preceding HF step by a factor of $n$ in the mod-$n$ approximation adopted here.\cite{Shimazaki2009,hirata_qp}

See Ref.\ \cite{Hirata2017} for a comprehensive treatise of MBGF($n$) at any $n$. 

\section{Results and Discussion} 

%
%
%
%

\subsection{Polyethylene, (CH$_2$)$_x$\label{sec:CH2}}

\begin{figure}
\includegraphics[width=0.85\columnwidth]{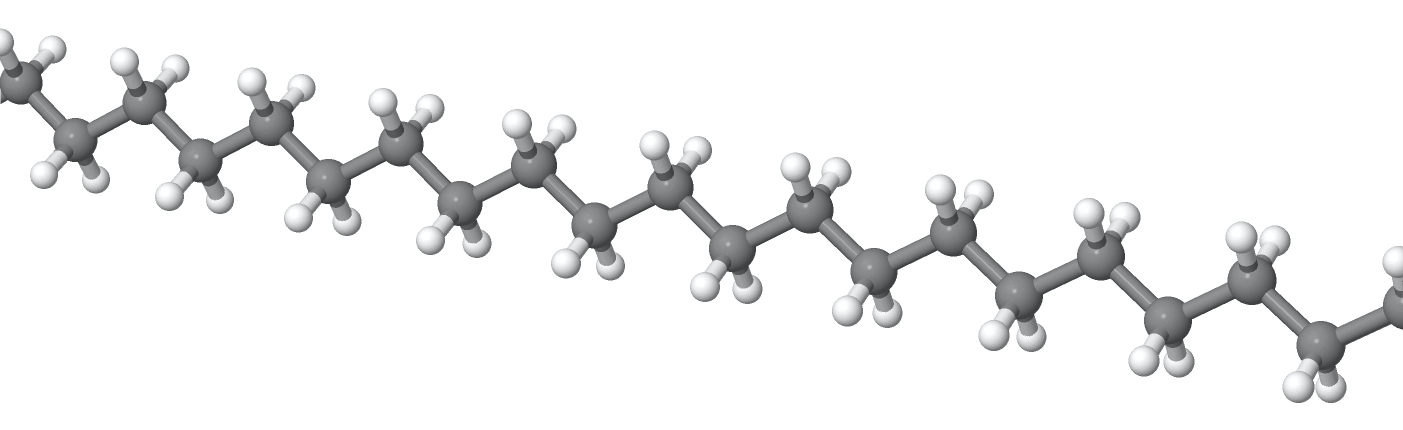}
\caption{Polyethylene, (CH$_2$)$_x$.\label{fig:CH2}}
\end{figure}

Polyethylene (Fig.\ \ref{fig:CH2}) is the one of the most produced plastics. It is thermodynamically and air stable. 
It is a linear zigzag chain with the C$_2$H$_4$ translational repeat unit or 
a $2/1$ helix with the CH$_2$ rototranslational (physical) repeat unit with the helical angle of $\varphi = 180^\circ$.\cite{Bower1989} 
Its electronic,\cite{Pireaux_PE_UPS,Ueno_PE_ARUP} structural,\cite{Shearer1956} and vibrational\cite{Krimm1956,Nielsen1957,Nielsen1961,Brown1963,Snyder1963,Parker_INS} properties have been throughly characterized experimentally. They
are also computationally reproduced accurately.\cite{HirataIwata_PE,sun_qp,hirata_qp} 

In this subsection, therefore, we briefly confirm the predictive accuracy of the B3LYP/cc-pVDZ and 6-31G** methods for structures, IR and Raman band positions, and phonon dispersions and phonon DOS as well as 
of MBGF(2)/cc-pVDZ for valence electronic bands and DOS. The B3LYP functional\cite{B3LYP} is the version that used the VWN5 functional.\cite{VWN}
All calculations were performed under the helical periodic boundary conditions
employing {\sc polymer},\cite{polymer} with $S = 8$, $L=12$, $K=48$ 
and the frozen-core and mod-6 approximations for MBGF(2)\cite{Shimazaki2009,hirata_qp} using a CH$_2$ group as the rototranslational (physical) repeat unit. 
During the latter calculation, we also obtained the MBPT(2) correction to the energy per unit cell.
See Sec.\ \ref{sec:Methods} for the details of these methods and the definitions of the computational parameters (such as $S$, $L$, and $K$).
Harmonic force constants were computed for up to the ninth nearest neighbor CH$_2$ groups by the supercell method,\cite{misc} using NWChem.\cite{nwchem}

The phonons of the 2/1 helix of polyethylene are optically active at the phase angles of $\theta = 0$ and $180^\circ$ in the phonon dispersion for a CH$_2$ rototranslational (physical) repeat unit\cite{Bower1989} (i.e., in the extended-zone scheme\cite{Piseri1973}). This may be understood as they all map onto $\theta = 0$ in the phonon dispersion for a  C$_2$H$_4$  translational repeat unit (in the true-zone scheme\cite{Piseri1973}). With these phase angles, the phonons have null linear momenta and can satisfy the momentum 
conservation law with photons with minuscule linear momenta during their optical (de)excitations. 
See Sec.\ \ref{sec:PTFE} for a more general discussion on the IR/Raman selection rules and various Brillouin-zone schemes of an infinite helical polymer. 

INS spectroscopy obeys  completely different selection rules.\cite{Marshall1971} 
Since thermal neutrons have linear momenta comparable to those of phonons, INS probes phonons of all momenta (phase angles). The scattering cross section 
is by far the largest for hydrogen motions because a momentum transfer is the most facile between particles with near-equal masses. 
Incoherent INS spectroscopy, therefore, measures the phonon DOS weighted by hydrogen amplitudes.
Coherent INS spectroscopy can further discern the momentum transfer of each transition and report the phonon dispersion curves as a function of the phonon momentum. 


\begin{table*}
\caption{Structural parameters of (CH$_2$)$_x$ and C$_2$H$_4$.  \label{PolyCH2_structure}}
\begin{tabular}{lcccccccc}
& \multicolumn{5}{c}{(CH$_2$)$_x$} &\multicolumn{3}{c}{C$_2$H$_4$\textsuperscript{\textit{a}}}  \\ \cline{2-6}\cline{7-9} 
Method & $r$(CC) & $r$(CH) & $a$(CCC) & $a$(HCH) & $\varphi$\textsuperscript{\textit{b}} & $r$(CC)  & $r$(CH) & $a$(HCH)  \\ \hline
B3LYP/6-31G** & 1.534\,\AA & 1.100\,\AA & $113.6^\circ$ & $105.9^\circ$ & $180.0^\circ$ & 1.330\,\AA & 1.087\,\AA & $116.3^\circ$ \\ 
B3LYP/cc-pVDZ & 1.533\,\AA & 1.107\,\AA & $113.7^\circ$ & $105.8^\circ$ & $180.0^\circ$  & 1.334\,\AA & 1.095\,\AA & $116.5^\circ$ \\ 
Observed\textsuperscript{\textit{c}} & 1.533\,\AA & 1.07\,\AA & $111.9^\circ$ & $107^\circ$ & $180.0^\circ$  & 1.339\,\AA & 1.086\,\AA & $117.6^\circ$ \\
\end{tabular} \\ 
\textsuperscript{\textit{a}}{Ethylene.}
\textsuperscript{\textit{b}}{The helical angle [Eqs.\ (\ref{X})--(\ref{Z})].}
\textsuperscript{\textit{c}}{The polyethylene structure from Teare.\cite{Teare1959} The ethylene structure from Herzberg.\cite{Herzberg}}
\end{table*}

Table \ref{PolyCH2_structure} compiles the internal coordinates of the equilibrium structure of polyethylene and its monomer, ethylene, determined by 
the B3LYP/cc-pVDZ and 6-31G** methods. The basis-set dependence is negligible and the bond lengths and angles are 
reproduced within a few hundredths of one \AA ngstrom and a few degrees, respectively, of the experimental data. The methods can, therefore, be considered predictive 
for structures.  

\begin{figure}
\includegraphics[width=1.0\columnwidth]{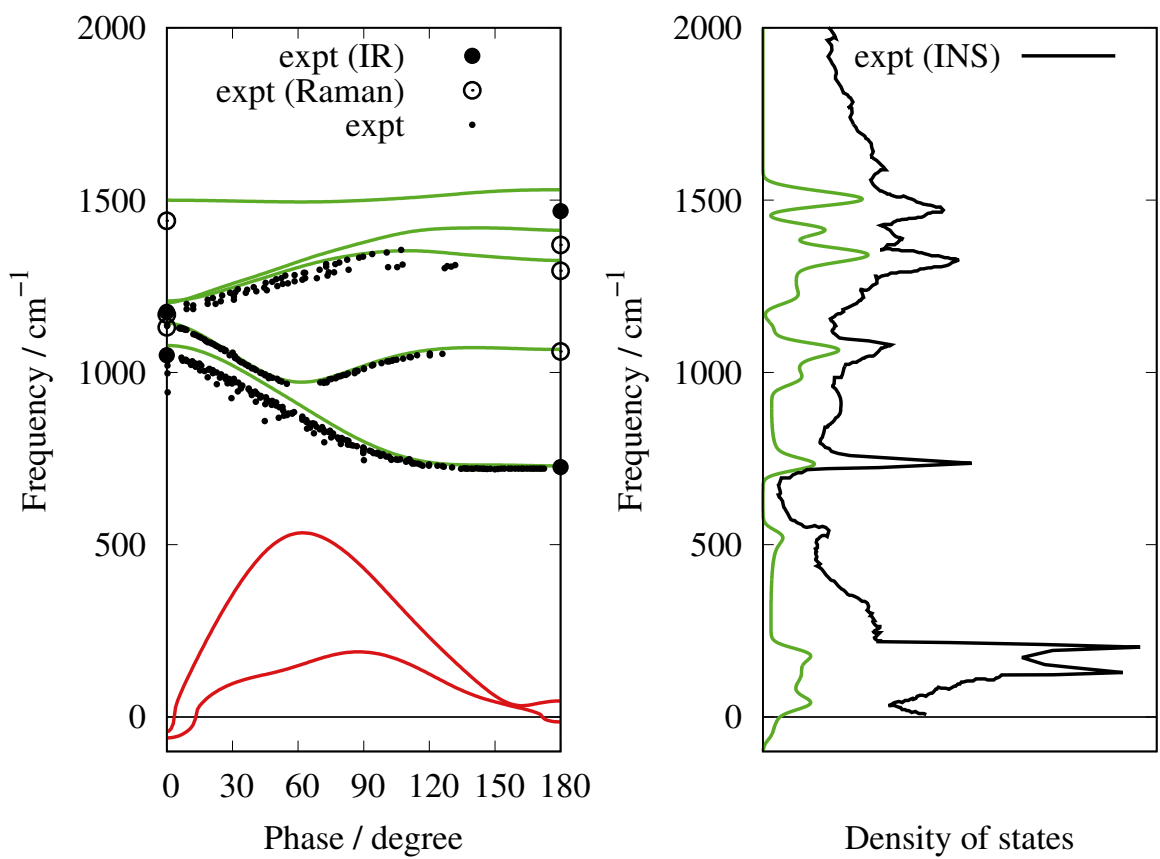}
\caption{\label{fig:CH2phonon}Phonon dispersion curves and DOS of (CH$_2$)$_x$ computed by B3LYP/6-31G** in the extended-zone scheme (i.e., corresponding to a CH$_2$ group as the repeat unit). The DOS is convoluted with a Gaussian of a FWHM of 40 cm$^{-1}$. 
The experimental phonon dispersion curves from Snyder and Schachtschneider.\cite{Snyder1963} 
The IR data from Krimm {\it et al.}\cite{Krimm1956} and from Nielsen and Holland.\cite{Nielsen1961} 
The Raman data from Nielsen and Woollett\cite{Nielsen1957} and  from Brown.\cite{Brown1963}
The incoherent inelastic neutron scattering (INS) spectrum from Parker.\cite{Parker_INS}}
\end{figure}

Figure \ref{fig:CH2phonon} compares the calculated phonon dispersion and phonon DOS with the observed IR and Raman band positions\cite{Krimm1956,Nielsen1961,Brown1963} and INS spectrum.\cite{Parker_INS} Phonon dispersion curves deduced from 
experimental data of $n$-alkanes are also overlaid.\cite{Snyder1963} As established already,\cite{HirataIwata_PE} B3LYP/6-31G** reproduces 
the nonmonotonic shapes of the phonon dispersions in the fingerprint region remarkably well.  
The calculated phonon DOS (without the hydrogen-amplitude weighting in this case) also provides  unmistakable assignments of all major peaks in
the observed INS spectrum.\cite{Parker_INS} The errors in the calculated peak positions are insignificant in light of the intrinsic difficulty of 
determining thermal neutron energies precisely. 

Note that the distance-based truncation of the force-constant matrices 
results in a loss of the strict translational and rotational invariance, causing the acoustic branches (the red curves in Fig.\ \ref{fig:CH2phonon}) to not converge exactly 
at zero frequencies at $\theta=0$ and 180$^\circ$ as they should theoretically. The errors can reach 61$i$ cm$^{-1}$ at $\theta=0$ 
and 47 cm$^{-1}$ at $\theta = 180^\circ$ despite the fact that force constants for up to the ninth nearest neighbor CH$_2$ groups on both sides
of the zeroth unit cell were taken into account. 
Since polyethylene is fully established to have a planar zigzag backbone in its equilibrium structure, an imaginary frequency of this size does not
imply a saddle point of the potential energy surface. 

\begin{table}
\caption{Vibrational frequencies (in cm$^{-1}$) of (CH$_2$)$_x$.  \label{PolyCH2_freqs}}
\begin{tabular}{lrr}
Irrep.\textsuperscript{\textit{a}}; phase; activity& \multicolumn{1}{c}{B3LYP/6-31G**} & \multicolumn{1}{c}{Observed\textsuperscript{\textit{b}}} \\ \hline
$A_g$; $\theta=0$; Raman & 3008.7  & 2848 \\ 
& 1499.7 & 1440 \\
& 1142.9 & 1131 \\
$B_{1g}$; $\theta=\varphi$; Raman & 1412.4 & 1370 \\
&  1066.6 & 1061 \\
$B_{2g}$; $\theta=\varphi$; Raman & 1324.7 & 1295 \\
$B_{3g}$; $\theta=0$; Raman & 3026.1  & 2883 \\
& 1207.2 & 1168 \\
$A_u$; $\theta=0$; IR & 1078.0 & 1050 \\
$B_{1u}$; $\theta=\varphi$; IR & 3072.8 & 2919 \\
& 728.6 & 725 \\
$B_{2u}$; $\theta=\varphi$; IR & 3023.6 & 2851 \\
&1530.0  & 1468 \\
$B_{3u}$; $\theta=0$; IR & 1202.6 & 1176 \\
\end{tabular} \\
\textsuperscript{\textit{a}}{Isomorphic to the $D_{2h}$ point group. $\varphi$ is the helical angle.}
\textsuperscript{\textit{b}}{The IR data from Krimm {\it et al.}\cite{Krimm1956} and from Nielsen and Holland.\cite{Nielsen1961} 
The Raman data from Nielsen and Woollett\cite{Nielsen1957} and  from Brown.\cite{Brown1963}}
\end{table}

Table \ref{PolyCH2_freqs} makes a more detailed comparison of the calculated and observed frequencies 
of the IR and Raman bands. Without scaling of frequencies,\cite{HirataIwata_PE} the calculated and observed data in the fingerprint region
agree with each other within 62 cm$^{-1}$ or usually much more closely. Since the observed bands are separated from one another more widely,
reliable assignments can be made on the basis of the calculated phonon frequencies and symmetry alone.
However, unsurprisingly, the C--H stretching modes at {\it ca.}\ 3000 cm$^{-1}$ suffer from much greater errors in excess of 100 cm$^{-1}$. 
This is well understood to be caused by anharmonicity,\cite{Qin2020} which also engenders numerous Fermi resonances, limiting the utility of
this spectral region for structural characterization. See Ref.\ \cite{Qin2020} for anharmonic phonon dispersion of polyethylene. 
For our purpose, the B3LYP/6-31G** method is considered predictive for vibrational spectra also.

\begin{figure}
\includegraphics[width=1.0\columnwidth]{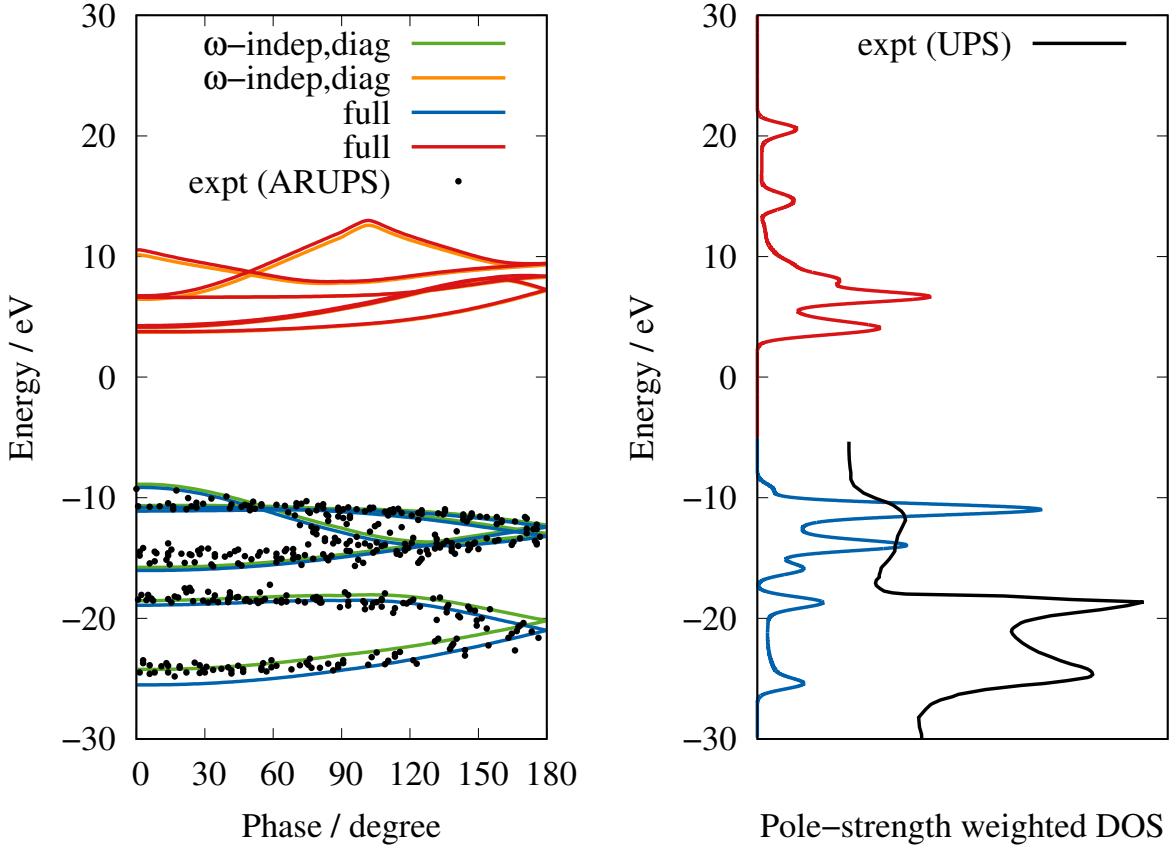}
\caption{\label{fig:CH2bands}Electronic energy bands and DOS of (C$_2$H$_4$)$_x$ computed by MBGF(2)/cc-pVDZ in the true-zone scheme (i.e., corresponding to a C$_2$H$_4$ group as the repeat unit). 
The phase angle is equal to $ka$ when $k$ is the wave vector and $a$ is the translational period. 
Here, `$\omega$-indep,diag' stands for the diagonal, frequency-independent approximation to the self-energy [Eq.\ (\ref{indep})], whereas
`full' for the use of the nondiagonal, frequency-dependent self-energy [Eq.\ (\ref{invDyson2})].
The DOS is weighted by the pole strength of Eq.\ (\ref{HellmannFeynman}) and convoluted with a Gaussian of a full-width-half-maximum (FWHM) of 1 eV. 
The ARUPS from Ueno {\it et al.}\cite{Ueno_PE_ARUP} 
The UPS from Pireaux {\it et al.}\cite{Pireaux_PE_UPS}}
\end{figure}

Figure \ref{fig:CH2bands} shows the calculated quasiparticle energy bands overlaid with the ones measured by angle-resolved ultraviolet 
photoelectron spectroscopy (ARUPS).\cite{Ueno_PE_ARUP} In the same figure, the electronic DOS is compared with the ultraviolet photoelectron spectrum (UPS).\cite{Pireaux_PE_UPS} The calculation was performed at the MBGF(2)/cc-pVDZ level with the diagonal and frequency-independent approximations (synonymous
with the $\Delta$MP2 method\cite{deltamp}) or without any approximation to the self-energy (labeled ``full'' in Fig.\ \ref{fig:CH2bands}).  The impact of the diagonal and frequency-independent approximations 
is minimal and the two types of calculated energy bands agree accurately with each other. Without shifting vertically, the calculated 
valence  bands agree with the ones obtained by ARUPS within the experimental error bars.
The electronic DOS explains every major
peak in the observed UPS quantitatively, although the peak intensities are not reproduced. The valence band edge at $-9.2$ eV is much lower than the air oxidation
potential of 5.2 eV,\cite{Rajendran2012} and thus polyethylene is air stable. For valence  bands, the MBGF(2)/cc-pVDZ method is predictive.

Experimentally, the fundamental band gap of polyethylene is 8.0 eV,\cite{Fujihira1972} with the valence and conduction band edges located at $-8.5$ and $-0.5$ eV, respectively. 
The valence band edge will be lowered to $-9.6$ eV for an isolated polyethylene chain in a vacuum, 
according to an extrapolation of the experimental data of oligomers (see p.169 of Ref.\ \cite{Seki1990}). (The lowering by 1.1 eV is a dielectric effect of the bulk material.)
The band gap of an isolated polyethylene chain in a vacuum is, therefore, estimated to be 9.1 eV (assuming the constancy of the conduction band edge at $-0.5$ eV). The MBGF(2)/cc-pVDZ method (with ``full'' self-energy) predicts the band gap of 13.1 eV with the valence band edge at $-9.1$ eV and conduction band edge at $4.1$ eV. Given the good agreement of the calculated and observed valence  bands, this large error is ascribed exclusively to the calculated conduction bands being too high. This, in turn, is likely  due (primarily) to the inadequacy of the basis 
set (cc-pVDZ) in describing diffuse orbitals. In our theoretical model of a single chain in a vacuum, the converged (i.e., infinite-basis-set) conduction band edge should be at least as low as zero.
If we assumed this, the calculated band gap would become 9.1 eV, which is more in line with the observed value of 9.1 eV. Nonetheless,
MBGF(2)/cc-pVDZ cannot be considered predictive for fundamental band gaps or conduction bands. 

\begin{table}
\caption{Binding energy (in kcal/mol) of (C$_2$H$_4$)$_x$.  \label{PolyCH2_binding}}
\begin{tabular}{lr}
Method & Binding energy\textsuperscript{\textit{a}} \\ \hline
B3LYP/cc-pVDZ &  23.6 \\ 
MBPT(2)/cc-pVDZ & 29.6  \\ 
\end{tabular} \\
\textsuperscript{\textit{a}}{The energy difference between the unit-cell energy of (C$_2$H$_4$)$_x$ and the energy of C$_2$H$_4$ 
in their respective B3LYP/cc-pVDZ-optimized geometries.}
\end{table}


\begin{figure}
\includegraphics[width=1.0\columnwidth]{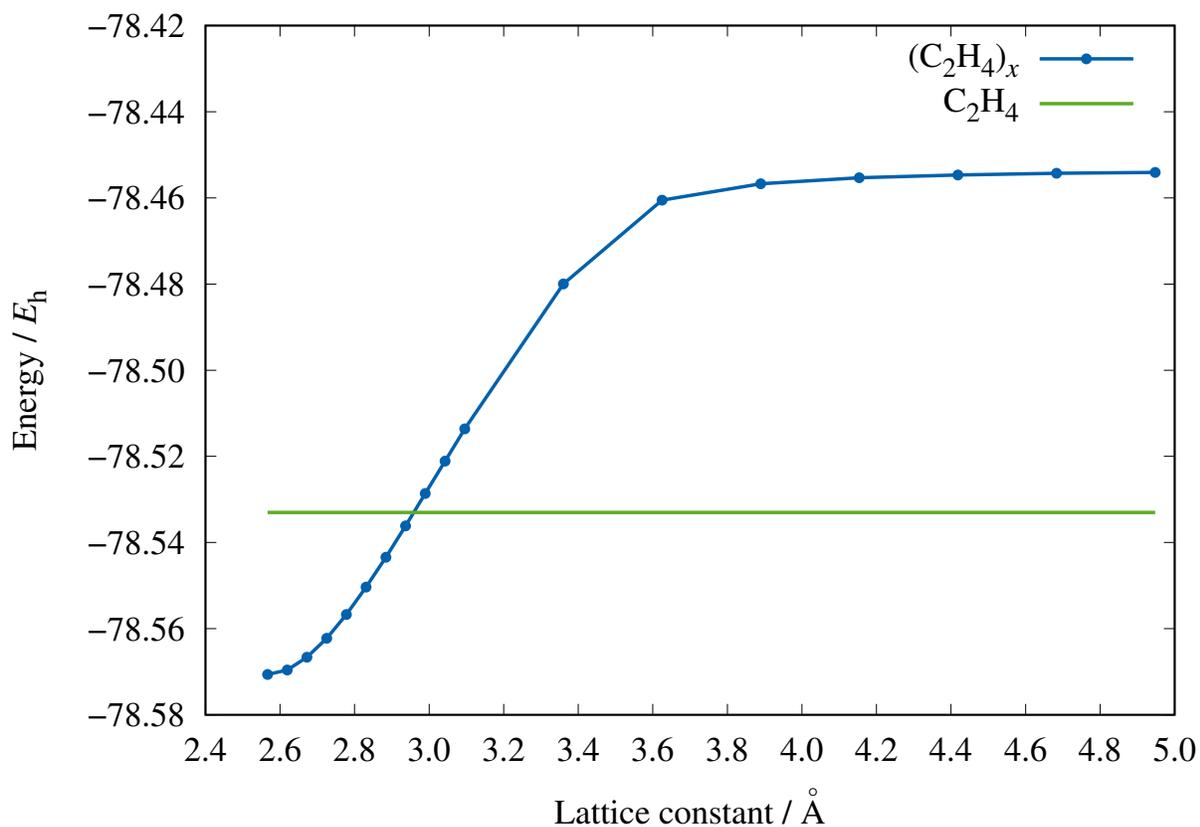}
\caption{\label{fig:PEdissociation}Unit-cell energy of (C$_2$H$_4$)$_x$ as a function of the lattice constant computed by B3LYP/cc-pVDZ.}
\end{figure}

Table \ref{PolyCH2_binding} underscores the thermodynamic stability of polyethylene; its binding energy per C$_2$H$_4$ group is 23.6 to 29.6 kcal/mol 
relative to its monomer, ethylene. Figure \ref{fig:PEdissociation} plots the energy per C$_2$H$_4$ group as a function of the lattice constant (translational
period). It may be viewed as a reaction energy profile of simultaneous dissociation of polyethylene into an infinite number of isolated ethylene molecules without 
geometrical relaxation in the latter. The energy of ethylene in its equilibrium geometry is indicated by a green line. While such a reaction pathway is unrealistic, 
the curve serves as a benchmark of highly stable polymers in comparison with other less stable ones discussed later. The figure shows that not only is the dissociation of
polyethylene  highly endothermic, it also faces a steep activation barrier partly because the unit-cell structure of (C$_2$H$_4$)$_x$ differs significantly
from the planar ethylene molecules.

%
%
%
%

\subsection{Polyacetylene, (C$_2$H$_2$)$_x$\label{sec:C2H2}}

\begin{figure}
\includegraphics[width=0.85\columnwidth]{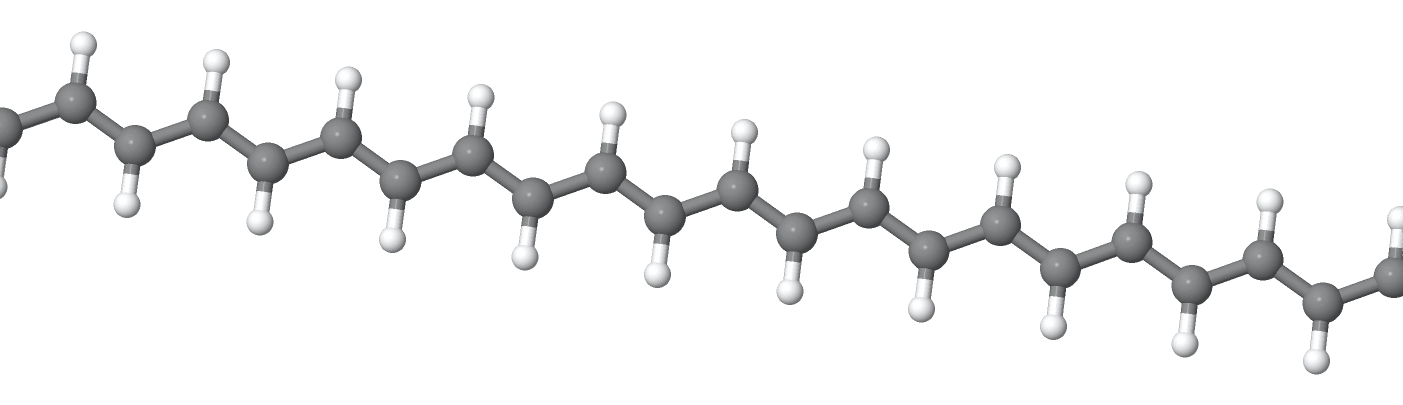}
\caption{Polyacetylene, (C$_2$H$_2$)$_x$.\label{fig:C2H2}}
\end{figure}

Polyacetylene\cite{Shirakawa1971,Shirakawa1973,Chiang1977,Chien1984,Heeger1988} (Fig.\ \ref{fig:C2H2}) is a paradigm of organic conductive polymers,\cite{Rajendran2012} and its geometrical, vibrational, and electronic structures are thoroughly studied.
It is thermodynamically stable, although unsubstituted linear oligoenes are increasingly unstable with chain length.\cite{Yoshida1988,Hirata1995Deca} 
It is also unstable to air because of its low ionization potential.
There are four stereoregular isomers of polyacetylene:\cite{Hirata1998} {\it trans}-transoid (all-{\it trans}), {\it cis}-transoid, {\it trans}-cisoid, and {\it cis}-cisoid, with the former two being a linear planar structure, whereas the latter two being a helix.\cite{Hirata1996} 
Only the linear planar {\it trans}-transoid\cite{Shirakawa1971,Lieser1980_trans,Shimamura1981,Fincher1982} and {\it cis}-transoid\cite{Baughman1978,Lieser1980,Chien1982_1,Chien1982_2} isomers have been observed. The {\it cis}-transoid form thermally isomerizes to the more stable {\it trans}-transoid form. In this study, we focus on the {\it trans}-transoid isomer, which is most thoroughly characterized experimentally.\cite{Shirakawa1971,Shirakawa1973,Kuzmany1980,Yannoni1983,Takeuchi1987,Kamiya_UPS,Hirata_INS} 

There are many DFT and {\it ab initio} studies of polyacetylene.\cite{Falk1975,Grant1979,Kasowski1980,Mintmire1983,Springborg1986,Mintmire1987,vonBoehm1987,Liegener1988,Vogl1989,Ashkenazi1989,Springborg1991,Paloheimo1992,SuhaiCPL,suhai_qp,Suhai1995,Hirata1995,SunBartlett1996,Hirata1996,Hirata1997,HirataIwata1998,Hirata1998,Ayala2001,HirataCC2004,hirata_qp} 
Some of them revealed that nonhybrid DFT (i.e., not hybridized with the HF exchange) is incapable of reproducing Peierls' distortion, predicting 
too small a bond-length alternation and nearly metallic electronic structure without doping.\cite{Mintmire1987,Vogl1989,Ashkenazi1989,Paloheimo1992,Suhai1995,Hirata1998}
HF theory, in contrast, exaggerates the bond-length alternation and fundamental band gap, whereas hybrid DFT and MBPT(2) tend to strike a good balance.\cite{Suhai1995,SunBartlett1996,Hirata1998,hirata_qp}

Here, we applied B3LYP/cc-pVDZ and 6-31G** for structures and phonon dispersion and DOS with $S=4$, $L=6$, and $K=24$ using a C$_2$H$_2$ group as the translational repeat unit.
We also performed MBGF(2)/cc-pVDZ calculations with the frozen-core and mod-3 approximations.\cite{Shimazaki2009,hirata_qp}  


\begin{table*}
\caption{Structural parameters of (C$_2$H$_2$)$_x$ and C$_2$H$_2$.  \label{PolyCH_structure}}
\begin{tabular}{lccccccc}
& \multicolumn{5}{c}{(C$_2$H$_2$)$_x$} &\multicolumn{2}{c}{C$_2$H$_2$\textsuperscript{\textit{a}}}  \\ \cline{2-6}\cline{7-8} 
Method & $r$(C=C)\textsuperscript{\textit{b}} & $r$(C--C)\textsuperscript{\textit{b}} & $r$(CH) & $a$(CCC) & $a$(C=CH) & $r$(C$\equiv$C)  & $r$(CH)  \\ \hline
B3LYP/6-31G** & 1.368\,\AA & 1.427\,\AA & 1.091\,\AA & $124.5^\circ$ & $118.5^\circ$ & 1.206\,\AA & 1.066\,\AA \\ 
B3LYP/cc-pVDZ & 1.371\,\AA & 1.429\,\AA & 1.097\,\AA & $124.5^\circ$ & $118.4^\circ$ & 1.210\,\AA & 1.073\,\AA \\ 
Observed\textsuperscript{\textit{c}} & 1.36\,\AA & 1.44\,\AA & $\dots$ & $\dots$ & $\dots$ & 1.203\,\AA & 1.063\,\AA \\
\end{tabular} \\ 
\textsuperscript{\textit{a}}{Acetylene.}
\textsuperscript{\textit{b}}{Peierls' distortion.}
\textsuperscript{\textit{c}}{The polyacetylene structure from Yannoni and Clarke.\cite{Yannoni1983} The acetylene structure from Kuchitsu.\cite{Kuchitsu}}
\end{table*}

The calculated equilibrium structures of all-{\it trans} polyacetylene and acetylene are compared with the observed\cite{Kuchitsu,Yannoni1983} in Table \ref{PolyCH_structure}. The B3LYP hybrid functional with either basis set correctly predicts the bond-length alternation, satisfying Peierls' theorem.\cite{Suhai1995,Hirata1998} All measured bond lengths are reproduced by the calculation within 0.01\,\AA. The B3LYP/cc-pVDZ or 6-31G** methods are, therefore,
predictive for the structures of Peierls' systems, while nonhybrid DFT models are not.\cite{Mintmire1987,Vogl1989,Ashkenazi1989,Paloheimo1992,Suhai1995,Hirata1998}

\begin{figure}
\includegraphics[width=1.0\columnwidth]{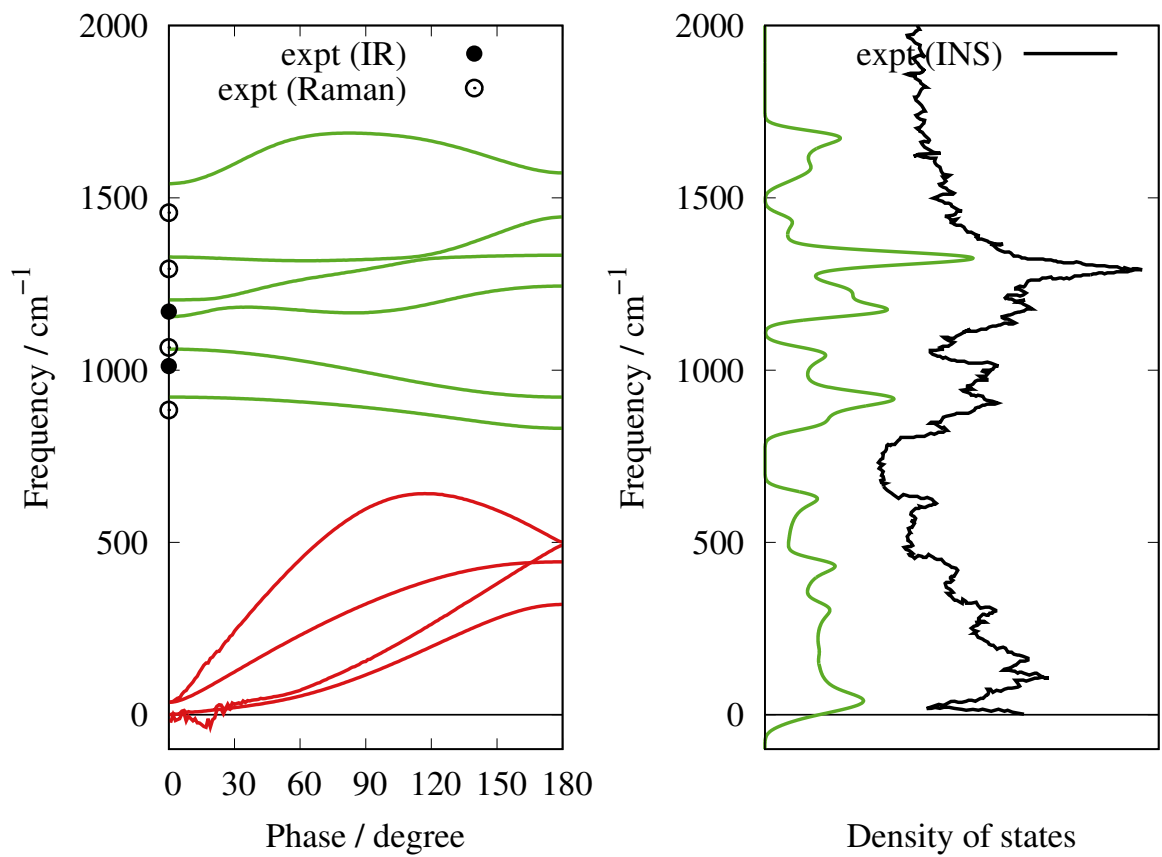}
\caption{\label{fig:polyCHdisp}Phonon dispersion curves and DOS of (C$_2$H$_2$)$_x$ computed by B3LYP/6-31G**. The DOS is convoluted with a Gaussian of a FWHM of 40 cm$^{-1}$. 
The IR data from Shirakawa and Ikeda\cite{Shirakawa1971} and from Takeuchi {\it et al.}\cite{Takeuchi1987} 
The Raman data from Kuzmany\cite{Kuzmany1980} and  from Takeuchi {\it et al.}\cite{Takeuchi1987} 
The INS spectrum from Hirata {\it et al.}\cite{Hirata_INS}}
\end{figure}

\begin{table}
\caption{Vibrational frequencies (in cm$^{-1}$) of (C$_2$H$_2$)$_x$.  \label{PolyCH_freqs}}
\begin{tabular}{lrr}
Irrep.\textsuperscript{\textit{a}}; phase; activity& \multicolumn{1}{c}{B3LYP/6-31G**} & \multicolumn{1}{c}{Observed\textsuperscript{\textit{b}}} \\ \hline
$A_g$; $\theta=0$; Raman &3138.8  & 2990 \\ 
& 1541.0 & 1457 \\
& 1328.1 & 1294 \\
&  1155.0 & 1066 \\
$B_g$; $\theta=0$; Raman & 921.8 & 884 \\
$A_u$; $\theta=0$; IR & 1060.9 & 1012 \\
$B_u$; $\theta=0$; IR & 3151.9 & 3013 \\
& 1203.7 & 1170 \\
\end{tabular} \\ 
\textsuperscript{\textit{a}}{Isomorphic to the $C_{2h}$ point group.}
\textsuperscript{\textit{b}}{The IR data from Shirakawa and Ikeda\cite{Shirakawa1971} and from Takeuchi {\it et al.}\cite{Takeuchi1987} 
The Raman data from Kuzmany\cite{Kuzmany1980} and  from Takeuchi {\it et al.}\cite{Takeuchi1987} }
\end{table}

Figure \ref{fig:polyCHdisp} plots the calculated phonon dispersion curves and phonon DOS in comparison with the IR and Raman band positions\cite{Shirakawa1971,Kuzmany1980,Takeuchi1987} and INS spectrum.\cite{Hirata_INS} Table \ref{PolyCH_freqs} compiles
the calculated and observed frequencies of the optically active phonons, which occur at the phase angle $\theta=0$. 
While the calculated frequencies are systematically higher than the observed by a few percent, the level of agreement is good enough
for reliable band assignments for all modes. They include the C=C and C--C stretching modes at 1457 and 1066 cm$^{-1}$, respectively, whose  
correct description must be predicated on a proper account of Peierls' distortion.
The overall shapes of the phonon dispersion curves are in excellent agreement with the ones calculated with an accurate force field determined 
by a hybrid experimental-computational approach.\cite{Hirata1995} 
All major peaks in the INS spectrum can also be unambiguously ascribed to the peaks in the phonon DOS. 
The B3LYP/6-31G** method is, therefore, predictive for the vibrational spectra of Peierls' systems.
The numerical noise in the acoustic-mode frequencies at $\theta=0$ 
is no more than 36 cm$^{-1}$, when up to the fifth nearest neighbor C$_2$H$_2$ groups are included in the normal-mode analysis.

\begin{figure}
\includegraphics[width=1.0\columnwidth]{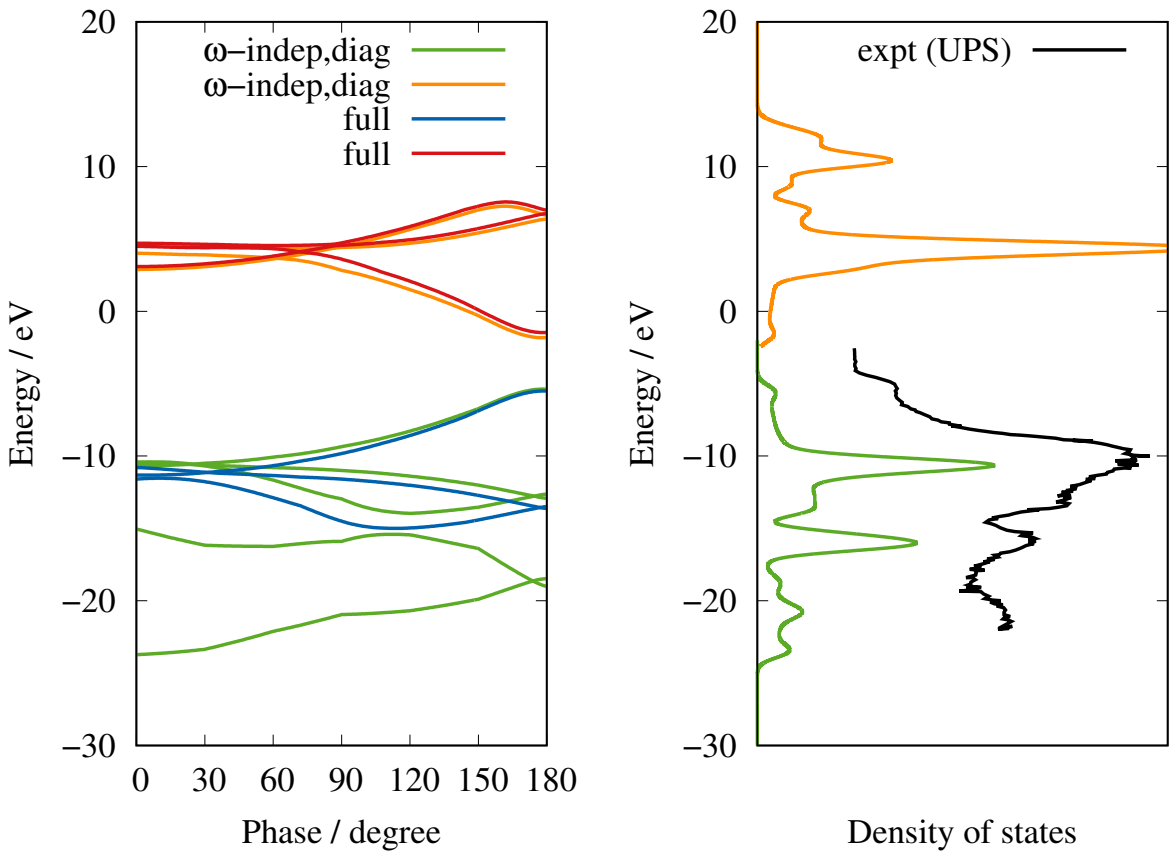}
\caption{\label{fig:polyCHbands}Electronic energy bands and DOS of (C$_2$H$_2$)$_x$ computed by MBGF(2)/cc-pVDZ. The DOS is convoluted with a Gaussian of a FWHM of 1 eV. 
The UPS from Kamiya {\it et al.}\cite{Kamiya_UPS}}
\end{figure}

Figure \ref{fig:polyCHbands} draws the quasiparticle energy bands calculated by MBGF(2)/cc-pVDZ using two types of the self-energy. 
The impact of the frequency-independent and diagonal approximations to the self-energy is generally small, but it is greater for lower-lying valence bands 
and higher-lying conduction bands. This is understood by Eq.\ (\ref{mbgf2}): In a higher-lying conduction 
band, the denominator of the first term can approach zero, where approximating $\omega$ by $\epsilon_i(k_i)$ can have a large effect. 
In a lower-lying valence band, the second term has the same propensity for a near-zero denominator. In more physical terms, the higher-lying (lower-lying)
the conduction (valence) band, the greater the overlap between multi-electron and one-electron bands and the involvements of non-Koopmans' states,
which cannot be described by the frequency-independent, diagonal approximation. The two major peaks in the calculated DOS occur at $-10.6$ and $-16.0$ eV. 
They correspond well with the observed peaks in the UPS at $-10.0$ and $-15.9$ eV. The MBGF(2)/cc-pVDZ method is predictive for valence  bands 
with accuracy of a few tenths of an electronvolt.

The calculated valence band edge is located at $-5.5$ eV. Its proximity to the air oxidation potential of 5.2 eV is consistent with the air instability of polyacetylene and its oligomers. 
The measured values of the fundamental band gap fall in the range of 1.4 --1.8 eV,\cite{Aulbur2000} as compared with the calculated value of 4.1 eV. 
The overestimation is again chiefly due to conduction bands being too high. In polyethylene (Sec.\ \ref{sec:CH2}), the calculated conduction band edge is overestimated by {\it ca.}\ 4.6 eV.
In polyacetylene, it is too high by {\it ca.}\ 2.2--2.6 eV only. In both cases, this is because of the inability of the compact basis set in describing diffuse conduction bands. 
The degree of overestimation is less in polyacetylene since its lower-lying conduction bands are made of conjugated $\pi^*$ orbitals, which are relatively less diffuse. 
Nonetheless, the MBGF(2)/cc-pVDZ method is not predictive for band gaps or conduction bands. 

The nonhybrid DFT models are well known\cite{Mintmire1987,Vogl1989,Ashkenazi1989,Paloheimo1992,Suhai1995,Hirata1998} to predict zero or near-zero fundamental band gap
for polyacetylene, violating Peierls' theorem, owing to the incomplete cancellation of self-interaction. Hybridizing the exchange functional with a HF exchange alleviates this problem, sometimes
yielding accurate band gaps for a variety of solids. However, the predicted band gaps sensitively depend on the weight of the admixed HF exchange and their agreement with the experimental values should be judged more or less arbitrary. 
We, therefore, relied on MBGF(2)/cc-pVDZ for energy bands, which is predictive for valence bands (but not for conduction bands or band gaps due to the inadequate basis set), while we used 
B3LYP for structures, vibrations, and thermodynamics only. 

\begin{table}
\caption{Binding energy (in kcal/mol) of (C$_2$H$_2$)$_x$.  \label{PolyCH_binding}}
\begin{tabular}{lr}
Method & Binding energy\textsuperscript{\textit{a}} \\ \hline
B3LYP/cc-pVDZ &  49.5  \\ 
MBPT(2)/cc-pVDZ & 47.5  \\ 
\end{tabular} \\ 
\textsuperscript{\textit{a}}{The energy difference between the unit-cell energy of (C$_2$H$_2$)$_x$ and the energy of C$_2$H$_2$ 
in their respective B3LYP/cc-pVDZ-optimized geometries.}
\end{table}


\begin{figure}
\includegraphics[width=1.0\columnwidth]{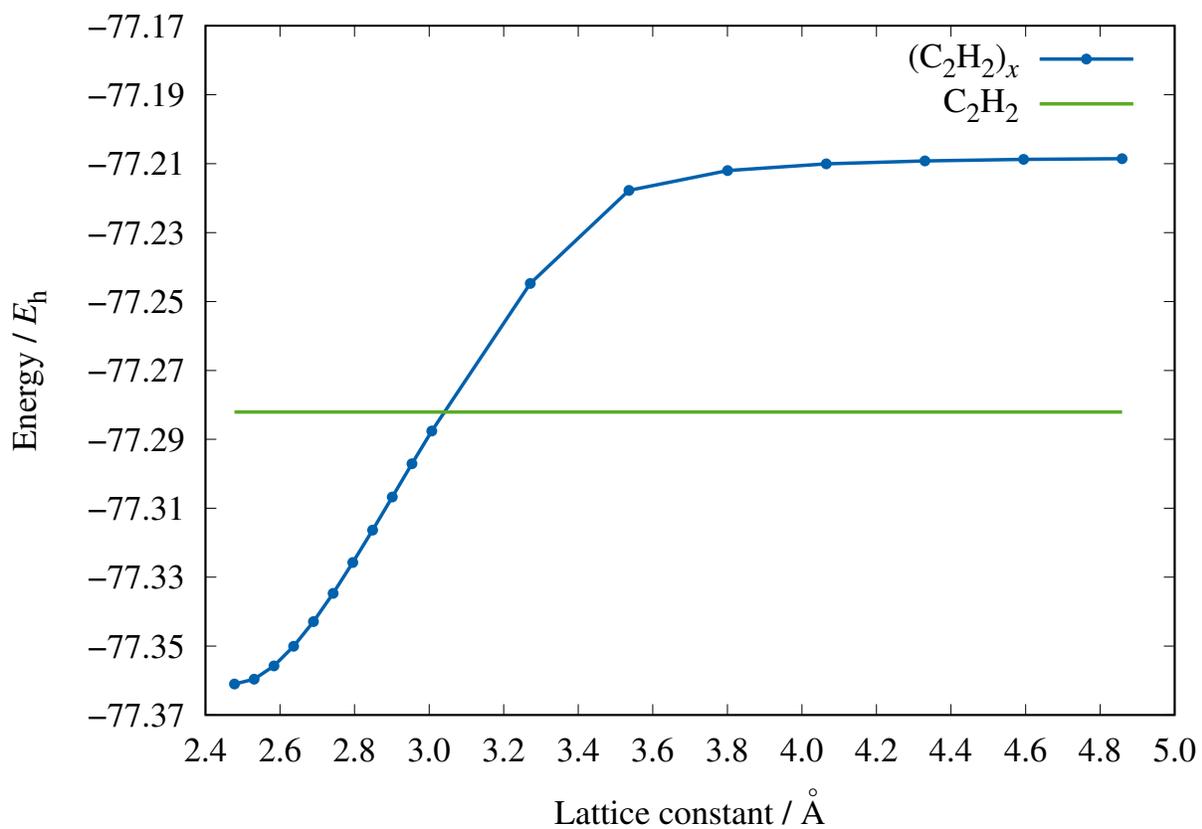}
\caption{\label{fig:CHdissociation}Unit-cell energy of (C$_2$H$_2$)$_x$ as a function of the lattice constant computed by B3LYP/cc-pVDZ.}
\end{figure}

Table \ref{PolyCH_binding} indicates 
that polyacetylene is thermodynamically stable with the binding energy per C$_2$H$_2$ unit cell of 47.5--49.6 kcal/mol relative to its monomer, acetylene. Figure \ref{fig:CHdissociation} suggests that a complete dissociation of polyacetylene into acetylenes 
will likely face an activation barrier higher than the energy difference between the product and reactant because 
of a large change in the structure from the polymer unit cell to the monomer.

%
%
%
%

\subsection{Polytetrafluoroethylene, (CF$_2$)$_x$\label{sec:PTFE}}

\begin{figure}
\includegraphics[width=0.85\columnwidth]{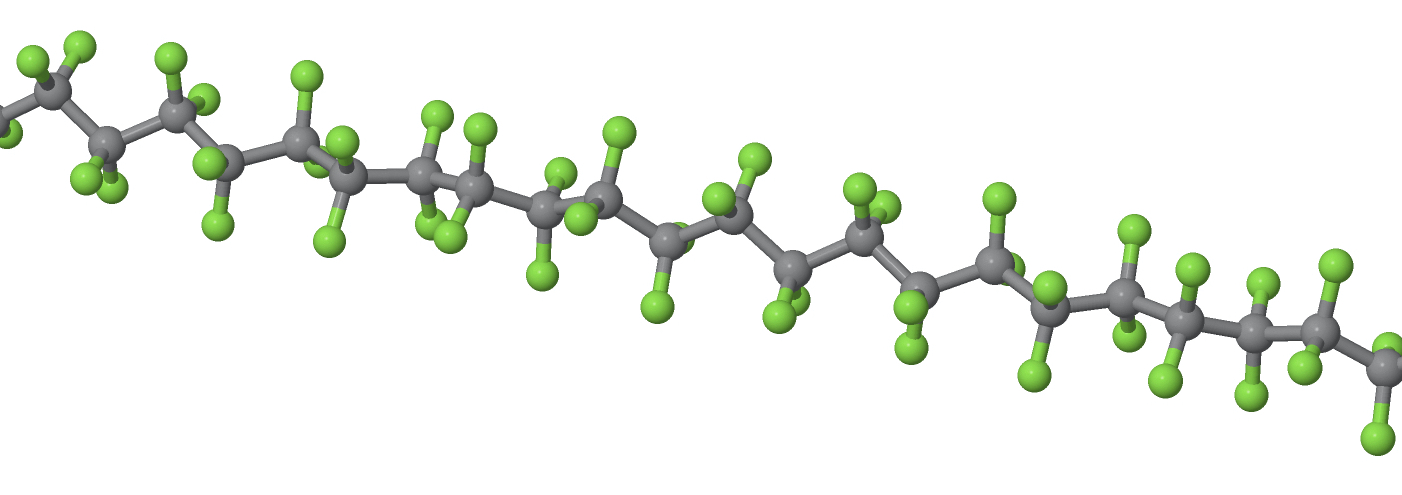}
\caption{Polytetrafluoroethylene, (CF$_2$)$_x$. \label{fig:CF2}}
\end{figure}

Polytetrafluoroethylene (Fig.\ \ref{fig:CF2}) adopts a $13/6$ helical structure (phase II) at low temperature and pressure,\cite{Bunn_Nature} meaning that 13 CF$_2$ 
groups make six complete turns with the helical angle of $\varphi = 6\times360^\circ/13 = 166.2^\circ$. 
At 19$^\circ$C, it undergoes a phase transition to a slightly more relaxed $15/7$ helix (phase IV) with helical angle of $\varphi = 7\times360^\circ/15 = 168.0^\circ$.\cite{Clark1999}
Crystal structures of various phases were interrogated experimentally\cite{Bunn_Nature,Weeks1981,Nakafuku1975,Lorenzen2003} and computationally.\cite{Farmer1981,DAmore2004,DAmore2006,Fatti2019}
Bunn and Howells,\cite{Bunn_Nature} who first determined the helical conformation of this polymer, argued that its cylindrical profile permits 
axial rotation and translation of chains relative to one another far below melting points, contributing to its super low friction.
The primary source of the low friction is, however, its low polarizability and thus weak dispersion forces of the system containing ``hard'' fluorine atoms. 
It is thermodynamically and chemically stable.\cite{Grainger2001}


\begin{table*}
\begin{scriptsize}
\caption{Structural parameters of (CF$_2$)$_x$ and C$_2$F$_4$.  \label{PolyCF2_structure}}
\begin{tabular}{lccccccccc}
& \multicolumn{6}{c}{(CF$_2$)$_x$} &\multicolumn{3}{c}{C$_2$F$_4$\textsuperscript{\textit{a}}}  \\ \cline{2-7}\cline{8-10} 
Method & $r$(CC) & $r$(CF) & $a$(CCC) & $a$(FCF) & $d$(CCCC) & $\varphi$\textsuperscript{\textit{b}} & $r$(CC)  & $r$(CF) & $a$(FCF)  \\ \hline
B3LYP/6-31G** & 1.562\,\AA & 1.353\,\AA & $113.3^\circ$ & $109.4^\circ$ & $163.0^\circ$ & $165.8^\circ$ & 1.326\,\AA & 1.325\,\AA & $113.7^\circ$ \\ 
B3LYP/cc-pVDZ & 1.567\,\AA & 1.353\,\AA & $113.0^\circ$ & $109.1^\circ$  & $161.7^\circ$  & $164.7^\circ$& 1.329\,\AA & 1.324\,\AA & $113.3^\circ$ \\ 
Observed (13/6)\textsuperscript{\textit{c}} & (1.53\,\AA) & (1.33\,\AA) & $116^\circ$\textsuperscript{\textit{d}} & ($108^\circ$)  & & $166.2^\circ$\textsuperscript{\textit{d}} & 1.311\,\AA & 1.319\,\AA & $112.4^\circ$ \\
\end{tabular} \\ 
\textsuperscript{\textit{a}}{Tetrafluoroethylene.}
\textsuperscript{\textit{b}}{The helical angle [Eqs.\ (\ref{X})--(\ref{Z})].}
\textsuperscript{\textit{c}}{The polytetrafluoroethylene structure (in parentheses) consistent with the X-ray diffraction data of Weeks {\it et al.}\cite{Weeks1981} The tetrafluoroethylene structure from Hellwege and Hellwege.\cite{Hellwege}}
\textsuperscript{\textit{d}}{Bunn and Howells.\cite{Bunn_Nature}}
\end{scriptsize}
\end{table*}

X-ray photoelectron spectroscopic, UPS, and
electron energy loss spectroscopic studies were reported.\cite{Clark1971,Pireaux_UPS1974,Falk1975,Delhalle_UPS1977,Seki_UPS1990,Seki1990,Miyamae2000,Ono2005,Wang2014} Miyamae {\it et al.}\cite{Miyamae2000} obtained the valence band structures 
by ARUPS, but the observed topmost valence band was in poor agreement with the HF/STO-3G calculation of Seki {\it et al.},\cite{Seki_UPS1990} 
which assumed a planar zigzag conformation. Yoshimura {\it et al.}\cite{Yoshimura_ARUP} reevaluated the ARUPS data to bring it into more accurate agreement with a simulation. 
See also Refs.\ \cite{Morokuma1971,Mccubbin1971,Pireaux_UPS1974,Delhalle1974,Falk1975,Delhalle_UPS1977,Kasowski1980,Otto1985,Springborg1989,Seki_UPS1990,Seki1990,Cain1992,Wang2014} for 
other band-structure calculations. 

Generally, three types of Brillouin zone are considered for a helix:\cite{Piseri1973}\ In the true-zone scheme, the translational repeat unit (C$_{13}$F$_{26}$ in the case of 13/6-helical polytetrafluoroethylene) is taken as the real-space unit cell. In the extended-zone scheme, the rototranslational (physical) repeat unit (CF$_2$) is the real-space unit cell.\cite{Bower1989} 
In the semi-extended-zone scheme, two rototranslational repeat units (C$_2$F$_4$) forms the real-space unit cell. The last scheme is introduced here because of the structural 
similarity of polytetrafluoroethylene ($\varphi \approx 166^\circ$) with polyethylene ($\varphi = 180^\circ$), whose true-zone scheme is based on a C$_2$H$_4$ group.

The selection rules of the IR and Raman transitions in an infinite helical polymer were deduced by Higgs.\cite{Higgs} 
Phonons in a helix are optically active if the phase angle ($\theta$) between adjacent unit-cell oscillators in the extended-zone scheme is equal to zero, $\varphi$, or $2\varphi$,
where $\varphi$ is the helical angle. More detailed IR and Raman activities can be decided on the basis of isomorphism with a point group. For example, the line group of the 13/6 chain of polytetrafluoroethylene
is isomorphic to the $D_{13}$ point group. Its phonons with vibrational phase angle of $\theta = 0$ transform as $A_1$ or $A_2$, which are Raman or IR active, respectively.
The phonons at $\theta = \varphi$ transform as $E_1$, which are both IR and Raman active, whereas those at $\theta = 2\varphi$ transform as $E_2$, which are Raman active.

The IR spectra of polytetrafluoroethylene was measured by Liang and Krimm.\cite{Liang1956} Koenig and Boerio\cite{Koenig1969} and Peacock {\it et al.}\cite{Peacock1970}
reported the Raman spectra. Coherent INS spectra were observed by Twisleton and White,\cite{Twisleton1972} following a measurement by LaGarde {\it et al.}\cite{LaGarde1969} 
Hannon {\it et al.}\cite{Hannon1969}\ performed a normal-mode analysis with empirical force fields, using three helix models: 13/6, 15/7, and planar zigzag. 
The most complete vibrational analysis was carried out by Piseri {\it et al.},\cite{Piseri1973} who also reported coherent INS spectra and 
established IR and Raman band assignments.

In this study, we ran the B3LYP/cc-pVDZ and 6-31G** calculations for structures, phonon dispersions, and phonon DOS and the MBGF(2)/cc-pVDZ calculations in the frozen-core and mod-6 approximations\cite{Shimazaki2009,hirata_qp}  
for quasiparticle energy bands and electronic DOS with $S=8$, $L=12$, and $K=48$.  A rototranslational (physical) repeat unit used was a CF$_2$ group.
Up to the eighth nearest neighbor CF$_2$ groups were considered in the phonon dispersion calculation.

Table \ref{PolyCF2_structure} compares the optimized geometries of polytetrafluoroethylene and its monomer, tetrafluoroethylene, with the observed.\cite{Bunn_Nature,Weeks1981,Hellwege} The B3LYP method with either basis set reproduces bond lengths and bond angles within a few hundredths of
one \AA ngstrom and a few degrees, respectively, of the experimental values, and is deemed predictive for structures. The helical angle ($\varphi$) is calculated
to be 164.7--165.8$^\circ$, which are closer to 166.2$^\circ$ of the 13/6 helix (low-temperature conformation) than to 168.0$^\circ$ of the 15/7 helix (high-temperature conformation). The calculated $\varphi$, however, does not precisely correspond to any commensurable structure. 

In the $13/6$-helical conformation, the fluorine atoms are not farthest apart from one another.\cite{Bunn_Nature} 
It was argued that the steric repulsion between the fluorine atoms is compensated for by
having a larger CCC angle of 116$^\circ$ than the corresponding angle of 112$^\circ$ in polyethylene since it is energetically harder to stretch the C--C bond.\cite{Bunn_Nature} 
This argument is not supported by our calculations:\ The CCC angles are roughly the same (113$^\circ$) between polyethylene and polytetrafluoroethylene, but the C--C bond is noticeably longer in the latter. Rather, the helical angle of 166$^\circ$ may be chosen by nature to stagger (at least slightly) the fluorine atoms not only with the first but also 
with the second nearest neighbor CF$_2$ groups (see Fig.\ 1 of Ref.\ \cite{Piseri1973}). 

\begin{figure}
\includegraphics[width=1.0\columnwidth]{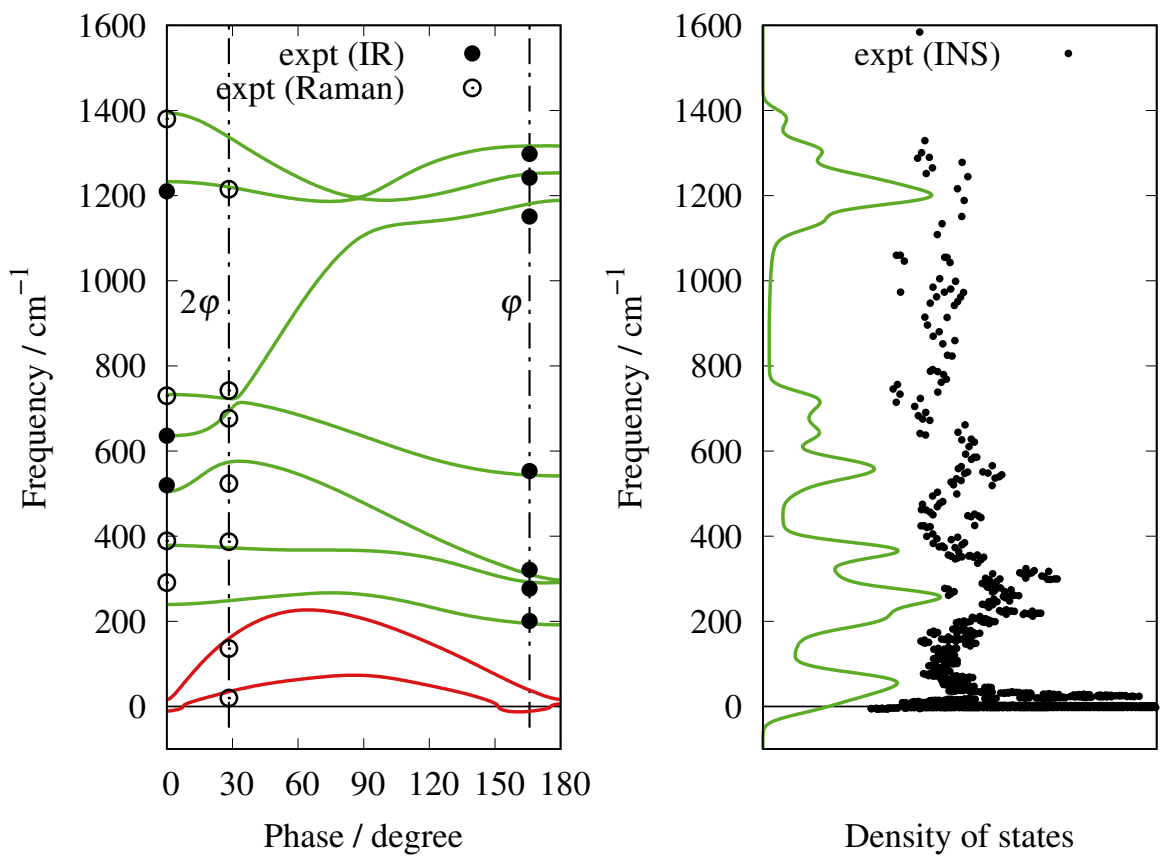}
\caption{\label{fig:PTFE}Phonon dispersion curves and DOS of (CF$_2$)$_x$ computed by B3LYP/6-31G** in the extended-zone scheme (i.e., corresponding to a CF$_2$ group as the repeat unit). 
Phonons are optically active when the phase angle of adjacent CF$_2$ oscillators is zero, $\varphi$ (helical angle), or $2\varphi$, which are indicated by dotted-dashed lines.
The DOS is convoluted with a Gaussian of a FWHM of 40 cm$^{-1}$.
The coherent INS spectrum (the scattering angle of 36$^\circ$) from Twisleton and White.\cite{Twisleton1972}}
\end{figure}

\begin{table}
\caption{Vibrational frequencies (in cm$^{-1}$) of (CF$_2$)$_x$.  \label{PolyCF2_freqs}}
\begin{tabular}{lrr}
Irrep.\textsuperscript{\textit{a}}; phase; activity& \multicolumn{1}{c}{B3LYP/6-31G**} & \multicolumn{1}{c}{Observed\textsuperscript{\textit{b}}} \\ \hline
$A_1$; $\theta=0$; Raman & 1394.2 & 1380 \\ 
& 733.0 & 730 \\
& 378.5 & 389 \\
& 239.8 & 291 \\
$A_2$; $\theta=0$; IR & 1232.8 & 1210 \\
 &635.8 & 636 \\
 & 504.7 & 520 \\ 
$E_1$; $\theta=\varphi$; IR, Raman & 1316.6 & 1298 \\
&1251.0& 1242 \\
&1180.7& 1151 \\
&543.5& 553 \\
&310.5& 321 \\
&292.1& 277 \\
&194.8& 201 \\
&38.3 & $\dots$ \\
$E_2$; $\theta=2\varphi$; Raman & 1337.4 & $\dots$ \\
& 1220.3 & 1215 \\
& 723.1 & 742 \\
& 693.2 & 677 \\
& 573.9 & 524 \\
& 373.0 & 387 \\
& 248.6 & $\dots$ \\
& 160.0 & 136 \\
& 34.7 & 20 \\
\end{tabular} \\ 
\textsuperscript{\textit{a}}{Isomorphic to the $D_{13}$ point group. $\varphi$ is the helical angle.}
\textsuperscript{\textit{b}}{Based on the assignment {\it F}-II of Piseri {\it et al.}\cite{Piseri1973}
The IR data ($A_2$, $E_1$) from Liang and Krimm.\cite{Liang1956} The Raman data ($A_1$, $E_2$) from Koenig and Boerio.\cite{Koenig1969}
}
\end{table}

Comparison of the calculated and observed frequencies of the IR and Raman bands\cite{Liang1956,Koenig1969} is made in Table \ref{PolyCF2_freqs}.
 Without any alteration to the assignment {\it F}-II of  Piseri {\it et al.},\cite{Piseri1973} the calculated frequencies agree excellently with the observed with the mean absolute deviation of 17 cm$^{-1}$. 
Figure \ref{fig:PTFE} plots the phonon dispersion curves and phonon DOS alongside the IR and Raman band positions and INS spectrum.\cite{Twisleton1972} 
Note that the INS spectrum is due to coherent scattering with a specific scattering angle of 36$^\circ$ and does not necessarily reflect the phonon DOS if the dispersions
are large. Nonetheless, the calculated DOS is consistent with the INS spectrum, both displaying broad peaks in the same frequency domains. 
The B3LYP/6-31G** method is predictive for vibrations. 

In a helical polymer, three zero-frequency acoustic modes are expected:\ two at $\theta=0$ (longitudinal and spinning) and one at $\theta=\varphi$ (transverse).  
Their frequencies are calculated to be 11$i$, 16, and 11$i$ cm$^{-1}$, respectively. The imaginary frequencies of these sizes
are numerical errors caused by the distance-based truncation of the force-constant matrices and do not imply a saddle point on the potential energy surface. 
They are relatively small for polytetrafluoroethylene likely because of the absence
of hydrogens, whose small masses may amplify numerical errors in the reciprocal-mass-weighted force constants.

\begin{figure}
\includegraphics[width=1.0\columnwidth]{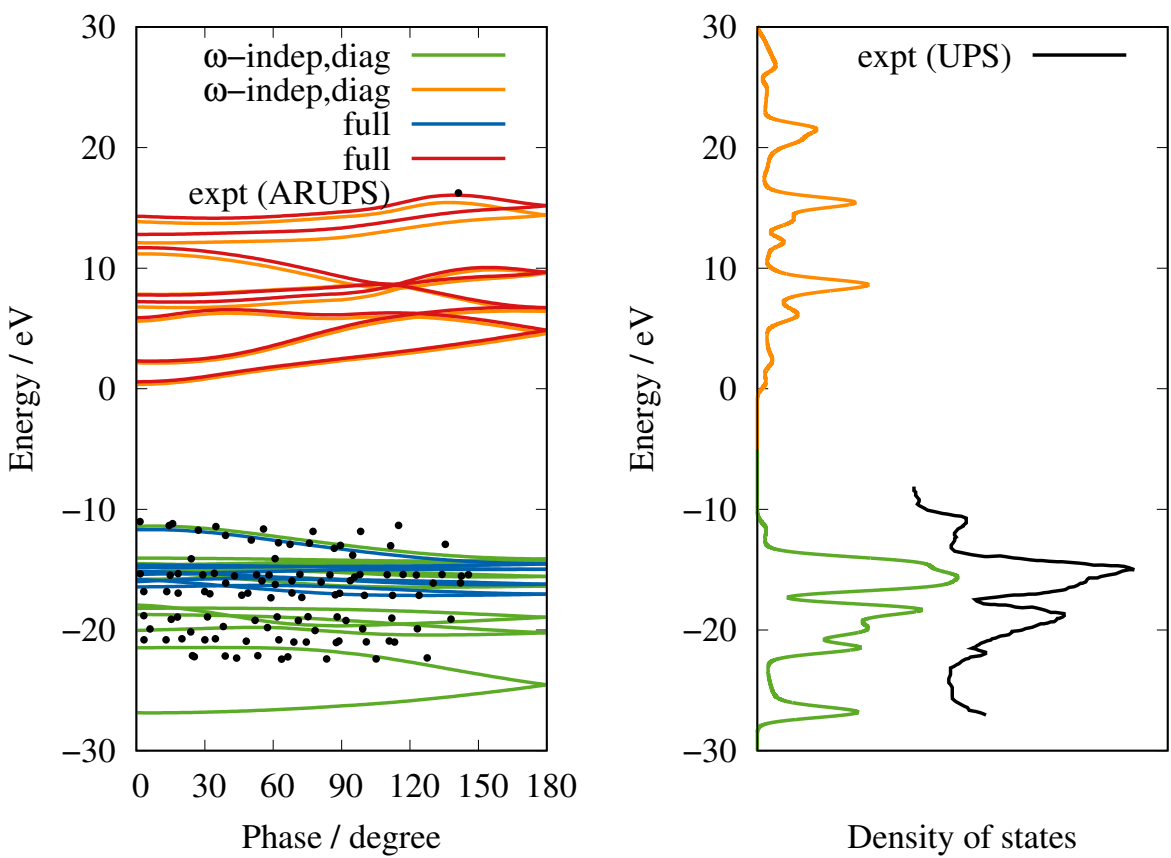}
\caption{\label{fig:PTFEbands}Electronic energy bands and DOS of (C$_2$F$_4$)$_x$ computed by MBGF(2)/cc-pVDZ in the semi-extended-zone scheme (i.e., corresponding to a C$_2$F$_4$ group as the repeat unit).
The DOS is convoluted with a Gaussian of a FWHM of 1 eV.
The ARUPS and UPS were measured originally by Miyamae {\it et al.}\cite{Miyamae2000} and then recalibrated by Yoshimura {\it et al.}\cite{Yoshimura_ARUP}}
\end{figure}

Figure \ref{fig:PTFEbands} compares the MBGF(2)/cc-pVDZ quasiparticle energy bands and DOS with the experimental data obtained by ARUPS and UPS.\cite{Miyamae2000,Yoshimura_ARUP}
They are in accurate agreement with each other, regardless of the approximations used in the self-energy. The two highest peaks in the UPS are observed at $-15.0$ and $-18.8$ eV,
which are in good accord with the calculated peaks in the convoluted DOS at $-15.6$ and $-18.3$ eV, respectively. Therefore, the MBGF(2)/cc-pVDZ method is predictive for valence  bands
with accuracy of {\it ca.}\ 0.5 eV. The reinterpretation of the experimental data by Yoshimura {\it et al.}\cite{Yoshimura_ARUP}\ is supported by our calculations. 

The fundamental band gap of an isolated polytetrafluoroethylene chain in a vacuum is estimated to be 11.9 eV with the valence band edge at $-11.7$ eV and 
conduction band edge at $0.2$ eV.\cite{Seki1990} These values were deduced from an extrapolation of the UPS and vacuum UV absorption spectra of oligomers, and 
differ from the ones in the bulk material (e.g., the valence band edge of the bulk is located at $-10.6\pm0.1$ eV).\cite{Seki1990}
The large band gap is consistent with the good insulation and low friction of polytetrafluoroethylene,
and the deep valence band edge explains its chemical and air stability.
In our MBGF(2)/cc-pVDZ calculation using the ``full'' self-energy, the valence and conduction band edges are located 
at $-11.7$ and $0.6$ eV, respectively, with the predicted band gap of 12.3 eV, which is in reasonable agreement with the observed (11.9 eV). 
As in polyethylene, the valence band edge is accurately reproduced (within 0.1 eV of the observed in this case).
The conduction band edge of polytetrafluoroethylene is overestimated by only 0.4 eV in contrast to an overestimation by 4.7 eV in polyethylene.
This is understandable because the electron-attached states of fluorine-rich molecules should be describable reasonably well by the cc-pVDZ basis set.
Nevertheless, the predictive accuracy of the MBGF(2)/cc-pVDZ method for conduction bands and band gaps cannot be claimed generally.

\begin{table}
\caption{Binding energy (in kcal/mol) of (C$_2$F$_4$)$_x$.  \label{PolyCF2_binding}}
\begin{tabular}{lr}
Method & Binding energy\textsuperscript{\textit{a}} \\ \hline
B3LYP/cc-pVDZ &  32.7  \\ 
MBPT(2)/cc-pVDZ & 41.9  \\
\end{tabular} \\ 
\textsuperscript{\textit{a}}{The energy difference between the unit-cell energy of (C$_2$F$_4$)$_x$ and  the energy of C$_2$F$_4$ 
in their respective B3LYP/cc-pVDZ-optimized geometries.}
\end{table}


\begin{figure}
\includegraphics[width=1.0\columnwidth]{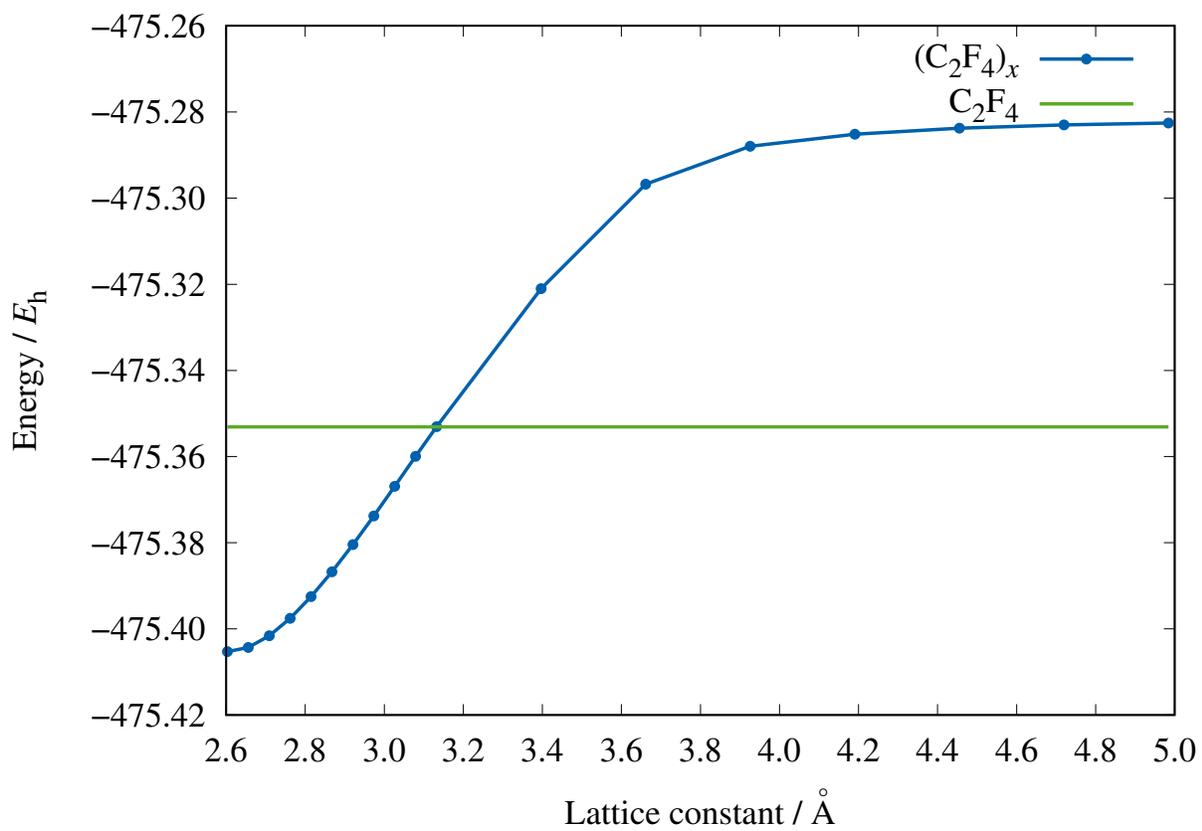}
\caption{\label{fig:PTFEdiss}Unit-cell energy of (C$_2$F$_4$)$_x$ as a function of the lattice constant computed by B3LYP/cc-pVDZ.}
\end{figure}

Table \ref{PolyCF2_binding} attests to the thermodynamic stability of polytetrafluoroethylene against dissociation into its monomers. 
An unrelaxed dissociation reaction energy profile illustrated in Fig.\ \ref{fig:PTFEdiss} also suggests the existence of a barrier to dissociation that is higher 
than the energy difference between product and reactant, adding to its kinetic stability. This, in turn, may be partly 
due to the large structural change from the $sp^3$-bonded C$_2$F$_4$ in the polymer to $sp^2$-bonded planar tetrafluoroethylene. 

%
%
%
%

\subsection{Polyazene, (N$_2$)$_x$\label{sec:N2}}

\begin{figure}
\includegraphics[width=0.85\columnwidth]{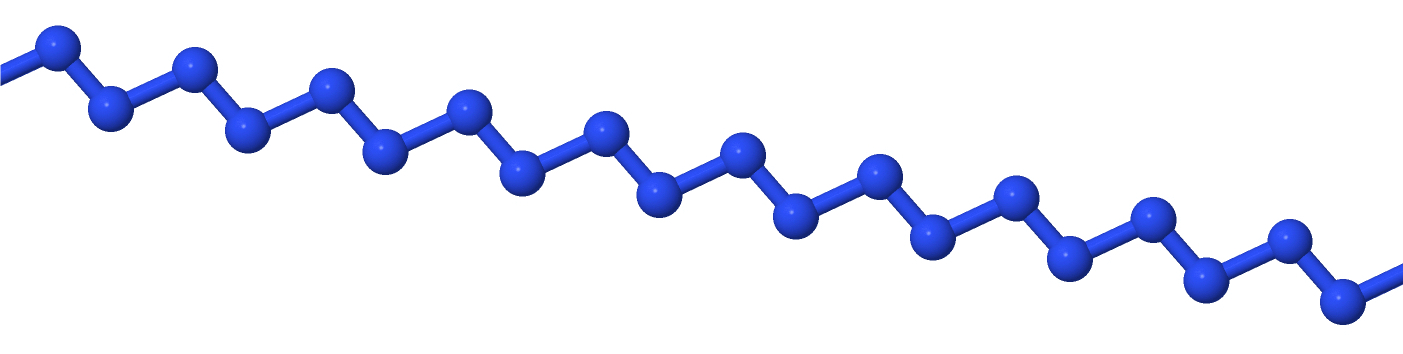}
\caption{Polyazene, (N$_2$)$_x$. \label{fig:N2}}
\end{figure}

As an infinitely catenated chain of nitrogen, we consider all-{\it trans} polyazene (Fig.\ \ref{fig:N2}). Since it is isoelectronic with all-{\it trans} polyacetylene (Sec.\ \ref{sec:C2H2}), we sought its possible metastable structure in a planar zigzag conformation with alternating N--N and N=N bonds. We also examined other conformations with no avail (see below).
The B3LYP/cc-pVDZ and 6-31G** calculations were performed for the structures, phonon dispersions, and phonon DOS with $S=4$, $L=6$, and $K=24$ using 
a N$_2$ group as the translational repeat unit. The MBGF(2)/cc-pVDZ calculations for 
the quasiparticle energy bands and electronic DOS additionally used the frozen-core and mod-3 approximations.\cite{Shimazaki2009,hirata_qp} Quantitative accuracy of these methods has been established 
for the organic polymers in Secs.\ \ref{sec:CH2}--\ref{sec:PTFE}.

There are two approaches to realizing longer catenated chains of nitrogen under ambient conditions.
One is the chemistry (bottom-up) approach, which enlarges the ring or elongates the chain with careful design of its protective terminal groups.\cite{Bartlett2000,Steele_bookchapter2019,ReviewN,AQC2014} 
The other is the physics (top-down) approach, which aims at high-pressure polymerization of solid or liquid N$_2$, followed by lifting of the pressure to kinetically trap the metastable polymers.\cite{Mailhiot1992,Eremets2004JCP,Ma_polymericN_2009,Pickard_polymericN_2009,Steele_bookchapter2019} 
In this context, ``polymer'' refers to a nonmolecular solid including two- or three-dimensional covalently-bonded networks.
In both approaches, synthetic and computational studies played essential and complementary roles.

In the chemistry approach, successful syntheses\cite{Christe1999,Vij2001,Cacace2002,Vij2002,Bi2010,Li2010,Tan2012,Tang2013,Barazov2016,Zhang2017,Zhang2018,Huang2018} of nitrogen allotropes 
were piloted by computations such as 
those on N$_4$,\cite{Lauderdale1992,Nguyen1996,Nguyen2003} N$_5$,\cite{Nguyen2001,Nguyen2003} N$_6$,\cite{Huber1982,Saxe1983,Schleyer1992,Ha1992,Lauderdale1992,Nguyen1996,Nguyen2001,Tobita2001,Wilson2001,Greschner2016}  N$_8$,\cite{Lauderdale1992,Nguyen1996,Fau2001,Hirshberg2014}  N$_{10}$,\cite{Fau2002} 
N$_{12}$,\cite{Olah2001} and even N$_{20}$,\cite{N20_1999} often using such highly accurate methods as CCSD(T) (coupled-cluster singles and doubles with 
noniterative triples).\cite{Raghavachari1989,Watts1993}
Christe {\it et al.}\cite{Christe1999} reported the synthesis and detection of a bent open-chain structure of N$_5^+$. See also Vij {\it et al.}\cite{Vij2001} Cacace {\it et al.}\cite{Cacace2002} synthesized an open-chain N$_4$.  
These were followed by the synthesis and isolation of the pentazole anion or cyclo-N$_5^-$ by Vij {\it et al.},\cite{Vij2002} by Barazov {\it et al.},\cite{Barazov2016} and by Zhang {\it et al.}\cite{Zhang2017} 
Subsequently, Bi {\it et al.}\cite{Bi2010} synthesized N$_6$H$_2$, consisting of cyclo-N$_5$ appended with another nitrogen atom. 
Zhang {\it et al.}\cite{Zhang2018} constructed stable zeolitic clusters composed of Na$^+$ and cyclo-N$_5^-$, containing as many as 60 nitrogen atoms. 

Among these, the most salient may be hexazine, all nitrogen analog\cite{Dewar1975} of benzene; hexazine is to polyazene as benzene is to polyacetylene.
Previous computational studies\cite{Huber1982,Schleyer1992,Ha1992,Lauderdale1992,Tobita2001} (with the exception of Ref.\ \cite{Saxe1983})
indicated that its planar $D_{6h}$ structure is a saddle point on its potential energy surface and is, therefore, unstable either towards  dissociation into N$_2$ (without an activation barrier\cite{Ha1981})
or towards structural relaxations. Lauderdale {\it et al.}\cite{Lauderdale1992} showed that it has an out-of-plane bending vibration with an imaginary frequency and thus
spontaneously transforms to either a twist or boat structure and eventually to hexaazaprismane (which is bound by N--N single bonds). 
The most stable isomer, however, seems to be a (twisted) open chain,\cite{Huber1983,Schleyer1992,Ha1992,Tobita2001,Olah2001} which may be further stabilized in condensed phase.\cite{Greschner2016}
These findings suggest that polyazene (either in the planar {\it trans}-transoid, {\it cis}-transoid, or {\it trans}-cisoid conformation) may not be sufficiently stabilized by aromaticity  
to withstand an out-of-plane structural distortion or even dissociation. It was also pointed out that B3LYP and CCSD(T) give 
largely consistent structural predictions for N$_6$.\cite{Tobita2001} 

Li {\it et al.}\cite{Li2010} reported the synthesis of a molecule containing a catena-8 nitrogen chain in the all-{\it trans} polyazene conformation. Its termini
are protected by five-membered rings, in much the same way all-{\it trans} polyenes need to be stabilized by terminal phenyl rings.\cite{Kim2002}
They furthermore observed reversible {\it trans}-{\it cis} photoisomerization, although the same isomerization 
in polyacetylene primarily occurs thermally or by doping.\cite{Tanaka1983} Tang {\it et al.}\cite{Tan2012,Tang2013} extended the all-{\it trans} polyazene motif to catena-10 and 11 nitrogen chains.
Unlike hexazine, that these can be synthesized and isolated may be taken as supportive evidence for the potential existence of infinite all-{\it trans} polyazene.
C$_2$N$_{14}$ (1-diazidocarbamoyl-5-azidotetrazole) is one of the most sensitive explosives ever made.\cite{Klapotke_AAA2011,Klapotke_AAA,Banert_AAA}
Its central motif is a catenated N$_5$ chain in a {\it cis}-polyazene-like conformation, also supporting the metastable existence of {\it cis}-transoid or {\it trans}-cisoid polyazene (see below for
potential high-pressure polymerization of {\it cis}-polyazene).
Open-chain N$_7$ detected by Huang {\it et al.},\cite{Huang2018} in contrast, is wire-like and looks more like polyynes.

Turning to the physics approach, Nellis {\it et al.}\cite{Nellis1984}\ is the first to computationally explore the possibility of infinitely catenated nitrogen under high pressure, predicting 
a polymeric liquid nitrogen. 
McMahan and LeSar\cite{McMahan1985} proposed the {\it cubic gauche} (cg) form of solid nitrogen under 1 Mbar, the pressure
achievable by diamond anvil cells, spurring both experimental and computational searches. 
Mailhoit {\it et al.}\cite{Mailhiot1992}\ refined the computational characterization of the cg form and also discovered other polymeric forms including a two-dimensional 
black phosphorous structure (stable above 210 GPa) and one-dimensional bond-alternating chain. The latter was predicted to be metastable, likely transforming to a tri-coordinated solid.
A one-dimensional chain was also studied with crystal-orbital theory by Pohl {\it et al.},\cite{Springborg1994} using DFT in the local density approximation (LDA).
The method displayed a severe underestimation of the bond-length alternation, also predicting a metallic electronic structure. 
This is well established to be due to the self-interaction error in LDA, causing the identical problem for isoelectronic {\it trans}-polyacetylene\cite{Mintmire1987,Vogl1989,Ashkenazi1989,Paloheimo1992,Suhai1995,Hirata1998} 
 (see Sec.\ \ref{sec:C2H2}).
These results clearly indicate that DFT without a hybrid of the HF exchange is fundamentally incapable of describing Peierls' systems, which are bond-alternating insulators at 0\,K. 
Subsequent DFT studies by Alemany and Martin\cite{Alemany2003} and by Mattson {\it et al.}\cite{Mattson2004}\ seem to suffer from the same qualitative error of LDA,
predicting a metallic zigzag chain.

The first experimental report of infinitely catenated nitrogen under high pressure (150 GPa) came from Goncharov {\it et al.},\cite{Goncharov2000} who observed the disappearance of the N$\equiv$N stretching vibration
of 2300--2500 cm$^{-1}$ and concomitant emergence of lattice vibrations in the range of 200--600 cm$^{-1}$. This phase is found to be a semiconductor and believed to be amorphous. 
Eremets {\it et al.}\cite{Eremets2001} were able to release pressure and bring this phase to a metastable state under ambient conditions. Gregoryanz {\it et al.}\cite{Gregoryanz2001,Gregoryanz2002}\ determined
the phase diagrams of the relevant phases spectroscopically. The first crystalline phase of catenated nitrogen was achieved by Eremet {\it et al.}\cite{Eremets2004,Eremets2004JCP,Eremets2007}\ in the cg form 
at high temperature and pressure. Whereas this phase could not be quenched at ambient conditions, this synthetic achievement spurred further computational predictions
of new polymeric phases.\cite{Pickard_polymericN_2009,Ma_polymericN_2009} 
Tomasino {\it et al.}\cite{Tomasino2014}\ synthesized a two-dimensional layered-polymeric form of nitrogen at 120--180 GPa, well above the stability range of the cg form. 
See Sontising and Beran\cite{Sontising2019,Sontising2020} for more recent, high-accuracy computational studies of solid nitrogen phases.

The past decade has seen a combination of the chemistry and physics approaches:\ high-pressure polymerization of shorter catenated nitrogen rings or chains.
Li {\it et al.}\cite{Li2013}\ computationally predicted high-pressure condensation of KN$_3$ into benzene-like rings or polymer chains. 
Prasad {\it et al.}\cite{Prasad2013}\ performed an evolutionary crystal structure exploration of LiN$_3$ under high pressure up to 300 GPa, and found the formation
of benzene-like N$_6$ rings and {\it cis}-polyacetylene-like infinite one-dimensional chains, which are furthermore predicted to be metastable under ambient conditions. 
They are both metallic as they are negatively charged (the authors used a hybrid DFT functional to guard against the aforementioned problem\cite{Mintmire1987,Vogl1989,Ashkenazi1989,Paloheimo1992,Suhai1995,Hirata1998}). 
This is perhaps the most supportive study yet of the existence of polyazene, albeit in the {\it cis} conformation.
Shen {\it et al.}\cite{Shen2015}\ reported a similar study, exploring a larger pool of high-pressure structures of LiN$_x$ including {\it cis}-polyacetylene-like chains.
In their DFT-based crystal-structure search, Peng {\it et al.}\cite{PengSR2015}\ discovered that, under high pressure, CsN$_2$ transforms 
to infinite helical chains of nitrogen. Since the chain is doped with Cs, it is said to adopt a similar conformation as helical polyoxane\cite{Oganov2006} (see Sec.\ \ref{sec:O}),
although it contradicts the observed effect of chemical doping on the structure of polyacetylene.\cite{Shirakawa1971,Shirakawa1973,Chiang1977,Chien1984,Heeger1988} 
Steele and Oleynik,\cite{Steele2017} on the other hand, found the chain structure of nitrogen in the high-pressure phase of KN$_x$ to be unstable. 
Williams {\it et al.}\cite{Williams2017}\ studied complexes of rubidium and catenated nitrogen under high pressures, again causing the nitrogen atoms to form {\it cis}-polyazene chains. 
These ideas have been recently realized synthetically by Bykov {\it et al.}\cite{Bykov2018,Bykov2020,Bykov2021} Some of the high-pressure structures indeed feature 
{\it cis}-polyazene chains chelating metals, which are metastable under ambient conditions. 


\begin{table}
\caption{Structural parameters of (N$_2$)$_x$ and N$_2$.  \label{PolyN2_structure}}
\begin{tabular}{lcccc}
& \multicolumn{3}{c}{(N$_2$)$_x$} & N$_2$  \\ \cline{2-4}\cline{5-5} 
Method & $r$(N=N)\textsuperscript{\textit{a}} & $r$(N--N)\textsuperscript{\textit{a}} & $a$(NNN) & $r$(N$\equiv$N) \\ \hline
B3LYP/6-31G** & 1.249\,\AA & 1.432\,\AA & 106.1$^\circ$ & 1.106\,\AA\\ 
B3LYP/cc-pVDZ & 1.248\,\AA & 1.434\,\AA & 106.0$^\circ$ & 1.105\,\AA \\ 
Observed\textsuperscript{\textit{b}} & (1.205\,\AA) & (1.429\,\AA) & (109.2$^\circ$) & 1.098\,\AA \\
\end{tabular} \\ 
\textsuperscript{\textit{a}}{Peierls' distortion.}
\textsuperscript{\textit{b}}{The {\it trans}-tetrazene-(2) structure (in parentheses) from Veith and Schlemmer.\cite{Veith1982}
The N$_2$ structure from Huber and Herzberg.\cite{HuberHerzberg}}
\end{table}

Table \ref{PolyN2_structure} lists the calculated structural parameters of a planar zigzag chain of all-{\it trans} polyazene.
It is predicted to display a bond-length alternation of 0.18--0.19\,\AA, obeying Peierls' theorem. The same B3LYP method reproduces
the observed N$_2$ bond length within 0.01\,\AA. The calculated N=N and N--N bond lengths of polyazene are much longer and slightly longer, respectively, than
the corresponding bond lengths of {\it trans}-tetrazene-(2), H$_2$N--N=N--NH$_2$, which is not conjugated (see also Sec.\ \ref{sec:NH}).
That the N--N bond in polyazene is not shorter, but instead longer than the nonconjugated N--N bond in tetrazene underscores the severity
of the lone-pair-lone-pair repulsion, destabilizing polyazene.

\begin{figure}
\includegraphics[width=1.0\columnwidth]{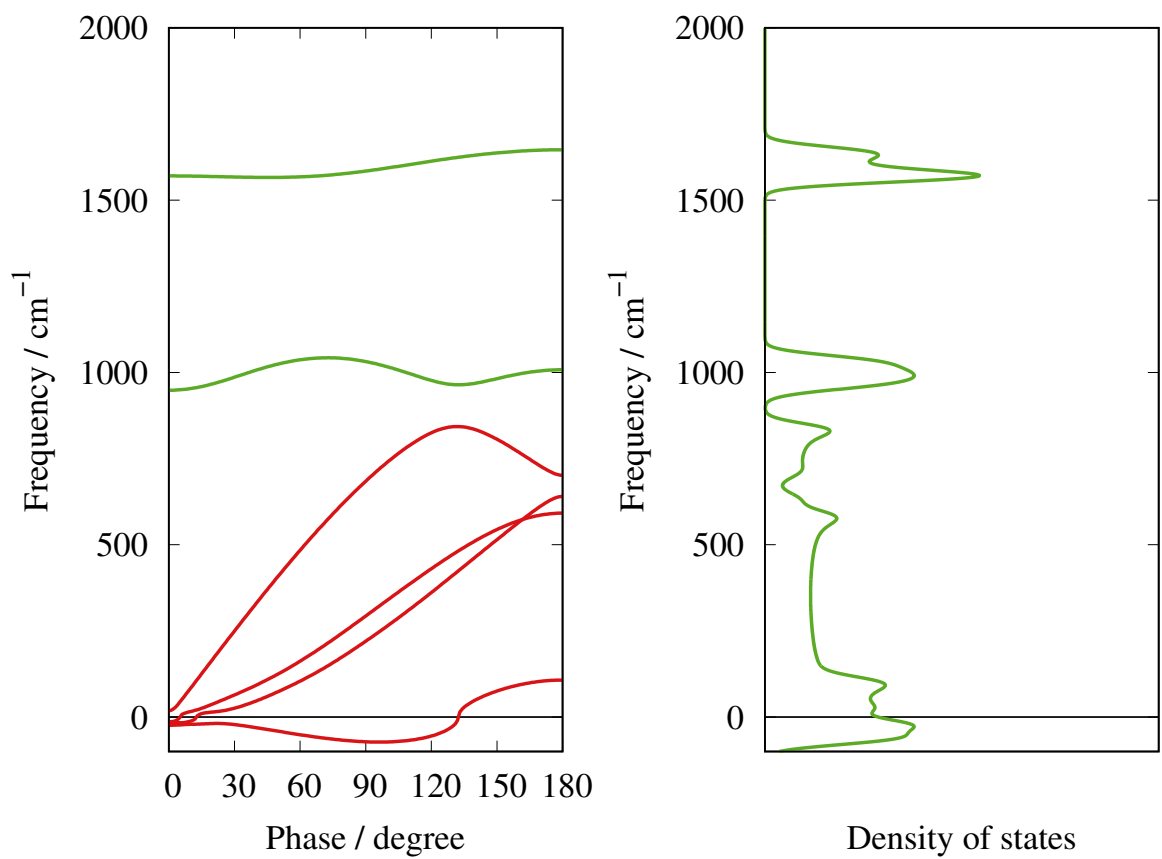}
\caption{\label{fig:phononN2}Phonon dispersion curves and DOS of (N$_2$)$_x$ computed by B3LYP/6-31G**. The DOS is convoluted with a Gaussian of a FWHM of 40 cm$^{-1}$.}
\end{figure}

\begin{table}
\caption{Vibrational frequencies (in cm$^{-1}$) of (N$_2$)$_x$.  \label{PolyN2_freqs}}
\begin{tabular}{lr}
Irrep.\textsuperscript{\textit{a}}; phase; activity& \multicolumn{1}{c}{B3LYP/6-31G**} \\ \hline
$A_g$; $\theta=0$; Raman & 948.3  \\ 
& 1570.2 \\
\end{tabular} \\ 
\textsuperscript{\textit{a}}{Isomorphic to the $C_{2h}$ point group.}
\end{table}

Figure \ref{fig:phononN2} plots the phonon dispersion curves and phonon DOS, which consist of two optical (green) and four acoustic (red) branches.
Table \ref{PolyN2_freqs} gives the frequencies of the optical phonons at the zone center ($\theta=0$),
which are predicted to be Raman active. The high-pressure (150-GPa) synthesis\cite{Goncharov2000} of a catenated form of nitrogen was accompanied by 
the disappearance of IR and Raman bands at 2300--2500 cm$^{-1}$ and the concomitant emergence of bands in the range of 200--600 cm$^{-1}$.  
While the disappearance of the N$\equiv$N stretching bands at 2300--2500 cm$^{-1}$ clearly indicates an extended chain formation and is not inconsistent with 
all-{\it trans} polyazene, both the number and frequencies of the lattice vibrations contradict such an interpretation, even after 
considering the immense pressure difference between theory and experiment. 

However, more alarming is the fact that the lowest acoustic branch has an imaginary frequency (reaching $73i$ cm$^{-1}$ at $\theta=96^\circ$) 
in a wide range of the phase angle $\theta$. 
This cannot be ascribed to an inevitable numerical error caused by the lack of strict translational and rotational
invariance as it occurs at phase angles far from $\theta=0$ (the calculation included force-constant matrices up to the fifth nearest neighbor N$_2$ units). 
It indicates that the planar zigzag structure is a saddle point on the potential energy surface, 
which is reminiscent of the planar hexagonal structure of hexazine being also a saddle point.\cite{Huber1982,Schleyer1992,Ha1992,Lauderdale1992,Tobita2001}

Hence, we sought other stable structures including the planar {\it cis}-transoid, planar {\it trans}-cisoid, 
helical {\it trans}-transoid,
corrugated {\it trans}-transoid, Ni-chelated {\it cis}-transoid, and Ni-chelated {\it trans}-cisoid structures. All of them tended to rearrange into separate 
N$_2$ fragments during the course
of geometry optimization. This by no means rules out the existence of local minima for these structures, but it strongly suggests that they are too unstable to exist 
as metastable species under ambient conditions. 

\begin{table}
\caption{Binding energy (in kcal/mol) of (N$_2$)$_x$.  \label{PolyN2_binding}}
\begin{tabular}{lr}
Method & Binding energy\textsuperscript{\textit{a}} \\ \hline
B3LYP/cc-pVDZ & $-58.4$ \\
MBPT(2)/cc-pVDZ & $-70.3$  \\ 
\end{tabular} \\ 
\textsuperscript{\textit{a}}{The energy difference between the unit-cell energy of (N$_2$)$_x$ and the energy of N$_2$ 
in their respective B3LYP/cc-pVDZ-optimized geometries.}
\end{table}

\begin{figure}
\includegraphics[width=1.0\columnwidth]{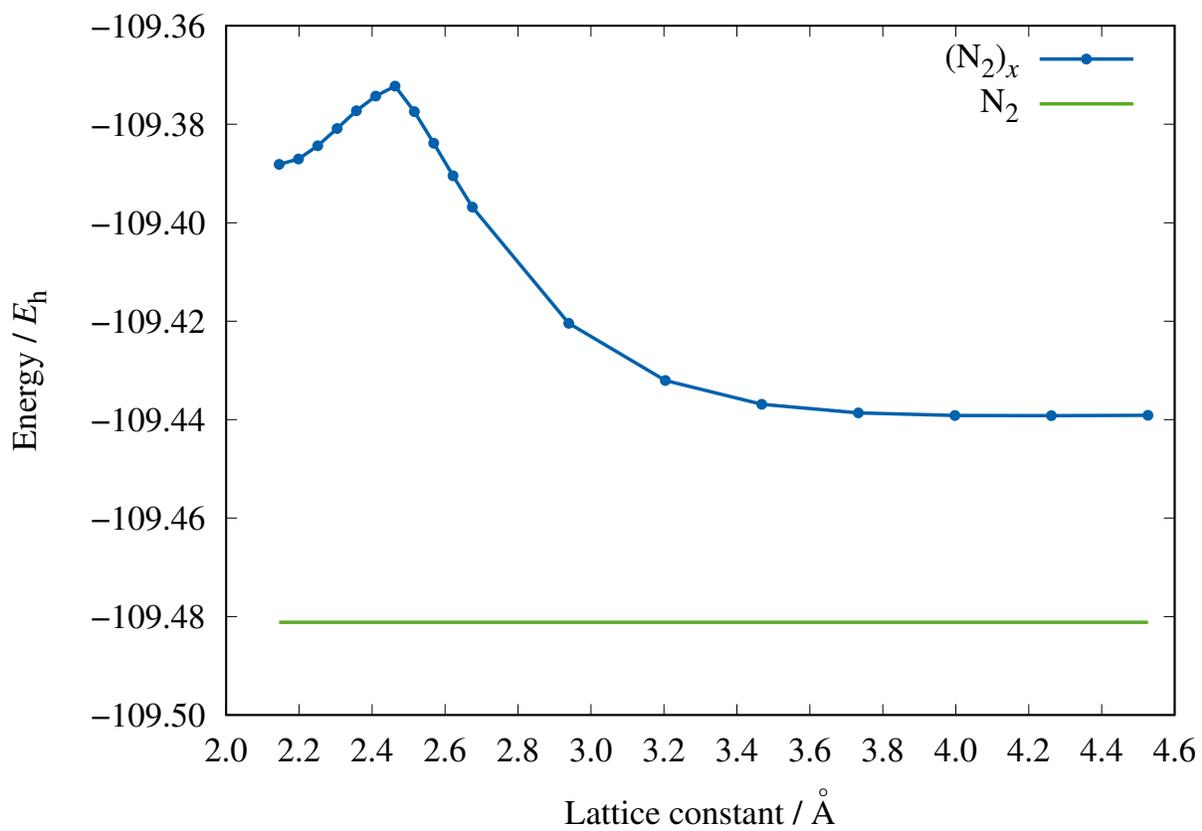}
\caption{\label{fig:N2diss}Unit-cell energy of (N$_2$)$_x$ as a function of the lattice constant computed by B3LYP/cc-pVDZ.}
\end{figure}

The large negative binding energies in Table \ref{PolyN2_binding} indicate that all-{\it trans} polyazene is unstable towards
dissociation into N$_2$. The unrelaxed dissociation reaction energy profile in Fig.\ \ref{fig:N2diss} exhibits a low activation barrier of {\it ca.}\ 12.5 kcal/mol
for the exothermic reaction. The barrier may not exist once the N$_2$ structure is relaxed along the reaction coordinate, 
but even if the barrier persists, it may be low enough
for the polymer to dissociate thermally. The low barrier may, in turn, be partly due to the similarity of the unit-cell and monomer structures. 
The energy profile is completely different from those of the stable organic polymers (Figs.\ \ref{fig:PEdissociation}, \ref{fig:CHdissociation}, and \ref{fig:PTFEdiss}).

We, therefore, conclude that all-{\it trans} polyazene is unlikely to exist under ambient conditions or even at low temperatures.


\begin{figure}
\includegraphics[width=1.0\columnwidth]{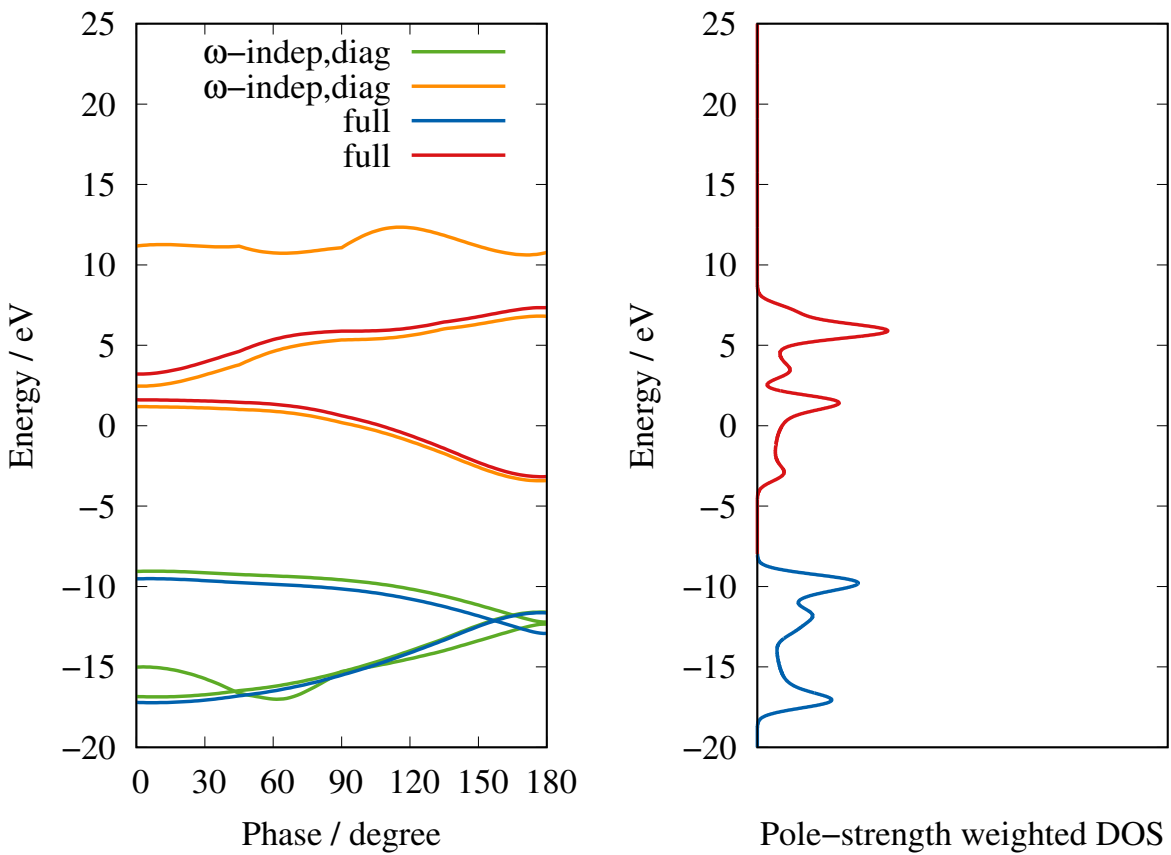}
\caption{\label{fig:N2bands} Electronic energy bands and DOS of (N$_2$)$_x$ computed by MBGF(2)/cc-pVDZ. 
The DOS is weighted by the pole strength of Eq.\ (\ref{HellmannFeynman}) and convoluted with a Gaussian of a FWHM of 1 eV.}
\end{figure}

For completeness, we document in Fig.\ \ref{fig:N2bands} the quasiparticle energy bands and electronic DOS calculated by the MBGF(2)/cc-pVDZ method.
The energy bands of polyazene differ strikingly from those of isoelectronic polyacetylene. 
In polyazene, the valence band is roughly parallel with the conduction band, and has the opposite dispersion from the valence band of polyacetylene. 
This is due to the fact that the valence band of polyazene consists of in-plane $\sigma$ orbitals accommodating lone pairs, 
whereas the valence and conduction bands of polyacetylene are $\pi$ and $\pi^*$ orbitals, respectively. 
The predicted UPS from polyazene has a peak at the ionization onset owing to the nearly flat valence band at its top.
The HF energy bands (not shown) are qualitatively the same as the MBGF(2) energy bands (which is a prerequisite for smooth energy-band interpolation in the mod-$n$ approximation\cite{Shimazaki2009,hirata_qp}). Quantitatively,
HF valence (conduction) bands are lower (higher) by a few electronvolts than the MBGF(2) counterparts, and the correlation effect has a greater impact on the $\pi$ and $\pi^*$ bands than on the $\sigma$ band.

%
%
%
%

\subsection{Isotactic polyazane, (NH)$_x$\label{sec:NH}}

\begin{figure}
\includegraphics[width=0.85\columnwidth]{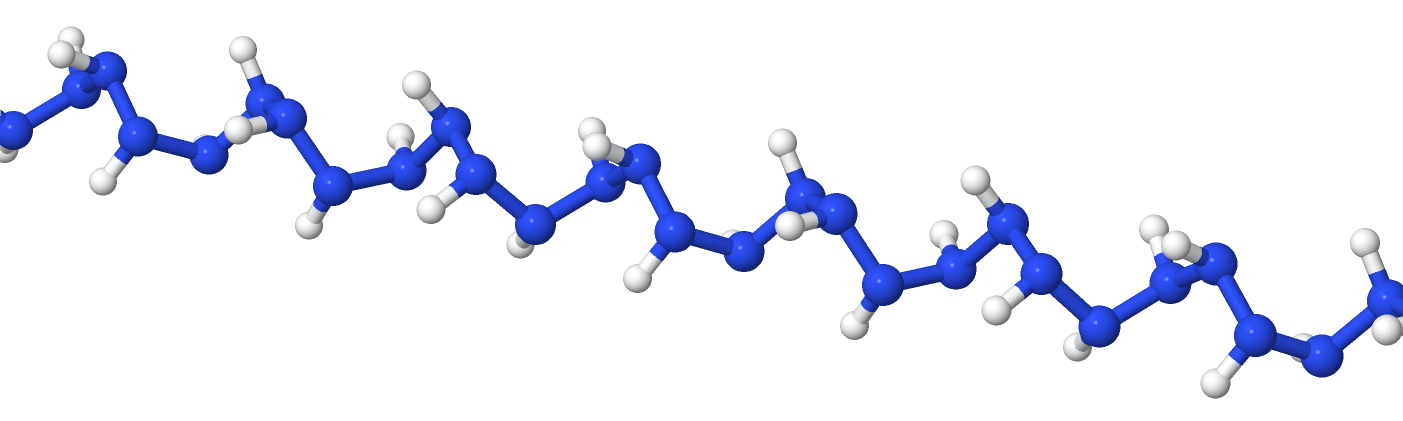}
\caption{Isotactic polyazane, (NH)$_x$. \label{fig:NH}}
\end{figure}

While polyazene (Sec.\ \ref{sec:N2}) may be deemed too unstable to exist under ambient conditions, stability of a catenated nitrogen chain 
can  be enhanced generally by introducing heteroatoms in its structure.\cite{Bartlett2000}
For instance, hexazine (see also Sec.\ \ref{sec:N2}) becomes stable in the planar hexagonal structure when one or more oxygen atoms are attached to nitrogens by dative bonds.\cite{Wilson2001}  
The repulsions between the lone pairs of adjacent nitrogen atoms, which are responsible for the instability, are alleviated by the lone pairs being consumed by these dative bonds.

In this subsection, we explore the stabilization of catenated nitrogen bonds by hydrogenation. We thus consider isotactic polyazane (NH)$_x$ (Fig.\ \ref{fig:NH}), which is helical and
isoelectronic with polyethylene (Sec.\ \ref{sec:CH2}). 
Intrachain hydrogen bonds might further stabilize it,\cite{Schlegel1993} although this has been disputed.\cite{Zhao1994} 
We performed B3LYP/cc-pVDZ and 6-31G** 
geometry optimization with $S=12$, $L=24$, and $K=48$ using a NH group as the rototranslational (physical) repeat unit. 
This was followed by phonon dispersion and phonon DOS calculations in the ninth nearest neighbor approximation.
We then carried out the MBGF(2)/cc-pVDZ calculations for 
the quasiparticle energy bands and electronic DOS in the frozen-core and mod-6 approximations.\cite{Shimazaki2009,hirata_qp} 

There are five stable isomers of N$_3$H$_3$.\cite{Magers1988} The most stable {\it trans}-triazene is already too unstable towards rapid acid-catalyzed decomposition to be isolated (see F\"{o}rstel {\it et al.}\cite{Forstel2016N3H3}
for time-of-flight mass-spectrometry detection of its isomers), despite the utility of synthetic reagents containing the triazene moiety.\cite{Kimball2002} Cyclo-N$_3$H$_3$ or triaziridine is isoelectronic with cyclopropane; cyclo-N$_3$H$_3$ is to polyazane as cyclopropane is to polyethylene. It is predicted\cite{Magers1988} to be the least stable isomer, yet has been isolated for structural determination by Kim {\it et al.}\cite{Kim1977}\ and by Heo {\it et al.},\cite{Heo2016} hinting at an unknown
factor stabilizing it. 
Kim {\it et al.}\cite{Kim1977}\ and Heo {\it et al.}\cite{Heo2016}\ also isolated triazane N$_3$H$_5$ in its complexes with Ag$^+$, whereas F\"{o}rstel {\it et al.}\cite{Forstel2015}\ detected it in the gas phase.
See Richard and Ball,\cite{Richard2008} and Dana {\it et al.}\cite{TriazaneTetrazane} for computation studies of triazane.

Tetrazene,\cite{ReviewN} H$_2$N--N=N--NH$_2$, first isolated by Wiberg {\it et al.}\cite{IsolationofTetrazene}\ in 1975, was shown\cite{Veith1982} to adopt a planar {\it trans} conformation. 
It is unstable and prone to thermolysis into NH$_4$ and N$_3$ or into N$_2$ and hydrazine (N$_2$H$_4$) above 0$^\circ$C, which may be taken as a more discouraging finding for polyazene (Sec.\ \ref{sec:N2})
 than for polyazane. 
Hydrazine itself is also thermodynamically unstable, but it is kinetically stable, and because of that, it is 
an excellent rocket fuel with the second highest thrust per mass per time (the highest is liquid hydrogen).\cite{TriazaneTetrazane} 
Cyclo-(NH)$_4$ (tetrazetidine) and N$_4$H$_6$ (tetrazane) are structurally more salient to polyazane and have been studied computationally.\cite{RitterN4H4_1989,tetrazane_calc,cyclo-tetrazane_calc}
Ritter {\it et al.}\cite{RitterN4H4_1989}\ concluded that despite its instability towards diazene dissociation, ``the tetrazetidine seems to be trapped kinetically by rather high energy barriers.''
Larger (NH)$_x$ rings and longer H(NH)$_x$H chains were studied by Zhao and Gimarc,\cite{Zhao1994} focusing on their strain energies. 
Experimentally, triazane N$_3$H$_5$ and terazane N$_4$H$_6$ were detected in a microwave discharge of hydazine.\cite{Fujii2002} 
A molecule containing the triazane moiety was also formed in an environment of astrophysical relevance.\cite{Forstel_astro2016} 
These are supportive of the metastable existence of polyazane. 


\begin{table*}
\begin{scriptsize}
\caption{Structural parameters of (NH)$_x$ and N$_2$H$_2$.  \label{PolyNH_structure}}
\begin{tabular}{lcccccccccc}
& \multicolumn{7}{c}{(NH)$_x$} &\multicolumn{3}{c}{{\it trans}-N$_2$H$_2$\textsuperscript{\textit{a}}}  \\ \cline{2-8}\cline{9-11} 
Method & $r$(NN) & $r$(NH) & $a$(NNN) & $a$(HNN) & $d$(NNNN) & $d$(HNNH) & $\varphi$\textsuperscript{\textit{b}} & $r$(NN)  & $r$(NH) & $a$(HNN)  \\ \hline
B3LYP/6-31G** & 1.431\,\AA & 1.025\,\AA & $109.9^\circ$ & $105.9^\circ$ & $73.7^\circ$ & $75.5^\circ$ & $98.1^\circ$& 1.246\,\AA & 1.040\,\AA & $106.1^\circ$ \\ 
B3LYP/cc-pVDZ & 1.429\,\AA & 1.029\,\AA & $110.0^\circ$ & $105.8^\circ$ & $73.7^\circ$ & $75.5^\circ$ & $98.1^\circ$& 1.245\,\AA & 1.045\,\AA & $105.8^\circ$ \\ 
Observed\textsuperscript{\textit{a}} &&&&&&& ($98.2^\circ$)\textsuperscript{\textit{c}} & 1.252\,\AA & 1.028\,\AA & $106.9^\circ$  \\
\end{tabular} \\ 
\textsuperscript{\textit{a}}{The {\it trans}-diazene structure from Carlotti {\it et al.}\cite{Carlotti}}
\textsuperscript{\textit{b}}{The helical angle [Eqs.\ (\ref{X})--(\ref{Z})].}
\textsuperscript{\textit{c}}{The helical angle for a 11/3 helix (not observed) is given in parentheses.}
\end{scriptsize}
\end{table*}

Table \ref{PolyNH_structure} compiles the predicted structural parameters of isotactic polyazane. Given the good agreement between the calculated and observed structures 
of its monomer ({\it trans}-diazene), we expect them to be within a few hundredths of one \AA ngstrom and a few degrees of the exact values. 
The optimized helical angle is 98.1$^\circ$. A commensurable helix that has a similar helical angle is 11/3 ($\varphi =98.2^\circ$). 
The N--N bond length in cyclo-N$_3$H$_3$ was measured to be 1.49\,\AA.\cite{Kim1977} That this is slightly longer than the predicted N--N bond length (1.43\,\AA) in polyazane
is reasonable given the presence of ring strain in cyclo-N$_3$H$_3$. The observed N--N bond length (1.46\,\AA) of hydrazine\cite{Collin1951} is closer to the calculated N--N bond length of polyazane.

The NNNN dihedral angle in polyazane is predicted to be 73.7$^\circ$, which is less than half of
the CCCC dihedral angle of 180$^\circ$ in isoelectronic polyethylene. 
It is smaller still than the corresponding angle of 166$^\circ$ in polytetrafluoroethylene, which suffers from the fluorine-fluorine repulsion.
Zhao and Gimarc\cite{Zhao1994} also observed the dihedral angle of  about 90$^\circ$ subtended by adjacent lone pairs in the (NH)$_x$ rings or long H(NH)$_x$H chains.
They invoked the {\it gauche} effect\cite{Mizushima_book,Mizushima1954,Gaucheeffect} to rationalize the conformation qualitatively.
The predicted structure furthermore suggests that there is no extra stabilization
due to intrachain hydrogen bonds between adjacent NH bonds because they are not oriented appropriately. 

\begin{figure}
\includegraphics[width=1.0\columnwidth]{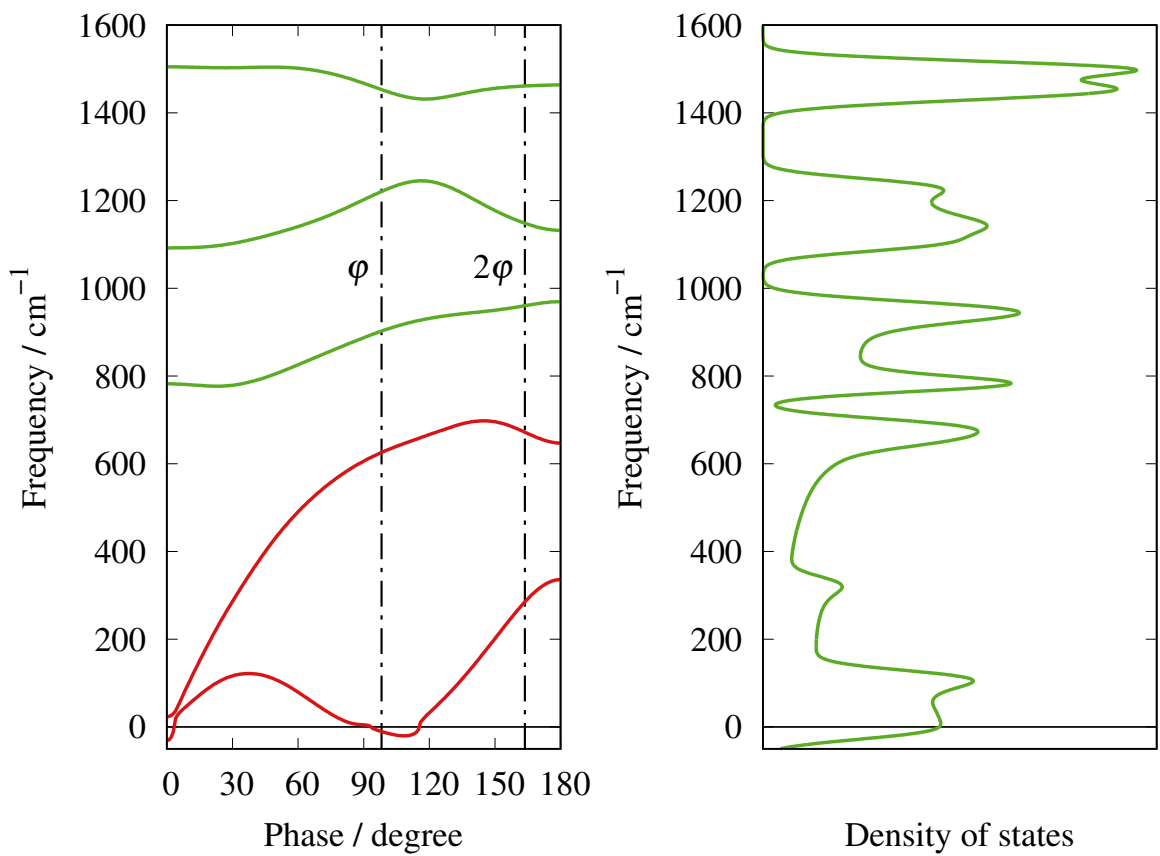}
\caption{\label{fig:NHdisp} Phonon dispersion curves and DOS of (NH)$_x$ computed by B3LYP/6-31G**. The DOS is convoluted with a Gaussian of a FWHM of 40 cm$^{-1}$.}
\end{figure}

\begin{table}
\caption{Vibrational frequencies (in cm$^{-1}$) of (NH)$_x$.  \label{PolyNH_freqs}}
\begin{tabular}{lr}
Irrep.\textsuperscript{\textit{a}}; phase; activity& \multicolumn{1}{c}{B3LYP/6-31G**} \\ \hline
$A$; $\theta=0$; IR, Raman & 3425.7  \\
& 1504.9 \\
& 1092.1 \\
& 782.2 \\
$E_1$; $\theta=\varphi$; IR, Raman & 3419.2 \\
& 1453.1 \\
& 1220.7 \\
& 902.4 \\
& 625.7 \\
$E_2$; $\theta=2\varphi$; Raman & 3420.7 \\
& 1461.2 \\
& 1148.3 \\
& 960.3 \\
& 671.8 \\
& 285.1 \\
\end{tabular} \\ 
\textsuperscript{\textit{a}}{Isomorphic to the $C_{11}$ point group. $\varphi$ is the helical angle.}
\end{table}

Figure \ref{fig:NHdisp} plots the phonon dispersion curves and phonon DOS. They consist of four optical (green) and two acoustic (red) branches.
The N--H stretching optical modes above 3000 cm$^{-1}$ are not shown in this figure. 
Since this is a helix, two acoustic (longitudinal and spinning) modes have zero frequencies at $\theta=0$ and one acoustic (transverse) mode also has null energy
at $\theta=\varphi$. This expected behavior is reproduced with the maximum deviation from zero frequencies being $31i$ cm$^{-1}$. 
These imaginary-frequency modes are observed only at $\theta\approx 0$ and $\varphi$ and their magnitudes are small, indicating that they are 
numerical errors due to the distance-based truncation of the force-constant matrices. The structure given in Table \ref{PolyNH_structure} is, therefore, a local minimum, 
and isotactic polyazane can exist at zero temperature and pressure (ignoring the zero-point vibrational energy). 
The predicted frequencies of the IR and Raman bands in Table \ref{PolyNH_freqs}
are expected to overestimate the (unavailable) experimental results by a few percent, judging from the results for the organic polymers (Secs.\ \ref{sec:CH2}--\ref{sec:PTFE}). 

\begin{table}
\caption{Binding energy (in kcal/mol) of (N$_2$H$_2$)$_x$.  \label{PolyNH_binding}}
\begin{tabular}{lrrr}
Method & {\it trans}-N$_2$H$_2$\textsuperscript{\textit{a}} & {\it cis}-N$_2$H$_2$\textsuperscript{\textit{b}} & N$_2$+NH$_3$\textsuperscript{\textit{c}} \\ \hline
B3LYP/cc-pVDZ & 7.5  & 12.4  & $-42.1$  \\ 
MBPT(2)/cc-pVDZ & 7.7 & 13.2 & $-50.1$ \\ 
\end{tabular} \\ 
\textsuperscript{\textit{a}}{The energy difference between  the unit-cell energy of (N$_2$H$_2$)$_x$ and the energy of {\it trans}-N$_2$H$_2$ 
in their respective B3LYP/cc-pVDZ-optimized geometries.}
\textsuperscript{\textit{b}}{The energy difference between  the unit-cell energy of (N$_2$H$_2$)$_x$ and the energy of {\it cis}-N$_2$H$_2$ 
in their respective B3LYP/cc-pVDZ-optimized geometries.}
\textsuperscript{\textit{c}}{The energy difference between  the unit-cell energy of (N$_2$H$_2$)$_x$ and two thirds of the energy of N$_2$+NH$_3$ 
in their respective B3LYP/cc-pVDZ-optimized geometries.}
\end{table}

\begin{figure}
\includegraphics[width=1.0\columnwidth]{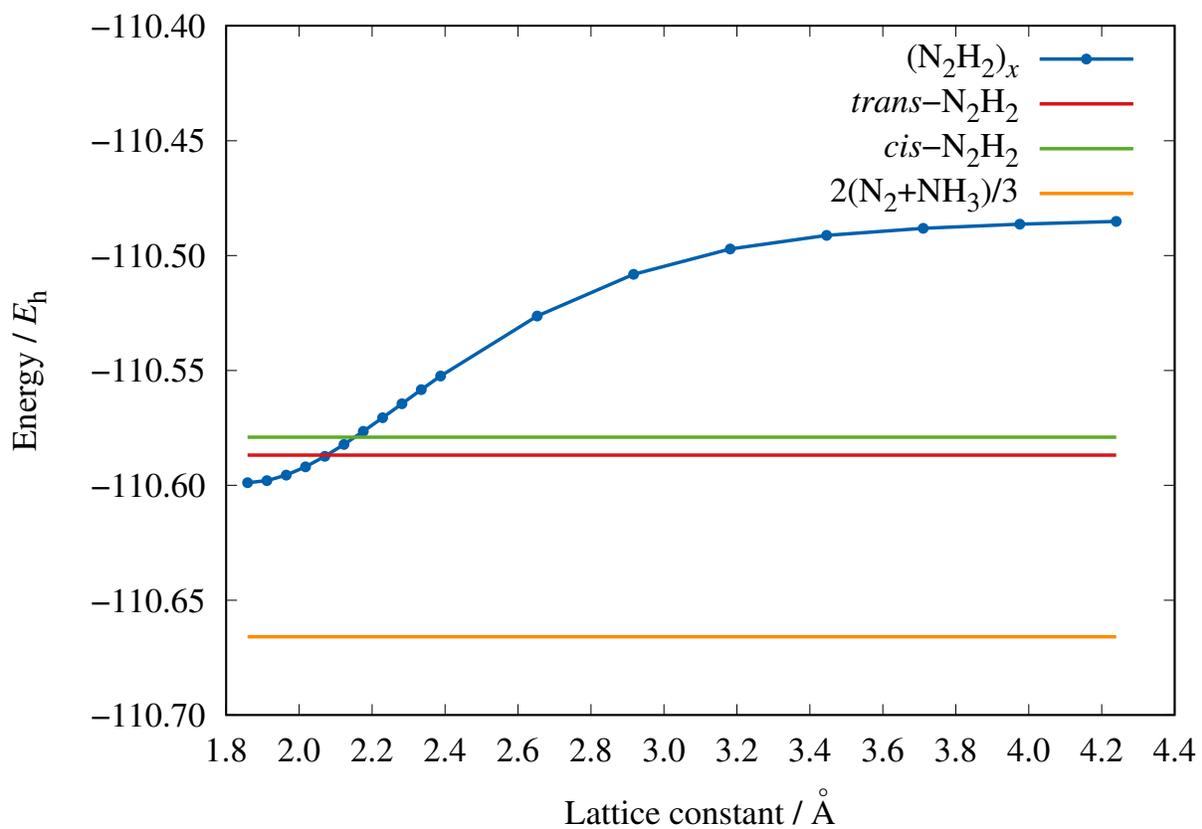}
\caption{\label{fig:NHdissociation}Unit-cell energy of (N$_2$H$_2$)$_x$ as a function of the lattice constant computed by B3LYP/cc-pVDZ.}
\end{figure}

Table \ref{PolyNH_binding} summarizes the binding energy of polyazane relative to its monomer (diazene) or to N$_2$ and NH$_3$. 
The diazene molecule exists in either the {\it trans} or {\it cis} isomers with the former being slightly more stable. 
The positive binding energies mean that polyazane is stable against dissociation into diazenes. In contrast, 
the dissociation into N$_2$ and NH$_3$ is highly exothermic. 
This underscores the disproportionate strength of the N$\equiv$N bond as compared with the N=N and N--N bonds. 

Figure \ref{fig:NHdissociation} plots the unrelaxed reaction energy profile of polyazane's dissociation into monomers. 
The unit-cell structure has the {\it gauche} conformation\cite{Mizushima_book,Mizushima1954} with the predicted HNNH dihedral angle of 75.5$^\circ$. 
Therefore, an considerable amount of internal rotation must take place during the dissociation into either {\it trans}- or {\it cis}-diazenes, causing 
the energy profile to rise to a much higher level than that of either diazene isomer, adding to the kinetic stability of polyazane. 
The dissociation into N$_2$ and NH$_3$ should involve even more significant rearrangements of atoms, including N--H bond breaking and formation,
likely creating a steeper activation barrier. We, therefore, expect polyazane to be kinetically stable, while unstable toward thermolysis. Theoretically, it 
can exist as a metastable species at low temperatures if not under ambient conditions. 


\begin{figure}
\includegraphics[width=1.0\columnwidth]{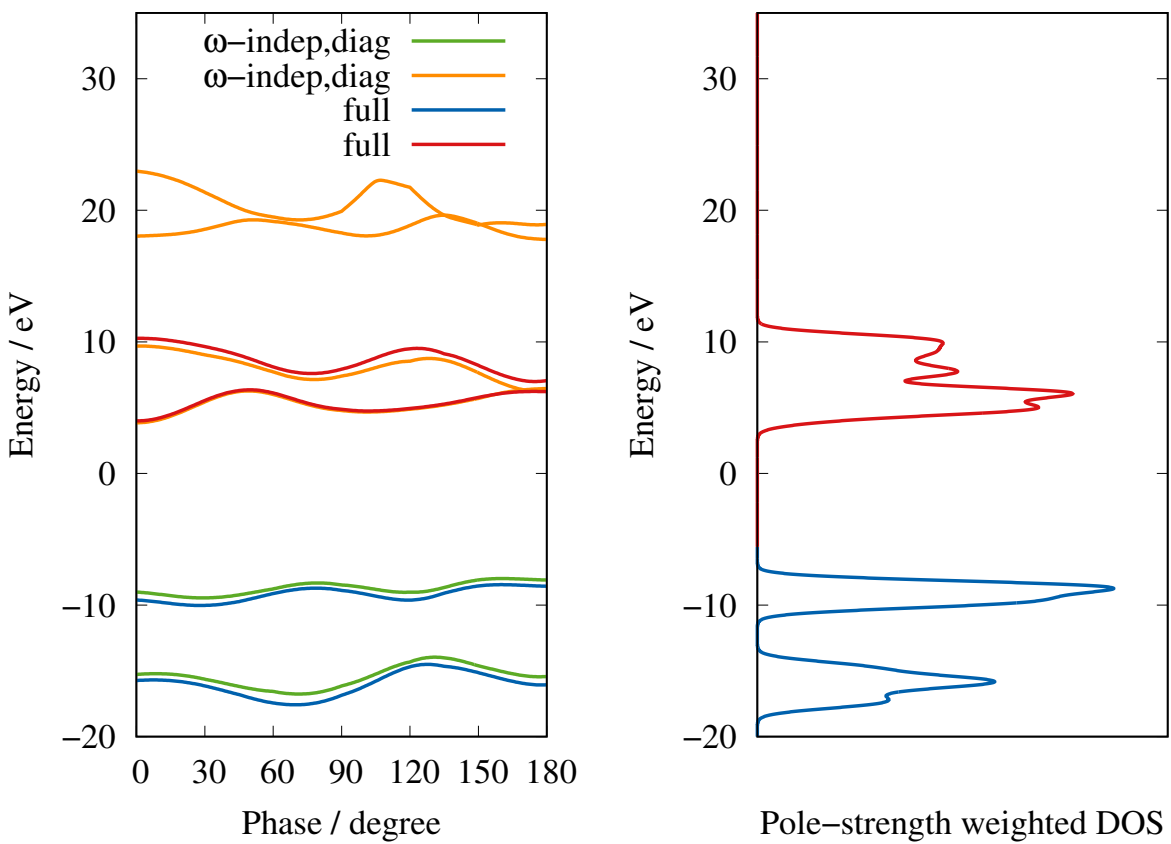}
\caption{\label{fig:NHbands}Electronic energy bands and DOS of (NH)$_x$ computed by MBGF(2)/cc-pVDZ. 
The DOS is weighted by the pole strength of Eq.\ (\ref{diagpole}) and convoluted with a Gaussian of a FWHM of 1 eV.}
\end{figure}

The calculated energy bands and electronic DOS are shown in Fig.\ \ref{fig:NHbands}. The top two valence bands have small dispersions and, therefore, the UPS 
of polyazane is predicted to have two sharp peaks in the range of 0--20 eV. The valence band edge occurs at $-8.5$ eV at $\theta = 160^\circ$, which should be 
 within a few tenths of an electronvolt of the exact value. Owing to this relatively large ionization energy, polyazane is expected to be stable in the air, but 
 will likely undergo an acid-catalyzed decomposition in solutions.

%
%
%
%

\subsection{Isotactic polyfluoroazane, (NF)$_x$\label{sec:NF}}

\begin{figure}
\includegraphics[width=0.85\columnwidth]{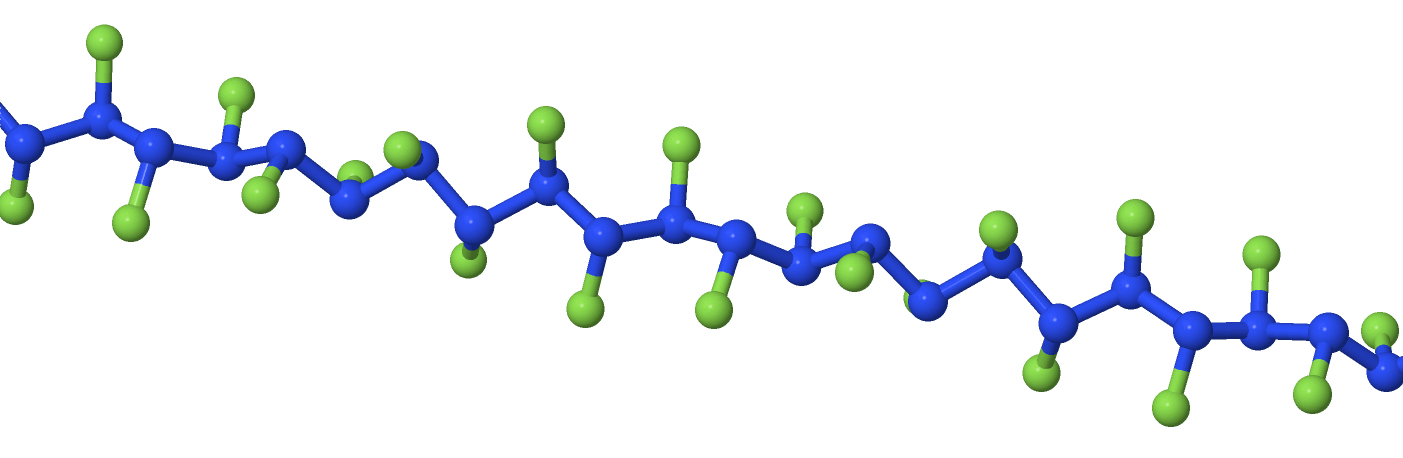}
\caption{Isotactic polyfluoroazane, (NF)$_x$. \label{fig:NF}}
\end{figure}

In addition to hydrogenation (Sec.\ \ref{sec:NH}), we consider fluorination to stabilize the catenated nitrogen bonds,\cite{Kirchmeier1992} leading to
isotactic polyfluoroazane (NF)$_x$ (Fig.\ \ref{fig:NF}). It is a nitrogen analog of polytetrafluoroethylene (Sec.\ \ref{sec:PTFE}) and adopts a helical conformation. 
The logic underlying the expected stability of fluorinated nitrogen chains is concisely summarized by Krumm {\it et al.},\cite{Krumm1995}
who synthesized hexakis(trifluoromethyl)tetrazane N$_4$(CF$_3$)$_6$:\ ``While no element can compete with carbon in the number of contiguous atoms in saturated chemical compounds, stable straight-chain species composed of other elements are known, especially if fluorine atoms, fluorinated groups, or other electronegative species are present in the molecule, for example CF$_3$(O)$_n$CF$_3$ ($n = 1-3$) and O$_n$F$_2$ ($n = 1, 2, 4$). Fluorinated N$_n$ compounds do exist and do exhibit surprising hydrolytic, thermal, and chemical stabilities. In contrast, the hydrogen-substituted analogues such as diazane, triazane, and tetrazane are increasingly unstable with increasing chain length.''\cite{Krumm1995} Employing the same strategy, Criton {\it et al.}\cite{Criton_cispolyN2021}\ synthesized helical polyazanes up to pentazane substituted by electron-withdrawing carbamate groups. Other substituted triazane\cite{Egger1983,Kanzian2010} and tetrazane\cite{Hope1967,Pirkle1978,Martin2017} were also reported.

In this study, we applied B3LYP/cc-pVDZ and 6-31G** for geometry optimization, phonon dispersion and phonon DOS calculations with $S=8$, $L=12$, and $K=48$ using a NF group as the rototranslational (physical) repeat unit.
The subsequent MBGF(2)/cc-pVDZ calculation for the quasiparticle energy bands and electronic DOS invoked the frozen-core and mod-6 approximations.
The force-constant matrices extending to the ninth nearest neighbor NF groups were included in the normal-mode analysis.


\begin{table*}
\begin{scriptsize}
\caption{Structural parameters of (NF)$_x$ and N$_2$F$_2$.  \label{PolyNF_structure}}
\begin{tabular}{lcccccccccc}
& \multicolumn{7}{c}{(NF)$_x$} &\multicolumn{3}{c}{{\it cis}-N$_2$F$_2$}  \\ \cline{2-8}\cline{9-11} 
Method & $r$(NN) & $r$(NF) & $a$(NNN) & $a$(FNN) & $d$(NNNN) & $d$(FNNF) & $\varphi$\textsuperscript{\textit{b}} & $r$(NN)  & $r$(NF) & $a$(FNN)   \\ \hline
B3LYP/6-31G** & 1.509\,\AA & 1.396\,\AA & $102.0^\circ$ & $103.0^\circ$ & $153.7^\circ$ & $156.6^\circ$ & $159.6^\circ$ & 1.220\,\AA & 1.392\,\AA & $114.3^\circ$  \\ 
B3LYP/cc-pVDZ & 1.510\,\AA & 1.395\,\AA & $102.0^\circ$ & $103.4^\circ$ & $152.0^\circ$ & $155.3^\circ$ & $158.3^\circ$ & 1.219\,\AA & 1.392\,\AA & $114.6^\circ$  \\
Observed\textsuperscript{\textit{a}} &&&&&&& ($160.0^\circ$)\textsuperscript{\textit{c}} &  1.214\,\AA & 1.384\,\AA & $114.5^\circ$
\end{tabular} \\ 
\textsuperscript{\textit{a}}{The {\it cis}-difluorodiazene structure from Hellwege and Hellwege.\cite{Hellwege}}
\textsuperscript{\textit{b}}{The helical angle [Eqs.\ (\ref{X})--(\ref{Z})].}
\textsuperscript{\textit{c}}{The helical angle for a 9/4 helix (not observed) is given in parentheses.}
\end{scriptsize}
\end{table*}

The optimized equilibrium structure of isotactic polyfluoroazane is given in Table \ref{PolyNF_structure}. 
The calculated helical angle is 158--160$^\circ$. A commensurable helix with a similar helical angle (160$^\circ$) is 9/4. 
It may be recalled that polytetrafluoroethylene also has a similar helical angle of 166$^\circ$. 
As before, we can safely expect the predicted structural parameters to be accurate to within
a few hundredths of one \AA ngstrom and a few degrees. However, the observed N--N bond length of 
hexakis(trifluoromethyl)tetrazane, N$_4$(CF$_3$)$_6$, is 1.379\,\AA,\cite{Krumm1995} which is considerably shorter than the calculated N--N bond length (1.51\,\AA) of polyfluoroazane. This is in spite of the bulkier CF$_3$ substituent on every nitrogen in the former. Thus, 
these two sets of results may be considered incompatible with each other. Note that the observed N--N bond lengths in cyclo-N$_3$H$_3$\cite{Kim1977} and hydrazine\cite{Collin1951} are 1.49\,\AA\ and 1.46\,\AA, respectively, which seem more in line with our calculated N--N bond length (1.51\,\AA) of polyfluoroazane. 

The calculated NNNN and FNNF dihedral angles fall in the range of 152--157$^\circ$. The same dihedral angle is subtended by the 
adjacent lone pairs. This is consistent with the  rationalization in terms of the {\it gauche} effect,\cite{Mizushima_book,Mizushima1954,Gaucheeffect} although such a
qualitative explanation may be
less pertinent in light of the present quantitative calculations. The measured NNNN dihedral angle in hexakis(trifluoromethyl)tetrazane, N$_4$(CF$_3$)$_6$, is 95.2$^\circ$,\cite{Krumm1995} which is considerably smaller than the corresponding predicted value (152--154$^\circ$) in polyfluoroazane. The cause of this disparity is also unknown. 

\begin{figure}
\includegraphics[width=1.0\columnwidth]{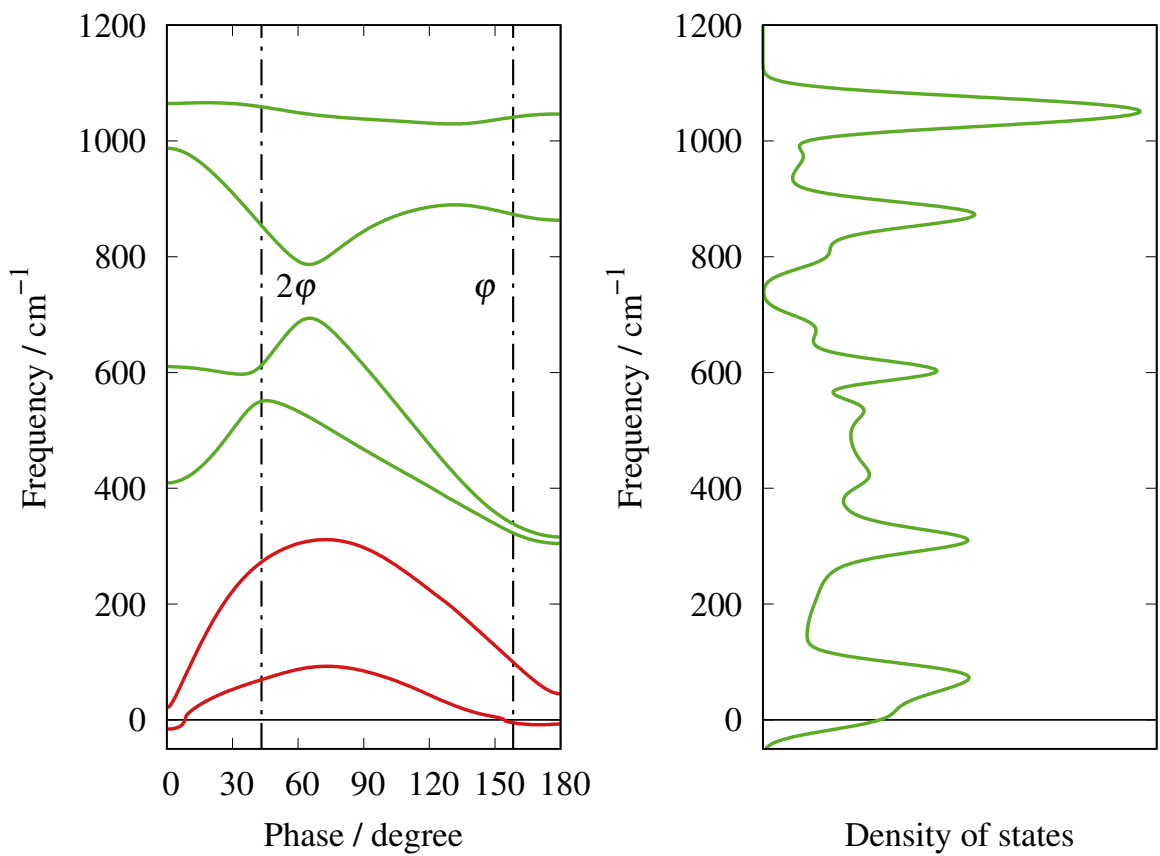}
\caption{\label{fig:NFdispersion}Phonon dispersion curves and DOS of (NF)$_x$ computed by B3LYP/6-31G**. The DOS is convoluted with a Gaussian of a FWHM of 40 cm$^{-1}$.}
\end{figure}

\begin{table}
\caption{Vibrational frequencies (in cm$^{-1}$) of (NF)$_x$.  \label{PolyNF_freqs}}
\begin{tabular}{lr}
Irrep.\textsuperscript{\textit{a}}; phase; activity& \multicolumn{1}{c}{B3LYP/6-31G**} \\ \hline
$A$; $\theta=0$; IR, Raman &  1064.5 \\
&  987.4 \\
&  610.2 \\
&  409.3 \\
$E_1$; $\theta=\varphi$; IR, Raman & 1041.0 \\
& 873.1  \\
& 338.6 \\
& 322.8 \\
& 99.6 \\
$E_2$; $\theta=2\varphi$; Raman & 1059.0 \\
& 854.0 \\
& 612.5 \\
& 550.3 \\
& 272.6 \\
& 69.5 \\
\end{tabular} \\ 
\textsuperscript{\textit{a}}{Isomorphic to the $C_{9}$ point group. $\varphi$ is the helical angle.}
\end{table}

Figure \ref{fig:NFdispersion} presents the calculated phonon dispersion curves and phonon DOS.  There are four optical (green) and two acoustic (red) branches.
The frequencies of the latter should become zero at $\theta=0$ (longitudinal and spinning) and $\theta=\varphi$ (transverse), which is
borne out numerically with errors of only $16i$, $21$, and $5i$ cm$^{-1}$. Clearly, they are caused by 
the distance-based truncation of force-constant matrices. Imaginary-frequency modes occurring in the range of $160^\circ \leq \theta \leq180^\circ$
are due to the zero-frequency acoustic mode at $\theta=\varphi$ and the periodic nature of the phonon dispersion, and do not 
imply an instability of the structure. In fact, the same numerical errors with similar magnitude
are detected in the phonon dispersion of polytetrafluoroethylene (Fig.\ \ref{fig:PTFE}). Therefore, the structure of Table \ref{PolyNF_structure} is a local minimum, and polyfluoroazane exists at zero temperature in a vacuum, ignoring the zero-point vibrations.

The frequencies of the IR- and/or Raman-active modes,
occurring at $\theta=0$, $\varphi$, and $2\varphi$, are compiled in Table \ref{PolyNF_freqs}.
They should be reliable with a tendency of systematic overestimation by a few percent.

\begin{table}
\caption{Binding energy (in kcal/mol) of (N$_2$F$_2$)$_x$.  \label{PolyNF_binding}}
\begin{tabular}{lrrr}
Method & {\it trans}-N$_2$F$_2$\textsuperscript{\textit{a}} & {\it cis}-N$_2$F$_2$\textsuperscript{\textit{b}} & N$_2$+NF$_3$\textsuperscript{\textit{c}} \\ \hline
B3LYP/cc-pVDZ & $-30.1$ & $-33.0$ & $-64.5$ \\ 
MBPT(2)/cc-pVDZ & $-31.4$ & $-34.1$ & $-73.8$ \\
\end{tabular} \\ 
\textsuperscript{\textit{a}}{The energy difference between  the unit-cell energy of (N$_2$F$_2$)$_x$ and the energy of {\it trans}-N$_2$F$_2$ 
in their respective B3LYP/cc-pVDZ-optimized geometries.}
\textsuperscript{\textit{b}}{The energy difference between  the unit-cell energy of (N$_2$F$_2$)$_x$ and the energy of {\it cis}-N$_2$F$_2$ 
in their respective B3LYP/cc-pVDZ-optimized geometries.}
\textsuperscript{\textit{c}}{The energy difference between  the unit-cell energy of (N$_2$F$_2$)$_x$ and two thirds of the energy of N$_2$+NF$_3$ 
in their respective B3LYP/cc-pVDZ-optimized geometries.}
\end{table}

\begin{figure}
\includegraphics[width=1.0\columnwidth]{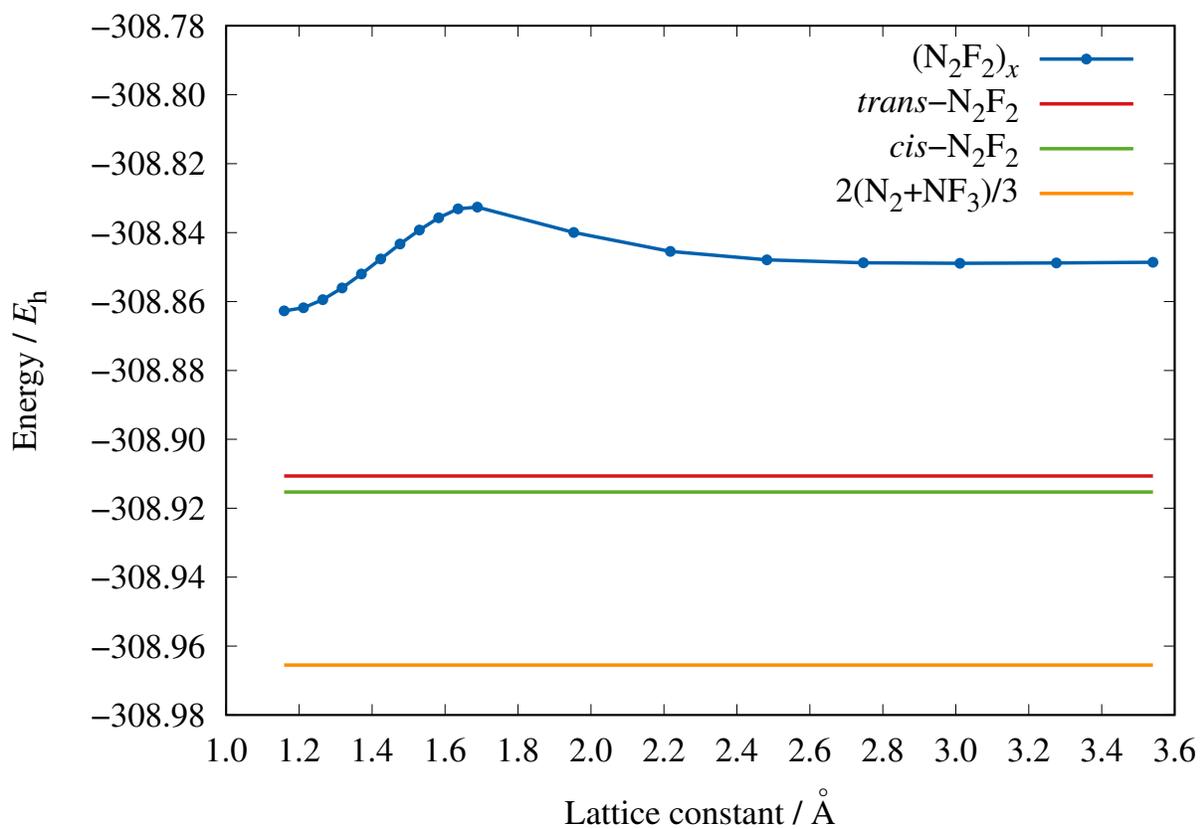}
\caption{\label{fig:NFdissociation}Unit-cell energy of (N$_2$F$_2$)$_x$ as a function of the lattice constant computed by B3LYP/cc-pVDZ.}
\end{figure}

Table \ref{PolyNF_binding} compiles the binding energies of polyfluoroazane relative to dissociation into its monomers ({\it trans}- or {\it cis}-difluorodiazenes)
or into dinitrogen and nitrogen trifluoride. 
Note that, unlike diazenes, the {\it cis} isomer of difluorodiazene is more stable than the {\it trans} isomer by 1.4 kcal/mol at 298 K.\cite{ChristieDixon2010} 
The negative values across the board in Table \ref{PolyNF_binding} mean that polyfluoroazane is thermodynamically unstable and 
seems even less stable than polyazane (Sec.\ \ref{sec:NH}). Contrary 
to the expectation that fluorination will stabilize the catenated nitrogen bonds,\cite{Krumm1995}  it seems to have the opposite effect. This may be because perfluorination stabilizes the short nitrogen chains 
to a greater degree than it does longer ones. 

The unrelaxed energy profile in Fig.\ \ref{fig:NFdissociation} corresponds to the dissociation into highly stretched, slightly twisted {\it trans}-difluorodiazenes. 
Unlike the corresponding energy profile for polyazane (Fig.\ \ref{fig:NHdissociation}), it is not a monotonically increasing function of the lattice constant, but
displays a low activation barrier of {\it ca.}\ 20 kcal/mol, which will be made even lower by geometry optimization of the transition state. Polyfluoroazane is, therefore,
expected to possess limited kinetic stability if any, and it may readily undergo thermolysis. Therefore, polyfluoroazane is unlikely to exist under ambient conditions, contrary
to the theoretical expectation and experimental findings for short catenated nitrogen molecules.\cite{Krumm1995} 


\begin{figure}
\includegraphics[width=1.0\columnwidth]{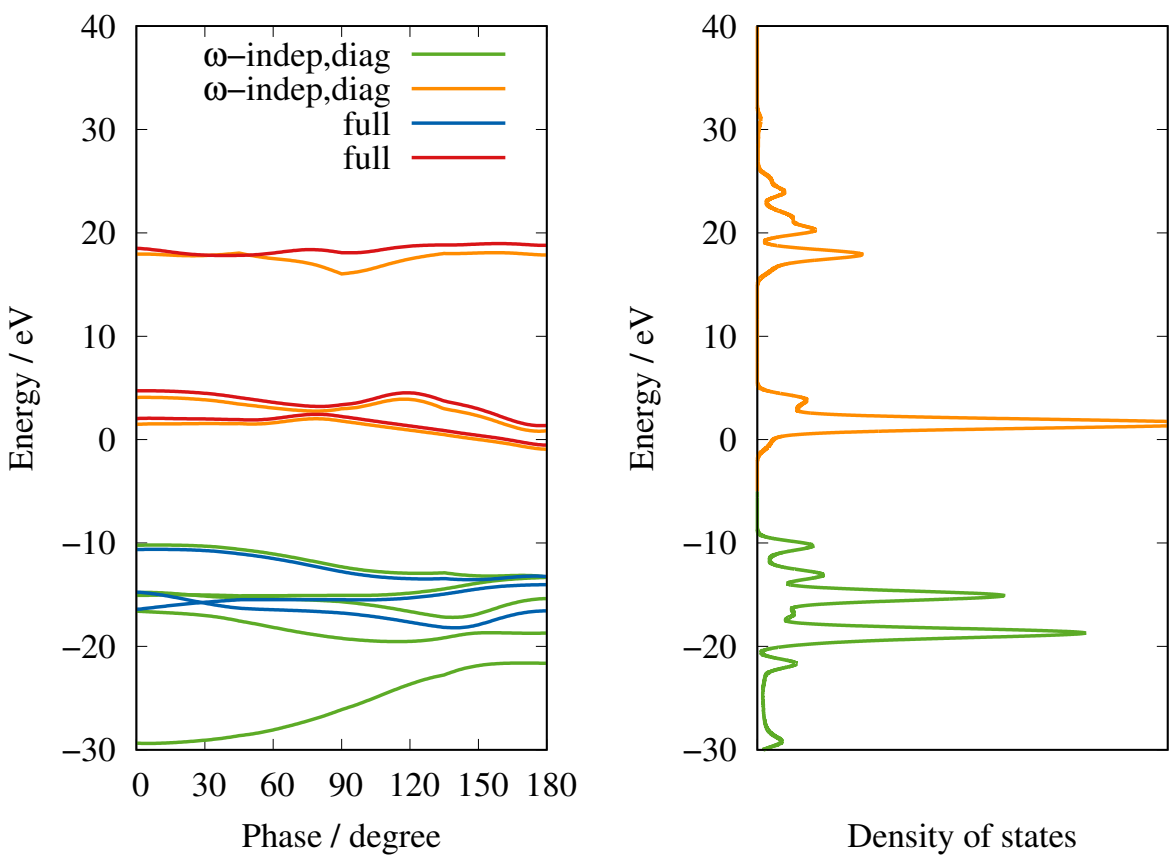}
\caption{\label{fig:NFbands}Electronic energy bands and DOS of (NF)$_x$ computed by MBGF(2)/cc-pVDZ. The DOS is convoluted with a Gaussian of a FWHM of 1 eV.}
\end{figure}

Figure \ref{fig:NFbands} shows the quasiparticle energy bands and electronic DOS of polyfluoroazane computed by MBGF(2)/cc-pVDZ. 
The convoluted DOS due to valence  bands
is expected to be an accurate prediction of UPS with errors in peak positions being no more than a few tenths of an electronvolt. 

%
%
%
%

\subsection{Polyoxane, (O)$_x$\label{sec:O}}

\begin{figure}
\includegraphics[width=0.85\columnwidth]{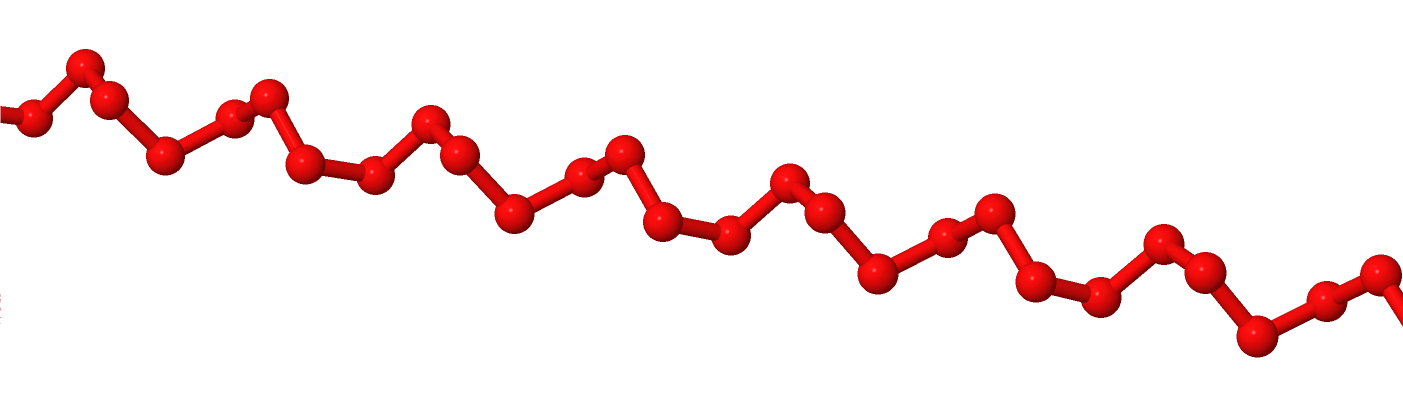}
\caption{Polyoxane, (O)$_x$.\label{fig:O}}
\end{figure}

A helical polymeric form of oxygen, polyoxane (Fig.\ \ref{fig:O}), may be implied by its stable sulfur homologue, helical polysulfane.\cite{SpringborgSulfur1986} 
Computational and experimental searches 
for long catenated forms of oxygen have been carried out using both the chemistry and physics approaches. 

In the chemistry approach, oxygen allotropes of various structures  were computationally characterized. They include O$_4$,\cite{Seidl1988,Seidl1992,Peterka1999,Politzer2000,RamirezSolis2010,Gadzhiev2013} O$_6$,\cite{Xie_O6_1992,Politzer2000,Gadzhiev2013} O$_{8}$,\cite{Politzer2000} and O$_{12}$.\cite{Politzer2000}
Among these, cyclo-O$_6$ is found to be stable in the cyclohexane-like chair conformation.\cite{Gadzhiev2013} This may be auspicious for polyoxane since cyclo-(O)$_6$
is to polyoxane as cyclohexane is to polyethylene.

McKay and Wright\cite{Howlong} computationally explored longer members of the hydrogen peroxide series H(O)$_n$H, whose
sulfur homologues H(S)$_n$H are known to exist. They found\cite{Howlong} that the bond dissociation energy reaches a minimum at H$_2$O$_6$, and then turns to increase for a longer chain, suggesting that an infinite chain may be relatively stable.
Martins-Costa {\it et al.}\cite{MartinsCostaCPL2009,MartinsCosta2011}\ computationally studied H(O)$_n$H with up to $n=10$, revealing that longer chains
adopt a helical conformation with bond-length alternation. They also showed that the heats of formation become negative for H$_2$O$_6$ and longer, which
are, therefore, metastable. See Refs.\ \cite{XuGoddard2002,Denis2009} for other computational studies of H$_2$O$_3$ and H$_2$O$_4$.

Experimentally, H$_2$O$_3$ has been detected by IR (Ref.\ \cite{Engdahl2002}) and microwave spectroscopies\cite{Suma2005} and H$_2$O$_3$ and H$_2$O$_4$ by Raman spectroscopy.\cite{Levanov2011} Substitution with electron-withdrawing fluorine terminal groups will enhance the stability considerably.\cite{Kirchmeier1992} Hence, the syntheses of perfluorinated polyoxanes as long as F$_2$O$_5$ and F$_2$O$_6$ were reported by Streng and Grosse,\cite{F2O6} hinting at an intrinsic ability of oxygens to form longer consecutive bonds.
 
In the physics approach, Gorelli {\it et al.}\cite{Gorelli1999,Gorelli2001}\ observed a pressure-induced transition to the $\epsilon$ phase of 
solid oxygen. This was accompanied by
a dramatic color change with strong IR bands emerging at 300--600 cm$^{-1}$, suggesting condensation
of multiple O$_2$ molecules into a longer catenated form. These authors ruled out polymeric forms with more than four atoms.\cite{Gorelli2001}
Neaton and Ashcroft\cite{Neaton2002} computationally proposed a linear herringbone (zigzag chain) structure in the  
pressure range corresponding to the $\epsilon$ phase.  
Their predicted IR and Raman band positions were said to be consistent with the observed.\cite{Gorelli1999,Gorelli2001} 
However, Goncharov {\it et al.}\cite{Goncharov2003}\ measured the Raman spectra of the $\epsilon$ phase, and concluded that the results were inconsistent with the herringbone structure. 
A subsequent computational study by Bartolomei {\it et al.}\cite{Bartolomei2011}\ also suggested that van-der-Waals clusters of the type (O$_2$)$_4$ for the $\epsilon$  phase.
Despite the well-defined O$_2$ molecular constituents, this structure is 
shown to have a singlet spin multiplicity and is consistent with the observed nonmagnetic character of this phase.\cite{Goncharenko2005}
This interpretation was also compatible with an inelastic X-ray scattering study of Meng {\it et al.},\cite{MengPNAS2008} once again ruling out the herringbone structure for the $\epsilon$ phase in favor of the O$_8$ structure.

Oganov and Glass,\cite{Oganov2006} using their pioneering evolutionary techniques, determined a helical polymeric form of oxygen as the most stable structure at 25 GPa,
which is similar to the herringbone structure of Neaton and Ashcroft.\cite{Neaton2002} This is not surprising given the similarity of DFT models (which consist in LDA or generalized gradient approximation) underlying both groups' simulations. 
Zhu {\it et al.}\cite{Zhu2012}\ also 
computationally predicted a new ($\theta$) phase of solid oxygen made of 4/1-helical polymers 
at 2 TPa, which is structurally analogous to phase III of solid sulfur. 

More recently, Hagiwara {\it et al.}\cite{Hagiwara2014}\ fashioned an antiferromagnetic one-dimensional chain of weakly interacting, non-catenated O$_2$ molecules in a single-walled carbon nanotube (SWCNT).
This may be contrasted with a long catenated metallic chain of sulfur formed in a SWCNT.\cite{Fujimori2013}  
 
In this study, we determined the structure, phonon dispersion, and phonon DOS of helical polyoxane at the B3LYP/cc-pVDZ or 6-31G** level 
with $S=8$, $L=12$, and $K=48$. We then computed the quasiparticle energy bands and electronic DOS at the MBGF(2)/cc-pVDZ level with 
the frozen-core and mod-4 approximations, using a single oxygen atom as the rototranslational (physical) repeat unit. We employed the ninth-nearest-neighbor
approximation in the normal-mode analysis.


\begin{table*}
\caption{Structural parameters of (O)$_x$ and O$_2$.  \label{PolyO_structure}}
\begin{tabular}{lcccccc}
& \multicolumn{4}{c}{(O)$_x$} & O$_2$ ($^1\Delta_\text{g}$)  & O$_2$ ($^3\Sigma_\text{g}^-$)  \\ \cline{2-5}\cline{6-6} \cline{7-7} 
Method & $r$(OO) & $a$(OOO) & $d$(OOOO) & $\varphi$\textsuperscript{\textit{a}} & $r$(OO) & $r$(OO) \\ \hline
B3LYP/6-31G** & 1.422\,\AA & 108.1$^\circ$ & 79.4$^\circ$ & $102.9^\circ$ & 1.216\,\AA & 1.215\,\AA \\ 
B3LYP/cc-pVDZ & 1.420\,\AA & 108.3$^\circ$ & 78.9$^\circ$ & $102.5^\circ$ & 1.210\,\AA  & 1.209\,\AA \\ 
PBE (2 TPa)\cite{Zhu2012} & 1.153\,\AA & $98.8^\circ$ & & $90.0^\circ$\\
Observed\textsuperscript{\textit{b}} & & & & ($102.9^\circ$)\textsuperscript{\textit{c}} & 1.216\,\AA & 1.208\,\AA \\
\end{tabular} \\ 
\textsuperscript{\textit{a}}{The helical angle.}
\textsuperscript{\textit{b}}{The O$_2$ structures from Huber and Herzberg.\cite{HuberHerzberg}}
\textsuperscript{\textit{c}}{The helical angle for a 7/2 helix (not observed) is given in parentheses.}
\end{table*}

The calculated equilibrium structure of polyoxane is given in Table \ref{PolyO_structure}.  It is predicted to have a 7/2-helical structure
with the helical angle of $\varphi=103^\circ$. This angle is not dissimilar from the one ($\varphi=90^\circ$) in the 4/1-helical structure predicted\cite{Zhu2012} 
under an immense pressure of 2 TPa. However, the O--O bond lengths are vastly different between these two pressures:\ it is 1.42\,\AA\ in our calculation under 
zero pressure, while 
1.15\,\AA\ under 2 TPa,\cite{Zhu2012} the latter being even shorter than the O=O bond in the singlet or triplet O$_2$ molecule in the gas phase. 

\begin{figure}
\includegraphics[width=1.0\columnwidth]{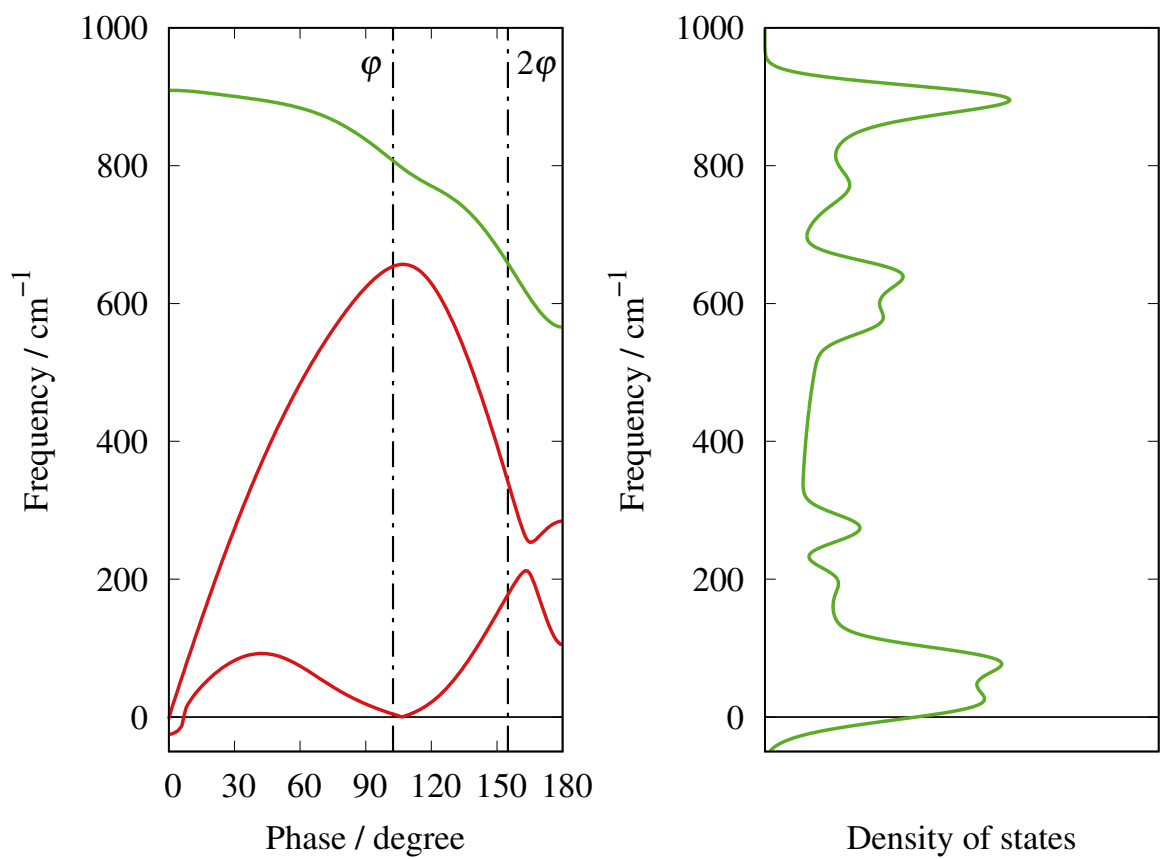}
\caption{\label{fig:Odisp} Phonon dispersion curves and DOS of (O)$_x$ computed by B3LYP/6-31G**. The DOS is convoluted with a Gaussian of a FWHM of 40 cm$^{-1}$.}
\end{figure}

\begin{table}
\caption{Vibrational frequencies (in cm$^{-1}$) of (O)$_x$.  \label{PolyO_freqs}}
\begin{tabular}{lr}
Irrep.\textsuperscript{\textit{a}}; phase; activity& \multicolumn{1}{c}{B3LYP/6-31G**} \\ \hline
$A_1$; $\theta=0$; Raman &  909.4 \\
$E_1$; $\theta=\varphi$; IR, Raman & 806.1 \\
& 654.4 \\
$E_2$; $\theta=2\varphi$; Raman & 662.5 \\
& 350.7 \\
& 172.9 \\
\end{tabular} \\ 
\textsuperscript{\textit{a}}{Isomorphic to the $D_{7}$ point group. $\varphi$ is the helical angle.}
\end{table}

The phonon dispersion curves and phonon DOS are drawn in Fig.\ \ref{fig:Odisp}. 
With only one atom in the rototranslational unit, there are one optical (green) and two acoustic (red) branches. 
The acoustic branches exhibit the expected behavior of having zero frequencies at $\theta=0$ and $\varphi$
with the maximum error of $25i$ cm$^{-1}$. Therefore, the structure of Table \ref{PolyO_structure} is unambiguously a local minimum, and
there is no bond-length alternation\cite{MartinsCostaCPL2009,MartinsCosta2011} according to our calculation.
The Raman bands from H$_2$O$_3$ are observed at 500, 756, and 878 cm$^{-1}$ and those from H$_2$O$_4$ at 
449, 586, 624, 827, and 865 cm$^{-1}$.\cite{Levanov2011} They mostly fall on the optical branch (some possibly on the upper acoustic branch), and are consistent
with our calculation. 

According to Higgs' selection rules,\cite{Higgs} six modes are optically active: Two are IR active and all six are Raman active. 
Their calculated frequencies are given in Table \ref{PolyO_freqs}, which are hoped to assist in future experimental identification.
Goncharov {\it et al.}\cite{Goncharov2003}\ detected seven Raman bands from the $\epsilon$ phase at 260, 290, 580, 600, 625, 695, and 1750 cm$^{-1}$ under 65 GPa. 
While these frequencies must be considerably elevated by the high pressure, they are clearly incompatible with the predicted Raman band positions 
of polyoxane in Table \ref{PolyO_freqs}.
The phase transition to the $\epsilon$ phase is accompanied by a strong IR absorption at 300 cm$^{-1}$ at 20 GPa.\cite{Gorelli1999,Gorelli2001} 
This cannot be explained by the predicted IR band positions for polyoxane, either.  These comparisons confirm the earlier conclusion\cite{Gorelli1999,Gorelli2001,Goncharov2003} ruling out the polyoxane-like structure
for the $\epsilon$ phase. 
The observed Raman bands\cite{Goncharov2003} from the $\zeta$ phase at 260, 440, 640, 700, 760, 830, and 1750 cm$^{-1}$ under 135 GPa
are also inconsistent with the polyoxane structure.

\begin{figure}
\includegraphics[width=1.0\columnwidth]{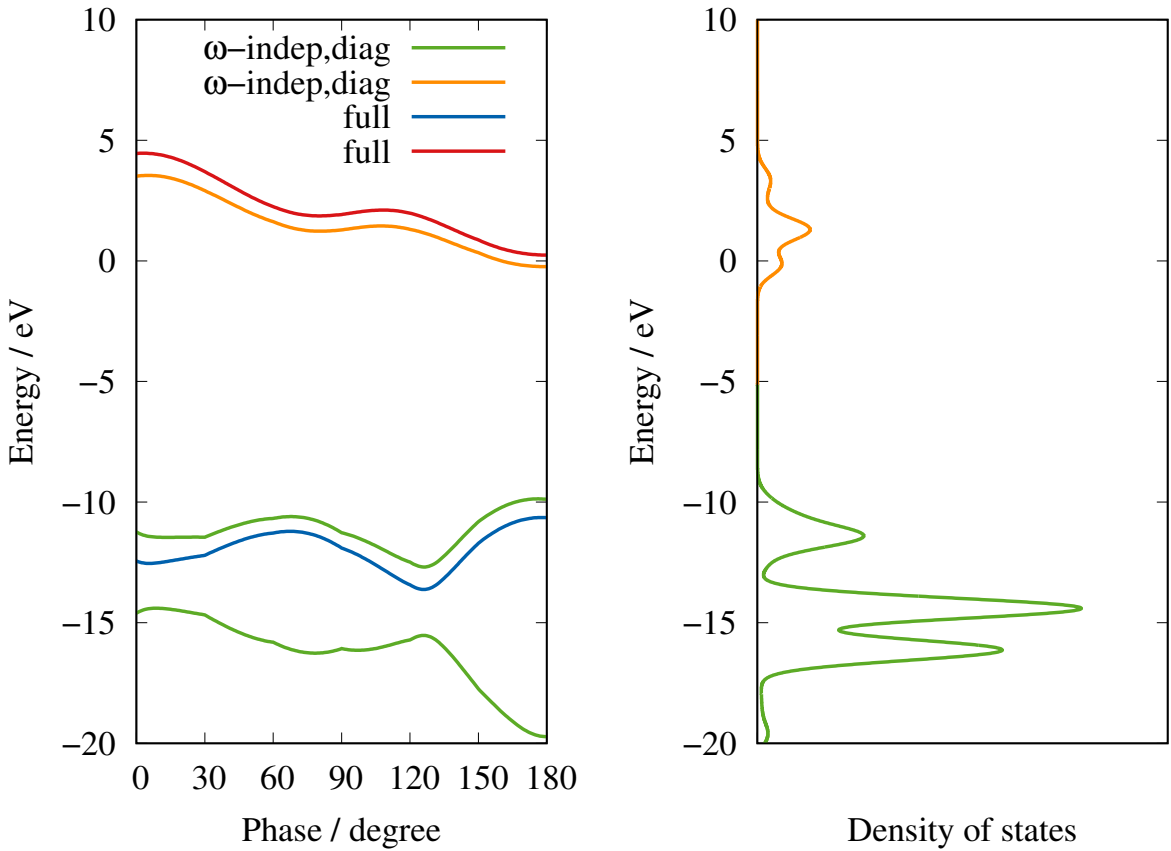}
\caption{\label{fig:Obands}Electronic energy bands and DOS of (O)$_x$ computed by MBGF(2)/cc-pVDZ. The DOS is weighted by the pole strength of Eq.\ (\ref{HellmannFeynman}) and convoluted with a Gaussian of a FWHM of 1 eV.}
\end{figure}

Figure \ref{fig:Obands} shows the calculated quasiparticle energy bands and electronic DOS.
The valence band edge is located at $-10.6$ eV according to MBGF(2)/cc-pVDZ with the ``full'' self-energy,
and this value should be accurate to within a few tenths of an electronvolt. 
The conduction band edge is 0.2 eV, which is too high. The amount of overestimation in the conduction band edge 
was 0.4 eV in electronegative polytetrafluoroethylene
and 4.6 eV in polyethylene. The corresponding value for moderately electronegative polyoxane may fall somewhere in between these two values. 
The fundamental band gap is, therefore, direct and in the range of 6.1--10.8 eV. 
In comparison, the ultra-high-pressure $\theta$ phase (a 4/1 helix) was predicted\cite{Zhu2012} to be an insulator with a band gap of 
3.2 eV or 5.9 eV according to the nonhybrid PBE or hybrid HSE functional, respectively.
The valence DOS in Fig. \ref{fig:Obands} should be predictive of the UPS of polyoxane without an energy shift.

\begin{table}
\caption{Binding energy (in kcal/mol) of (O$_2$)$_x$.  \label{PolyO_binding}}
\begin{tabular}{lrr}
Method & O$_2$ ($^1\Delta_\text{g}$)\textsuperscript{\textit{a}} & O$_2$ ($^3\Sigma_\text{g}^-$)\textsuperscript{\textit{b}} \\ \hline
B3LYP/cc-pVDZ & 13.8 & $-25.7$ \\ 
MBPT(2)/cc-pVDZ & 3.3 & $-29.3$ \\
\end{tabular} \\ 
\textsuperscript{\textit{a}}{The energy difference between  the unit-cell energy of (O$_2$)$_x$ and the energy of the singlet O$_2$ 
in their respective B3LYP/cc-pVDZ-optimized geometries.}
\textsuperscript{\textit{b}}{The energy difference between  the unit-cell energy of (O$_2$)$_x$ and the energy of the triplet O$_2$ 
in their respective B3LYP/cc-pVDZ-optimized geometries.}
\end{table}


\begin{figure}
\includegraphics[width=1.0\columnwidth]{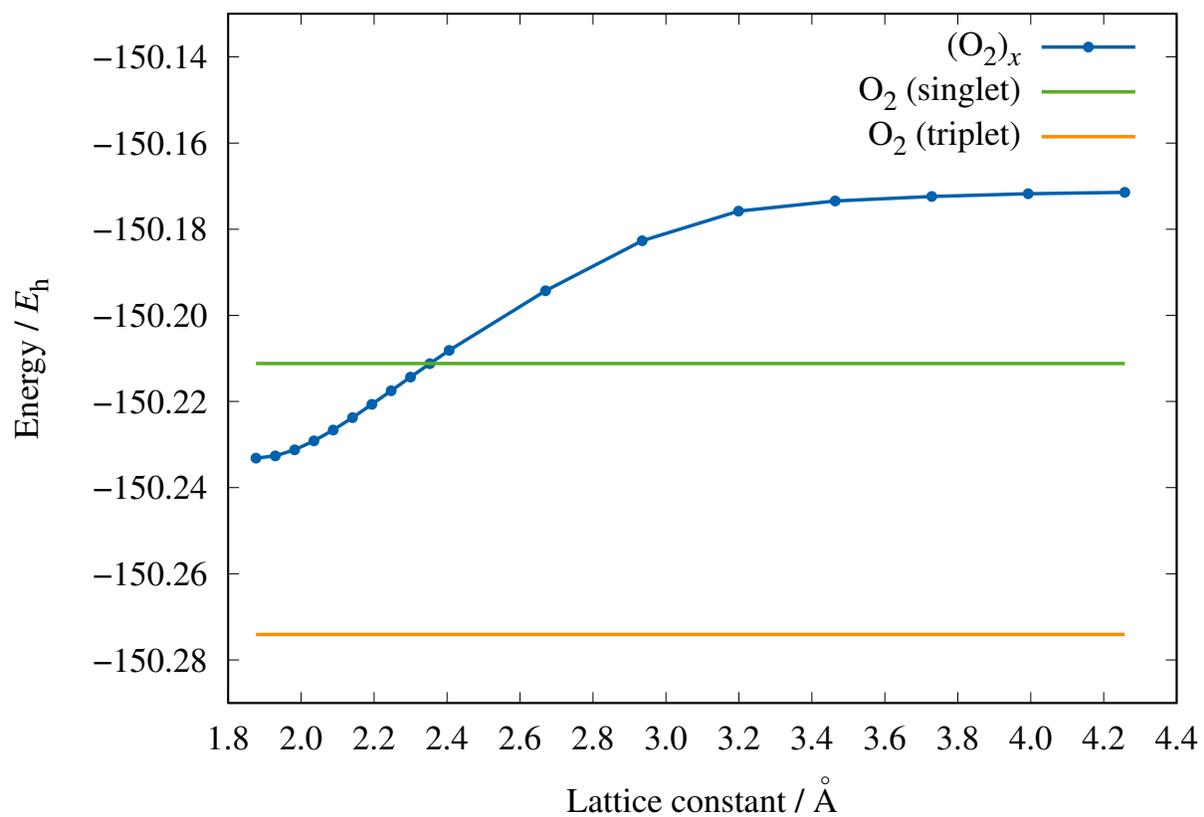}
\caption{\label{fig:Odissociation}Unit-cell energy of (O$_2$)$_x$ as a function of the lattice constant computed by B3LYP/cc-pVDZ.}
\end{figure}

Table \ref{PolyO_binding} compiles the binding energy of polyoxane relative to its monomer O$_2$. 
Polyoxane is metastable with respect to triplet O$_2$. Remarkably, it is thermodynamically
stable against dissociation into singlet O$_2$ according to the electron-correlated treatments.
Figure \ref{fig:Odissociation} indicates that the unrelaxed dissociation reaction energy profile is a monotonically increasing function
of the lattice constant, although this may a computational artifact because the spin-restricted formulation compels 
the dissociation products to be singlet. Comparison with similar energy profiles of the organic polymers (Secs.\ \ref{sec:CH2}--\ref{sec:PTFE}) 
nonetheless suggests that polyoxane may also be kinetically stable.

\section{Conclusions}

In this article, we have reported an implementation of the {\it ab initio} Gaussian-basis-set MBGF(2) method for an infinite helical polymer. 
It can compute the quasiparticle energy bands and pole strengths in the extended-zone scheme with or without the diagonal and frequency-independent approximations
to the self-energy. It simultaneously evaluates the MBPT(2) correction to the energy per the smallest rototranslational (physical) repeat unit. Together with the Gaussian-basis-set DFT method\cite{Hirata_MP2022} for
energy and analytical gradients with respect to atomic positions, translational period, and helical angle, it can simulate 
ARUPS, UPS, IR and Raman band positions, phonon dispersion curves, and INS spectra of infinite helical polymers even with incommensurable structures, which cannot be handled well  
by either a molecular or solid-state method.

Through applications to several well-characterized organic polymers, i.e., polyethylene, polyacetylene, and polytetrafluoroethylene,
we have demonstrated the predictive accuracy of B3LYP/cc-pVDZ or 6-31G** for structures, phonon dispersion curves, and phonon DOS
and of MBGF(2)/cc-pVDZ for valence  bands and valence DOS, if not for conduction  bands or fundamental band gaps. 
The measured bond lengths and angles of these organic polymers are reproduced within a few hundredths of one \AA ngstrom
and a few degrees. The vibrational frequencies are systematically overestimated by a few percent, but they are accurate enough for reliable band assignments
of their IR, Raman, and INS spectral bands. The valence  bands and DOS have negligible dependence on the diagonal and frequency-independent approximations
and are within a few tenths of an electronvolt or a typical experimental error bar of the observed energy bands or photoelectron spectral peaks. 
The band gaps are overestimated by 0.4--4.1 eV, depending on the nature of the conduction bands.

Armed with these predictive methods for helical polymers, we have explored the possibilities of the kinetically-trapped metastable existence 
of second-row-element inorganic polymers under ambient conditions, notwithstanding the obvious difficulties of their synthesis and isolation.
All-{\it trans} polyazene (N$_2$)$_x$ is isoelectronic with all-{\it trans} polyacetylene and an insulator with bond-length alternation. Various isomers including 
the {\it trans}-transoid are both thermodynamically and mechanically unstable with a propensity to spontaneously dissociate into N$_2$ fragments during geometry optimizations. 
On the other hand, isotactic polyazane (NH)$_x$, which
is a nitrogen analog of polyethylene, is thermodynamically stable in its 11/3-helical conformation against dissociation into its monomers. It is not thermodynamically stable 
against dissociation into N$_2$ and NH$_3$, but such dissociation reaction involves N--H bond breaking and formation and likely has a high activation barrier. 
It may, therefore, exist at low temperatures, although it is likely unstable in solutions.
Contrary to theoretical expectations and experimental findings for short chains, perfluorinated polyazane or isotactic polyfluoroazane (NF)$_x$ in its 9/4-helical conformation seems no more
stable than polyazane either thermodynamically or kinetically. This may be because the fluorination stabilizes the smaller dissociation products (difluorodiazenes or trifluoroamine) more than it does the polymer. The 7/2-helical polyoxane (O)$_x$ is thermodynamically stable with respect to singlet O$_2$. It 
also seems kinetically stable against dissociation into triplet O$_2$. Echoing with previous computational predictions of the existence of (helical) polyoxane under high pressures,\cite{Neaton2002,Oganov2006,Zhu2012} our calculation supports its possible metastable existence at low pressures and temperatures. 
The IR and Raman band positions, INS spectra, and (AR)UPS of these polymers have been predicted in order to assist in their experimental detections.

\section{Acknowledgements}

SH is indebted to The Late Professor Mitsuo Tasumi of the University of Tokyo for decades of encouragements and tutelage especially in the area of polymer spectroscopy.
He was supported by the U.S. Department of Energy (DoE), Office of Science, Office of Basic Energy Sciences under Grant No.\ DE-SC0006028. 
SH is a Guggenheim Fellow of the John Simon Guggenheim Memorial Foundation. He also thanks the hospitality of University of Tsukuba, where the initial phase of this study was conducted.

SSX was supported by the Center for Scalable Predictive methods for Excitations and Correlated phenomena (SPEC), which is funded by the U.S. DoE, Office of Science, Office of Basic Energy Sciences, Division of Chemical Sciences, Geosciences and Biosciences as part of the Computational Chemical Sciences (CCS) program at Pacific Northwest National Laboratory (PNNL) under FWP 70942. PNNL is a multi-program national laboratory operated by Battelle Memorial Institute for the U.S. DoE.

RJB was supported by the Air Force Office of Scientific Research under AFOSR (Award No.\ FA9550-19-1-0091).

This research used resources of the National Energy Research Scientific Computing Center (NERSC), a U.S. DoE Office of Science User Facility located at Lawrence Berkeley National Laboratory, operated under Contract No.\ DE-AC02-05CH11231 under NERSC award m3196 (2022).

\begin{tocentry}
      \includegraphics[width=0.9\linewidth]{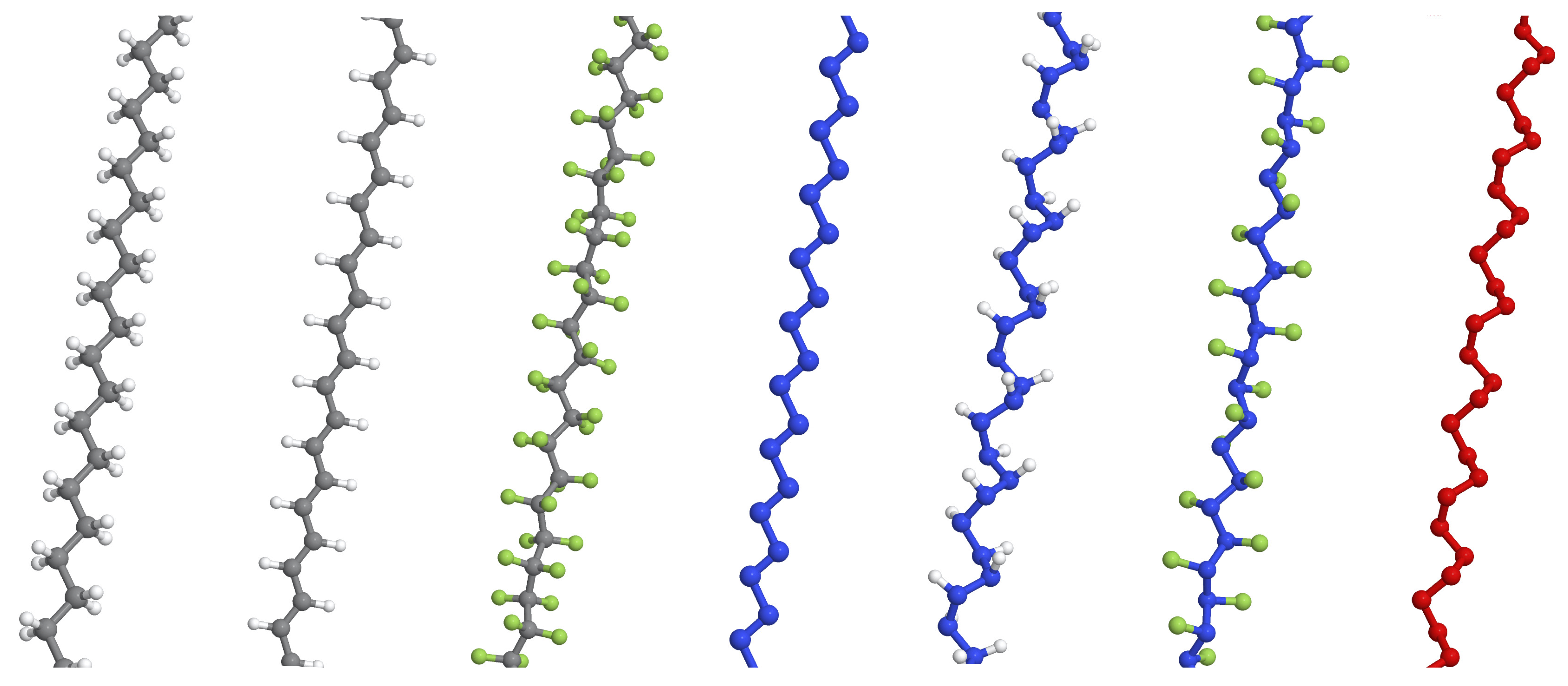}
\end{tocentry}

\bibliography{library.bib}

\providecommand{\latin}[1]{#1}
\makeatletter
\providecommand{\doi}
  {\begingroup\let\do\@makeother\dospecials
  \catcode`\{=1 \catcode`\}=2 \doi@aux}
\providecommand{\doi@aux}[1]{\endgroup\texttt{#1}}
\makeatother
\providecommand*\mcitethebibliography{\thebibliography}
\csname @ifundefined\endcsname{endmcitethebibliography}
  {\let\endmcitethebibliography\endthebibliography}{}
\begin{mcitethebibliography}{306}
\providecommand*\natexlab[1]{#1}
\providecommand*\mciteSetBstSublistMode[1]{}
\providecommand*\mciteSetBstMaxWidthForm[2]{}
\providecommand*\mciteBstWouldAddEndPuncttrue
  {\def\EndOfBibitem{\unskip.}}
\providecommand*\mciteBstWouldAddEndPunctfalse
  {\let\EndOfBibitem\relax}
\providecommand*\mciteSetBstMidEndSepPunct[3]{}
\providecommand*\mciteSetBstSublistLabelBeginEnd[3]{}
\providecommand*\EndOfBibitem{}
\mciteSetBstSublistMode{f}
\mciteSetBstMaxWidthForm{subitem}{(\alph{mcitesubitemcount})}
\mciteSetBstSublistLabelBeginEnd
  {\mcitemaxwidthsubitemform\space}
  {\relax}
  {\relax}

\bibitem[Glukhovtsev \latin{et~al.}(1996)Glukhovtsev, Jiao, and
  Schleyer]{HEDM1996}
Glukhovtsev,~M.~N.; Jiao,~H.~J.; Schleyer,~P.~v. Besides {N}$_2$, what is the
  most stable molecule composed only of nitrogen atoms? \emph{Inorg. Chem.}
  \textbf{1996}, \emph{35}, 7124--7133\relax
\mciteBstWouldAddEndPuncttrue
\mciteSetBstMidEndSepPunct{\mcitedefaultmidpunct}
{\mcitedefaultendpunct}{\mcitedefaultseppunct}\relax
\EndOfBibitem
\bibitem[{McKay} and Wright(1998){McKay}, and Wright]{Howlong}
{McKay},~D.~J.; Wright,~J.~S. How long can you make an oxygen chain? \emph{J.
  Am. Chem. Soc.} \textbf{1998}, \emph{120}, 1003--1013\relax
\mciteBstWouldAddEndPuncttrue
\mciteSetBstMidEndSepPunct{\mcitedefaultmidpunct}
{\mcitedefaultendpunct}{\mcitedefaultseppunct}\relax
\EndOfBibitem
\bibitem[Bartlett(2000)]{Bartlett2000}
Bartlett,~R.~J. Exploding the Mysteries of Nitrogen. \emph{Chemistry \&
  Industry} \textbf{2000}, \emph{4}, 140--143\relax
\mciteBstWouldAddEndPuncttrue
\mciteSetBstMidEndSepPunct{\mcitedefaultmidpunct}
{\mcitedefaultendpunct}{\mcitedefaultseppunct}\relax
\EndOfBibitem
\bibitem[Steele and Oleynik(2019)Steele, and Oleynik]{Steele_bookchapter2019}
Steele,~B.~A.; Oleynik,~I.~I. Computational Discovery of New High-Nitrogen
  Energetic Materials. \emph{Computational Approaches for Chemistry under
  Extreme Conditions} \textbf{2019}, \emph{28}, 25--52\relax
\mciteBstWouldAddEndPuncttrue
\mciteSetBstMidEndSepPunct{\mcitedefaultmidpunct}
{\mcitedefaultendpunct}{\mcitedefaultseppunct}\relax
\EndOfBibitem
\bibitem[O'Sullivan and Zdilla(2020)O'Sullivan, and Zdilla]{ReviewN}
O'Sullivan,~O.~T.; Zdilla,~M.~J. Properties and Promise of Catenated Nitrogen
  Systems As High-Energy-Density Materials. \emph{Chem. Rev.} \textbf{2020},
  \emph{120}, 5682--5744\relax
\mciteBstWouldAddEndPuncttrue
\mciteSetBstMidEndSepPunct{\mcitedefaultmidpunct}
{\mcitedefaultendpunct}{\mcitedefaultseppunct}\relax
\EndOfBibitem
\bibitem[AQC(2014)]{AQC2014}
Sabin,~J.~R., Ed. \emph{Advances in Quantum Chemistry}; Elsevier, 2014;
  Vol.~69\relax
\mciteBstWouldAddEndPuncttrue
\mciteSetBstMidEndSepPunct{\mcitedefaultmidpunct}
{\mcitedefaultendpunct}{\mcitedefaultseppunct}\relax
\EndOfBibitem
\bibitem[Shimizu \latin{et~al.}(1998)Shimizu, Suhara, Ikumo, Eremets, and
  Amaya]{Shimizu1998}
Shimizu,~K.; Suhara,~K.; Ikumo,~M.; Eremets,~M.~I.; Amaya,~K. Superconductivity
  in oxygen. \emph{Nature} \textbf{1998}, \emph{393}, 767--769\relax
\mciteBstWouldAddEndPuncttrue
\mciteSetBstMidEndSepPunct{\mcitedefaultmidpunct}
{\mcitedefaultendpunct}{\mcitedefaultseppunct}\relax
\EndOfBibitem
\bibitem[Neaton and Ashcroft(2002)Neaton, and Ashcroft]{Neaton2002}
Neaton,~J.~B.; Ashcroft,~N.~W. Low-energy linear structures in dense oxygen:
  Implications for the $\epsilon$ phase. \emph{Phys. Rev. Lett.} \textbf{2002},
  \emph{88}, 205503\relax
\mciteBstWouldAddEndPuncttrue
\mciteSetBstMidEndSepPunct{\mcitedefaultmidpunct}
{\mcitedefaultendpunct}{\mcitedefaultseppunct}\relax
\EndOfBibitem
\bibitem[Ma \latin{et~al.}(2007)Ma, Oganov, and Glass]{Ma2007}
Ma,~Y.~M.; Oganov,~A.~R.; Glass,~C.~W. Structure of the metallic $\zeta$-phase
  of oxygen and isosymmetric nature of the $\epsilon$--$\zeta$ phase
  transition: {\it Ab initio} simulations. \emph{Phys. Rev. B} \textbf{2007},
  \emph{76}, 064101\relax
\mciteBstWouldAddEndPuncttrue
\mciteSetBstMidEndSepPunct{\mcitedefaultmidpunct}
{\mcitedefaultendpunct}{\mcitedefaultseppunct}\relax
\EndOfBibitem
\bibitem[Meng \latin{et~al.}(2008)Meng, Eng, Tse, Shaw, Hu, Shu, Gramsch, Kao,
  Hemley, and Mao]{MengPNAS2008}
Meng,~Y.; Eng,~P.~J.; Tse,~J.~S.; Shaw,~D.~M.; Hu,~M.~Y.; Shu,~J.~F.;
  Gramsch,~S.~A.; Kao,~C.; Hemley,~R.~J.; Mao,~H.~K. Inelastic x-ray scattering
  of dense solid oxygen: Evidence for intermolecular bonding. \emph{Proc. Natl.
  Acad. Sci. USA} \textbf{2008}, \emph{105}, 11640--11644\relax
\mciteBstWouldAddEndPuncttrue
\mciteSetBstMidEndSepPunct{\mcitedefaultmidpunct}
{\mcitedefaultendpunct}{\mcitedefaultseppunct}\relax
\EndOfBibitem
\bibitem[Kimball and Haley(2002)Kimball, and Haley]{Kimball2002}
Kimball,~D.~B.; Haley,~M.~M. Triazenes: A versatile tool in organic synthesis.
  \emph{Angew. Chem. Int. Ed.} \textbf{2002}, \emph{41}, 3338--3351\relax
\mciteBstWouldAddEndPuncttrue
\mciteSetBstMidEndSepPunct{\mcitedefaultmidpunct}
{\mcitedefaultendpunct}{\mcitedefaultseppunct}\relax
\EndOfBibitem
\bibitem[Wright and McKay(1999)Wright, and McKay]{Wright1999}
Wright,~J.~S.; McKay,~D.~J. Polymeric oxygen and nitrogen: why not? \emph{Sci.
  Prog.} \textbf{1999}, \emph{82}, 151--170\relax
\mciteBstWouldAddEndPuncttrue
\mciteSetBstMidEndSepPunct{\mcitedefaultmidpunct}
{\mcitedefaultendpunct}{\mcitedefaultseppunct}\relax
\EndOfBibitem
\bibitem[Manners(1996)]{Manners1996}
Manners,~I. Polymers and the periodic table: Recent developments in inorganic
  polymer science. \emph{Angew. Chem. Int. Ed.} \textbf{1996}, \emph{35},
  1602--1621\relax
\mciteBstWouldAddEndPuncttrue
\mciteSetBstMidEndSepPunct{\mcitedefaultmidpunct}
{\mcitedefaultendpunct}{\mcitedefaultseppunct}\relax
\EndOfBibitem
\bibitem[Wolfe(1972)]{Gaucheeffect}
Wolfe,~S. Gauche Effect. {Some} Stereochemical Consequences of Adjacent
  Electron Pairs and Polar Bonds. \emph{Acc. Chem. Res.} \textbf{1972},
  \emph{5}, 102--111\relax
\mciteBstWouldAddEndPuncttrue
\mciteSetBstMidEndSepPunct{\mcitedefaultmidpunct}
{\mcitedefaultendpunct}{\mcitedefaultseppunct}\relax
\EndOfBibitem
\bibitem[Mizushima(1954)]{Mizushima_book}
Mizushima,~S. \emph{Internal Rotation}; Academic Press: New York, 1954\relax
\mciteBstWouldAddEndPuncttrue
\mciteSetBstMidEndSepPunct{\mcitedefaultmidpunct}
{\mcitedefaultendpunct}{\mcitedefaultseppunct}\relax
\EndOfBibitem
\bibitem[Mizushima \latin{et~al.}(1954)Mizushima, Nakagawa, Ichishima, and
  Miyazawa]{Mizushima1954}
Mizushima,~S.; Nakagawa,~I.; Ichishima,~I.; Miyazawa,~T. Further Evidence for
  the Existence of the {\it Gauche}-Form of 1,2-Dichloroethane. \emph{J. Chem.
  Phys.} \textbf{1954}, \emph{22}\relax
\mciteBstWouldAddEndPuncttrue
\mciteSetBstMidEndSepPunct{\mcitedefaultmidpunct}
{\mcitedefaultendpunct}{\mcitedefaultseppunct}\relax
\EndOfBibitem
\bibitem[Criton \latin{et~al.}(2021)Criton, Vilona, Jacob, M\'{e}debielle,
  Dumont, Joucla, and Lac\^{o}te]{Criton_cispolyN2021}
Criton,~T.; Vilona,~D.; Jacob,~G.; M\'{e}debielle,~M.; Dumont,~E.; Joucla,~L.;
  Lac\^{o}te,~E. Synthesis and Properties of Higher Nuclearity Polyazanes.
  \emph{Chem. Eur. J.} \textbf{2021}, \emph{27}, 3670--3674\relax
\mciteBstWouldAddEndPuncttrue
\mciteSetBstMidEndSepPunct{\mcitedefaultmidpunct}
{\mcitedefaultendpunct}{\mcitedefaultseppunct}\relax
\EndOfBibitem
\bibitem[Martins-Costa \latin{et~al.}(2009)Martins-Costa, Anglada, and
  Ruiz-Lopez]{MartinsCostaCPL2009}
Martins-Costa,~M.; Anglada,~J.~M.; Ruiz-Lopez,~M.~F. Hyperconjugation in
  adjacent {OO} bonds: Remarkable odd/even effects. \emph{Chem. Phys. Lett.}
  \textbf{2009}, \emph{481}, 180--182\relax
\mciteBstWouldAddEndPuncttrue
\mciteSetBstMidEndSepPunct{\mcitedefaultmidpunct}
{\mcitedefaultendpunct}{\mcitedefaultseppunct}\relax
\EndOfBibitem
\bibitem[Zhu \latin{et~al.}(2012)Zhu, Wang, Wang, Zou, Mao, and Ma]{Zhu2012}
Zhu,~L.; Wang,~Z.~W.; Wang,~Y.~C.; Zou,~G.~T.; Mao,~H.~K.; Ma,~Y.~M. Spiral
  chain {O}$_4$ form of dense oxygen. \emph{Proc. Natl. Acad. Sci. USA}
  \textbf{2012}, \emph{109}, 751--753\relax
\mciteBstWouldAddEndPuncttrue
\mciteSetBstMidEndSepPunct{\mcitedefaultmidpunct}
{\mcitedefaultendpunct}{\mcitedefaultseppunct}\relax
\EndOfBibitem
\bibitem[McMahan and LeSar(1985)McMahan, and LeSar]{McMahan1985}
McMahan,~A.~K.; LeSar,~R. Pressure Dissociation of Solid Nitrogen under 1
  {Mbar}. \emph{Phys. Rev. Lett.} \textbf{1985}, \emph{54}, 1929--1932\relax
\mciteBstWouldAddEndPuncttrue
\mciteSetBstMidEndSepPunct{\mcitedefaultmidpunct}
{\mcitedefaultendpunct}{\mcitedefaultseppunct}\relax
\EndOfBibitem
\bibitem[Mailhiot \latin{et~al.}(1992)Mailhiot, Yang, and
  {McMahan}]{Mailhiot1992}
Mailhiot,~C.; Yang,~L.~H.; {McMahan},~A.~K. Polymeric Nitrogen. \emph{Phys.
  Rev. B} \textbf{1992}, \emph{46}, 14419--14435\relax
\mciteBstWouldAddEndPuncttrue
\mciteSetBstMidEndSepPunct{\mcitedefaultmidpunct}
{\mcitedefaultendpunct}{\mcitedefaultseppunct}\relax
\EndOfBibitem
\bibitem[Eremets \latin{et~al.}(2004)Eremets, Gavriliuk, Trojan, Dzivenko, and
  Boehler]{Eremets2004}
Eremets,~M.~I.; Gavriliuk,~A.~G.; Trojan,~I.~A.; Dzivenko,~D.~A.; Boehler,~R.
  Single-bonded cubic form of nitrogen. \emph{Nature Mater.} \textbf{2004},
  \emph{3}, 558--563\relax
\mciteBstWouldAddEndPuncttrue
\mciteSetBstMidEndSepPunct{\mcitedefaultmidpunct}
{\mcitedefaultendpunct}{\mcitedefaultseppunct}\relax
\EndOfBibitem
\bibitem[Eremets \latin{et~al.}(2007)Eremets, Gavriliuk, and
  Trojan]{Eremets2007}
Eremets,~M.~I.; Gavriliuk,~A.~G.; Trojan,~I.~A. Single-crystalline polymeric
  nitrogen. \emph{Appl. Phys. Lett.} \textbf{2007}, \emph{90}\relax
\mciteBstWouldAddEndPuncttrue
\mciteSetBstMidEndSepPunct{\mcitedefaultmidpunct}
{\mcitedefaultendpunct}{\mcitedefaultseppunct}\relax
\EndOfBibitem
\bibitem[Schlegel and Skancke(1993)Schlegel, and Skancke]{Schlegel1993}
Schlegel,~H.~B.; Skancke,~A. Thermochemistry, Energy Comparisons, and
  Conformational-Analysis of Hydrazine, Triazane, and Triaminoammonia. \emph{J.
  Am. Chem. Soc.} \textbf{1993}, \emph{115}, 7465--7471\relax
\mciteBstWouldAddEndPuncttrue
\mciteSetBstMidEndSepPunct{\mcitedefaultmidpunct}
{\mcitedefaultendpunct}{\mcitedefaultseppunct}\relax
\EndOfBibitem
\bibitem[Zhao and Gimarc(1994)Zhao, and Gimarc]{Zhao1994}
Zhao,~M.; Gimarc,~B.~M. Strain Energies of ({NH})$_n$ Rings, $n=3-8$. \emph{J.
  Phys. Chem.} \textbf{1994}, \emph{98}, 7497--7503\relax
\mciteBstWouldAddEndPuncttrue
\mciteSetBstMidEndSepPunct{\mcitedefaultmidpunct}
{\mcitedefaultendpunct}{\mcitedefaultseppunct}\relax
\EndOfBibitem
\bibitem[Del~Re \latin{et~al.}(1967)Del~Re, Ladik, and Bicz\'o]{DelRe1967}
Del~Re,~G.; Ladik,~J.; Bicz\'o,~G. Self-Consistent-Field Tight-Binding
  Treatment of Polymers. {I}. Infinite three-Dimensional Case. \emph{Phys.
  Rev.} \textbf{1967}, \emph{155}, 997\relax
\mciteBstWouldAddEndPuncttrue
\mciteSetBstMidEndSepPunct{\mcitedefaultmidpunct}
{\mcitedefaultendpunct}{\mcitedefaultseppunct}\relax
\EndOfBibitem
\bibitem[Andr\.e(1969)]{Andre1969}
Andr\.e,~J.~M. Self-Consistent Field Theory for Electronic Structure of
  Polymers. \emph{J. Chem. Phys.} \textbf{1969}, \emph{50}, 1536\relax
\mciteBstWouldAddEndPuncttrue
\mciteSetBstMidEndSepPunct{\mcitedefaultmidpunct}
{\mcitedefaultendpunct}{\mcitedefaultseppunct}\relax
\EndOfBibitem
\bibitem[Mintmire and Sabin(1980)Mintmire, and Sabin]{MintmireSabin1980}
Mintmire,~J.~W.; Sabin,~J.~R. Local Density Functional Methods in
  Two-Dimensionally Periodic Systems. {I}. {The} Atomic Hydrogen Monolayer.
  \emph{Int. J. Quantum Chem.} \textbf{1980}, \emph{17}, 707--713\relax
\mciteBstWouldAddEndPuncttrue
\mciteSetBstMidEndSepPunct{\mcitedefaultmidpunct}
{\mcitedefaultendpunct}{\mcitedefaultseppunct}\relax
\EndOfBibitem
\bibitem[Mintmire \latin{et~al.}(1982)Mintmire, Sabin, and
  Trickey]{MintmireSabin1982}
Mintmire,~J.~W.; Sabin,~J.~R.; Trickey,~S.~B. Local Density Functional Methods
  in two-Dimensionally Periodic-Systems: Hydrogen and Beryllium Monolayers.
  \emph{Phys. Rev. B} \textbf{1982}, \emph{26}, 1743--1753\relax
\mciteBstWouldAddEndPuncttrue
\mciteSetBstMidEndSepPunct{\mcitedefaultmidpunct}
{\mcitedefaultendpunct}{\mcitedefaultseppunct}\relax
\EndOfBibitem
\bibitem[Kert\'esz(1982)]{Kertesz1982}
Kert\'esz,~M. Electronic Structure of Polymers. \emph{Adv. Quantum Chem.}
  \textbf{1982}, \emph{15}, 161--214\relax
\mciteBstWouldAddEndPuncttrue
\mciteSetBstMidEndSepPunct{\mcitedefaultmidpunct}
{\mcitedefaultendpunct}{\mcitedefaultseppunct}\relax
\EndOfBibitem
\bibitem[Sun and Bartlett(1999)Sun, and Bartlett]{Sun1999}
Sun,~J.~Q.; Bartlett,~R.~J. Modern correlation theories for extended, periodic
  systems. \emph{Top. Curr. Chem.} \textbf{1999}, \emph{203}, 121--145\relax
\mciteBstWouldAddEndPuncttrue
\mciteSetBstMidEndSepPunct{\mcitedefaultmidpunct}
{\mcitedefaultendpunct}{\mcitedefaultseppunct}\relax
\EndOfBibitem
\bibitem[Hirata(2009)]{Hirata2009_PCCP}
Hirata,~S. Quantum chemistry of macromolecules and solids. \emph{Phys. Chem.
  Chem. Phys.} \textbf{2009}, \emph{11}, 8397--8412\relax
\mciteBstWouldAddEndPuncttrue
\mciteSetBstMidEndSepPunct{\mcitedefaultmidpunct}
{\mcitedefaultendpunct}{\mcitedefaultseppunct}\relax
\EndOfBibitem
\bibitem[Nakano and Okamoto(2001)Nakano, and Okamoto]{Helicalreview1}
Nakano,~T.; Okamoto,~Y. Synthetic helical polymers: Conformation and function.
  \emph{Chem. Rev.} \textbf{2001}, \emph{101}, 4013--4038\relax
\mciteBstWouldAddEndPuncttrue
\mciteSetBstMidEndSepPunct{\mcitedefaultmidpunct}
{\mcitedefaultendpunct}{\mcitedefaultseppunct}\relax
\EndOfBibitem
\bibitem[Yashima \latin{et~al.}(2009)Yashima, Maeda, Iida, Furusho, and
  Nagai]{Helicalreview2}
Yashima,~E.; Maeda,~K.; Iida,~H.; Furusho,~Y.; Nagai,~K. Helical Polymers:
  Synthesis, Structures, and Functions. \emph{Chem. Rev.} \textbf{2009},
  \emph{109}, 6102--6211\relax
\mciteBstWouldAddEndPuncttrue
\mciteSetBstMidEndSepPunct{\mcitedefaultmidpunct}
{\mcitedefaultendpunct}{\mcitedefaultseppunct}\relax
\EndOfBibitem
\bibitem[Imamura(1970)]{Imamura1970}
Imamura,~A. Electronic Structures of Polymers Using Tight-Binding
  Approximation. {I}. {Polyethylene} by Extended {H\"{u}ckel} Method. \emph{J.
  Chem. Phys.} \textbf{1970}, \emph{52}, 3168--3175\relax
\mciteBstWouldAddEndPuncttrue
\mciteSetBstMidEndSepPunct{\mcitedefaultmidpunct}
{\mcitedefaultendpunct}{\mcitedefaultseppunct}\relax
\EndOfBibitem
\bibitem[Fujita and Imamura(1970)Fujita, and Imamura]{Fujita1970}
Fujita,~H.; Imamura,~A. Electronic Structures of Polymers Using Tight-Binding
  Approximation. {II}. Polyethylene and Polyglycine by {CNDO} Method. \emph{J.
  Chem. Phys.} \textbf{1970}, \emph{53}, 4555\relax
\mciteBstWouldAddEndPuncttrue
\mciteSetBstMidEndSepPunct{\mcitedefaultmidpunct}
{\mcitedefaultendpunct}{\mcitedefaultseppunct}\relax
\EndOfBibitem
\bibitem[Blumen and Merkel(1977)Blumen, and Merkel]{Blumen1977}
Blumen,~A.; Merkel,~C. Energy-Band Calculations on Helical Systems. \emph{Phys.
  Stat. Solidi B} \textbf{1977}, \emph{83}, 425--431\relax
\mciteBstWouldAddEndPuncttrue
\mciteSetBstMidEndSepPunct{\mcitedefaultmidpunct}
{\mcitedefaultendpunct}{\mcitedefaultseppunct}\relax
\EndOfBibitem
\bibitem[Teramae \latin{et~al.}(1983)Teramae, Yamabe, Satoko, and
  Imamura]{Teramae1983}
Teramae,~H.; Yamabe,~T.; Satoko,~C.; Imamura,~A. Energy Gradient in the Ab
  initio {Hartree--Fock} Crystal-Orbital Formalism of One-Dimensional Infinite
  Polymers. \emph{Chem. Phys. Lett.} \textbf{1983}, \emph{101}, 149--152\relax
\mciteBstWouldAddEndPuncttrue
\mciteSetBstMidEndSepPunct{\mcitedefaultmidpunct}
{\mcitedefaultendpunct}{\mcitedefaultseppunct}\relax
\EndOfBibitem
\bibitem[Karpfen and Beyer(1984)Karpfen, and Beyer]{Karpfen1984}
Karpfen,~A.; Beyer,~A. Ab Initio Studies on Polymers. {VI}. {Torsional}
  Potential in Regular Polyethylene Chains. \emph{J. Comput. Chem.}
  \textbf{1984}, \emph{5}, 11--18\relax
\mciteBstWouldAddEndPuncttrue
\mciteSetBstMidEndSepPunct{\mcitedefaultmidpunct}
{\mcitedefaultendpunct}{\mcitedefaultseppunct}\relax
\EndOfBibitem
\bibitem[Andr\'{e} \latin{et~al.}(1984)Andr\'{e}, Vercauteren, Bodart, and
  Fripiat]{Andre1984}
Andr\'{e},~J.~M.; Vercauteren,~D.~P.; Bodart,~V.~P.; Fripiat,~J.~G. Ab Initio
  Calculations of the Electronic Structure of Helical Polymers. \emph{J.
  Comput. Chem.} \textbf{1984}, \emph{5}, 535--547\relax
\mciteBstWouldAddEndPuncttrue
\mciteSetBstMidEndSepPunct{\mcitedefaultmidpunct}
{\mcitedefaultendpunct}{\mcitedefaultseppunct}\relax
\EndOfBibitem
\bibitem[Teramae \latin{et~al.}(1984)Teramae, Yamabe, and Imamura]{Teramea1984}
Teramae,~H.; Yamabe,~T.; Imamura,~A. Ab Initio Studies on the Geometrical and
  Vibrational Structures of Polymers. \emph{J. Chem. Phys.} \textbf{1984},
  \emph{81}, 3564--3572\relax
\mciteBstWouldAddEndPuncttrue
\mciteSetBstMidEndSepPunct{\mcitedefaultmidpunct}
{\mcitedefaultendpunct}{\mcitedefaultseppunct}\relax
\EndOfBibitem
\bibitem[Springborg and Jones(1986)Springborg, and Jones]{SpringborgSulfur1986}
Springborg,~M.; Jones,~R.~O. Energy Surfaces of Polymeric Sulfur: Structure and
  Electronic Properties. \emph{Phys. Rev. Lett.} \textbf{1986}, \emph{57},
  1145--1148\relax
\mciteBstWouldAddEndPuncttrue
\mciteSetBstMidEndSepPunct{\mcitedefaultmidpunct}
{\mcitedefaultendpunct}{\mcitedefaultseppunct}\relax
\EndOfBibitem
\bibitem[Teramae and Takeda(1989)Teramae, and Takeda]{Teramae1989}
Teramae,~H.; Takeda,~K. Ab Initio Studies on Silicon Compounds. {II}. On the
  Gauche Structure of the Parent Polysilane. \emph{J. Am. Chem. Soc.}
  \textbf{1989}, \emph{111}, 1281--1285\relax
\mciteBstWouldAddEndPuncttrue
\mciteSetBstMidEndSepPunct{\mcitedefaultmidpunct}
{\mcitedefaultendpunct}{\mcitedefaultseppunct}\relax
\EndOfBibitem
\bibitem[Mintmire(1991)]{Mintmire1991}
Mintmire,~J.~W. In \emph{Density Functional Methods in Chemistry};
  Labanowski,~J.~K., Andzelm,~J.~W., Eds.; Springer-Verlag, 1991; pp
  125--137\relax
\mciteBstWouldAddEndPuncttrue
\mciteSetBstMidEndSepPunct{\mcitedefaultmidpunct}
{\mcitedefaultendpunct}{\mcitedefaultseppunct}\relax
\EndOfBibitem
\bibitem[Mintmire \latin{et~al.}(1992)Mintmire, Dunlap, and
  White]{Mintmire1992}
Mintmire,~J.~W.; Dunlap,~B.~I.; White,~C.~T. Are Fullerene Tubules Metallic?
  \emph{Phys. Rev. Lett.} \textbf{1992}, \emph{68}, 631--634\relax
\mciteBstWouldAddEndPuncttrue
\mciteSetBstMidEndSepPunct{\mcitedefaultmidpunct}
{\mcitedefaultendpunct}{\mcitedefaultseppunct}\relax
\EndOfBibitem
\bibitem[Mintmire \latin{et~al.}(1993)Mintmire, Robertson, and
  White]{Mintmire1993}
Mintmire,~J.~W.; Robertson,~D.~H.; White,~C.~T. Properties of Fullerene
  Nanotubules. \emph{J. Phys. Chem. Solids} \textbf{1993}, \emph{54},
  1835--1840\relax
\mciteBstWouldAddEndPuncttrue
\mciteSetBstMidEndSepPunct{\mcitedefaultmidpunct}
{\mcitedefaultendpunct}{\mcitedefaultseppunct}\relax
\EndOfBibitem
\bibitem[Hirata and Iwata(1997)Hirata, and Iwata]{Hirata1997}
Hirata,~S.; Iwata,~S. Density functional crystal orbital study on the normal
  vibrations of polyacetylene and polymethineimine. \emph{J. Chem. Phys.}
  \textbf{1997}, \emph{107}, 10075--10084\relax
\mciteBstWouldAddEndPuncttrue
\mciteSetBstMidEndSepPunct{\mcitedefaultmidpunct}
{\mcitedefaultendpunct}{\mcitedefaultseppunct}\relax
\EndOfBibitem
\bibitem[Hirata \latin{et~al.}(1998)Hirata, Torii, and Tasumi]{Hirata1998}
Hirata,~S.; Torii,~H.; Tasumi,~M. Density-functional crystal orbital study on
  the structures and energetics of polyacetylene isomers. \emph{Phys. Rev. B}
  \textbf{1998}, \emph{57}, 11994--12001\relax
\mciteBstWouldAddEndPuncttrue
\mciteSetBstMidEndSepPunct{\mcitedefaultmidpunct}
{\mcitedefaultendpunct}{\mcitedefaultseppunct}\relax
\EndOfBibitem
\bibitem[Zhang \latin{et~al.}(1999)Zhang, Miao, Van~Doren, Ladik, and
  Mintmire]{ZhangMintmire1999}
Zhang,~M.~L.; Miao,~M.~S.; Van~Doren,~V.~E.; Ladik,~J.~J.; Mintmire,~J.~W.
  Calculation of the total energy per unit cell and of the band structures of
  the five nucleotide base stacks using the local-density approximation.
  \emph{J. Chem. Phys.} \textbf{1999}, \emph{111}, 8696--8700\relax
\mciteBstWouldAddEndPuncttrue
\mciteSetBstMidEndSepPunct{\mcitedefaultmidpunct}
{\mcitedefaultendpunct}{\mcitedefaultseppunct}\relax
\EndOfBibitem
\bibitem[Elizondo and Mintmire(2006)Elizondo, and Mintmire]{Mintmire2006}
Elizondo,~S.~L.; Mintmire,~J.~W. Ab initio study of helical silver single-wall
  nanotubes and nanowires. \emph{Phys. Rev. B} \textbf{2006}, \emph{73}\relax
\mciteBstWouldAddEndPuncttrue
\mciteSetBstMidEndSepPunct{\mcitedefaultmidpunct}
{\mcitedefaultendpunct}{\mcitedefaultseppunct}\relax
\EndOfBibitem
\bibitem[Mintmire(2022)]{Mintmire2022}
Mintmire,~J.~W. Density-functional methods for extended helical systems.
  \emph{Adv. Quantum Chem.} \textbf{2022}, \emph{85}, 177--196\relax
\mciteBstWouldAddEndPuncttrue
\mciteSetBstMidEndSepPunct{\mcitedefaultmidpunct}
{\mcitedefaultendpunct}{\mcitedefaultseppunct}\relax
\EndOfBibitem
\bibitem[Hirata(2022)]{Hirata_MP2022}
Hirata,~S. Nonvanishing quadrature derivatives in the analytical gradients of
  density functional energies in crystals and helices. \emph{Mol. Phys.}
  \textbf{2022}, e2086500\relax
\mciteBstWouldAddEndPuncttrue
\mciteSetBstMidEndSepPunct{\mcitedefaultmidpunct}
{\mcitedefaultendpunct}{\mcitedefaultseppunct}\relax
\EndOfBibitem
\bibitem[Jovanovic and Michl(2019)Jovanovic, and Michl]{Jovanovic2019}
Jovanovic,~M.; Michl,~J. Effect of Conformation on Electron Localization and
  Delocalization in Infinite Helical Chains [{X}({CH}$_3$)$_2$]$_\infty$ ({X} =
  {Si}, {Ge}, {Sn}, and {Pb}). \emph{J. Am. Chem. Soc.} \textbf{2019},
  \emph{141}, 13101--13113\relax
\mciteBstWouldAddEndPuncttrue
\mciteSetBstMidEndSepPunct{\mcitedefaultmidpunct}
{\mcitedefaultendpunct}{\mcitedefaultseppunct}\relax
\EndOfBibitem
\bibitem[Bunn and Howells(1954)Bunn, and Howells]{Bunn_Nature}
Bunn,~C.~W.; Howells,~E.~R. Structures of Molecules and Crystals of
  Fluorocarbons. \emph{Nature} \textbf{1954}, \emph{174}, 549--551\relax
\mciteBstWouldAddEndPuncttrue
\mciteSetBstMidEndSepPunct{\mcitedefaultmidpunct}
{\mcitedefaultendpunct}{\mcitedefaultseppunct}\relax
\EndOfBibitem
\bibitem[Clark(1999)]{Clark1999}
Clark,~E.~S. The molecular conformations of polytetrafluoroethylene: forms {II}
  and {IV}. \emph{Polymer} \textbf{1999}, \emph{40}, 4659--4665\relax
\mciteBstWouldAddEndPuncttrue
\mciteSetBstMidEndSepPunct{\mcitedefaultmidpunct}
{\mcitedefaultendpunct}{\mcitedefaultseppunct}\relax
\EndOfBibitem
\bibitem[D'Amore \latin{et~al.}(2006)D'Amore, Talarico, and Barone]{DAmore2006}
D'Amore,~M.; Talarico,~G.; Barone,~V. Periodic and high-temperature disordered
  conformations of polytetrafluoroethylene chains: An ab initio modeling.
  \emph{J. Am. Chem. Soc.} \textbf{2006}, \emph{128}, 1099--1108\relax
\mciteBstWouldAddEndPuncttrue
\mciteSetBstMidEndSepPunct{\mcitedefaultmidpunct}
{\mcitedefaultendpunct}{\mcitedefaultseppunct}\relax
\EndOfBibitem
\bibitem[Weeks \latin{et~al.}(1981)Weeks, Clark, and Eby]{Weeks1981}
Weeks,~J.~J.; Clark,~E.~S.; Eby,~R.~K. Crystal Structure of the Low-Temperature
  Phase {({II})} of Polytetrafluoroethylene. \emph{Polymer} \textbf{1981},
  \emph{22}, 1480--1486\relax
\mciteBstWouldAddEndPuncttrue
\mciteSetBstMidEndSepPunct{\mcitedefaultmidpunct}
{\mcitedefaultendpunct}{\mcitedefaultseppunct}\relax
\EndOfBibitem
\bibitem[Linderberg and {\"{O}}hrn(1965)Linderberg, and
  {\"{O}}hrn]{Linderberg65}
Linderberg,~J.; {\"{O}}hrn,~Y. Improved Single-Particle Propagators in Theory
  of Conjugated Systems. \emph{Proc. Roy. Soc. (London)} \textbf{1965},
  \emph{A285}, 445\relax
\mciteBstWouldAddEndPuncttrue
\mciteSetBstMidEndSepPunct{\mcitedefaultmidpunct}
{\mcitedefaultendpunct}{\mcitedefaultseppunct}\relax
\EndOfBibitem
\bibitem[Hedin(1965)]{Hedin}
Hedin,~L. New Method for Calculating one-Particle {Green's} Function with
  Application to Electron-Gas Problem. \emph{Phys. Rev.} \textbf{1965},
  \emph{139}, A796\relax
\mciteBstWouldAddEndPuncttrue
\mciteSetBstMidEndSepPunct{\mcitedefaultmidpunct}
{\mcitedefaultendpunct}{\mcitedefaultseppunct}\relax
\EndOfBibitem
\bibitem[{\"{O}}hrn and Linderberg(1965){\"{O}}hrn, and Linderberg]{Ohrn65}
{\"{O}}hrn,~Y.; Linderberg,~J. Propagators for Alternant Hydrocarbon Molecules.
  \emph{Phys. Rev.} \textbf{1965}, \emph{139}, A1063\relax
\mciteBstWouldAddEndPuncttrue
\mciteSetBstMidEndSepPunct{\mcitedefaultmidpunct}
{\mcitedefaultendpunct}{\mcitedefaultseppunct}\relax
\EndOfBibitem
\bibitem[Linderberg and {\"{O}}hrn(1967)Linderberg, and
  {\"{O}}hrn]{Linderberg67}
Linderberg,~J.; {\"{O}}hrn,~Y. Improved decoupling procedure for {Green}
  function. \emph{Chem. Phys. Lett.} \textbf{1967}, \emph{1}, 295--296\relax
\mciteBstWouldAddEndPuncttrue
\mciteSetBstMidEndSepPunct{\mcitedefaultmidpunct}
{\mcitedefaultendpunct}{\mcitedefaultseppunct}\relax
\EndOfBibitem
\bibitem[Goscinski and Lukman(1970)Goscinski, and Lukman]{GoscinskiLukman}
Goscinski,~O.; Lukman,~B. Moment-conserving decoupling of {Green} functions via
  {Pad\'e} approximants. \emph{Chem. Phys. Lett.} \textbf{1970}, \emph{7},
  573--576\relax
\mciteBstWouldAddEndPuncttrue
\mciteSetBstMidEndSepPunct{\mcitedefaultmidpunct}
{\mcitedefaultendpunct}{\mcitedefaultseppunct}\relax
\EndOfBibitem
\bibitem[Doll and Reinhard(1972)Doll, and Reinhard]{Doll}
Doll,~J.~D.; Reinhard,~W.~P. Many-Body {Green's} Functions for Finite,
  Nonuniform Systems: Applications to Closed Shell Atoms. \emph{J. Chem. Phys.}
  \textbf{1972}, \emph{57}, 1169\relax
\mciteBstWouldAddEndPuncttrue
\mciteSetBstMidEndSepPunct{\mcitedefaultmidpunct}
{\mcitedefaultendpunct}{\mcitedefaultseppunct}\relax
\EndOfBibitem
\bibitem[Pickup and Goscinski(1973)Pickup, and Goscinski]{pickup}
Pickup,~B.~T.; Goscinski,~O. Direct Calculation of Ionization Energies. {I}.
  {Closed} Shells. \emph{Mol. Phys.} \textbf{1973}, \emph{26}, 1013--1035\relax
\mciteBstWouldAddEndPuncttrue
\mciteSetBstMidEndSepPunct{\mcitedefaultmidpunct}
{\mcitedefaultendpunct}{\mcitedefaultseppunct}\relax
\EndOfBibitem
\bibitem[Yarlagadda \latin{et~al.}(1973)Yarlagadda, Csanak, Taylor, Schneider,
  and Yaris]{Yaris}
Yarlagadda,~B.~S.; Csanak,~G.; Taylor,~H.~S.; Schneider,~B.; Yaris,~R.
  Application of Many-Body {Green's} Functions to Scattering and Bound-State
  Properties of Helium. \emph{Phys. Rev. A} \textbf{1973}, \emph{7},
  146--154\relax
\mciteBstWouldAddEndPuncttrue
\mciteSetBstMidEndSepPunct{\mcitedefaultmidpunct}
{\mcitedefaultendpunct}{\mcitedefaultseppunct}\relax
\EndOfBibitem
\bibitem[Linderberg and {\"{O}}hrn(1973)Linderberg, and
  {\"{O}}hrn]{linderbergohrn}
Linderberg,~J.; {\"{O}}hrn,~Y. \emph{Propagators in Quantum Chemistry};
  Academic Press: London, 1973\relax
\mciteBstWouldAddEndPuncttrue
\mciteSetBstMidEndSepPunct{\mcitedefaultmidpunct}
{\mcitedefaultendpunct}{\mcitedefaultseppunct}\relax
\EndOfBibitem
\bibitem[Tsui and Freed(1974)Tsui, and Freed]{Freed74}
Tsui,~F. S.~M.; Freed,~K.~F. Relationship between One-Electron {Green's}
  Function and Quantum Chemical Theories. \emph{Chem. Phys.} \textbf{1974},
  \emph{5}, 337--349\relax
\mciteBstWouldAddEndPuncttrue
\mciteSetBstMidEndSepPunct{\mcitedefaultmidpunct}
{\mcitedefaultendpunct}{\mcitedefaultseppunct}\relax
\EndOfBibitem
\bibitem[Paldus and \v{C}\'{i}\v{z}ek(1975)Paldus, and
  \v{C}\'{i}\v{z}ek]{paldus}
Paldus,~J.; \v{C}\'{i}\v{z}ek,~J. Time-Independent Diagrammatic Approach to
  Perturbation Theory of Fermion Systems. \emph{Adv. Quantum Chem.}
  \textbf{1975}, \emph{9}, 105--197\relax
\mciteBstWouldAddEndPuncttrue
\mciteSetBstMidEndSepPunct{\mcitedefaultmidpunct}
{\mcitedefaultendpunct}{\mcitedefaultseppunct}\relax
\EndOfBibitem
\bibitem[Cederbaum(1975)]{Ceder}
Cederbaum,~L.~S. One-Body {Green's} Function for Atoms and Molecules: Theory
  and Application. \emph{J. Phys. B:\ At. Mol. Phys.} \textbf{1975}, \emph{8},
  290--303\relax
\mciteBstWouldAddEndPuncttrue
\mciteSetBstMidEndSepPunct{\mcitedefaultmidpunct}
{\mcitedefaultendpunct}{\mcitedefaultseppunct}\relax
\EndOfBibitem
\bibitem[Cederbaum and Domcke(1977)Cederbaum, and Domcke]{cederbaumacp}
Cederbaum,~L.~S.; Domcke,~W. Theoretical Aspects of Ionization Potentials and
  Photoelectron Spectroscopy: A {Green's} Function Approach. \emph{Adv. Chem.
  Phys.} \textbf{1977}, \emph{36}, 205--344\relax
\mciteBstWouldAddEndPuncttrue
\mciteSetBstMidEndSepPunct{\mcitedefaultmidpunct}
{\mcitedefaultendpunct}{\mcitedefaultseppunct}\relax
\EndOfBibitem
\bibitem[Simons(1977)]{simonsrev}
Simons,~J. Theoretical Studies of Negative Molecular Ions. \emph{Annu. Rev.
  Phys. Chem.} \textbf{1977}, \emph{28}, 15--45\relax
\mciteBstWouldAddEndPuncttrue
\mciteSetBstMidEndSepPunct{\mcitedefaultmidpunct}
{\mcitedefaultendpunct}{\mcitedefaultseppunct}\relax
\EndOfBibitem
\bibitem[Herman \latin{et~al.}(1978)Herman, Yeager, and Freed]{herman}
Herman,~M.~F.; Yeager,~D.~L.; Freed,~K.~F. Analysis of Third Order
  Contributions to Equations of Motion {Green's} Function Ionization
  Potentials: Application to {N}$_2$. \emph{Chem. Phys.} \textbf{1978},
  \emph{29}, 77--96\relax
\mciteBstWouldAddEndPuncttrue
\mciteSetBstMidEndSepPunct{\mcitedefaultmidpunct}
{\mcitedefaultendpunct}{\mcitedefaultseppunct}\relax
\EndOfBibitem
\bibitem[Baker and Pickup(1980)Baker, and Pickup]{BakerPickup}
Baker,~J.; Pickup,~B.~T. A Method for Molecular Ionization Potentials.
  \emph{Chem. Phys. Lett.} \textbf{1980}, \emph{76}, 537--541\relax
\mciteBstWouldAddEndPuncttrue
\mciteSetBstMidEndSepPunct{\mcitedefaultmidpunct}
{\mcitedefaultendpunct}{\mcitedefaultseppunct}\relax
\EndOfBibitem
\bibitem[\"{O}hrn and Born(1981)\"{O}hrn, and Born]{ohrnborn}
\"{O}hrn,~Y.; Born,~G. Molecular Electron Propagator Theory and Calculations.
  \emph{Adv. Quantum Chem.} \textbf{1981}, \emph{13}, 1--88\relax
\mciteBstWouldAddEndPuncttrue
\mciteSetBstMidEndSepPunct{\mcitedefaultmidpunct}
{\mcitedefaultendpunct}{\mcitedefaultseppunct}\relax
\EndOfBibitem
\bibitem[J{\o}rgensen and Simons(1981)J{\o}rgensen, and
  Simons]{jorgensensimons}
J{\o}rgensen,~P.; Simons,~J. \emph{Second Quantization-Based Methods in Quantum
  Chemistry}; Academic Press: New York, 1981\relax
\mciteBstWouldAddEndPuncttrue
\mciteSetBstMidEndSepPunct{\mcitedefaultmidpunct}
{\mcitedefaultendpunct}{\mcitedefaultseppunct}\relax
\EndOfBibitem
\bibitem[Schirmer(1982)]{schirmer1982}
Schirmer,~J. Beyond the Random-Phase Approximation: A New Approximation Scheme
  for the Polarization Propagator. \emph{Phys. Rev. A} \textbf{1982},
  \emph{26}, 2395--2416\relax
\mciteBstWouldAddEndPuncttrue
\mciteSetBstMidEndSepPunct{\mcitedefaultmidpunct}
{\mcitedefaultendpunct}{\mcitedefaultseppunct}\relax
\EndOfBibitem
\bibitem[Schirmer \latin{et~al.}(1983)Schirmer, Cederbaum, and
  Walter]{schirmer}
Schirmer,~J.; Cederbaum,~L.~S.; Walter,~O. New Approach to the One-Particle
  {Green's}-Function for Finite Fermi Systems. \emph{Phys. Rev. A}
  \textbf{1983}, \emph{28}, 1237--1259\relax
\mciteBstWouldAddEndPuncttrue
\mciteSetBstMidEndSepPunct{\mcitedefaultmidpunct}
{\mcitedefaultendpunct}{\mcitedefaultseppunct}\relax
\EndOfBibitem
\bibitem[von Niessen \latin{et~al.}(1984)von Niessen, Schirmer, and
  Cederbaum]{vonniessen}
von Niessen,~W.; Schirmer,~J.; Cederbaum,~L.~S. \emph{Comput. Phys. Reports}
  \textbf{1984}, \emph{1}, 57--125\relax
\mciteBstWouldAddEndPuncttrue
\mciteSetBstMidEndSepPunct{\mcitedefaultmidpunct}
{\mcitedefaultendpunct}{\mcitedefaultseppunct}\relax
\EndOfBibitem
\bibitem[Prasad \latin{et~al.}(1985)Prasad, Pal, and Mukherjee]{Prasad}
Prasad,~M.~D.; Pal,~S.; Mukherjee,~D. Some Aspects of Self-Consistent
  Propagator Theories. \emph{Phys. Rev. A} \textbf{1985}, \emph{31},
  1287--1298\relax
\mciteBstWouldAddEndPuncttrue
\mciteSetBstMidEndSepPunct{\mcitedefaultmidpunct}
{\mcitedefaultendpunct}{\mcitedefaultseppunct}\relax
\EndOfBibitem
\bibitem[Hybertsen and Louie(1986)Hybertsen, and Louie]{GWLouie}
Hybertsen,~M.~S.; Louie,~S.~G. Electron Correlation in Semiconductors and
  Insulators: Band-Gaps and Quasi-Particle Energies. \emph{Phys. Rev. B}
  \textbf{1986}, \emph{34}, 5390--5413\relax
\mciteBstWouldAddEndPuncttrue
\mciteSetBstMidEndSepPunct{\mcitedefaultmidpunct}
{\mcitedefaultendpunct}{\mcitedefaultseppunct}\relax
\EndOfBibitem
\bibitem[Oddershede(1987)]{oddershede}
Oddershede,~J. Propagator methods. \emph{Adv. Chem. Phys.} \textbf{1987},
  \emph{69}, 201--239\relax
\mciteBstWouldAddEndPuncttrue
\mciteSetBstMidEndSepPunct{\mcitedefaultmidpunct}
{\mcitedefaultendpunct}{\mcitedefaultseppunct}\relax
\EndOfBibitem
\bibitem[Kutzelnigg and Mukherjee(1997)Kutzelnigg, and Mukherjee]{Kutzelnigg}
Kutzelnigg,~W.; Mukherjee,~D. Normal order and extended {Wick} theorem for a
  multiconfiguration reference wave function. \emph{J. Chem. Phys.}
  \textbf{1997}, \emph{107}, 432--449\relax
\mciteBstWouldAddEndPuncttrue
\mciteSetBstMidEndSepPunct{\mcitedefaultmidpunct}
{\mcitedefaultendpunct}{\mcitedefaultseppunct}\relax
\EndOfBibitem
\bibitem[Aryasetiawan and Gunnarsson(1998)Aryasetiawan, and Gunnarsson]{GW1}
Aryasetiawan,~F.; Gunnarsson,~O. The {GW} method. \emph{Rep. Prog. Phys.}
  \textbf{1998}, \emph{61}, 237--312\relax
\mciteBstWouldAddEndPuncttrue
\mciteSetBstMidEndSepPunct{\mcitedefaultmidpunct}
{\mcitedefaultendpunct}{\mcitedefaultseppunct}\relax
\EndOfBibitem
\bibitem[Ortiz(1999)]{ortiz_aqc}
Ortiz,~J.~V. Toward an exact one-electron picture of chemical bonding.
  \emph{Adv. Quantum Chem.} \textbf{1999}, \emph{35}, 33--52\relax
\mciteBstWouldAddEndPuncttrue
\mciteSetBstMidEndSepPunct{\mcitedefaultmidpunct}
{\mcitedefaultendpunct}{\mcitedefaultseppunct}\relax
\EndOfBibitem
\bibitem[Onida \latin{et~al.}(2002)Onida, Reining, and Rubio]{GW2}
Onida,~G.; Reining,~L.; Rubio,~A. Electronic excitations: density-functional
  versus many-body {Green's}-function approaches. \emph{Rev. Mod. Phys.}
  \textbf{2002}, \emph{74}, 601--659\relax
\mciteBstWouldAddEndPuncttrue
\mciteSetBstMidEndSepPunct{\mcitedefaultmidpunct}
{\mcitedefaultendpunct}{\mcitedefaultseppunct}\relax
\EndOfBibitem
\bibitem[Ortiz(2013)]{Wire}
Ortiz,~J.~V. Electron propagator theory: An approach to prediction and
  interpretation in quantum chemistry. \emph{WIREs Comput. Mol. Sci.}
  \textbf{2013}, \emph{3}, 123--142\relax
\mciteBstWouldAddEndPuncttrue
\mciteSetBstMidEndSepPunct{\mcitedefaultmidpunct}
{\mcitedefaultendpunct}{\mcitedefaultseppunct}\relax
\EndOfBibitem
\bibitem[Szabo and Ostlund(1982)Szabo, and Ostlund]{szabo}
Szabo,~A.; Ostlund,~N.~S. \emph{Modern Quantum Chemistry}; MacMillan: New York,
  NY, 1982\relax
\mciteBstWouldAddEndPuncttrue
\mciteSetBstMidEndSepPunct{\mcitedefaultmidpunct}
{\mcitedefaultendpunct}{\mcitedefaultseppunct}\relax
\EndOfBibitem
\bibitem[Hirata \latin{et~al.}(2017)Hirata, Doran, Knowles, and
  Ortiz]{Hirata2017}
Hirata,~S.; Doran,~A.~E.; Knowles,~P.~J.; Ortiz,~J.~V. One-particle many-body
  {Green's} function theory: Algebraic recursive definitions, linked-diagram
  theorem, irreducible-diagram theorem, and general-order algorithms. \emph{J.
  Chem. Phys.} \textbf{2017}, \emph{147}, 044108\relax
\mciteBstWouldAddEndPuncttrue
\mciteSetBstMidEndSepPunct{\mcitedefaultmidpunct}
{\mcitedefaultendpunct}{\mcitedefaultseppunct}\relax
\EndOfBibitem
\bibitem[Bower and Maddams(1989)Bower, and Maddams]{Bower1989}
Bower,~D.~I.; Maddams,~W.~F. \emph{The Vibrational Spectroscopy of Polymers};
  Cambridge University Press: Cambridge, 1989\relax
\mciteBstWouldAddEndPuncttrue
\mciteSetBstMidEndSepPunct{\mcitedefaultmidpunct}
{\mcitedefaultendpunct}{\mcitedefaultseppunct}\relax
\EndOfBibitem
\bibitem[Hauser \latin{et~al.}(2017)Hauser, He, Garcia-Diaz, Simmerling, and
  Coutsias]{Hauser}
Hauser,~K.; He,~Y.~Q.; Garcia-Diaz,~M.; Simmerling,~C.; Coutsias,~E.
  Characterization of Biomolecular Helices and Their Complementarity Using
  Geometric Analysis. \emph{J. Chem. Inf. Model.} \textbf{2017}, \emph{57},
  864--874\relax
\mciteBstWouldAddEndPuncttrue
\mciteSetBstMidEndSepPunct{\mcitedefaultmidpunct}
{\mcitedefaultendpunct}{\mcitedefaultseppunct}\relax
\EndOfBibitem
\bibitem[Kittel(1963)]{Kittel}
Kittel,~C. \emph{Quantum Theory of Solids}, 2nd ed.; Wiley: Hoboken, 1963\relax
\mciteBstWouldAddEndPuncttrue
\mciteSetBstMidEndSepPunct{\mcitedefaultmidpunct}
{\mcitedefaultendpunct}{\mcitedefaultseppunct}\relax
\EndOfBibitem
\bibitem[Delhalle \latin{et~al.}(1980)Delhalle, Piela, Br\'{e}das, and
  Andr\'{e}]{Delhalle}
Delhalle,~J.; Piela,~L.; Br\'{e}das,~J.~L.; Andr\'{e},~J.-M. Multipole
  Expansion in Tight-Binding {Hartree--Fock} Calculations for Infinite-Model
  Polymers. \emph{Phys. Rev. B} \textbf{1980}, \emph{22}, 6254--6267\relax
\mciteBstWouldAddEndPuncttrue
\mciteSetBstMidEndSepPunct{\mcitedefaultmidpunct}
{\mcitedefaultendpunct}{\mcitedefaultseppunct}\relax
\EndOfBibitem
\bibitem[Obara and Saika(1986)Obara, and Saika]{Obara1986}
Obara,~S.; Saika,~A. Efficient recursive computation of molecular integrals
  over {Cartesian Gaussian} functions. \emph{J. Chem. Phys.} \textbf{1986},
  \emph{84}, 3963\relax
\mciteBstWouldAddEndPuncttrue
\mciteSetBstMidEndSepPunct{\mcitedefaultmidpunct}
{\mcitedefaultendpunct}{\mcitedefaultseppunct}\relax
\EndOfBibitem
\bibitem[Ortiz(2020)]{OrtizDyson}
Ortiz,~J.~V. Dyson-orbital concepts for description of electrons in molecules.
  \emph{J. Chem. Phys.} \textbf{2020}, \emph{153}, 070902\relax
\mciteBstWouldAddEndPuncttrue
\mciteSetBstMidEndSepPunct{\mcitedefaultmidpunct}
{\mcitedefaultendpunct}{\mcitedefaultseppunct}\relax
\EndOfBibitem
\bibitem[Hirata and Bartlett(2000)Hirata, and Bartlett]{Hirata_GF2}
Hirata,~S.; Bartlett,~R.~J. Many-body {Green's}-function calculations on the
  electronic excited states of extended systems. \emph{J. Chem. Phys.}
  \textbf{2000}, \emph{112}, 7339--7344\relax
\mciteBstWouldAddEndPuncttrue
\mciteSetBstMidEndSepPunct{\mcitedefaultmidpunct}
{\mcitedefaultendpunct}{\mcitedefaultseppunct}\relax
\EndOfBibitem
\bibitem[Suhai(1983)]{suhai_qp}
Suhai,~S. Quasiparticle Energy-Band Structures in Semiconducting Polymers:
  Correlation-Effects on the Band-Gap in Polyacetylene. \emph{Phys. Rev. B}
  \textbf{1983}, \emph{27}, 3506--3518\relax
\mciteBstWouldAddEndPuncttrue
\mciteSetBstMidEndSepPunct{\mcitedefaultmidpunct}
{\mcitedefaultendpunct}{\mcitedefaultseppunct}\relax
\EndOfBibitem
\bibitem[Sun and Bartlett(1996)Sun, and Bartlett]{SunBartlett1996}
Sun,~J.~Q.; Bartlett,~R.~J. Second-order many-body perturbation-theory
  calculations in extended systems. \emph{J. Chem. Phys.} \textbf{1996},
  \emph{104}, 8553--8565\relax
\mciteBstWouldAddEndPuncttrue
\mciteSetBstMidEndSepPunct{\mcitedefaultmidpunct}
{\mcitedefaultendpunct}{\mcitedefaultseppunct}\relax
\EndOfBibitem
\bibitem[Sun and Bartlett(1996)Sun, and Bartlett]{sun_qp}
Sun,~J.-Q.; Bartlett,~R.~J. Correlated prediction of the photoelectron spectrum
  of polyethylene: Explanation of {XPS} and {UPS} measurements. \emph{Phys.
  Rev. Lett.} \textbf{1996}, \emph{77}, 3669--3672\relax
\mciteBstWouldAddEndPuncttrue
\mciteSetBstMidEndSepPunct{\mcitedefaultmidpunct}
{\mcitedefaultendpunct}{\mcitedefaultseppunct}\relax
\EndOfBibitem
\bibitem[Hirata \latin{et~al.}(2015)Hirata, Hermes, Simons, and Ortiz]{deltamp}
Hirata,~S.; Hermes,~M.~R.; Simons,~J.; Ortiz,~J.~V. General-Order Many-Body
  {Green's} Function Method. \emph{J. Chem. Theory Comput.} \textbf{2015},
  \emph{11}, 1595--1606\relax
\mciteBstWouldAddEndPuncttrue
\mciteSetBstMidEndSepPunct{\mcitedefaultmidpunct}
{\mcitedefaultendpunct}{\mcitedefaultseppunct}\relax
\EndOfBibitem
\bibitem[Kunz(1972)]{Kunz1972}
Kunz,~A.~B. Electronic Polarons in Nonmetals. \emph{Phys. Rev. B}
  \textbf{1972}, \emph{6}, 606--615\relax
\mciteBstWouldAddEndPuncttrue
\mciteSetBstMidEndSepPunct{\mcitedefaultmidpunct}
{\mcitedefaultendpunct}{\mcitedefaultseppunct}\relax
\EndOfBibitem
\bibitem[Liegener(1988)]{Liegener1988}
Liegener,~C.-M. {\it Ab initio} Calculations of Correlation Effects in {\it
  Trans}-Polyacetylene. \emph{J. Chem. Phys.} \textbf{1988}, \emph{88},
  6999--7004\relax
\mciteBstWouldAddEndPuncttrue
\mciteSetBstMidEndSepPunct{\mcitedefaultmidpunct}
{\mcitedefaultendpunct}{\mcitedefaultseppunct}\relax
\EndOfBibitem
\bibitem[Suhai(1983)]{SuhaiCPL}
Suhai,~S. Bond Alternation in Infinite Polyene:\ {Peierls} Distortion Reduced
  by Electron Correlation. \emph{Chem. Phys. Lett.} \textbf{1983}, \emph{96},
  619--625\relax
\mciteBstWouldAddEndPuncttrue
\mciteSetBstMidEndSepPunct{\mcitedefaultmidpunct}
{\mcitedefaultendpunct}{\mcitedefaultseppunct}\relax
\EndOfBibitem
\bibitem[Ye \latin{et~al.}(1993)Ye, F\"orner, and Ladik]{Ye1993}
Ye,~Y.~J.; F\"orner,~W.; Ladik,~J. Numerical Application of the Coupled-Cluster
  Theory with Localized Orbitals to Polymers. {I}. {Total} Correlation Energy
  Per Unit Cell. \emph{Chem. Phys.} \textbf{1993}, \emph{178}, 1--23\relax
\mciteBstWouldAddEndPuncttrue
\mciteSetBstMidEndSepPunct{\mcitedefaultmidpunct}
{\mcitedefaultendpunct}{\mcitedefaultseppunct}\relax
\EndOfBibitem
\bibitem[Suhai(1994)]{SuhaiJCP1994}
Suhai,~S. Cooperative Effects in Hydrogen Bonding: Fourth-Order Many-Body
  Perturbation Theory Studies of Water Oligomers and of an Infinite Water Chain
  as a Model for Ice. \emph{J. Chem. Phys.} \textbf{1994}, \emph{101},
  9766--9782\relax
\mciteBstWouldAddEndPuncttrue
\mciteSetBstMidEndSepPunct{\mcitedefaultmidpunct}
{\mcitedefaultendpunct}{\mcitedefaultseppunct}\relax
\EndOfBibitem
\bibitem[Suhai(1994)]{SuhaiPRB1994}
Suhai,~S. Electron Correlation in Extended Systems: Fourth-Order Many-Body
  Perturbation Theory and Density-Functional Methods Applied to an Infinite
  Chain of Hydrogen Atoms. \emph{Phys. Rev. B} \textbf{1994}, \emph{50},
  14791--14801\relax
\mciteBstWouldAddEndPuncttrue
\mciteSetBstMidEndSepPunct{\mcitedefaultmidpunct}
{\mcitedefaultendpunct}{\mcitedefaultseppunct}\relax
\EndOfBibitem
\bibitem[Suhai(1995)]{Suhai1995}
Suhai,~S. Electron Correlation and Dimerization in {\it trans}-polyacetylene:
  Many-Body Perturbation Theory Versus Density-Functional Methods. \emph{Phys.
  Rev. B} \textbf{1995}, \emph{51}, 16553--16567\relax
\mciteBstWouldAddEndPuncttrue
\mciteSetBstMidEndSepPunct{\mcitedefaultmidpunct}
{\mcitedefaultendpunct}{\mcitedefaultseppunct}\relax
\EndOfBibitem
\bibitem[Hirata and Iwata(1998)Hirata, and Iwata]{HirataIwata1998}
Hirata,~S.; Iwata,~S. Analytical energy gradients in second-order {M}\o
  ller--{P}lesset perturbation theory for extended systems. \emph{J. Chem.
  Phys.} \textbf{1998}, \emph{109}, 4147--4155\relax
\mciteBstWouldAddEndPuncttrue
\mciteSetBstMidEndSepPunct{\mcitedefaultmidpunct}
{\mcitedefaultendpunct}{\mcitedefaultseppunct}\relax
\EndOfBibitem
\bibitem[Ayala \latin{et~al.}(2001)Ayala, Kudin, and Scuseria]{Ayala2001}
Ayala,~P.~Y.; Kudin,~K.~N.; Scuseria,~G.~E. Atomic orbital
  {Laplace}-transformed second-order {M}\o ller--{P}lesset theory for periodic
  systems. \emph{J. Chem. Phys.} \textbf{2001}, \emph{115}, 9698--9707\relax
\mciteBstWouldAddEndPuncttrue
\mciteSetBstMidEndSepPunct{\mcitedefaultmidpunct}
{\mcitedefaultendpunct}{\mcitedefaultseppunct}\relax
\EndOfBibitem
\bibitem[Hirata \latin{et~al.}(2004)Hirata, Podeszwa, Tobita, and
  Bartlett]{HirataCC2004}
Hirata,~S.; Podeszwa,~R.; Tobita,~M.; Bartlett,~R.~J. Coupled-cluster singles
  and doubles for extended systems. \emph{J. Chem. Phys.} \textbf{2004},
  \emph{120}, 2581--2592\relax
\mciteBstWouldAddEndPuncttrue
\mciteSetBstMidEndSepPunct{\mcitedefaultmidpunct}
{\mcitedefaultendpunct}{\mcitedefaultseppunct}\relax
\EndOfBibitem
\bibitem[Shimazaki and Hirata(2009)Shimazaki, and Hirata]{Shimazaki2009}
Shimazaki,~T.; Hirata,~S. On the {Brillouin}-Zone Integrations in Second-Order
  Many-Body Perturbation Calculations for Extended Systems of One-Dimensional
  Periodicity. \emph{Int. J. Quantum Chem.} \textbf{2009}, \emph{109},
  2953--2959\relax
\mciteBstWouldAddEndPuncttrue
\mciteSetBstMidEndSepPunct{\mcitedefaultmidpunct}
{\mcitedefaultendpunct}{\mcitedefaultseppunct}\relax
\EndOfBibitem
\bibitem[Hirata and Shimazaki(2009)Hirata, and Shimazaki]{hirata_qp}
Hirata,~S.; Shimazaki,~T. Fast second-order many-body perturbation method for
  extended systems. \emph{Phys. Rev. B} \textbf{2009}, \emph{80}, 085118\relax
\mciteBstWouldAddEndPuncttrue
\mciteSetBstMidEndSepPunct{\mcitedefaultmidpunct}
{\mcitedefaultendpunct}{\mcitedefaultseppunct}\relax
\EndOfBibitem
\bibitem[Ohnishi and Hirata(2010)Ohnishi, and Hirata]{Ohnishi2010}
Ohnishi,~Y.~Y.; Hirata,~S. Logarithm second-order many-body perturbation method
  for extended systems. \emph{J. Chem. Phys.} \textbf{2010}, \emph{133},
  034106\relax
\mciteBstWouldAddEndPuncttrue
\mciteSetBstMidEndSepPunct{\mcitedefaultmidpunct}
{\mcitedefaultendpunct}{\mcitedefaultseppunct}\relax
\EndOfBibitem
\bibitem[Pireaux \latin{et~al.}(1976)Pireaux, Svensson, Basilier, Malmqvist,
  Gelius, Gaudano, and Siegbahn]{Pireaux_PE_UPS}
Pireaux,~J.~J.; Svensson,~S.; Basilier,~E.; Malmqvist,~P.-{\AA}.; Gelius,~U.;
  Gaudano,~R.; Siegbahn,~K. Core-Electron Relaxation Energies and Valence-Band
  Formation of Linear Alkanes Studied in Gas-Phase by Means of Electron
  Spectroscopy. \emph{Phys. Rev. A} \textbf{1976}, \emph{14}, 2133--2145\relax
\mciteBstWouldAddEndPuncttrue
\mciteSetBstMidEndSepPunct{\mcitedefaultmidpunct}
{\mcitedefaultendpunct}{\mcitedefaultseppunct}\relax
\EndOfBibitem
\bibitem[Ueno \latin{et~al.}(1990)Ueno, Seki, Sato, Fujimoto, Kuramochi,
  Sugita, and Inokuchi]{Ueno_PE_ARUP}
Ueno,~N.; Seki,~K.; Sato,~N.; Fujimoto,~H.; Kuramochi,~T.; Sugita,~K.;
  Inokuchi,~H. Energy-Band Dispersion in Oriented Thin-Films of
  Pentatriacontan-18-One by Angle-Resolved Photoemission with Synchrotron
  Radiation. \emph{Phys. Rev. B} \textbf{1990}, \emph{41}, 1176--1183\relax
\mciteBstWouldAddEndPuncttrue
\mciteSetBstMidEndSepPunct{\mcitedefaultmidpunct}
{\mcitedefaultendpunct}{\mcitedefaultseppunct}\relax
\EndOfBibitem
\bibitem[Shearer and Vand(1956)Shearer, and Vand]{Shearer1956}
Shearer,~H. M.~M.; Vand,~V. The Crystal Structure of the Monoclinic Form of
  Normal-Hexatriacontane. \emph{Acta Crystallogr.} \textbf{1956}, \emph{9},
  379--384\relax
\mciteBstWouldAddEndPuncttrue
\mciteSetBstMidEndSepPunct{\mcitedefaultmidpunct}
{\mcitedefaultendpunct}{\mcitedefaultseppunct}\relax
\EndOfBibitem
\bibitem[Krimm \latin{et~al.}(1956)Krimm, Liang, and Sutherland]{Krimm1956}
Krimm,~S.; Liang,~C.~Y.; Sutherland,~G. B. B.~M. Infrared Spectra of High
  Polymers. {II}. {Polyethylene}. \emph{J. Chem. Phys.} \textbf{1956},
  \emph{25}, 549--562\relax
\mciteBstWouldAddEndPuncttrue
\mciteSetBstMidEndSepPunct{\mcitedefaultmidpunct}
{\mcitedefaultendpunct}{\mcitedefaultseppunct}\relax
\EndOfBibitem
\bibitem[Nielsen and Woollett(1957)Nielsen, and Woollett]{Nielsen1957}
Nielsen,~J.~R.; Woollett,~A.~H. Vibrational Spectra of Polyethylenes and
  Related Substances. \emph{J. Chem. Phys.} \textbf{1957}, \emph{26},
  1391--1400\relax
\mciteBstWouldAddEndPuncttrue
\mciteSetBstMidEndSepPunct{\mcitedefaultmidpunct}
{\mcitedefaultendpunct}{\mcitedefaultseppunct}\relax
\EndOfBibitem
\bibitem[Nielsen and Holland(1961)Nielsen, and Holland]{Nielsen1961}
Nielsen,~J.~R.; Holland,~R.~F. Dichroism and interpretation of the infrared
  bands of oriented crystalline polyethylene. \emph{J. Mol. Spectrosc.}
  \textbf{1961}, \emph{6}, 394--418\relax
\mciteBstWouldAddEndPuncttrue
\mciteSetBstMidEndSepPunct{\mcitedefaultmidpunct}
{\mcitedefaultendpunct}{\mcitedefaultseppunct}\relax
\EndOfBibitem
\bibitem[Brown(1963)]{Brown1963}
Brown,~R.~G. Raman Spectra of Polyethylenes. \emph{J. Chem. Phys.}
  \textbf{1963}, \emph{38}, 221\relax
\mciteBstWouldAddEndPuncttrue
\mciteSetBstMidEndSepPunct{\mcitedefaultmidpunct}
{\mcitedefaultendpunct}{\mcitedefaultseppunct}\relax
\EndOfBibitem
\bibitem[Snyder and Schachtschneider(1963)Snyder, and
  Schachtschneider]{Snyder1963}
Snyder,~R.~G.; Schachtschneider,~J.~H. Vibrational Analysis of the {\it
  n}-Paraffins. {I}. {Assignments} of Infrared Bands in the Spectra of
  {C}$_3${H}$_8$ through {\it n}-{C}$_{19}${H}$_{40}$. \emph{Spectrochim. Acta}
  \textbf{1963}, \emph{19}, 85--116\relax
\mciteBstWouldAddEndPuncttrue
\mciteSetBstMidEndSepPunct{\mcitedefaultmidpunct}
{\mcitedefaultendpunct}{\mcitedefaultseppunct}\relax
\EndOfBibitem
\bibitem[Parker(1996)]{Parker_INS}
Parker,~S.~F. Inelastic neutron scattering spectra of polyethylene. \emph{J.
  Chem. Soc., Faraday Trans.} \textbf{1996}, \emph{92}, 1941--1946\relax
\mciteBstWouldAddEndPuncttrue
\mciteSetBstMidEndSepPunct{\mcitedefaultmidpunct}
{\mcitedefaultendpunct}{\mcitedefaultseppunct}\relax
\EndOfBibitem
\bibitem[Hirata and Iwata(1998)Hirata, and Iwata]{HirataIwata_PE}
Hirata,~S.; Iwata,~S. Density functional crystal orbital study on the normal
  vibrations and phonon dispersion curves of all-{\it trans} polyethylene.
  \emph{J. Chem. Phys.} \textbf{1998}, \emph{108}, 7901--7908\relax
\mciteBstWouldAddEndPuncttrue
\mciteSetBstMidEndSepPunct{\mcitedefaultmidpunct}
{\mcitedefaultendpunct}{\mcitedefaultseppunct}\relax
\EndOfBibitem
\bibitem[Becke(1993)]{B3LYP}
Becke,~A.~D. Density-Functional Thermochemistry. {III}. {The} Role of Exact
  Exchange. \emph{J. Chem. Phys.} \textbf{1993}, \emph{98}, 5648--5652\relax
\mciteBstWouldAddEndPuncttrue
\mciteSetBstMidEndSepPunct{\mcitedefaultmidpunct}
{\mcitedefaultendpunct}{\mcitedefaultseppunct}\relax
\EndOfBibitem
\bibitem[Vosko \latin{et~al.}(1980)Vosko, Wilk, and Nusair]{VWN}
Vosko,~S.~H.; Wilk,~L.; Nusair,~M. Accurate Spin-Dependent Electron Liquid
  Correlation Energies for Local Spin-Density Calculations: A Critical
  Analysis. \emph{Can. J. Phys.} \textbf{1980}, \emph{58}, 1200--1211\relax
\mciteBstWouldAddEndPuncttrue
\mciteSetBstMidEndSepPunct{\mcitedefaultmidpunct}
{\mcitedefaultendpunct}{\mcitedefaultseppunct}\relax
\EndOfBibitem
\bibitem[Hirata(2022)]{polymer}
Hirata,~S. {\sc polymer 3.0}. 2022\relax
\mciteBstWouldAddEndPuncttrue
\mciteSetBstMidEndSepPunct{\mcitedefaultmidpunct}
{\mcitedefaultendpunct}{\mcitedefaultseppunct}\relax
\EndOfBibitem
\bibitem[mis()]{misc}
The vibrational frequencies at the phase angle $\theta=0$ of (NH)$_x$ and
  (O)$_x$ were obtained by the crystal-orbital method with {\sc
  polymer}\cite{polymer} using a large unit cell of 13 NH or 13 O groups and
  also by the supercell method of 19 NH or 19 O groups with
  NWChem.\cite{nwchem} They agree with each other within 6 cm$^{-1}$ except for
  the lowest optical mode whose frequencies agree within 26 cm$^{-1}$. Since
  phonons at $\theta \neq 0$ destroy periodic symmetry, the intrinsic
  efficiency of the crystal-orbital method is lost. Hence, we elected to use
  the supercell method and molecular software for the normal-mode
  analysis.\relax
\mciteBstWouldAddEndPunctfalse
\mciteSetBstMidEndSepPunct{\mcitedefaultmidpunct}
{}{\mcitedefaultseppunct}\relax
\EndOfBibitem
\bibitem[Apr\`a \latin{et~al.}(2020)Apr\`a, Bylaska, de~Jong, Govind, Kowalski,
  Straatsma, Valiev, van Dam, Alexeev, Anchell, Anisimov, Aquino, Atta-Fynn,
  Autschbach, Bauman, Becca, Bernholdt, Bhaskaran-Nair, Bogatko, Borowski,
  Boschen, Brabec, Bruner, Cauet, Chen, Chuev, Cramer, Daily, Deegan, Dunning,
  Dupuis, Dyall, Fann, Fischer, Fonari, Fr\"uchtl, Gagliardi, Garza, Gawande,
  Ghosh, Glaesemann, G\"otz, Hammond, Helms, Hermes, Hirao, Hirata, Jacquelin,
  Jensen, Johnson, J\'onsson, Kendall, Klemm, Kobayashi, Konkov,
  Krishnamoorthy, Krishnan, Lin, Lins, Littlefield, Logsdail, Lopata, Ma,
  Marenich, del Campo, Mejia-Rodriguez, Moore, Mullin, Nakajima, Nascimento,
  Nichols, Nichols, Nieplocha, Otero-de-la Roza, Palmer, Panyala, Pirojsirikul,
  Peng, Peverati, Pittner, Pollack, Richard, Sadayappan, Schatz, Shelton,
  Silverstein, Smith, Soares, Song, Swart, Taylor, Thomas, Tipparaju, Truhlar,
  Tsemekhman, Van~Voorhis, V\'azquez-Mayagoitia, Verma, Villa, Vishnu,
  Vogiatzis, Wang, Weare, Williamson, Windus, Woli\'nski, Wong, Wu, Yang, Yu,
  Zacharias, Zhang, Zhao, and Harrison]{nwchem}
Apr\`a,~E. \latin{et~al.}  {NWChem}: Past, present, and future. \emph{J. Chem.
  Phys.} \textbf{2020}, \emph{152}, 184102\relax
\mciteBstWouldAddEndPuncttrue
\mciteSetBstMidEndSepPunct{\mcitedefaultmidpunct}
{\mcitedefaultendpunct}{\mcitedefaultseppunct}\relax
\EndOfBibitem
\bibitem[Piseri \latin{et~al.}(1973)Piseri, Powell, and Dolling]{Piseri1973}
Piseri,~L.; Powell,~B.~M.; Dolling,~G. Lattice dynamics of
  Polytetrafluoroethylene. \emph{J. Chem. Phys.} \textbf{1973}, \emph{58},
  158--171\relax
\mciteBstWouldAddEndPuncttrue
\mciteSetBstMidEndSepPunct{\mcitedefaultmidpunct}
{\mcitedefaultendpunct}{\mcitedefaultseppunct}\relax
\EndOfBibitem
\bibitem[Marshall and Lovesey(1971)Marshall, and Lovesey]{Marshall1971}
Marshall,~W.; Lovesey,~S.~W. \emph{Theory of Thermal Neutron Scattering};
  Oxford University Press: London, 1971\relax
\mciteBstWouldAddEndPuncttrue
\mciteSetBstMidEndSepPunct{\mcitedefaultmidpunct}
{\mcitedefaultendpunct}{\mcitedefaultseppunct}\relax
\EndOfBibitem
\bibitem[Teare(1959)]{Teare1959}
Teare,~P.~W. The Crystal Structure of Orthorhombic Hexatriacontane
  {C$_{36}$H$_{74}$}. \emph{Acta Crystallogr.} \textbf{1959}, \emph{12},
  294--300\relax
\mciteBstWouldAddEndPuncttrue
\mciteSetBstMidEndSepPunct{\mcitedefaultmidpunct}
{\mcitedefaultendpunct}{\mcitedefaultseppunct}\relax
\EndOfBibitem
\bibitem[Herzberg(1966)]{Herzberg}
Herzberg,~G. \emph{Electronic Spectra and Electronic Structure of Polyatomic
  Molecules}; Van Nostrand: New York, 1966\relax
\mciteBstWouldAddEndPuncttrue
\mciteSetBstMidEndSepPunct{\mcitedefaultmidpunct}
{\mcitedefaultendpunct}{\mcitedefaultseppunct}\relax
\EndOfBibitem
\bibitem[Qin and Hirata(2020)Qin, and Hirata]{Qin2020}
Qin,~X.~Y.; Hirata,~S. Anharmonic Phonon Dispersion in Polyethylene. \emph{J.
  Phys. Chem. B} \textbf{2020}, \emph{124}, 10477--10485\relax
\mciteBstWouldAddEndPuncttrue
\mciteSetBstMidEndSepPunct{\mcitedefaultmidpunct}
{\mcitedefaultendpunct}{\mcitedefaultseppunct}\relax
\EndOfBibitem
\bibitem[Rajendran \latin{et~al.}(2012)Rajendran, Tsuchiya, Hirata, and
  Iordanov]{Rajendran2012}
Rajendran,~A.; Tsuchiya,~T.; Hirata,~S.; Iordanov,~T.~D. Predicting Properties
  of Organic Optoelectronic Materials: Asymptotically Corrected Density
  Functional Study. \emph{J. Phys. Chem. A} \textbf{2012}, \emph{116},
  12153--12162\relax
\mciteBstWouldAddEndPuncttrue
\mciteSetBstMidEndSepPunct{\mcitedefaultmidpunct}
{\mcitedefaultendpunct}{\mcitedefaultseppunct}\relax
\EndOfBibitem
\bibitem[Fujihira and Inokuchi(1972)Fujihira, and Inokuchi]{Fujihira1972}
Fujihira,~M.; Inokuchi,~H. Photoemission from Polyethylene. \emph{Chem. Phys.
  Lett.} \textbf{1972}, \emph{17}, 554--556\relax
\mciteBstWouldAddEndPuncttrue
\mciteSetBstMidEndSepPunct{\mcitedefaultmidpunct}
{\mcitedefaultendpunct}{\mcitedefaultseppunct}\relax
\EndOfBibitem
\bibitem[Seki \latin{et~al.}(1990)Seki, Tanaka, Ohta, Aoki, Imamura, Fujimoto,
  Yamamoto, and Inokuchi]{Seki1990}
Seki,~K.; Tanaka,~H.; Ohta,~T.; Aoki,~Y.; Imamura,~A.; Fujimoto,~H.;
  Yamamoto,~H.; Inokuchi,~H. Electronic Structure of Poly(Tetrafluoroethylene)
  Studied by {UPS}, {VUV} Absorption, and Band Calculations. \emph{Phys. Scr.}
  \textbf{1990}, \emph{41}, 167--171\relax
\mciteBstWouldAddEndPuncttrue
\mciteSetBstMidEndSepPunct{\mcitedefaultmidpunct}
{\mcitedefaultendpunct}{\mcitedefaultseppunct}\relax
\EndOfBibitem
\bibitem[Shirakawa and Ikeda(1971)Shirakawa, and Ikeda]{Shirakawa1971}
Shirakawa,~H.; Ikeda,~S. Infrared Spectra of Poly(Acetylene). \emph{Polymer J.}
  \textbf{1971}, \emph{2}, 231\relax
\mciteBstWouldAddEndPuncttrue
\mciteSetBstMidEndSepPunct{\mcitedefaultmidpunct}
{\mcitedefaultendpunct}{\mcitedefaultseppunct}\relax
\EndOfBibitem
\bibitem[Shirakawa \latin{et~al.}(1973)Shirakawa, Ito, and
  Ikeda]{Shirakawa1973}
Shirakawa,~H.; Ito,~T.; Ikeda,~S. Raman Scattering and Electronic Spectra of
  Poly(Acetylene). \emph{Polym. J.} \textbf{1973}, \emph{4}, 460--462\relax
\mciteBstWouldAddEndPuncttrue
\mciteSetBstMidEndSepPunct{\mcitedefaultmidpunct}
{\mcitedefaultendpunct}{\mcitedefaultseppunct}\relax
\EndOfBibitem
\bibitem[Chiang \latin{et~al.}(1977)Chiang, Fincher, Park, Heeger, Shirakawa,
  Louis, Gau, and MacDiarmid]{Chiang1977}
Chiang,~C.~K.; Fincher,~C.~R.; Park,~Y.~W.; Heeger,~A.~J.; Shirakawa,~H.;
  Louis,~E.~J.; Gau,~S.~C.; MacDiarmid,~A.~G. Electrical Conductivity in Doped
  Polyacetylene. \emph{Phys. Rev. Lett.} \textbf{1977}, \emph{39},
  1098--1101\relax
\mciteBstWouldAddEndPuncttrue
\mciteSetBstMidEndSepPunct{\mcitedefaultmidpunct}
{\mcitedefaultendpunct}{\mcitedefaultseppunct}\relax
\EndOfBibitem
\bibitem[Chien(1984)]{Chien1984}
Chien,~J. C.~W. \emph{Polyacetylene: Chemistry, Physics, and Material Science};
  Academic: Orlando, 1984\relax
\mciteBstWouldAddEndPuncttrue
\mciteSetBstMidEndSepPunct{\mcitedefaultmidpunct}
{\mcitedefaultendpunct}{\mcitedefaultseppunct}\relax
\EndOfBibitem
\bibitem[Heeger \latin{et~al.}(1988)Heeger, Kivelson, Schrieffer, and
  Su]{Heeger1988}
Heeger,~A.~J.; Kivelson,~S.; Schrieffer,~J.~R.; Su,~W.~P. Solitons in
  Conducting Polymers. \emph{Rev. Mod. Phys.} \textbf{1988}, \emph{60},
  781--850\relax
\mciteBstWouldAddEndPuncttrue
\mciteSetBstMidEndSepPunct{\mcitedefaultmidpunct}
{\mcitedefaultendpunct}{\mcitedefaultseppunct}\relax
\EndOfBibitem
\bibitem[Yoshida and Tasumi(1988)Yoshida, and Tasumi]{Yoshida1988}
Yoshida,~H.; Tasumi,~M. Infrared and {Raman} Spectra of {\it
  Trans,Trans}-1,3,5,7-Octatetraene and Normal-Coordinate Analysis Based on
  {\it Ab initio} Molecular Orbital Calculations. \emph{J. Chem. Phys.}
  \textbf{1988}, \emph{89}, 2803--2809\relax
\mciteBstWouldAddEndPuncttrue
\mciteSetBstMidEndSepPunct{\mcitedefaultmidpunct}
{\mcitedefaultendpunct}{\mcitedefaultseppunct}\relax
\EndOfBibitem
\bibitem[Hirata \latin{et~al.}(1995)Hirata, Yoshida, Torii, and
  Tasumi]{Hirata1995Deca}
Hirata,~S.; Yoshida,~H.; Torii,~H.; Tasumi,~M. Vibrational Analyses of {\it
  Trans,Trans}-1,3,5,7-Octatetraene and All-{\it Trans}-1,3,5,7,9-Decapentaene
  Based on {\it Ab Initio} Molecular Orbital Calculations and Observed Infrared
  and {Raman} Spectra. \emph{J. Chem. Phys.} \textbf{1995}, \emph{103},
  8955--8963\relax
\mciteBstWouldAddEndPuncttrue
\mciteSetBstMidEndSepPunct{\mcitedefaultmidpunct}
{\mcitedefaultendpunct}{\mcitedefaultseppunct}\relax
\EndOfBibitem
\bibitem[Hirata \latin{et~al.}(1996)Hirata, Torii, and Tasumi]{Hirata1996}
Hirata,~S.; Torii,~H.; Tasumi,~M. Stereostructural and vibrational analyses of
  cis-polyacetylene based on density functional calculations of oligoenes.
  \emph{Bull. Chem. Soc. Jpn.} \textbf{1996}, \emph{69}, 3089--3106\relax
\mciteBstWouldAddEndPuncttrue
\mciteSetBstMidEndSepPunct{\mcitedefaultmidpunct}
{\mcitedefaultendpunct}{\mcitedefaultseppunct}\relax
\EndOfBibitem
\bibitem[Lieser \latin{et~al.}(1980)Lieser, Wegner, M\"uller, Enkelmann, and
  Meyer]{Lieser1980_trans}
Lieser,~G.; Wegner,~G.; M\"uller,~W.; Enkelmann,~V.; Meyer,~W.~H. The Structure
  and Morphology of {\it Trans}-Poly(Acetylene). \emph{Makromol. Chem. Rapid
  Commun.} \textbf{1980}, \emph{1}, 627--632\relax
\mciteBstWouldAddEndPuncttrue
\mciteSetBstMidEndSepPunct{\mcitedefaultmidpunct}
{\mcitedefaultendpunct}{\mcitedefaultseppunct}\relax
\EndOfBibitem
\bibitem[Shimamura \latin{et~al.}(1981)Shimamura, Karasz, Hirsch, and
  Chien]{Shimamura1981}
Shimamura,~K.; Karasz,~F.~E.; Hirsch,~J.~A.; Chien,~J. C.~W. Crystal Structure
  of {\it Trans}-Polyacetylene. \emph{Makromol. Chem. Rapid Commun.}
  \textbf{1981}, \emph{2}, 473--480\relax
\mciteBstWouldAddEndPuncttrue
\mciteSetBstMidEndSepPunct{\mcitedefaultmidpunct}
{\mcitedefaultendpunct}{\mcitedefaultseppunct}\relax
\EndOfBibitem
\bibitem[Fincher \latin{et~al.}(1982)Fincher, Chen, Heeger, Macdiarmid, and
  Hastings]{Fincher1982}
Fincher,~C.~R.; Chen,~C.~E.; Heeger,~A.~J.; Macdiarmid,~A.~G.; Hastings,~J.~B.
  Structural Determination of the Symmetry-Breaking Parameter in {\it
  Trans}-({CH)}$_x$. \emph{Phys. Rev. Lett.} \textbf{1982}, \emph{48},
  100--104\relax
\mciteBstWouldAddEndPuncttrue
\mciteSetBstMidEndSepPunct{\mcitedefaultmidpunct}
{\mcitedefaultendpunct}{\mcitedefaultseppunct}\relax
\EndOfBibitem
\bibitem[Baughman \latin{et~al.}(1978)Baughman, Hsu, Pez, and
  Signorelli]{Baughman1978}
Baughman,~R.~H.; Hsu,~S.~L.; Pez,~G.~P.; Signorelli,~A.~J. Structures of {\it
  Cis}-Polyacetylene and Highly Conducting Derivatives. \emph{J. Chem. Phys.}
  \textbf{1978}, \emph{68}, 5405--5409\relax
\mciteBstWouldAddEndPuncttrue
\mciteSetBstMidEndSepPunct{\mcitedefaultmidpunct}
{\mcitedefaultendpunct}{\mcitedefaultseppunct}\relax
\EndOfBibitem
\bibitem[Lieser \latin{et~al.}(1980)Lieser, Wegner, M\"uller, and
  Enkelmann]{Lieser1980}
Lieser,~G.; Wegner,~G.; M\"uller,~W.; Enkelmann,~V. On the Morphology and
  Structure of {\it Cis}-Poly(Acetylene). \emph{Makromol. Chem. Rapid Commun.}
  \textbf{1980}, \emph{1}, 621--626\relax
\mciteBstWouldAddEndPuncttrue
\mciteSetBstMidEndSepPunct{\mcitedefaultmidpunct}
{\mcitedefaultendpunct}{\mcitedefaultseppunct}\relax
\EndOfBibitem
\bibitem[Chien \latin{et~al.}(1982)Chien, Karasz, and Shimamura]{Chien1982_1}
Chien,~J. C.~W.; Karasz,~F.~E.; Shimamura,~K. Electron-Diffraction Study of
  Pristine and iodine-Doped and Iodine-Doped Poly({\it Cis}-Acetylene).
  \emph{J. Polym. Sci. Polym. Lett. Ed.} \textbf{1982}, \emph{20},
  97--102\relax
\mciteBstWouldAddEndPuncttrue
\mciteSetBstMidEndSepPunct{\mcitedefaultmidpunct}
{\mcitedefaultendpunct}{\mcitedefaultseppunct}\relax
\EndOfBibitem
\bibitem[Chien \latin{et~al.}(1982)Chien, Karasz, and Shimamura]{Chien1982_2}
Chien,~J. C.~W.; Karasz,~F.~E.; Shimamura,~K. Crystal Structure of Pristine and
  Iodine-Doped {\it Cis}-Polyacetylene. \emph{Macromolecules} \textbf{1982},
  \emph{15}, 1012--1017\relax
\mciteBstWouldAddEndPuncttrue
\mciteSetBstMidEndSepPunct{\mcitedefaultmidpunct}
{\mcitedefaultendpunct}{\mcitedefaultseppunct}\relax
\EndOfBibitem
\bibitem[Kuzmany(1980)]{Kuzmany1980}
Kuzmany,~H. Resonance {Raman} Scattering from Neutral and Doped Polyacetylene.
  \emph{Phys. Status Solidi B} \textbf{1980}, \emph{97}, 521--531\relax
\mciteBstWouldAddEndPuncttrue
\mciteSetBstMidEndSepPunct{\mcitedefaultmidpunct}
{\mcitedefaultendpunct}{\mcitedefaultseppunct}\relax
\EndOfBibitem
\bibitem[Yannoni and Clarke(1983)Yannoni, and Clarke]{Yannoni1983}
Yannoni,~C.~S.; Clarke,~T.~C. Molecular Geometry of {\it Cis}-Polyacetylene and
  {\it Trans}-Polyacetylene by Nutation {NMR} Spectroscopy. \emph{Phys. Rev.
  Lett.} \textbf{1983}, \emph{51}, 1191--1193\relax
\mciteBstWouldAddEndPuncttrue
\mciteSetBstMidEndSepPunct{\mcitedefaultmidpunct}
{\mcitedefaultendpunct}{\mcitedefaultseppunct}\relax
\EndOfBibitem
\bibitem[Takeuchi \latin{et~al.}(1987)Takeuchi, Arakawa, Furukawa, Harada, and
  Shirakawa]{Takeuchi1987}
Takeuchi,~H.; Arakawa,~T.; Furukawa,~Y.; Harada,~I.; Shirakawa,~H. Density of
  Vibrational States in {\it Trans}-Polyene: Comparison with the Infrared,
  {Raman} and Neutron Spectra of {\it Trans}-polyacetylene. \emph{J. Mol.
  Struct.} \textbf{1987}, \emph{158}, 179--193\relax
\mciteBstWouldAddEndPuncttrue
\mciteSetBstMidEndSepPunct{\mcitedefaultmidpunct}
{\mcitedefaultendpunct}{\mcitedefaultseppunct}\relax
\EndOfBibitem
\bibitem[Kamiya \latin{et~al.}(1996)Kamiya, Miyamae, Oku, Seki, Inokuchi,
  Tanaka, and Tanaka]{Kamiya_UPS}
Kamiya,~K.; Miyamae,~T.; Oku,~M.; Seki,~K.; Inokuchi,~H.; Tanaka,~C.;
  Tanaka,~J. Ultraviolet photoemission spectra of perchlorate-doped {\it cis}-
  and {\it trans}-polyacetylene. \emph{J. Phys. Chem.} \textbf{1996},
  \emph{100}, 16213--16217\relax
\mciteBstWouldAddEndPuncttrue
\mciteSetBstMidEndSepPunct{\mcitedefaultmidpunct}
{\mcitedefaultendpunct}{\mcitedefaultseppunct}\relax
\EndOfBibitem
\bibitem[Hirata \latin{et~al.}(1996)Hirata, Torii, Furukawa, Tasumi, and
  Tomkinson]{Hirata_INS}
Hirata,~S.; Torii,~H.; Furukawa,~Y.; Tasumi,~M.; Tomkinson,~J. Inelastic
  neutron scattering from trans-polyacetylene. \emph{Chem. Phys. Lett.}
  \textbf{1996}, \emph{261}, 241--245\relax
\mciteBstWouldAddEndPuncttrue
\mciteSetBstMidEndSepPunct{\mcitedefaultmidpunct}
{\mcitedefaultendpunct}{\mcitedefaultseppunct}\relax
\EndOfBibitem
\bibitem[Falk and Fleming(1975)Falk, and Fleming]{Falk1975}
Falk,~J.~E.; Fleming,~R.~J. Study of Electronic Band Structure of Polyacetylene
  and Polyfluoroethylenes. \emph{J. Phys. C: Solid State Phys.} \textbf{1975},
  \emph{8}, 627--646\relax
\mciteBstWouldAddEndPuncttrue
\mciteSetBstMidEndSepPunct{\mcitedefaultmidpunct}
{\mcitedefaultendpunct}{\mcitedefaultseppunct}\relax
\EndOfBibitem
\bibitem[Grant and Batra(1979)Grant, and Batra]{Grant1979}
Grant,~P.~M.; Batra,~I.~P. Band Structure of Polyacetylene, ({CH})$_x$.
  \emph{Solid State Commun.} \textbf{1979}, \emph{29}, 225--229\relax
\mciteBstWouldAddEndPuncttrue
\mciteSetBstMidEndSepPunct{\mcitedefaultmidpunct}
{\mcitedefaultendpunct}{\mcitedefaultseppunct}\relax
\EndOfBibitem
\bibitem[Kasowski \latin{et~al.}(1980)Kasowski, Hsu, and
  Caruthers]{Kasowski1980}
Kasowski,~R.~V.; Hsu,~W.~Y.; Caruthers,~E.~B. Electronic Properties of
  Polyacetylene, Polyethylene, and Polytetrafluoroethylene. \emph{J. Chem.
  Phys.} \textbf{1980}, \emph{72}, 4896--4900\relax
\mciteBstWouldAddEndPuncttrue
\mciteSetBstMidEndSepPunct{\mcitedefaultmidpunct}
{\mcitedefaultendpunct}{\mcitedefaultseppunct}\relax
\EndOfBibitem
\bibitem[Mintmire and White(1983)Mintmire, and White]{Mintmire1983}
Mintmire,~J.~W.; White,~C.~T. Theoretical Treatment of the Dielectric Response
  of All-{\it Trans}-Polyacetylene. \emph{Phys. Rev. B} \textbf{1983},
  \emph{27}, 1447--1449\relax
\mciteBstWouldAddEndPuncttrue
\mciteSetBstMidEndSepPunct{\mcitedefaultmidpunct}
{\mcitedefaultendpunct}{\mcitedefaultseppunct}\relax
\EndOfBibitem
\bibitem[Springborg(1986)]{Springborg1986}
Springborg,~M. Self-Consistent Electronic Structures of Polyacetylene.
  \emph{Phys. Rev. B} \textbf{1986}, \emph{33}, 8475--8489\relax
\mciteBstWouldAddEndPuncttrue
\mciteSetBstMidEndSepPunct{\mcitedefaultmidpunct}
{\mcitedefaultendpunct}{\mcitedefaultseppunct}\relax
\EndOfBibitem
\bibitem[Mintmire and White(1987)Mintmire, and White]{Mintmire1987}
Mintmire,~J.~W.; White,~C.~T. Local-Density-Functional Results for the
  Dimerization of {\it Trans}-polyacetylene: Relationship to the Band Gap
  Problem. \emph{Phys. Rev. B} \textbf{1987}, \emph{35}, 4180--4183\relax
\mciteBstWouldAddEndPuncttrue
\mciteSetBstMidEndSepPunct{\mcitedefaultmidpunct}
{\mcitedefaultendpunct}{\mcitedefaultseppunct}\relax
\EndOfBibitem
\bibitem[von Boehm \latin{et~al.}(1987)von Boehm, Kuivalainen, and
  Calais]{vonBoehm1987}
von Boehm,~J.; Kuivalainen,~P.; Calais,~J.-L. Self-Consistent
  Linear-Combination-of-{Gaussian}-Orbitals Approach for Polymers: Application
  to {\it Trans}-({CH})$_x$. \emph{Phys. Rev. B} \textbf{1987}, \emph{35},
  8177--8183\relax
\mciteBstWouldAddEndPuncttrue
\mciteSetBstMidEndSepPunct{\mcitedefaultmidpunct}
{\mcitedefaultendpunct}{\mcitedefaultseppunct}\relax
\EndOfBibitem
\bibitem[Vogl and Campbell(1989)Vogl, and Campbell]{Vogl1989}
Vogl,~P.; Campbell,~D.~K. Three-Dimensional Structure and Intrinsic Defects in
  {\it Trans}-Polyacetylene. \emph{Phys. Rev. Lett.} \textbf{1989}, \emph{62},
  2012--2015\relax
\mciteBstWouldAddEndPuncttrue
\mciteSetBstMidEndSepPunct{\mcitedefaultmidpunct}
{\mcitedefaultendpunct}{\mcitedefaultseppunct}\relax
\EndOfBibitem
\bibitem[Ashkenazi \latin{et~al.}(1989)Ashkenazi, Pickett, Krakauer, Wang,
  Klein, and Chubb]{Ashkenazi1989}
Ashkenazi,~J.; Pickett,~W.~E.; Krakauer,~H.; Wang,~C.~S.; Klein,~B.~M.;
  Chubb,~S.~R. Ground State of {\it Trans}-Polyacetylene and the {Peierls}
  Mechanism. \emph{Phys. Rev. Lett.} \textbf{1989}, \emph{62}, 2016--2019\relax
\mciteBstWouldAddEndPuncttrue
\mciteSetBstMidEndSepPunct{\mcitedefaultmidpunct}
{\mcitedefaultendpunct}{\mcitedefaultseppunct}\relax
\EndOfBibitem
\bibitem[Springborg \latin{et~al.}(1991)Springborg, Calais, Goscinski, and
  Eriksson]{Springborg1991}
Springborg,~M.; Calais,~J.-L.; Goscinski,~O.; Eriksson,~L.~A.
  Linear-Muffin-Tin-Orbital Method for Helical Polymers: A Detailed Study of
  {\it Trans}-polyacetylene. \emph{Phys. Rev. B} \textbf{1991}, \emph{44},
  12713--12736\relax
\mciteBstWouldAddEndPuncttrue
\mciteSetBstMidEndSepPunct{\mcitedefaultmidpunct}
{\mcitedefaultendpunct}{\mcitedefaultseppunct}\relax
\EndOfBibitem
\bibitem[Paloheimo and von Boehm(1992)Paloheimo, and von Boehm]{Paloheimo1992}
Paloheimo,~J.; von Boehm,~J. Density-Functional Study of the Dimerization of
  {\it Trans}-polyacetylene. \emph{Phys. Rev. B} \textbf{1992}, \emph{46},
  4304--4307\relax
\mciteBstWouldAddEndPuncttrue
\mciteSetBstMidEndSepPunct{\mcitedefaultmidpunct}
{\mcitedefaultendpunct}{\mcitedefaultseppunct}\relax
\EndOfBibitem
\bibitem[Hirata \latin{et~al.}(1995)Hirata, Torii, and Tasumi]{Hirata1995}
Hirata,~S.; Torii,~H.; Tasumi,~M. Vibrational Analyses of {\it
  Trans}-polyacetylene Based on {\it Ab Initio} Second-Order {M}\o
  ller--{P}lesset Perturbation Calculations of {\it Trans}-Oligoenes. \emph{J.
  Chem. Phys.} \textbf{1995}, \emph{103}, 8964--8979\relax
\mciteBstWouldAddEndPuncttrue
\mciteSetBstMidEndSepPunct{\mcitedefaultmidpunct}
{\mcitedefaultendpunct}{\mcitedefaultseppunct}\relax
\EndOfBibitem
\bibitem[Kuchitsu(1998)]{Kuchitsu}
Kuchitsu,~K. \emph{Structure Data of Free Polyatomic Molecules: Basic Data};
  Springer: Berlin, 1998\relax
\mciteBstWouldAddEndPuncttrue
\mciteSetBstMidEndSepPunct{\mcitedefaultmidpunct}
{\mcitedefaultendpunct}{\mcitedefaultseppunct}\relax
\EndOfBibitem
\bibitem[Aulbur \latin{et~al.}(2000)Aulbur, Jonsson, and Wilkins]{Aulbur2000}
Aulbur,~W.~G.; Jonsson,~L.; Wilkins,~J.~W. Quasiparticle calculations in
  solids. \emph{Solid State Phys.} \textbf{2000}, \emph{54}, 1--218\relax
\mciteBstWouldAddEndPuncttrue
\mciteSetBstMidEndSepPunct{\mcitedefaultmidpunct}
{\mcitedefaultendpunct}{\mcitedefaultseppunct}\relax
\EndOfBibitem
\bibitem[Nakafuku and Takemura(1975)Nakafuku, and Takemura]{Nakafuku1975}
Nakafuku,~C.; Takemura,~T. Crystal Structure of High Pressure Phase of
  Polytetrafluoroethylene. \emph{Jpn. J. Appl. Phys.} \textbf{1975}, \emph{14},
  599--602\relax
\mciteBstWouldAddEndPuncttrue
\mciteSetBstMidEndSepPunct{\mcitedefaultmidpunct}
{\mcitedefaultendpunct}{\mcitedefaultseppunct}\relax
\EndOfBibitem
\bibitem[Lorenzen \latin{et~al.}(2003)Lorenzen, Hanfland, and
  Mermet]{Lorenzen2003}
Lorenzen,~M.; Hanfland,~M.; Mermet,~A. Poly(tetrafluoroethylene) under
  pressure: X-diffraction studies. \emph{Nucl. Instr. Meth. Phys. Res. B}
  \textbf{2003}, \emph{200}, 416--420\relax
\mciteBstWouldAddEndPuncttrue
\mciteSetBstMidEndSepPunct{\mcitedefaultmidpunct}
{\mcitedefaultendpunct}{\mcitedefaultseppunct}\relax
\EndOfBibitem
\bibitem[Farmer and Eby(1981)Farmer, and Eby]{Farmer1981}
Farmer,~B.~L.; Eby,~R.~K. Energy Calculations of the Crystal Structure of the
  Low-Temperature Phase ({II}) of Polytetrafluoroethylene. \emph{Polymer}
  \textbf{1981}, \emph{22}, 1487--1495\relax
\mciteBstWouldAddEndPuncttrue
\mciteSetBstMidEndSepPunct{\mcitedefaultmidpunct}
{\mcitedefaultendpunct}{\mcitedefaultseppunct}\relax
\EndOfBibitem
\bibitem[D'Amore \latin{et~al.}(2004)D'Amore, Auriemma, De~Rosa, and
  Barone]{DAmore2004}
D'Amore,~M.; Auriemma,~F.; De~Rosa,~C.; Barone,~V. Disordered chain
  conformations of poly(tetrafluoroethylene) in the high-temperature
  crystalline form {I}. \emph{Macromolecules} \textbf{2004}, \emph{37},
  9473--9480\relax
\mciteBstWouldAddEndPuncttrue
\mciteSetBstMidEndSepPunct{\mcitedefaultmidpunct}
{\mcitedefaultendpunct}{\mcitedefaultseppunct}\relax
\EndOfBibitem
\bibitem[Fatti \latin{et~al.}(2019)Fatti, Righi, Dini, and Ciniero]{Fatti2019}
Fatti,~G.; Righi,~M.~C.; Dini,~D.; Ciniero,~A. First-Principles Insights into
  the Structural and Electronic Properties of Polytetrafluoroethylene in Its
  High-Pressure Phase (Form {III}). \emph{J. Phys. Chem. C} \textbf{2019},
  \emph{123}, 6250--6255\relax
\mciteBstWouldAddEndPuncttrue
\mciteSetBstMidEndSepPunct{\mcitedefaultmidpunct}
{\mcitedefaultendpunct}{\mcitedefaultseppunct}\relax
\EndOfBibitem
\bibitem[Grainger and Stewart(2001)Grainger, and Stewart]{Grainger2001}
Grainger,~D.~W.; Stewart,~C.~W. \emph{Fluorinated Surfaces, Coatings, and
  Films}; American Chemical Society, 2001; Vol. 787; pp 1--14\relax
\mciteBstWouldAddEndPuncttrue
\mciteSetBstMidEndSepPunct{\mcitedefaultmidpunct}
{\mcitedefaultendpunct}{\mcitedefaultseppunct}\relax
\EndOfBibitem
\bibitem[Hellwege and Hellwege(1976)Hellwege, and Hellwege]{Hellwege}
Hellwege,~K.~H.; Hellwege,~A.~M. \emph{Landolt-Bornstein: Group II: Atomic and
  Molecular Physics Volume 7: Structure Data of Free Polyatomic Molecules};
  Springer-Verlag: Berlin, 1976\relax
\mciteBstWouldAddEndPuncttrue
\mciteSetBstMidEndSepPunct{\mcitedefaultmidpunct}
{\mcitedefaultendpunct}{\mcitedefaultseppunct}\relax
\EndOfBibitem
\bibitem[Clark and Kilcast(1971)Clark, and Kilcast]{Clark1971}
Clark,~D.~T.; Kilcast,~D. Study of Core and Valence Energy Levels of {PTFE}.
  \emph{Nature Phys. Sci.} \textbf{1971}, \emph{233}, 77--79\relax
\mciteBstWouldAddEndPuncttrue
\mciteSetBstMidEndSepPunct{\mcitedefaultmidpunct}
{\mcitedefaultendpunct}{\mcitedefaultseppunct}\relax
\EndOfBibitem
\bibitem[Pireaux \latin{et~al.}(1974)Pireaux, Riga, Caudano, Verbist,
  Andr\'{e}, Delhalle, and Delhalle]{Pireaux_UPS1974}
Pireaux,~J.~J.; Riga,~J.; Caudano,~R.; Verbist,~J.~J.; Andr\'{e},~J.~M.;
  Delhalle,~J.; Delhalle,~S. Electronic Structure of Fluoropolymers: Theory and
  {ESCA} Measurements. \emph{J. Electron Spectrosc. Relat. Phenom.}
  \textbf{1974}, \emph{5}, 531--550\relax
\mciteBstWouldAddEndPuncttrue
\mciteSetBstMidEndSepPunct{\mcitedefaultmidpunct}
{\mcitedefaultendpunct}{\mcitedefaultseppunct}\relax
\EndOfBibitem
\bibitem[Delhalle \latin{et~al.}(1977)Delhalle, Delhalle, Andr\'{e}, Pireaux,
  Riga, Caudano, and Verbist]{Delhalle_UPS1977}
Delhalle,~J.; Delhalle,~S.; Andr\'{e},~J.~M.; Pireaux,~J.~J.; Riga,~J.;
  Caudano,~R.; Verbist,~J.~J. Electronic Structure of Linear Fluoropolymers:
  Theory and {ESCA} Measurements Revisited. \emph{J. Electron Spectrosc. Relat.
  Phenom.} \textbf{1977}, \emph{12}, 293--303\relax
\mciteBstWouldAddEndPuncttrue
\mciteSetBstMidEndSepPunct{\mcitedefaultmidpunct}
{\mcitedefaultendpunct}{\mcitedefaultseppunct}\relax
\EndOfBibitem
\bibitem[Seki \latin{et~al.}(1990)Seki, Tanaka, Ohta, Aoki, Imamura, Fujimoto,
  Yamamoto, and Inokuchi]{Seki_UPS1990}
Seki,~K.; Tanaka,~H.; Ohta,~T.; Aoki,~Y.; Imamura,~A.; Fujimoto,~H.;
  Yamamoto,~H.; Inokuchi,~H. Electronic Structure of Poly(Tetrafluoroethylene)
  Studied by {UPS}, {VUV} Absorption, and Band Calculations. \emph{Phys. Scr.}
  \textbf{1990}, \emph{41}, 167--171\relax
\mciteBstWouldAddEndPuncttrue
\mciteSetBstMidEndSepPunct{\mcitedefaultmidpunct}
{\mcitedefaultendpunct}{\mcitedefaultseppunct}\relax
\EndOfBibitem
\bibitem[Miyamae \latin{et~al.}(2000)Miyamae, Hasegawa, Yoshimura, Ishii, Ueno,
  and Seki]{Miyamae2000}
Miyamae,~T.; Hasegawa,~S.; Yoshimura,~D.; Ishii,~H.; Ueno,~N.; Seki,~K.
  Intramolecular energy-band dispersion in oriented thin films of {\it
  n}-{CF}$_3$({CF}$_2$)$_{22}${CF}$_3$ observed by angle-resolved photoemission
  with synchrotron radiation. \emph{J. Chem. Phys.} \textbf{2000}, \emph{112},
  3333--3338\relax
\mciteBstWouldAddEndPuncttrue
\mciteSetBstMidEndSepPunct{\mcitedefaultmidpunct}
{\mcitedefaultendpunct}{\mcitedefaultseppunct}\relax
\EndOfBibitem
\bibitem[Ono \latin{et~al.}(2005)Ono, Yamane, Fukagawa, Kera, Yoshimura,
  Okudaira, Morikawa, Seki, and Ueno]{Ono2005}
Ono,~M.; Yamane,~H.; Fukagawa,~H.; Kera,~S.; Yoshimura,~D.; Okudaira,~K.~K.;
  Morikawa,~E.; Seki,~K.; Ueno,~N. {UPS} study of {VUV}-photodegradation of
  polytetrafluoroethylene ({PTFE}) ultrathin film by using synchrotron
  radiation. \emph{Nucl. Inst. Meth. Phys. Res. B} \textbf{2005}, \emph{236},
  377--382\relax
\mciteBstWouldAddEndPuncttrue
\mciteSetBstMidEndSepPunct{\mcitedefaultmidpunct}
{\mcitedefaultendpunct}{\mcitedefaultseppunct}\relax
\EndOfBibitem
\bibitem[Wang \latin{et~al.}(2014)Wang, Duscher, and Paddison]{Wang2014}
Wang,~C.; Duscher,~G.; Paddison,~S.~J. Electron energy loss spectroscopy of
  polytetrafluoroethylene: Experiment and first principles calculations.
  \emph{Microscopy} \textbf{2014}, \emph{63}, 73--83\relax
\mciteBstWouldAddEndPuncttrue
\mciteSetBstMidEndSepPunct{\mcitedefaultmidpunct}
{\mcitedefaultendpunct}{\mcitedefaultseppunct}\relax
\EndOfBibitem
\bibitem[Yoshimura \latin{et~al.}(2004)Yoshimura, Ishii, Ouchi, Miyamae,
  Hasegawa, Okudaira, Ueno, and Seki]{Yoshimura_ARUP}
Yoshimura,~D.; Ishii,~H.; Ouchi,~Y.; Miyamae,~T.; Hasegawa,~S.;
  Okudaira,~K.~K.; Ueno,~N.; Seki,~K. Simulation study of angle-resolved
  photoemission spectra and intramolecular energy-band dispersion of a
  poly(tetrafluoroethylene) oligomer film. \emph{J. Chem. Phys.} \textbf{2004},
  \emph{120}, 10753--10762\relax
\mciteBstWouldAddEndPuncttrue
\mciteSetBstMidEndSepPunct{\mcitedefaultmidpunct}
{\mcitedefaultendpunct}{\mcitedefaultseppunct}\relax
\EndOfBibitem
\bibitem[Morokuma(1971)]{Morokuma1971}
Morokuma,~K. Electronic Structures of Linear Polymers. {II}. {Formulation} and
  {CNDO}/2 Calculation for Polyethylene and Poly(Tetrafluoroethylene). \emph{J.
  Chem. Phys.} \textbf{1971}, \emph{54}, 962--971\relax
\mciteBstWouldAddEndPuncttrue
\mciteSetBstMidEndSepPunct{\mcitedefaultmidpunct}
{\mcitedefaultendpunct}{\mcitedefaultseppunct}\relax
\EndOfBibitem
\bibitem[McCubbin(1971)]{Mccubbin1971}
McCubbin,~W.~L. Assessment of Polymer Band Structure Calculations. \emph{Chem.
  Phys. Lett.} \textbf{1971}, \emph{8}, 507--512\relax
\mciteBstWouldAddEndPuncttrue
\mciteSetBstMidEndSepPunct{\mcitedefaultmidpunct}
{\mcitedefaultendpunct}{\mcitedefaultseppunct}\relax
\EndOfBibitem
\bibitem[Delhalle(1974)]{Delhalle1974}
Delhalle,~J. Influence of Chemical Substitution on Energy Band Structure of
  Polyfluoroethylenes. \emph{Chem. Phys.} \textbf{1974}, \emph{5},
  306--314\relax
\mciteBstWouldAddEndPuncttrue
\mciteSetBstMidEndSepPunct{\mcitedefaultmidpunct}
{\mcitedefaultendpunct}{\mcitedefaultseppunct}\relax
\EndOfBibitem
\bibitem[Otto \latin{et~al.}(1985)Otto, Ladik, and F\"orner]{Otto1985}
Otto,~P.; Ladik,~J.; F\"orner,~W. The Energy Band Structure of
  Polyfluoroethylene: Influence of Chemical Substitution and Conformation.
  \emph{Chem. Phys.} \textbf{1985}, \emph{95}, 365--372\relax
\mciteBstWouldAddEndPuncttrue
\mciteSetBstMidEndSepPunct{\mcitedefaultmidpunct}
{\mcitedefaultendpunct}{\mcitedefaultseppunct}\relax
\EndOfBibitem
\bibitem[Springborg and Lev(1989)Springborg, and Lev]{Springborg1989}
Springborg,~M.; Lev,~M. Electronic Structures of Polyethylene and
  Polytetrafluoroethylene. \emph{Phys. Rev. B} \textbf{1989}, \emph{40},
  3333--3339\relax
\mciteBstWouldAddEndPuncttrue
\mciteSetBstMidEndSepPunct{\mcitedefaultmidpunct}
{\mcitedefaultendpunct}{\mcitedefaultseppunct}\relax
\EndOfBibitem
\bibitem[Cain and Matienzo(1992)Cain, and Matienzo]{Cain1992}
Cain,~S.~R.; Matienzo,~L.~J. Charge Iterated Parameters for Extended {H}\"uckel
  Calculations on Polytetrafluoroethylene: Relevance to {X}-Ray Photoelectron
  Spectroscopy. \emph{J. Electron Spectrosc. Relat. Phenom.} \textbf{1992},
  \emph{58}, 365--373\relax
\mciteBstWouldAddEndPuncttrue
\mciteSetBstMidEndSepPunct{\mcitedefaultmidpunct}
{\mcitedefaultendpunct}{\mcitedefaultseppunct}\relax
\EndOfBibitem
\bibitem[Higgs(1953)]{Higgs}
Higgs,~P.~W. The Vibration Spectra of Helical Molecules: Infra-Red and {Raman}
  Selection Rules, Intensities and Approximate Frequencies. \emph{Proc. R. Soc.
  A} \textbf{1953}, \emph{220}, 472--485\relax
\mciteBstWouldAddEndPuncttrue
\mciteSetBstMidEndSepPunct{\mcitedefaultmidpunct}
{\mcitedefaultendpunct}{\mcitedefaultseppunct}\relax
\EndOfBibitem
\bibitem[Liang and Krimm(1956)Liang, and Krimm]{Liang1956}
Liang,~C.~Y.; Krimm,~S. Infrared Spectra of High Polymers. {III}.
  {Polytetrafluoroethylene} and Polychlorotrifluoroethylene. \emph{J. Chem.
  Phys.} \textbf{1956}, \emph{25}, 563--571\relax
\mciteBstWouldAddEndPuncttrue
\mciteSetBstMidEndSepPunct{\mcitedefaultmidpunct}
{\mcitedefaultendpunct}{\mcitedefaultseppunct}\relax
\EndOfBibitem
\bibitem[Koenig and Boerio(1969)Koenig, and Boerio]{Koenig1969}
Koenig,~J.~L.; Boerio,~F.~J. Raman Scattering and Band Assignments in
  Polytetrafluoroethylene. \emph{J. Chem. Phys.} \textbf{1969}, \emph{50},
  2823--2829\relax
\mciteBstWouldAddEndPuncttrue
\mciteSetBstMidEndSepPunct{\mcitedefaultmidpunct}
{\mcitedefaultendpunct}{\mcitedefaultseppunct}\relax
\EndOfBibitem
\bibitem[Peacock \latin{et~al.}(1970)Peacock, Hendra, Willis, and
  Cudby]{Peacock1970}
Peacock,~C.~J.; Hendra,~P.~J.; Willis,~H.~A.; Cudby,~M. E.~A. Raman Spectrum
  and Vibrational Assignment for Poly(Tetrafluoro-Ethylene). \emph{J. Chem.
  Soc. A} \textbf{1970}, \emph{1970}, 2943--2947\relax
\mciteBstWouldAddEndPuncttrue
\mciteSetBstMidEndSepPunct{\mcitedefaultmidpunct}
{\mcitedefaultendpunct}{\mcitedefaultseppunct}\relax
\EndOfBibitem
\bibitem[Twisleton and White(1972)Twisleton, and White]{Twisleton1972}
Twisleton,~J.~F.; White,~J.~W. Interchain Force Field and Elastic-Constants of
  Polytetrafluoroethylene. \emph{Polymer} \textbf{1972}, \emph{13},
  40--42\relax
\mciteBstWouldAddEndPuncttrue
\mciteSetBstMidEndSepPunct{\mcitedefaultmidpunct}
{\mcitedefaultendpunct}{\mcitedefaultseppunct}\relax
\EndOfBibitem
\bibitem[{LaGarde} \latin{et~al.}(1969){LaGarde}, Prask, and
  Trevino]{LaGarde1969}
{LaGarde},~V.; Prask,~H.; Trevino,~S. Vibrations in Teflon. \emph{Discuss.
  Faraday Soc.} \textbf{1969}, \emph{48}, 15--18\relax
\mciteBstWouldAddEndPuncttrue
\mciteSetBstMidEndSepPunct{\mcitedefaultmidpunct}
{\mcitedefaultendpunct}{\mcitedefaultseppunct}\relax
\EndOfBibitem
\bibitem[Hannon \latin{et~al.}(1969)Hannon, Boerio, and Koenig]{Hannon1969}
Hannon,~M.~J.; Boerio,~F.~J.; Koenig,~J.~L. Vibrational analysis of
  polytetrafluoroethylene. \emph{J. Chem. Phys.} \textbf{1969}, \emph{50},
  2829--2836\relax
\mciteBstWouldAddEndPuncttrue
\mciteSetBstMidEndSepPunct{\mcitedefaultmidpunct}
{\mcitedefaultendpunct}{\mcitedefaultseppunct}\relax
\EndOfBibitem
\bibitem[Eremets \latin{et~al.}(2004)Eremets, Gavriliuk, Serebryanaya, Trojan,
  Dzivenko, Boehler, Mao, and Hemley]{Eremets2004JCP}
Eremets,~M.~I.; Gavriliuk,~A.~G.; Serebryanaya,~N.~R.; Trojan,~I.~A.;
  Dzivenko,~D.~A.; Boehler,~R.; Mao,~H.~K.; Hemley,~R.~J. Structural
  transformation of molecular nitrogen to a single-bonded atomic state at high
  pressures. \emph{J. Chem. Phys.} \textbf{2004}, \emph{121},
  11296--11300\relax
\mciteBstWouldAddEndPuncttrue
\mciteSetBstMidEndSepPunct{\mcitedefaultmidpunct}
{\mcitedefaultendpunct}{\mcitedefaultseppunct}\relax
\EndOfBibitem
\bibitem[Ma \latin{et~al.}(2009)Ma, Oganov, Li, Xie, and
  Kotakoski]{Ma_polymericN_2009}
Ma,~Y.~M.; Oganov,~A.~R.; Li,~Z.~W.; Xie,~Y.; Kotakoski,~J. Novel High Pressure
  Structures of Polymeric Nitrogen. \emph{Phys. Rev. Lett.} \textbf{2009},
  \emph{102}, 065501\relax
\mciteBstWouldAddEndPuncttrue
\mciteSetBstMidEndSepPunct{\mcitedefaultmidpunct}
{\mcitedefaultendpunct}{\mcitedefaultseppunct}\relax
\EndOfBibitem
\bibitem[Pickard and Needs(2009)Pickard, and Needs]{Pickard_polymericN_2009}
Pickard,~C.~J.; Needs,~R.~J. High-Pressure Phases of Nitrogen. \emph{Phys. Rev.
  Lett.} \textbf{2009}, \emph{102}, 125702\relax
\mciteBstWouldAddEndPuncttrue
\mciteSetBstMidEndSepPunct{\mcitedefaultmidpunct}
{\mcitedefaultendpunct}{\mcitedefaultseppunct}\relax
\EndOfBibitem
\bibitem[Christe \latin{et~al.}(1999)Christe, Wilson, Sheehy, and
  Boatz]{Christe1999}
Christe,~K.~O.; Wilson,~W.~W.; Sheehy,~J.~A.; Boatz,~J.~A. {N$_5^+$}: A novel
  homoleptic polynitrogen ion as a high energy density material. \emph{Angew.
  Chem. Int. Ed.} \textbf{1999}, \emph{38}, 2004--2009\relax
\mciteBstWouldAddEndPuncttrue
\mciteSetBstMidEndSepPunct{\mcitedefaultmidpunct}
{\mcitedefaultendpunct}{\mcitedefaultseppunct}\relax
\EndOfBibitem
\bibitem[Vij \latin{et~al.}(2001)Vij, Wilson, Vij, Tham, Sheehy, and
  Christe]{Vij2001}
Vij,~A.; Wilson,~W.~W.; Vij,~V.; Tham,~F.~S.; Sheehy,~J.~A.; Christe,~K.~O.
  Polynitrogen chemistry: Synthesis, characterization, and crystal structure of
  surprisingly stable fluoroantimonate salts of {N}$_5^+$. \emph{J. Am. Chem.
  Soc.} \textbf{2001}, \emph{123}, 6308--6313\relax
\mciteBstWouldAddEndPuncttrue
\mciteSetBstMidEndSepPunct{\mcitedefaultmidpunct}
{\mcitedefaultendpunct}{\mcitedefaultseppunct}\relax
\EndOfBibitem
\bibitem[Cacace \latin{et~al.}(2002)Cacace, de~Petris, and Troiani]{Cacace2002}
Cacace,~F.; de~Petris,~G.; Troiani,~A. Experimental detection of tetranitrogen.
  \emph{Science} \textbf{2002}, \emph{295}, 480--481\relax
\mciteBstWouldAddEndPuncttrue
\mciteSetBstMidEndSepPunct{\mcitedefaultmidpunct}
{\mcitedefaultendpunct}{\mcitedefaultseppunct}\relax
\EndOfBibitem
\bibitem[Vij \latin{et~al.}(2002)Vij, Pavlovich, Wilson, Vij, and
  Christe]{Vij2002}
Vij,~A.; Pavlovich,~J.~G.; Wilson,~W.~W.; Vij,~V.; Christe,~K.~O. Experimental
  detection of the pentaazacyclopentadienide (pentazolate) anion, {\it
  cyclo}-{N$_5^-$}. \emph{Angew. Chem. Int. Ed.} \textbf{2002}, \emph{41},
  3051--3054\relax
\mciteBstWouldAddEndPuncttrue
\mciteSetBstMidEndSepPunct{\mcitedefaultmidpunct}
{\mcitedefaultendpunct}{\mcitedefaultseppunct}\relax
\EndOfBibitem
\bibitem[Bi \latin{et~al.}(2010)Bi, Liao, Xu, Deng, Wang, Wu, Gao, and
  Zhang]{Bi2010}
Bi,~Y.~F.; Liao,~W.~P.; Xu,~G.~C.; Deng,~R.~P.; Wang,~M.~Y.; Wu,~Z.~J.;
  Gao,~S.; Zhang,~H.~J. Three {\it p-tert}-Butylthiacalix[4]arene-Supported
  Cobalt Compounds Obtained in One Pot Involving In Situ Formation of
  {N}$_6${H}$_2$ Ligand. \emph{Inorg. Chem.} \textbf{2010}, \emph{49},
  7735--7740\relax
\mciteBstWouldAddEndPuncttrue
\mciteSetBstMidEndSepPunct{\mcitedefaultmidpunct}
{\mcitedefaultendpunct}{\mcitedefaultseppunct}\relax
\EndOfBibitem
\bibitem[Li \latin{et~al.}(2010)Li, Qi, Li, Zhang, Sun, Yu, and Pang]{Li2010}
Li,~Y.~C.; Qi,~C.; Li,~S.~H.; Zhang,~H.~J.; Sun,~C.~H.; Yu,~Y.~Z.; Pang,~S.~P.
  1,1$^\prime$-{A}zobis-1,2,3-triazole: A High-Nitrogen Compound with Stable
  {N}$_8$ Structure and Photochromism. \emph{J. Am. Chem. Soc.} \textbf{2010},
  \emph{132}, 12172--12173\relax
\mciteBstWouldAddEndPuncttrue
\mciteSetBstMidEndSepPunct{\mcitedefaultmidpunct}
{\mcitedefaultendpunct}{\mcitedefaultseppunct}\relax
\EndOfBibitem
\bibitem[Tang \latin{et~al.}(2012)Tang, Yang, Shen, Wu, Ju, Lu, and
  Cheng]{Tan2012}
Tang,~Y.~X.; Yang,~H.~W.; Shen,~J.~H.; Wu,~B.; Ju,~X.~H.; Lu,~C.~X.;
  Cheng,~G.~B. Synthesis and characterization of
  1,1$^\prime$-azobis(5-methyltetrazole). \emph{New J. Chem.} \textbf{2012},
  \emph{36}, 2447--2450\relax
\mciteBstWouldAddEndPuncttrue
\mciteSetBstMidEndSepPunct{\mcitedefaultmidpunct}
{\mcitedefaultendpunct}{\mcitedefaultseppunct}\relax
\EndOfBibitem
\bibitem[Tang \latin{et~al.}(2013)Tang, Yang, Wu, Ju, Lu, and Cheng]{Tang2013}
Tang,~Y.~X.; Yang,~H.~W.; Wu,~B.; Ju,~X.~H.; Lu,~C.~X.; Cheng,~G.~B. Synthesis
  and Characterization of a Stable, Catenated {N}$_{11}$ Energetic Salt.
  \emph{Angew. Chem. Int. Ed.} \textbf{2013}, \emph{52}, 4875--4877\relax
\mciteBstWouldAddEndPuncttrue
\mciteSetBstMidEndSepPunct{\mcitedefaultmidpunct}
{\mcitedefaultendpunct}{\mcitedefaultseppunct}\relax
\EndOfBibitem
\bibitem[Bazanov \latin{et~al.}(2016)Bazanov, Geiger, Carmieli, Grinstein,
  Welner, and Haas]{Barazov2016}
Bazanov,~B.; Geiger,~U.; Carmieli,~R.; Grinstein,~D.; Welner,~S.; Haas,~Y.
  Detection of Cyclo-{N$_5^-$} in {THF} Solution. \emph{Angew. Chem. Int. Ed.}
  \textbf{2016}, \emph{55}, 13233--13235\relax
\mciteBstWouldAddEndPuncttrue
\mciteSetBstMidEndSepPunct{\mcitedefaultmidpunct}
{\mcitedefaultendpunct}{\mcitedefaultseppunct}\relax
\EndOfBibitem
\bibitem[Zhang \latin{et~al.}(2017)Zhang, Sun, Hu, Yu, and Lu]{Zhang2017}
Zhang,~C.; Sun,~C.~G.; Hu,~B.~C.; Yu,~C.~M.; Lu,~M. Synthesis and
  characterization of the pentazolate anion cyclo-{N}$_5^-$ in
  ({N}$_5$)$_6$({H}$_3${O})$_3$({NH}$_4$)$_4${Cl}. \emph{Science}
  \textbf{2017}, \emph{355}, 374--376\relax
\mciteBstWouldAddEndPuncttrue
\mciteSetBstMidEndSepPunct{\mcitedefaultmidpunct}
{\mcitedefaultendpunct}{\mcitedefaultseppunct}\relax
\EndOfBibitem
\bibitem[Zhang \latin{et~al.}(2018)Zhang, Wang, Li, Lin, Song, Huang, Liu, Nie,
  and Zhang]{Zhang2018}
Zhang,~W.; Wang,~K.; Li,~J.; Lin,~Z.; Song,~S.; Huang,~S.; Liu,~Y.; Nie,~F.;
  Zhang,~Q. Stabilization of the Pentazolate Anion in a Zeolitic Architecture
  with {Na$_{20}$N$_{60}$} and {Na$_{24}$N$_{60}$} Nanocages. \emph{Angew.
  Chem. Int. Ed.} \textbf{2018}, \emph{57}, 2592--2595\relax
\mciteBstWouldAddEndPuncttrue
\mciteSetBstMidEndSepPunct{\mcitedefaultmidpunct}
{\mcitedefaultendpunct}{\mcitedefaultseppunct}\relax
\EndOfBibitem
\bibitem[Huang \latin{et~al.}(2018)Huang, Zhang, Yan, Xiong, Xu, and
  Ren]{Huang2018}
Huang,~R.~Y.; Zhang,~C.; Yan,~D.; Xiong,~Z.; Xu,~H.; Ren,~X.~M.
  {Pb}$^\text{II}$-catalyzed transformation of aromatic nitriles to
  heptanitrogen anions {\it via} sodium azide: A combined experimental and
  theoretical study. \emph{RSC Adv.} \textbf{2018}, \emph{8},
  39929--39936\relax
\mciteBstWouldAddEndPuncttrue
\mciteSetBstMidEndSepPunct{\mcitedefaultmidpunct}
{\mcitedefaultendpunct}{\mcitedefaultseppunct}\relax
\EndOfBibitem
\bibitem[Lauderdale \latin{et~al.}(1992)Lauderdale, Stanton, and
  Bartlett]{Lauderdale1992}
Lauderdale,~W.~J.; Stanton,~J.~F.; Bartlett,~R.~J. Stability and Energetics of
  Metastable Molecules: Tetraazatetrahedrane ({N}$_4$), Hexaazabenzene
  ({N}$_6$), and Octaazacubane ({N}$_8$). \emph{J. Phys. Chem.} \textbf{1992},
  \emph{96}, 1173--1178\relax
\mciteBstWouldAddEndPuncttrue
\mciteSetBstMidEndSepPunct{\mcitedefaultmidpunct}
{\mcitedefaultendpunct}{\mcitedefaultseppunct}\relax
\EndOfBibitem
\bibitem[Nguyen and Ha(1996)Nguyen, and Ha]{Nguyen1996}
Nguyen,~M.~T.; Ha,~T.~K. Azidopentazole is probably the lowest-energy {N}$_8$
  species: A theoretical study. \emph{Chem. Ber.} \textbf{1996}, \emph{129},
  1157--1159\relax
\mciteBstWouldAddEndPuncttrue
\mciteSetBstMidEndSepPunct{\mcitedefaultmidpunct}
{\mcitedefaultendpunct}{\mcitedefaultseppunct}\relax
\EndOfBibitem
\bibitem[Nguyen(2003)]{Nguyen2003}
Nguyen,~M.~T. Polynitrogen compounds. {I}. Structure and stability of {N$_4$
  and N$_5$} systems. \emph{Coord. Chem. Rev.} \textbf{2003}, \emph{244},
  93--113\relax
\mciteBstWouldAddEndPuncttrue
\mciteSetBstMidEndSepPunct{\mcitedefaultmidpunct}
{\mcitedefaultendpunct}{\mcitedefaultseppunct}\relax
\EndOfBibitem
\bibitem[Nguyen and Ha(2001)Nguyen, and Ha]{Nguyen2001}
Nguyen,~M.~T.; Ha,~T.~K. Decomposition mechanism of the polynitrogen {N$_5$ and
  N$_6$} clusters and their ions. \emph{Chem. Phys. Lett.} \textbf{2001},
  \emph{335}, 311--320\relax
\mciteBstWouldAddEndPuncttrue
\mciteSetBstMidEndSepPunct{\mcitedefaultmidpunct}
{\mcitedefaultendpunct}{\mcitedefaultseppunct}\relax
\EndOfBibitem
\bibitem[Huber(1982)]{Huber1982}
Huber,~H. Is Hexazine Stable. \emph{Angew. Chem. Int. Ed.} \textbf{1982},
  \emph{21}, 64--65\relax
\mciteBstWouldAddEndPuncttrue
\mciteSetBstMidEndSepPunct{\mcitedefaultmidpunct}
{\mcitedefaultendpunct}{\mcitedefaultseppunct}\relax
\EndOfBibitem
\bibitem[Saxe and Schaefer~III(1983)Saxe, and Schaefer~III]{Saxe1983}
Saxe,~P.; Schaefer~III,~H.~F. Cyclic ${D}_{6h}$ Hexaazabenzene: A Relative
  Minimum on the {N}$_6$ Potential Energy Hypersurface. \emph{J. Am. Chem.
  Soc.} \textbf{1983}, \emph{105}, 1760--1764\relax
\mciteBstWouldAddEndPuncttrue
\mciteSetBstMidEndSepPunct{\mcitedefaultmidpunct}
{\mcitedefaultendpunct}{\mcitedefaultseppunct}\relax
\EndOfBibitem
\bibitem[Glukhovtsev and Schleyer(1992)Glukhovtsev, and Schleyer]{Schleyer1992}
Glukhovtsev,~M.~N.; Schleyer,~P.~v. Structures, Bonding and Energies of {N}$_6$
  Isomers. \emph{Chem. Phys. Lett.} \textbf{1992}, \emph{198}, 547--554\relax
\mciteBstWouldAddEndPuncttrue
\mciteSetBstMidEndSepPunct{\mcitedefaultmidpunct}
{\mcitedefaultendpunct}{\mcitedefaultseppunct}\relax
\EndOfBibitem
\bibitem[Ha and Nguyen(1992)Ha, and Nguyen]{Ha1992}
Ha,~T.~K.; Nguyen,~M.~T. The Identity of the six Nitrogen-Atoms ({N}$_6$)
  Species. \emph{Chem. Phys. Lett.} \textbf{1992}, \emph{195}, 179--183\relax
\mciteBstWouldAddEndPuncttrue
\mciteSetBstMidEndSepPunct{\mcitedefaultmidpunct}
{\mcitedefaultendpunct}{\mcitedefaultseppunct}\relax
\EndOfBibitem
\bibitem[Tobita and Bartlett(2001)Tobita, and Bartlett]{Tobita2001}
Tobita,~M.; Bartlett,~R.~J. Structure and stability of {N}$_6$ isomers and
  their spectroscopic characteristics. \emph{J. Phys. Chem. A} \textbf{2001},
  \emph{105}, 4107--4113\relax
\mciteBstWouldAddEndPuncttrue
\mciteSetBstMidEndSepPunct{\mcitedefaultmidpunct}
{\mcitedefaultendpunct}{\mcitedefaultseppunct}\relax
\EndOfBibitem
\bibitem[Wilson \latin{et~al.}(2001)Wilson, Perera, Bartlett, and
  Watts]{Wilson2001}
Wilson,~K.~J.; Perera,~S.~A.; Bartlett,~R.~J.; Watts,~J.~D. Stabilization of
  the pseudo-benzene {N}$_6$ ring with oxygen. \emph{J. Phys. Chem. A}
  \textbf{2001}, \emph{105}, 7693--7699\relax
\mciteBstWouldAddEndPuncttrue
\mciteSetBstMidEndSepPunct{\mcitedefaultmidpunct}
{\mcitedefaultendpunct}{\mcitedefaultseppunct}\relax
\EndOfBibitem
\bibitem[Greschner \latin{et~al.}(2016)Greschner, Zhang, Majumdar, Liu, Peng,
  Tse, and Yao]{Greschner2016}
Greschner,~M.~J.; Zhang,~M.; Majumdar,~A.; Liu,~H.~Y.; Peng,~F.; Tse,~J.~S.;
  Yao,~Y.~S. A New Allotrope of Nitrogen as High-Energy Density Material.
  \emph{J. Phys. Chem. A} \textbf{2016}, \emph{120}, 2920--2925\relax
\mciteBstWouldAddEndPuncttrue
\mciteSetBstMidEndSepPunct{\mcitedefaultmidpunct}
{\mcitedefaultendpunct}{\mcitedefaultseppunct}\relax
\EndOfBibitem
\bibitem[Fau and Bartlett(2001)Fau, and Bartlett]{Fau2001}
Fau,~S.; Bartlett,~R.~J. Possible products of the end-on addition of {N}$_3^-$
  to {N}$_5^+$ and their stability. \emph{J. Phys. Chem. A} \textbf{2001},
  \emph{105}, 4096--4106\relax
\mciteBstWouldAddEndPuncttrue
\mciteSetBstMidEndSepPunct{\mcitedefaultmidpunct}
{\mcitedefaultendpunct}{\mcitedefaultseppunct}\relax
\EndOfBibitem
\bibitem[Hirshberg \latin{et~al.}(2014)Hirshberg, Gerber, and
  Krylov]{Hirshberg2014}
Hirshberg,~B.; Gerber,~R.~B.; Krylov,~A.~I. Calculations predict a stable
  molecular crystal of {N}$_8$. \emph{Nature Chem.} \textbf{2014}, \emph{6},
  52--56\relax
\mciteBstWouldAddEndPuncttrue
\mciteSetBstMidEndSepPunct{\mcitedefaultmidpunct}
{\mcitedefaultendpunct}{\mcitedefaultseppunct}\relax
\EndOfBibitem
\bibitem[Fau \latin{et~al.}(2002)Fau, Wilson, and Bartlett]{Fau2002}
Fau,~S.; Wilson,~K.~J.; Bartlett,~R.~J. On the stability of {N}$_5^+${N}$_5^-$.
  \emph{J. Phys. Chem. A} \textbf{2002}, \emph{106}, 4639--4644\relax
\mciteBstWouldAddEndPuncttrue
\mciteSetBstMidEndSepPunct{\mcitedefaultmidpunct}
{\mcitedefaultendpunct}{\mcitedefaultseppunct}\relax
\EndOfBibitem
\bibitem[Olah \latin{et~al.}(2001)Olah, Prakash, and Rasul]{Olah2001}
Olah,~G.~A.; Prakash,~G. K.~S.; Rasul,~G. {N}$_6^{2+}$ and {N}$_4^{2+}$
  dications and their {N}$_{12}$ and {N}$_{10}$ azido derivatives:
  {DFT/GIAO-MP2} theoretical studies. \emph{J. Am. Chem. Soc.} \textbf{2001},
  \emph{123}, 3308--3310\relax
\mciteBstWouldAddEndPuncttrue
\mciteSetBstMidEndSepPunct{\mcitedefaultmidpunct}
{\mcitedefaultendpunct}{\mcitedefaultseppunct}\relax
\EndOfBibitem
\bibitem[Ha \latin{et~al.}(1999)Ha, Suleimenov, and Nguyen]{N20_1999}
Ha,~T.~K.; Suleimenov,~O.; Nguyen,~M.~T. A quantum chemical study of three
  isomers of {N$_{20}$}. \emph{Chem. Phys. Lett.} \textbf{1999}, \emph{315},
  327--334\relax
\mciteBstWouldAddEndPuncttrue
\mciteSetBstMidEndSepPunct{\mcitedefaultmidpunct}
{\mcitedefaultendpunct}{\mcitedefaultseppunct}\relax
\EndOfBibitem
\bibitem[Raghavachari \latin{et~al.}(1989)Raghavachari, Trucks, Pople, and
  Headgordon]{Raghavachari1989}
Raghavachari,~K.; Trucks,~G.~W.; Pople,~J.~A.; Headgordon,~M. A Fifth-Order
  Perturbation Comparison of Electron Correlation Theories. \emph{Chem. Phys.
  Lett.} \textbf{1989}, \emph{157}, 479--483\relax
\mciteBstWouldAddEndPuncttrue
\mciteSetBstMidEndSepPunct{\mcitedefaultmidpunct}
{\mcitedefaultendpunct}{\mcitedefaultseppunct}\relax
\EndOfBibitem
\bibitem[Watts \latin{et~al.}(1993)Watts, Gauss, and Bartlett]{Watts1993}
Watts,~J.~D.; Gauss,~J.; Bartlett,~R.~J. Coupled-Cluster Methods with
  Noniterative Triple Excitations for Restricted Open-Shell {Hartree}-{Fock}
  and Other General Single Determinant Reference Functions: Energies and
  Analytical Gradients. \emph{J. Chem. Phys.} \textbf{1993}, \emph{98},
  8718--8733\relax
\mciteBstWouldAddEndPuncttrue
\mciteSetBstMidEndSepPunct{\mcitedefaultmidpunct}
{\mcitedefaultendpunct}{\mcitedefaultseppunct}\relax
\EndOfBibitem
\bibitem[Dewar(1975)]{Dewar1975}
Dewar,~M. J.~S. {MO} Studies of Some Nonbenzenoid Aromatic Systems. \emph{Pure
  Appl. Chem.} \textbf{1975}, \emph{44}, 767--782\relax
\mciteBstWouldAddEndPuncttrue
\mciteSetBstMidEndSepPunct{\mcitedefaultmidpunct}
{\mcitedefaultendpunct}{\mcitedefaultseppunct}\relax
\EndOfBibitem
\bibitem[Ha \latin{et~al.}(1981)Ha, Cimiraglia, and Nguyen]{Ha1981}
Ha,~T.~K.; Cimiraglia,~R.; Nguyen,~M.~T. Can Hexazine ({N}$_6$) Be Stable.
  \emph{Chem. Phys. Lett.} \textbf{1981}, \emph{83}, 317--319\relax
\mciteBstWouldAddEndPuncttrue
\mciteSetBstMidEndSepPunct{\mcitedefaultmidpunct}
{\mcitedefaultendpunct}{\mcitedefaultseppunct}\relax
\EndOfBibitem
\bibitem[Huber \latin{et~al.}(1983)Huber, Ha, and Nguyen]{Huber1983}
Huber,~H.; Ha,~T.~K.; Nguyen,~M.~T. Is {N}$_6$ an Open-Chain Molecule. \emph{J.
  Mol. Struct. Theochem} \textbf{1983}, \emph{105}, 351--358\relax
\mciteBstWouldAddEndPuncttrue
\mciteSetBstMidEndSepPunct{\mcitedefaultmidpunct}
{\mcitedefaultendpunct}{\mcitedefaultseppunct}\relax
\EndOfBibitem
\bibitem[Kim \latin{et~al.}(2002)Kim, Furukawa, Sakamoto, and Tasumi]{Kim2002}
Kim,~J.~Y.; Furukawa,~Y.; Sakamoto,~A.; Tasumi,~M. Infrared and Raman studies
  of the radical anion of $\alpha$,$\omega$-diphenyl-1,3,5,7,9-decapentaene as
  a model compound in polyacetylene. \emph{Synth. Metals} \textbf{2002},
  \emph{129}, 235--238\relax
\mciteBstWouldAddEndPuncttrue
\mciteSetBstMidEndSepPunct{\mcitedefaultmidpunct}
{\mcitedefaultendpunct}{\mcitedefaultseppunct}\relax
\EndOfBibitem
\bibitem[Tanaka \latin{et~al.}(1983)Tanaka, Koike, Ohzeki, Yoshizawa, and
  Yamabe]{Tanaka1983}
Tanaka,~K.; Koike,~T.; Ohzeki,~K.; Yoshizawa,~K.; Yamabe,~T.
  Photo-Isomerization of {\it Cis}-Polyacetylene. \emph{Solid State Commun.}
  \textbf{1983}, \emph{47}, 127--129\relax
\mciteBstWouldAddEndPuncttrue
\mciteSetBstMidEndSepPunct{\mcitedefaultmidpunct}
{\mcitedefaultendpunct}{\mcitedefaultseppunct}\relax
\EndOfBibitem
\bibitem[Klap\"{o}tke \latin{et~al.}(2011)Klap\"{o}tke, Martin, and
  Stierstorfer]{Klapotke_AAA2011}
Klap\"{o}tke,~T.~M.; Martin,~F.~A.; Stierstorfer,~J. {C}$_2${N}$_{14}$: An
  Energetic and Highly Sensitive Binary Azidotetrazole. \emph{Angew. Chem. Int.
  Ed.} \textbf{2011}, \emph{50}, 4227--4229\relax
\mciteBstWouldAddEndPuncttrue
\mciteSetBstMidEndSepPunct{\mcitedefaultmidpunct}
{\mcitedefaultendpunct}{\mcitedefaultseppunct}\relax
\EndOfBibitem
\bibitem[Klap\"{o}tke \latin{et~al.}(2012)Klap\"{o}tke, Krumm, Martin, and
  Stierstorfer]{Klapotke_AAA}
Klap\"{o}tke,~T.~M.; Krumm,~B.; Martin,~F.~A.; Stierstorfer,~J. New
  Azidotetrazoles: Structurally Interesting and Extremely Sensitive.
  \emph{Chem. Asian J.} \textbf{2012}, \emph{7}, 214--224\relax
\mciteBstWouldAddEndPuncttrue
\mciteSetBstMidEndSepPunct{\mcitedefaultmidpunct}
{\mcitedefaultendpunct}{\mcitedefaultseppunct}\relax
\EndOfBibitem
\bibitem[Banert \latin{et~al.}(2013)Banert, Richter, Schaarschmidt, and
  Lang]{Banert_AAA}
Banert,~K.; Richter,~S.; Schaarschmidt,~D.; Lang,~H. Well Known or New?
  {Synthesis} and Structure Assignment of Binary {C}$_2${N}$_{14}$ Compounds
  Reinvestigated. \emph{Angew. Chem. Int. Ed.} \textbf{2013}, \emph{52},
  3499--3502\relax
\mciteBstWouldAddEndPuncttrue
\mciteSetBstMidEndSepPunct{\mcitedefaultmidpunct}
{\mcitedefaultendpunct}{\mcitedefaultseppunct}\relax
\EndOfBibitem
\bibitem[Nellis \latin{et~al.}(1984)Nellis, Holmes, Mitchell, and
  Vanthiel]{Nellis1984}
Nellis,~W.~J.; Holmes,~N.~C.; Mitchell,~A.~C.; Vanthiel,~M. Phase-Transition in
  Fluid Nitrogen at High-Densities and Temperatures. \emph{Phys. Rev. Lett.}
  \textbf{1984}, \emph{53}, 1661--1664\relax
\mciteBstWouldAddEndPuncttrue
\mciteSetBstMidEndSepPunct{\mcitedefaultmidpunct}
{\mcitedefaultendpunct}{\mcitedefaultseppunct}\relax
\EndOfBibitem
\bibitem[Pohl \latin{et~al.}(1994)Pohl, Meider, and Springborg]{Springborg1994}
Pohl,~A.; Meider,~H.; Springborg,~M. On the {Peierls} Distortion and the
  Local-Density Approximation. \emph{J. Mol. Struct. (Theochem)} \textbf{1994},
  \emph{111}, 165--173\relax
\mciteBstWouldAddEndPuncttrue
\mciteSetBstMidEndSepPunct{\mcitedefaultmidpunct}
{\mcitedefaultendpunct}{\mcitedefaultseppunct}\relax
\EndOfBibitem
\bibitem[Alemany and Martins(2003)Alemany, and Martins]{Alemany2003}
Alemany,~M. M.~G.; Martins,~J.~L. Density-functional study of nonmolecular
  phases of nitrogen: Metastable phase at low pressure. \emph{Phys. Rev. B}
  \textbf{2003}, \emph{68}, 024110\relax
\mciteBstWouldAddEndPuncttrue
\mciteSetBstMidEndSepPunct{\mcitedefaultmidpunct}
{\mcitedefaultendpunct}{\mcitedefaultseppunct}\relax
\EndOfBibitem
\bibitem[Mattson \latin{et~al.}(2004)Mattson, Sanchez-Portal, Chiesa, and
  Martin]{Mattson2004}
Mattson,~W.~D.; Sanchez-Portal,~D.; Chiesa,~S.; Martin,~R.~M. Prediction of new
  phases of nitrogen at high pressure from first-principles simulations.
  \emph{Phys. Rev. Lett.} \textbf{2004}, \emph{93}, 125501\relax
\mciteBstWouldAddEndPuncttrue
\mciteSetBstMidEndSepPunct{\mcitedefaultmidpunct}
{\mcitedefaultendpunct}{\mcitedefaultseppunct}\relax
\EndOfBibitem
\bibitem[Goncharov \latin{et~al.}(2000)Goncharov, Gregoryanz, Mao, Liu, and
  Hemley]{Goncharov2000}
Goncharov,~A.~F.; Gregoryanz,~E.; Mao,~H.~K.; Liu,~Z.~X.; Hemley,~R.~J. Optical
  evidence for a nonmolecular phase of nitrogen above 150 {GPa}. \emph{Phys.
  Rev. Lett.} \textbf{2000}, \emph{85}, 1262--1265\relax
\mciteBstWouldAddEndPuncttrue
\mciteSetBstMidEndSepPunct{\mcitedefaultmidpunct}
{\mcitedefaultendpunct}{\mcitedefaultseppunct}\relax
\EndOfBibitem
\bibitem[Eremets \latin{et~al.}(2001)Eremets, Hemley, Mao, and
  Gregoryanz]{Eremets2001}
Eremets,~M.~L.; Hemley,~R.~J.; Mao,~H.; Gregoryanz,~E. Semiconducting
  non-molecular nitrogen up to 240 {GPa} and its low-pressure stability.
  \emph{Nature} \textbf{2001}, \emph{411}, 170--174\relax
\mciteBstWouldAddEndPuncttrue
\mciteSetBstMidEndSepPunct{\mcitedefaultmidpunct}
{\mcitedefaultendpunct}{\mcitedefaultseppunct}\relax
\EndOfBibitem
\bibitem[Gregoryanz \latin{et~al.}(2001)Gregoryanz, Goncharov, Hemley, and
  Mao]{Gregoryanz2001}
Gregoryanz,~E.; Goncharov,~A.~F.; Hemley,~R.~J.; Mao,~H.~K. High-pressure
  amorphous nitrogen. \emph{Phys. Rev. B} \textbf{2001}, \emph{64},
  052103\relax
\mciteBstWouldAddEndPuncttrue
\mciteSetBstMidEndSepPunct{\mcitedefaultmidpunct}
{\mcitedefaultendpunct}{\mcitedefaultseppunct}\relax
\EndOfBibitem
\bibitem[Gregoryanz \latin{et~al.}(2002)Gregoryanz, Goncharov, Hemley, Mao,
  Somayazulu, and Shen]{Gregoryanz2002}
Gregoryanz,~E.; Goncharov,~A.~F.; Hemley,~R.~J.; Mao,~H.~K.; Somayazulu,~M.;
  Shen,~G.~Y. Raman, infrared, and x-ray evidence for new phases of nitrogen at
  high pressures and temperatures. \emph{Phys. Rev. B} \textbf{2002},
  \emph{66}, 224108\relax
\mciteBstWouldAddEndPuncttrue
\mciteSetBstMidEndSepPunct{\mcitedefaultmidpunct}
{\mcitedefaultendpunct}{\mcitedefaultseppunct}\relax
\EndOfBibitem
\bibitem[Tomasino \latin{et~al.}(2014)Tomasino, Kim, Smith, and
  Yoo]{Tomasino2014}
Tomasino,~D.; Kim,~M.; Smith,~J.; Yoo,~C.~S. Pressure-Induced Symmetry-Lowering
  Transition in Dense Nitrogen to Layered Polymeric Nitrogen ({LP-N}) with
  Colossal {Raman} Intensity. \emph{Phys. Rev. Lett.} \textbf{2014},
  \emph{113}, 205502\relax
\mciteBstWouldAddEndPuncttrue
\mciteSetBstMidEndSepPunct{\mcitedefaultmidpunct}
{\mcitedefaultendpunct}{\mcitedefaultseppunct}\relax
\EndOfBibitem
\bibitem[Sontising and Beran(2019)Sontising, and Beran]{Sontising2019}
Sontising,~W.; Beran,~G. J.~O. Theoretical assessment of the structure and
  stability of the $\lambda$ phase of nitrogen. \emph{Phys. Rev. Mater.}
  \textbf{2019}, \emph{3}, 095002\relax
\mciteBstWouldAddEndPuncttrue
\mciteSetBstMidEndSepPunct{\mcitedefaultmidpunct}
{\mcitedefaultendpunct}{\mcitedefaultseppunct}\relax
\EndOfBibitem
\bibitem[Sontising and Beran(2020)Sontising, and Beran]{Sontising2020}
Sontising,~W.; Beran,~G. J.~O. Combining crystal structure prediction and
  simulated spectroscopy in pursuit of the unknown nitrogen phase $\zeta$
  crystal structure. \emph{Phys. Rev. Mater.} \textbf{2020}, \emph{4},
  063601\relax
\mciteBstWouldAddEndPuncttrue
\mciteSetBstMidEndSepPunct{\mcitedefaultmidpunct}
{\mcitedefaultendpunct}{\mcitedefaultseppunct}\relax
\EndOfBibitem
\bibitem[Li \latin{et~al.}(2013)Li, Wang, Xu, Li, Wang, and Chen]{Li2013}
Li,~J.~F.; Wang,~X.~L.; Xu,~N.; Li,~D.~Y.; Wang,~D.~C.; Chen,~L.
  Pressure-induced polymerization of nitrogen in potassium azides. \emph{Epl}
  \textbf{2013}, \emph{104}, 16005\relax
\mciteBstWouldAddEndPuncttrue
\mciteSetBstMidEndSepPunct{\mcitedefaultmidpunct}
{\mcitedefaultendpunct}{\mcitedefaultseppunct}\relax
\EndOfBibitem
\bibitem[Prasad \latin{et~al.}(2013)Prasad, Ashcroft, and Hoffmann]{Prasad2013}
Prasad,~D. L. V.~K.; Ashcroft,~N.~W.; Hoffmann,~R. Evolving Structural
  Diversity and Metallicity in Compressed Lithium Azide. \emph{J. Phys. Chem.
  C} \textbf{2013}, \emph{117}, 20838--20846\relax
\mciteBstWouldAddEndPuncttrue
\mciteSetBstMidEndSepPunct{\mcitedefaultmidpunct}
{\mcitedefaultendpunct}{\mcitedefaultseppunct}\relax
\EndOfBibitem
\bibitem[Shen \latin{et~al.}(2015)Shen, Oganov, Qian, Zhang, Dong, Zhu, and
  Zhou]{Shen2015}
Shen,~Y.~Q.; Oganov,~A.~R.; Qian,~G.~R.; Zhang,~J.; Dong,~H.~F.; Zhu,~Q.;
  Zhou,~Z.~X. Novel lithium-nitrogen compounds at ambient and high pressures.
  \emph{Sci. Rep.} \textbf{2015}, \emph{5}, 14204\relax
\mciteBstWouldAddEndPuncttrue
\mciteSetBstMidEndSepPunct{\mcitedefaultmidpunct}
{\mcitedefaultendpunct}{\mcitedefaultseppunct}\relax
\EndOfBibitem
\bibitem[Peng \latin{et~al.}(2015)Peng, Han, Liu, and Yao]{PengSR2015}
Peng,~F.; Han,~Y.~X.; Liu,~H.~Y.; Yao,~Y.~S. Exotic stable cesium polynitrides
  at high pressure. \emph{Sci. Rep.} \textbf{2015}, \emph{5}, 16902\relax
\mciteBstWouldAddEndPuncttrue
\mciteSetBstMidEndSepPunct{\mcitedefaultmidpunct}
{\mcitedefaultendpunct}{\mcitedefaultseppunct}\relax
\EndOfBibitem
\bibitem[Oganov and Glass(2006)Oganov, and Glass]{Oganov2006}
Oganov,~A.~R.; Glass,~C.~W. Crystal structure prediction using ab initio
  evolutionary techniques: Principles and applications. \emph{J. Chem. Phys.}
  \textbf{2006}, \emph{124}, 244704\relax
\mciteBstWouldAddEndPuncttrue
\mciteSetBstMidEndSepPunct{\mcitedefaultmidpunct}
{\mcitedefaultendpunct}{\mcitedefaultseppunct}\relax
\EndOfBibitem
\bibitem[Steele and Oleynik(2017)Steele, and Oleynik]{Steele2017}
Steele,~B.~A.; Oleynik,~I.~I. Novel Potassium Polynitrides at High Pressures.
  \emph{J. Phys. Chem. A} \textbf{2017}, \emph{121}, 8955--8961\relax
\mciteBstWouldAddEndPuncttrue
\mciteSetBstMidEndSepPunct{\mcitedefaultmidpunct}
{\mcitedefaultendpunct}{\mcitedefaultseppunct}\relax
\EndOfBibitem
\bibitem[Williams \latin{et~al.}(2017)Williams, Steele, and
  Oleynik]{Williams2017}
Williams,~A.~S.; Steele,~B.~A.; Oleynik,~I.~I. Novel rubidium poly-nitrogen
  materials at high pressure. \emph{J. Chem. Phys.} \textbf{2017}, \emph{147},
  234701\relax
\mciteBstWouldAddEndPuncttrue
\mciteSetBstMidEndSepPunct{\mcitedefaultmidpunct}
{\mcitedefaultendpunct}{\mcitedefaultseppunct}\relax
\EndOfBibitem
\bibitem[Bykov \latin{et~al.}(2018)Bykov, Bykova, Aprilis, Glazyrin, Koemets,
  Chuvashova, Kupenko, McCammon, Mezouar, Prakapenka, Liermann, Tasnadi,
  Ponomareva, Abrikosov, Dubrovinskaia, and Dubrovinsky]{Bykov2018}
Bykov,~M. \latin{et~al.}  {Fe--N} system at high pressure reveals a compound
  featuring polymeric nitrogen chains. \emph{Nat. Commun.} \textbf{2018},
  \emph{9}, 2756\relax
\mciteBstWouldAddEndPuncttrue
\mciteSetBstMidEndSepPunct{\mcitedefaultmidpunct}
{\mcitedefaultendpunct}{\mcitedefaultseppunct}\relax
\EndOfBibitem
\bibitem[Bykov \latin{et~al.}(2020)Bykov, Chariton, Bykova, Khandarkhaeva,
  Fedotenko, Ponomareva, Tidholm, Tasnadi, Abrikosov, Sedmak, Prakapenka,
  Hanfland, Liermann, Mahmood, Goncharov, Dubrovinskaia, and
  Dubrovinsky]{Bykov2020}
Bykov,~M. \latin{et~al.}  High-Pressure Synthesis of Metal-Inorganic Frameworks
  with {Hf$_4$N$_{20}\cdot$N$_2$}, {WN$_8\cdot$N$_2$}, and
  {Os$_5$N$_{28}\cdot$3N$_2$} Polymeric Nitrogen Linkers. \emph{Angew. Chem.
  Int. Ed.} \textbf{2020}, \emph{59}, 10321--10326\relax
\mciteBstWouldAddEndPuncttrue
\mciteSetBstMidEndSepPunct{\mcitedefaultmidpunct}
{\mcitedefaultendpunct}{\mcitedefaultseppunct}\relax
\EndOfBibitem
\bibitem[Bykov \latin{et~al.}(2021)Bykov, Bykova, Chariton, Prakapenka,
  Batyrev, Mahmood, and Goncharov]{Bykov2021}
Bykov,~M.; Bykova,~E.; Chariton,~S.; Prakapenka,~V.~B.; Batyrev,~I.~G.;
  Mahmood,~M.~F.; Goncharov,~A.~F. Stabilization of pentazolate anions in the
  high-pressure compounds {Na$_2$N$_5$} and {NaN$_5$} and in the sodium
  pentazolate framework {NaN$_5\cdot$N$_2$}. \emph{Dalton Trans.}
  \textbf{2021}, \emph{50}, 7229--7237\relax
\mciteBstWouldAddEndPuncttrue
\mciteSetBstMidEndSepPunct{\mcitedefaultmidpunct}
{\mcitedefaultendpunct}{\mcitedefaultseppunct}\relax
\EndOfBibitem
\bibitem[Veith and Schlemmer(1982)Veith, and Schlemmer]{Veith1982}
Veith,~M.; Schlemmer,~G. Crystal and Molecular Structure of {\it
  Trans}-Tetrazene-(2) ({N}$_4${H}$_4$) at $-90^\circ$ {C}. \emph{Z. Anorg.
  Allg. Chem.} \textbf{1982}, \emph{494}, 7--19\relax
\mciteBstWouldAddEndPuncttrue
\mciteSetBstMidEndSepPunct{\mcitedefaultmidpunct}
{\mcitedefaultendpunct}{\mcitedefaultseppunct}\relax
\EndOfBibitem
\bibitem[Huber and Herzberg(1979)Huber, and Herzberg]{HuberHerzberg}
Huber,~K.~P.; Herzberg,~G. \emph{Molecular Spectra and Molecular Structure:
  Constants of Diatomic Molecules}; Van Nostrand Reinhold: New York, 1979\relax
\mciteBstWouldAddEndPuncttrue
\mciteSetBstMidEndSepPunct{\mcitedefaultmidpunct}
{\mcitedefaultendpunct}{\mcitedefaultseppunct}\relax
\EndOfBibitem
\bibitem[Magers \latin{et~al.}(1988)Magers, Salter, Bartlett, Salter, Hess, and
  Schaad]{Magers1988}
Magers,~D.~H.; Salter,~E.~A.; Bartlett,~R.~J.; Salter,~C.; Hess,~B.~A.;
  Schaad,~L.~J. Do Stable Isomers of {N}$_3${H}$_3$ Exist? \emph{J. Am. Chem.
  Soc.} \textbf{1988}, \emph{110}, 3435--3446\relax
\mciteBstWouldAddEndPuncttrue
\mciteSetBstMidEndSepPunct{\mcitedefaultmidpunct}
{\mcitedefaultendpunct}{\mcitedefaultseppunct}\relax
\EndOfBibitem
\bibitem[F\"{o}rstel \latin{et~al.}(2016)F\"{o}rstel, Tsegaw, Maksyutenko,
  Mebel, Sander, and Kaiser]{Forstel2016N3H3}
F\"{o}rstel,~M.; Tsegaw,~Y.~A.; Maksyutenko,~P.; Mebel,~A.~M.; Sander,~W.;
  Kaiser,~R.~I. On the Formation of {N}$_3${H}$_3$ Isomers in Irradiated
  Ammonia Bearing Ices: Triazene ({H}$_2${NNNH}) or Triimide ({HNHNNH}).
  \emph{ChemPhysChem} \textbf{2016}, \emph{17}, 2726--2735\relax
\mciteBstWouldAddEndPuncttrue
\mciteSetBstMidEndSepPunct{\mcitedefaultmidpunct}
{\mcitedefaultendpunct}{\mcitedefaultseppunct}\relax
\EndOfBibitem
\bibitem[Kim \latin{et~al.}(1977)Kim, Gilje, and Seff]{Kim1977}
Kim,~Y.; Gilje,~J.~W.; Seff,~K. Synthesis and Structures of Two New Hydrides of
  Nitrogen, Triazane ({N}$_3${H}$_5$) and Cyclotriazane ({N}$_3${H}$_3$):
  Crystallographic and Mass-Spectrometric Analyses of Vacuum-Dehydrated
  Partially Decomposed Fully {Ag}$^+$-Exchanged Zeolite a Treated with Ammonia.
  \emph{J. Am. Chem. Soc.} \textbf{1977}, \emph{99}, 7057--7059\relax
\mciteBstWouldAddEndPuncttrue
\mciteSetBstMidEndSepPunct{\mcitedefaultmidpunct}
{\mcitedefaultendpunct}{\mcitedefaultseppunct}\relax
\EndOfBibitem
\bibitem[Heo \latin{et~al.}(2016)Heo, Kim, Kim, and Seff]{Heo2016}
Heo,~N.~H.; Kim,~Y.; Kim,~J.~J.; Seff,~K. Surprising Intrazeolitic Chemistry of
  Silver. \emph{J. Phys. Chem. C} \textbf{2016}, \emph{120}, 5277--5287\relax
\mciteBstWouldAddEndPuncttrue
\mciteSetBstMidEndSepPunct{\mcitedefaultmidpunct}
{\mcitedefaultendpunct}{\mcitedefaultseppunct}\relax
\EndOfBibitem
\bibitem[F\"{o}rstel \latin{et~al.}(2015)F\"{o}rstel, Maksyutenko, Jones, Sun,
  Chen, Chang, and Kaiser]{Forstel2015}
F\"{o}rstel,~M.; Maksyutenko,~P.; Jones,~B.~M.; Sun,~B.-J.; Chen,~S.-H.;
  Chang,~A. H.-H.; Kaiser,~R.~I. Detection of the Elusive Triazane Molecule
  ({N}$_3${H}$_5$) in the Gas Phase. \emph{ChemPhysChem} \textbf{2015},
  \emph{16}, 3139--3142\relax
\mciteBstWouldAddEndPuncttrue
\mciteSetBstMidEndSepPunct{\mcitedefaultmidpunct}
{\mcitedefaultendpunct}{\mcitedefaultseppunct}\relax
\EndOfBibitem
\bibitem[Richard and Ball(2008)Richard, and Ball]{Richard2008}
Richard,~R.~M.; Ball,~D.~W. {G2}, {G3}, and complete basis set calculations on
  the thermodynamic properties of triazane. \emph{J. Mol. Model.}
  \textbf{2008}, \emph{14}, 29--37\relax
\mciteBstWouldAddEndPuncttrue
\mciteSetBstMidEndSepPunct{\mcitedefaultmidpunct}
{\mcitedefaultendpunct}{\mcitedefaultseppunct}\relax
\EndOfBibitem
\bibitem[Grinberg~Dana \latin{et~al.}(2019)Grinberg~Dana, Moore, Jasper, and
  Green]{TriazaneTetrazane}
Grinberg~Dana,~A.; Moore,~K.~B.; Jasper,~A.~W.; Green,~W.~H. Large
  Intermediates in Hydrazine Decomposition: A Theoretical Study of the
  {N}$_3${H}$_5$ and {N}$_4${H}$_6$ Potential Energy Surfaces. \emph{J. Phys.
  Chem. A} \textbf{2019}, \emph{123}, 4679--4692\relax
\mciteBstWouldAddEndPuncttrue
\mciteSetBstMidEndSepPunct{\mcitedefaultmidpunct}
{\mcitedefaultendpunct}{\mcitedefaultseppunct}\relax
\EndOfBibitem
\bibitem[Wiberg \latin{et~al.}(1975)Wiberg, Bayer, and
  Bachhuber]{IsolationofTetrazene}
Wiberg,~N.; Bayer,~H.; Bachhuber,~H. Isolation of Tetrazene, {N}$_4${H}$_4$.
  \emph{Angew. Chem. Int. Ed.} \textbf{1975}, \emph{14}, 177--178\relax
\mciteBstWouldAddEndPuncttrue
\mciteSetBstMidEndSepPunct{\mcitedefaultmidpunct}
{\mcitedefaultendpunct}{\mcitedefaultseppunct}\relax
\EndOfBibitem
\bibitem[Ritter \latin{et~al.}(1989)Ritter, H\"{a}felinger, L\"{u}ddecke, and
  Rau]{RitterN4H4_1989}
Ritter,~G.; H\"{a}felinger,~G.; L\"{u}ddecke,~E.; Rau,~H. Tetrazetidine: {\it
  Ab initio} Calculations and Experimental Approach. \emph{J. Am. Chem. Soc.}
  \textbf{1989}, \emph{111}, 4627--4635\relax
\mciteBstWouldAddEndPuncttrue
\mciteSetBstMidEndSepPunct{\mcitedefaultmidpunct}
{\mcitedefaultendpunct}{\mcitedefaultseppunct}\relax
\EndOfBibitem
\bibitem[Ball(2001)]{tetrazane_calc}
Ball,~D.~W. Tetrazane: {Hartree}--{Fock}, {Gaussian}-2 and-3, and complete
  basis set predictions of some thermochemical properties of {N}$_4${H}$_6$.
  \emph{J. Phys. Chem. A} \textbf{2001}, \emph{105}, 465--470\relax
\mciteBstWouldAddEndPuncttrue
\mciteSetBstMidEndSepPunct{\mcitedefaultmidpunct}
{\mcitedefaultendpunct}{\mcitedefaultseppunct}\relax
\EndOfBibitem
\bibitem[Ball(2002)]{cyclo-tetrazane_calc}
Ball,~D.~W. High-level ab initio calculations on hydrogen-nitrogen compounds.
  Thermochemistry of tetrazetidine, {N}$_4${H}$_4$. \emph{J. Mol. Struct.
  Theochem} \textbf{2002}, \emph{619}, 37--43\relax
\mciteBstWouldAddEndPuncttrue
\mciteSetBstMidEndSepPunct{\mcitedefaultmidpunct}
{\mcitedefaultendpunct}{\mcitedefaultseppunct}\relax
\EndOfBibitem
\bibitem[Fujii \latin{et~al.}(2002)Fujii, Selvin, Sablier, and
  Iwase]{Fujii2002}
Fujii,~T.; Selvin,~C.~P.; Sablier,~M.; Iwase,~K. Analysis of hydronitrogen
  species generated by a microwave discharge in ({N}$_2${H}$_4$)/{He}. \emph{J.
  Phys. Chem. A} \textbf{2002}, \emph{106}, 3102--3105\relax
\mciteBstWouldAddEndPuncttrue
\mciteSetBstMidEndSepPunct{\mcitedefaultmidpunct}
{\mcitedefaultendpunct}{\mcitedefaultseppunct}\relax
\EndOfBibitem
\bibitem[F\"{o}rstel \latin{et~al.}(2016)F\"{o}rstel, Maksyutenko, Jones, Sun,
  Lee, Chang, and Kaiser]{Forstel_astro2016}
F\"{o}rstel,~M.; Maksyutenko,~P.; Jones,~B.~M.; Sun,~B.~J.; Lee,~H.~C.;
  Chang,~A. H.~H.; Kaiser,~R.~I. On the Formation of Amide Polymers Via
  Carbonyl-Amino Group Linkages in Energetically Processed Ices of
  Astrophysical Relevance. \emph{Astrophys. J.} \textbf{2016}, \emph{820},
  117\relax
\mciteBstWouldAddEndPuncttrue
\mciteSetBstMidEndSepPunct{\mcitedefaultmidpunct}
{\mcitedefaultendpunct}{\mcitedefaultseppunct}\relax
\EndOfBibitem
\bibitem[Carlotti \latin{et~al.}(1974)Carlotti, Johns, and Trombetti]{Carlotti}
Carlotti,~M.; Johns,~J. W.~C.; Trombetti,~A. $\nu_5$ Fundamental Bands of
  {N$_2$H$_2$} and {N$_2$D$_2$}. \emph{Can. J. Phys.} \textbf{1974}, \emph{52},
  340--344\relax
\mciteBstWouldAddEndPuncttrue
\mciteSetBstMidEndSepPunct{\mcitedefaultmidpunct}
{\mcitedefaultendpunct}{\mcitedefaultseppunct}\relax
\EndOfBibitem
\bibitem[Collin and Lipscomb(1951)Collin, and Lipscomb]{Collin1951}
Collin,~R.~L.; Lipscomb,~W.~N. The Crystal Structure of Hydrazine. \emph{Acta
  Cryst.} \textbf{1951}, \emph{4}, 10--14\relax
\mciteBstWouldAddEndPuncttrue
\mciteSetBstMidEndSepPunct{\mcitedefaultmidpunct}
{\mcitedefaultendpunct}{\mcitedefaultseppunct}\relax
\EndOfBibitem
\bibitem[Kirchmeier \latin{et~al.}(1992)Kirchmeier, Shreeve, and
  Verma]{Kirchmeier1992}
Kirchmeier,~R.~L.; Shreeve,~J.~M.; Verma,~R.~D. Fluorinated Compounds That
  Contain Catenated Oxygen, Sulfur or Nitrogen Atoms. \emph{Coord. Chem. Rev.}
  \textbf{1992}, \emph{112}, 169--213\relax
\mciteBstWouldAddEndPuncttrue
\mciteSetBstMidEndSepPunct{\mcitedefaultmidpunct}
{\mcitedefaultendpunct}{\mcitedefaultseppunct}\relax
\EndOfBibitem
\bibitem[Krumm \latin{et~al.}(1995)Krumm, Vij, Kirchmeier, Shreeve, and
  Oberhammer]{Krumm1995}
Krumm,~B.; Vij,~A.; Kirchmeier,~R.~J.; Shreeve,~J.~M.; Oberhammer,~H.
  Hexakis(Trifluoromethyl)Tetrazane. \emph{Angew. Chem. Int. Ed.}
  \textbf{1995}, \emph{34}, 586--588\relax
\mciteBstWouldAddEndPuncttrue
\mciteSetBstMidEndSepPunct{\mcitedefaultmidpunct}
{\mcitedefaultendpunct}{\mcitedefaultseppunct}\relax
\EndOfBibitem
\bibitem[Egger \latin{et~al.}(1983)Egger, Hoesch, and Dreiding]{Egger1983}
Egger,~N.; Hoesch,~L.; Dreiding,~A.~S. 3,3-Dialkyltriazenecarboxylic
  Derivatives by Oxidative Hydro-Acyl-Elimination from
  3,3-Dialkyltriazane-1,2-Dicarboxylic Derivatives. \emph{Helv. Chim. Acta}
  \textbf{1983}, \emph{66}, 1416--1426\relax
\mciteBstWouldAddEndPuncttrue
\mciteSetBstMidEndSepPunct{\mcitedefaultmidpunct}
{\mcitedefaultendpunct}{\mcitedefaultseppunct}\relax
\EndOfBibitem
\bibitem[Kanzian and Mayr(2010)Kanzian, and Mayr]{Kanzian2010}
Kanzian,~T.; Mayr,~H. Electrophilic Reactivities of Azodicarboxylates.
  \emph{Chem. Eur. J.} \textbf{2010}, \emph{16}, 11670--11677\relax
\mciteBstWouldAddEndPuncttrue
\mciteSetBstMidEndSepPunct{\mcitedefaultmidpunct}
{\mcitedefaultendpunct}{\mcitedefaultseppunct}\relax
\EndOfBibitem
\bibitem[Hope and Wiles(1967)Hope, and Wiles]{Hope1967}
Hope,~P.; Wiles,~L.~A. Action of Sulphur Monochloride on
  2-Acyl-1,1-Dimethylhydrazines: Formation of Tetrazans and Oxadiazolines.
  \emph{J. Chem. Soc. C} \textbf{1967}, \emph{1967}, 2636--2638\relax
\mciteBstWouldAddEndPuncttrue
\mciteSetBstMidEndSepPunct{\mcitedefaultmidpunct}
{\mcitedefaultendpunct}{\mcitedefaultseppunct}\relax
\EndOfBibitem
\bibitem[Pirkle and Gravel(1978)Pirkle, and Gravel]{Pirkle1978}
Pirkle,~W.~H.; Gravel,~P.~L. Persistent Cyclic Diacylhydrazyl Radicals from
  Urazoles and Pyrazolidine-3,5-Diones. \emph{J. Org. Chem.} \textbf{1978},
  \emph{43}, 808--815\relax
\mciteBstWouldAddEndPuncttrue
\mciteSetBstMidEndSepPunct{\mcitedefaultmidpunct}
{\mcitedefaultendpunct}{\mcitedefaultseppunct}\relax
\EndOfBibitem
\bibitem[Martin and Breton(2017)Martin, and Breton]{Martin2017}
Martin,~K.~L.; Breton,~G.~W. Computational, $^1${H} {NMR}, and {X}-ray
  structural studies on 1-arylurazole tetrazane dimers. \emph{Acta Crystallogr.
  C} \textbf{2017}, \emph{73}, 660--666\relax
\mciteBstWouldAddEndPuncttrue
\mciteSetBstMidEndSepPunct{\mcitedefaultmidpunct}
{\mcitedefaultendpunct}{\mcitedefaultseppunct}\relax
\EndOfBibitem
\bibitem[Christe \latin{et~al.}(2010)Christe, Dixon, Grant, Haiges, Tham, Vij,
  Vij, Wang, and Wilson]{ChristieDixon2010}
Christe,~K.~O.; Dixon,~D.~A.; Grant,~D.~J.; Haiges,~R.; Tham,~F.~S.; Vij,~A.;
  Vij,~V.; Wang,~T.~H.; Wilson,~W.~W. Dinitrogen Difluoride Chemistry.
  {Improved} Syntheses of {\it cis}- and {\it trans}-{N}$_2${F}$_2$, Synthesis
  and Characterization of {N}$_2${F}$^+${Sn}$_2${F}$_9^-$, Ordered Crystal
  Structure of {N}$_2${F}$^+${Sb}$_2${F}$_{11}^-$, High-Level Electronic
  Structure Calculations of {\it cis}-{N}$_2${F}$_2$, {\it
  trans}-{N}$_2${F}$_2$, {F}$_2${N=N}, and {N}$_2${F}$^+$, and Mechanism of the
  {\it trans}-{\it cis} Isomerization of {N}$_2${F}$_2$. \emph{Inorg. Chem.}
  \textbf{2010}, \emph{49}, 6823--6833\relax
\mciteBstWouldAddEndPuncttrue
\mciteSetBstMidEndSepPunct{\mcitedefaultmidpunct}
{\mcitedefaultendpunct}{\mcitedefaultseppunct}\relax
\EndOfBibitem
\bibitem[Seidl and Schaefer(1988)Seidl, and Schaefer]{Seidl1988}
Seidl,~E.~T.; Schaefer,~H.~F. Theoretical Studies of Oxygen Rings:
  Cyclotetraoxygen, {O$_4$}. \emph{J. Chem. Phys.} \textbf{1988}, \emph{88},
  7043--7049\relax
\mciteBstWouldAddEndPuncttrue
\mciteSetBstMidEndSepPunct{\mcitedefaultmidpunct}
{\mcitedefaultendpunct}{\mcitedefaultseppunct}\relax
\EndOfBibitem
\bibitem[Seidl and Schaefer(1992)Seidl, and Schaefer]{Seidl1992}
Seidl,~E.~T.; Schaefer,~H.~F. Is There a Transition-State for the Unimolecular
  Dissociation of Cyclotetraoxygen ({O}$_4$). \emph{J. Chem. Phys.}
  \textbf{1992}, \emph{96}, 1176--1182\relax
\mciteBstWouldAddEndPuncttrue
\mciteSetBstMidEndSepPunct{\mcitedefaultmidpunct}
{\mcitedefaultendpunct}{\mcitedefaultseppunct}\relax
\EndOfBibitem
\bibitem[Peterka \latin{et~al.}(1999)Peterka, Ahmed, Suits, Wilson, Korkin,
  Nooijen, and Bartlett]{Peterka1999}
Peterka,~D.~S.; Ahmed,~M.; Suits,~A.~G.; Wilson,~K.~J.; Korkin,~A.;
  Nooijen,~M.; Bartlett,~R.~J. Unraveling the mysteries of metastable
  {O$_4^*$}. \emph{J. Chem. Phys.} \textbf{1999}, \emph{110}, 6095--6098\relax
\mciteBstWouldAddEndPuncttrue
\mciteSetBstMidEndSepPunct{\mcitedefaultmidpunct}
{\mcitedefaultendpunct}{\mcitedefaultseppunct}\relax
\EndOfBibitem
\bibitem[Politzer and Lane(2000)Politzer, and Lane]{Politzer2000}
Politzer,~P.; Lane,~P. Kohn--{Sham} studies of oxygen systems. \emph{Int. J.
  Quantum Chem.} \textbf{2000}, \emph{77}, 336--340\relax
\mciteBstWouldAddEndPuncttrue
\mciteSetBstMidEndSepPunct{\mcitedefaultmidpunct}
{\mcitedefaultendpunct}{\mcitedefaultseppunct}\relax
\EndOfBibitem
\bibitem[Ram\'{i}rez-Sol\'{i}s \latin{et~al.}(2010)Ram\'{i}rez-Sol\'{i}s,
  Jolibois, and Maron]{RamirezSolis2010}
Ram\'{i}rez-Sol\'{i}s,~A.; Jolibois,~F.; Maron,~L. {\it Ab initio} molecular
  dynamics studies on the ground singlet potential energy surface of the
  tetraoxygen molecule, {O$_4$}. \emph{Chem. Phys. Lett.} \textbf{2010},
  \emph{485}, 16--20\relax
\mciteBstWouldAddEndPuncttrue
\mciteSetBstMidEndSepPunct{\mcitedefaultmidpunct}
{\mcitedefaultendpunct}{\mcitedefaultseppunct}\relax
\EndOfBibitem
\bibitem[Gadzhiev \latin{et~al.}(2013)Gadzhiev, Ignatov, Kulikov, Feigin,
  Razuvaev, Sennikov, and Schrems]{Gadzhiev2013}
Gadzhiev,~O.~B.; Ignatov,~S.~K.; Kulikov,~M.~Y.; Feigin,~A.~M.;
  Razuvaev,~A.~G.; Sennikov,~P.~G.; Schrems,~O. Structure, Energy, and
  Vibrational Frequencies of Oxygen Allotropes {O}$_n$ ($n \leq 6$) in the
  Covalently Bound and van der Waals Forms: Ab Initio Study at the {CCSD(T)}
  Level. \emph{J. Chem. Theory Comput.} \textbf{2013}, \emph{9}, 247--262\relax
\mciteBstWouldAddEndPuncttrue
\mciteSetBstMidEndSepPunct{\mcitedefaultmidpunct}
{\mcitedefaultendpunct}{\mcitedefaultseppunct}\relax
\EndOfBibitem
\bibitem[Xie \latin{et~al.}(1992)Xie, Schaefer, Jang, Mhin, Kim, Yoon, and
  Kim]{Xie_O6_1992}
Xie,~Y.~M.; Schaefer,~H.~F.; Jang,~J.~H.; Mhin,~B.~J.; Kim,~H.~S.; Yoon,~C.~W.;
  Kim,~K.~S. Sulfur Clusters: Structure, Infrared, and {Raman} Spectra of
  Cyclo-{S}$_6$ and Comparison with the Hypothetical Cyclo-{O}$_6$ Molecule.
  \emph{Mol. Phys.} \textbf{1992}, \emph{76}, 537--546\relax
\mciteBstWouldAddEndPuncttrue
\mciteSetBstMidEndSepPunct{\mcitedefaultmidpunct}
{\mcitedefaultendpunct}{\mcitedefaultseppunct}\relax
\EndOfBibitem
\bibitem[Martins-Costa \latin{et~al.}(2011)Martins-Costa, Anglada, and
  Ruiz-Lopez]{MartinsCosta2011}
Martins-Costa,~M.; Anglada,~J.~M.; Ruiz-Lopez,~M.~F. Structure, Stability, and
  Dynamics of Hydrogen Polyoxides. \emph{Int. J. Quantum Chem.} \textbf{2011},
  \emph{111}, 1543--1554\relax
\mciteBstWouldAddEndPuncttrue
\mciteSetBstMidEndSepPunct{\mcitedefaultmidpunct}
{\mcitedefaultendpunct}{\mcitedefaultseppunct}\relax
\EndOfBibitem
\bibitem[Xu and Goddard(2002)Xu, and Goddard]{XuGoddard2002}
Xu,~X.; Goddard,~W.~A. Peroxone chemistry: Formation of {H$_2$O$_3$ and
  ring-(HO$_2$)(HO$_3$) from O$_3$/H$_2$O$_2$}. \emph{Proc. Nat. Acad. Sci.
  USA} \textbf{2002}, \emph{99}, 15308--15312\relax
\mciteBstWouldAddEndPuncttrue
\mciteSetBstMidEndSepPunct{\mcitedefaultmidpunct}
{\mcitedefaultendpunct}{\mcitedefaultseppunct}\relax
\EndOfBibitem
\bibitem[Denis and Ornellas(2009)Denis, and Ornellas]{Denis2009}
Denis,~P.~A.; Ornellas,~F.~R. Theoretical Characterization of Hydrogen
  Polyoxides: {HOOH, HOOOH, HOOOOH, and HOOO}. \emph{J. Phys. Chem. A}
  \textbf{2009}, \emph{113}, 499--506\relax
\mciteBstWouldAddEndPuncttrue
\mciteSetBstMidEndSepPunct{\mcitedefaultmidpunct}
{\mcitedefaultendpunct}{\mcitedefaultseppunct}\relax
\EndOfBibitem
\bibitem[Engdahl and Nelander(2002)Engdahl, and Nelander]{Engdahl2002}
Engdahl,~A.; Nelander,~B. The vibrational spectrum of {H}$_2${O}$_3$.
  \emph{Science} \textbf{2002}, \emph{295}, 482--483\relax
\mciteBstWouldAddEndPuncttrue
\mciteSetBstMidEndSepPunct{\mcitedefaultmidpunct}
{\mcitedefaultendpunct}{\mcitedefaultseppunct}\relax
\EndOfBibitem
\bibitem[Suma \latin{et~al.}(2005)Suma, Sumiyoshi, and Endo]{Suma2005}
Suma,~K.; Sumiyoshi,~Y.; Endo,~Y. The rotational spectrum and structure of
  {HOOOH}. \emph{J. Am. Chem. Soc.} \textbf{2005}, \emph{127},
  14998--14999\relax
\mciteBstWouldAddEndPuncttrue
\mciteSetBstMidEndSepPunct{\mcitedefaultmidpunct}
{\mcitedefaultendpunct}{\mcitedefaultseppunct}\relax
\EndOfBibitem
\bibitem[Levanov \latin{et~al.}(2011)Levanov, Sakharov, Dashkova, Antipenko,
  and Lunin]{Levanov2011}
Levanov,~A.~V.; Sakharov,~D.~V.; Dashkova,~A.~V.; Antipenko,~E.~E.;
  Lunin,~V.~V. Synthesis of Hydrogen Polyoxides {H$_2$O$_4$} and {H$_2$O$_3$}
  and Their Characterization by {Raman} Spectroscopy. \emph{Eur. J. Inorg.
  Chem.} \textbf{2011}, \emph{2011}, 5144--5150\relax
\mciteBstWouldAddEndPuncttrue
\mciteSetBstMidEndSepPunct{\mcitedefaultmidpunct}
{\mcitedefaultendpunct}{\mcitedefaultseppunct}\relax
\EndOfBibitem
\bibitem[Streng and Grosse(1966)Streng, and Grosse]{F2O6}
Streng,~A.~G.; Grosse,~A.~V. Two New Fluorides of Oxygen {O$_5$F$_2$} and
  {O$_6$F$_2$}. \emph{J. Am. Chem. Soc.} \textbf{1966}, \emph{88},
  169--170\relax
\mciteBstWouldAddEndPuncttrue
\mciteSetBstMidEndSepPunct{\mcitedefaultmidpunct}
{\mcitedefaultendpunct}{\mcitedefaultseppunct}\relax
\EndOfBibitem
\bibitem[Gorelli \latin{et~al.}(1999)Gorelli, Ulivi, Santoro, and
  Bini]{Gorelli1999}
Gorelli,~F.~A.; Ulivi,~L.; Santoro,~M.; Bini,~R. The $\epsilon$ phase of solid
  oxygen: Evidence of an {O}$_4$ molecule lattice. \emph{Phys. Rev. Lett.}
  \textbf{1999}, \emph{83}, 4093--4096\relax
\mciteBstWouldAddEndPuncttrue
\mciteSetBstMidEndSepPunct{\mcitedefaultmidpunct}
{\mcitedefaultendpunct}{\mcitedefaultseppunct}\relax
\EndOfBibitem
\bibitem[Gorelli \latin{et~al.}(2001)Gorelli, Ulivi, Santoro, and
  Bini]{Gorelli2001}
Gorelli,~F.~A.; Ulivi,~L.; Santoro,~M.; Bini,~R. Spectroscopic study of the
  $\epsilon$ phase of solid oxygen. \emph{Phys. Rev. B} \textbf{2001},
  \emph{63}, 104110\relax
\mciteBstWouldAddEndPuncttrue
\mciteSetBstMidEndSepPunct{\mcitedefaultmidpunct}
{\mcitedefaultendpunct}{\mcitedefaultseppunct}\relax
\EndOfBibitem
\bibitem[Goncharov \latin{et~al.}(2003)Goncharov, Gregoryanz, Hemley, and
  Mao]{Goncharov2003}
Goncharov,~A.~F.; Gregoryanz,~E.; Hemley,~R.~J.; Mao,~H.~K. Molecular character
  of the metallic high-pressure phase of oxygen. \emph{Phys. Rev. B}
  \textbf{2003}, \emph{68}, 100102\relax
\mciteBstWouldAddEndPuncttrue
\mciteSetBstMidEndSepPunct{\mcitedefaultmidpunct}
{\mcitedefaultendpunct}{\mcitedefaultseppunct}\relax
\EndOfBibitem
\bibitem[Bartolomei \latin{et~al.}(2011)Bartolomei, Carmona-Novillo,
  Hern\'{a}ndez, P\'{e}rez-R\'{i}os, Campos-Mart\'{i}nez, and
  Hern\'{a}ndez-Lamoneda]{Bartolomei2011}
Bartolomei,~M.; Carmona-Novillo,~E.; Hern\'{a}ndez,~M.~I.;
  P\'{e}rez-R\'{i}os,~J.; Campos-Mart\'{i}nez,~J.; Hern\'{a}ndez-Lamoneda,~R.
  Molecular oxygen tetramer {(O$_2$)$_4$}: Intermolecular interactions and
  implications for the $\epsilon$ solid phase. \emph{Phys. Rev. B}
  \textbf{2011}, \emph{84}, 092105\relax
\mciteBstWouldAddEndPuncttrue
\mciteSetBstMidEndSepPunct{\mcitedefaultmidpunct}
{\mcitedefaultendpunct}{\mcitedefaultseppunct}\relax
\EndOfBibitem
\bibitem[Goncharenko(2005)]{Goncharenko2005}
Goncharenko,~I.~N. Evidence for a magnetic collapse in the epsilon phase of
  solid oxygen. \emph{Phys. Rev. Lett.} \textbf{2005}, \emph{94}, 205701\relax
\mciteBstWouldAddEndPuncttrue
\mciteSetBstMidEndSepPunct{\mcitedefaultmidpunct}
{\mcitedefaultendpunct}{\mcitedefaultseppunct}\relax
\EndOfBibitem
\bibitem[Hagiwara \latin{et~al.}(2014)Hagiwara, Ikeda, Kida, Matsuda, Tadera,
  Kyakuno, Yanagi, Maniwa, and Okunishi]{Hagiwara2014}
Hagiwara,~M.; Ikeda,~M.; Kida,~T.; Matsuda,~K.; Tadera,~S.; Kyakuno,~H.;
  Yanagi,~K.; Maniwa,~Y.; Okunishi,~K. Haldane State Formed by Oxygen Molecules
  Encapsulated in Single-Walled Carbon Nanotubes. \emph{J. Phys. Soc. Jpn.}
  \textbf{2014}, \emph{83}, 113706\relax
\mciteBstWouldAddEndPuncttrue
\mciteSetBstMidEndSepPunct{\mcitedefaultmidpunct}
{\mcitedefaultendpunct}{\mcitedefaultseppunct}\relax
\EndOfBibitem
\bibitem[Fujimori \latin{et~al.}(2013)Fujimori, Morelos-Gomez, Zhu, Muramatsu,
  Futamura, Urita, Terrones, Hayashi, Endo, Hong, Choi, Tom\'{a}nek, and
  Kaneko]{Fujimori2013}
Fujimori,~T.; Morelos-Gomez,~A.; Zhu,~Z.; Muramatsu,~H.; Futamura,~R.;
  Urita,~K.; Terrones,~M.; Hayashi,~T.; Endo,~M.; Hong,~S.~Y.; Choi,~Y.~C.;
  Tom\'{a}nek,~D.; Kaneko,~K. Conducting linear chains of sulphur inside carbon
  nanotubes. \emph{Nature Comm.} \textbf{2013}, \emph{4}, 2162\relax
\mciteBstWouldAddEndPuncttrue
\mciteSetBstMidEndSepPunct{\mcitedefaultmidpunct}
{\mcitedefaultendpunct}{\mcitedefaultseppunct}\relax
\EndOfBibitem
\end{mcitethebibliography}

\end{document}